\documentclass[11pt]{article}
\usepackage[margin=1in]{geometry}
\usepackage{amsmath,amssymb,amsthm,mathtools,microtype}
\usepackage[colorlinks=true,linkcolor=blue,urlcolor=blue]{hyperref}

\newtheorem{theorem}{Theorem}
\newtheorem{lemma}{Lemma}[section]
\newtheorem{proposition}[lemma]{Proposition}
\newtheorem{corollary}[lemma]{Corollary}
\newtheorem{definition}[lemma]{Definition}

\theoremstyle{definition}
\newtheorem{remark}[lemma]{Remark}

\usepackage{enumitem}
\theoremstyle{plain}
\newtheorem{fact}{Fact}
\newcommand{\E}{\mathbb E}

\newcommand{\R}{\mathbb R}
\newcommand{\TV}{\operatorname{TV}}

\newcommand{\op}{\mathrm{op}}

\newcommand{\Alg}{\mathrm{Alg}}

\usepackage{tikz}
\usepackage{booktabs,multirow}
\providecommand{\QC}{q^{\star}}  
\usepackage{authblk}

\usepackage{pgfplots}
\pgfplotsset{compat=1.18}
\usetikzlibrary{decorations.pathreplacing,calc}
\usetikzlibrary{patterns}
\usepackage{array}
\usepackage{xcolor}
\usepackage{annotate-equations}
\definecolor{cGate}{HTML}{3366CC}
\definecolor{cTrunc}{HTML}{CC6600}
\definecolor{cCut}{HTML}{2E8B57}
\definecolor{cField}{HTML}{8A2BE2}

\usepackage[most]{tcolorbox}
\newtcolorbox{defband}[1]{enhanced,colback=#1!8,frame hidden,boxrule=0pt,
  sharp corners,left=2pt,right=2pt,top=1pt,bottom=1pt,boxsep=2pt,
  before skip=6pt,after skip=6pt}

\usepackage{xcolor}
\usepackage{hyperref}

\hypersetup{
  colorlinks=true,
  citecolor=blue
}

\usepackage{tocloft}

\usepackage{thmtools}

\tcbset{
  theoremshade/.style={
    enhanced,
    breakable,             
    colback=gray!15,        
    frame hidden,
    boxrule=0pt,
    sharp corners,
    boxsep=0pt,
    left=6pt,
    right=6pt,
    top=4pt,
    bottom=4pt,
    before skip=8pt,
    after skip=8pt
  }
}

\usepackage{thm-restate}
\usepackage{etoolbox}
\usepackage{makecell}

\newcommand{\shadeenv}[1]{%
  \expandafter\pretocmd\csname #1\endcsname
    {\begin{tcolorbox}[theoremshade]}{}{}%
  \expandafter\apptocmd\csname end#1\endcsname
    {\end{tcolorbox}}{}{}%
}

\shadeenv{theorem}
\shadeenv{lemma}
\shadeenv{proposition}
\shadeenv{corollary}
\shadeenv{definition}
\shadeenv{remark}
\shadeenv{fact}
\shadeenv{assumption}

\allowdisplaybreaks
\title{Towards the Query Complexity of Sampling under Strong Log-Concavity and Log-Sobolev Inequalities}
\author[1]{Huanjian Zhou\thanks{zhou-huanjian185@g.ecc.u-tokyo.ac.jp}}
\author[1,2]{Taiji Suzuki\thanks{taiji@mist.i.u-tokyo.ac.jp }}
\affil[1]{The University of Tokyo}
\affil[2]{RIKEN Center for Advanced Intelligence Project}
\begin{document}

\pagenumbering{gobble}
\maketitle

\begin{abstract}
We study the query complexity of sampling from smooth distributions on
$\mathbb R^d$ using first-order oracles.
For $d$-dimensional strongly log-concave targets with condition number $\kappa=2$, we prove that any randomized sampler requires
$\exp({\Omega(\log^{1/3} d)})$ queries even from a constant warm start.
This rules out polylogarithmic dimension dependence and strictly separates strongly log-concave sampling from Gaussian sampling.
For targets with $L$-smooth potentials satisfying an $\alpha$-log-Sobolev
inequality with $\kappa= L/\alpha$, we prove an $\Omega(\kappa)$ query
lower bound for $d\gtrsim\log^2\kappa$.
This shows that log-Sobolev sampling is strictly harder than  strongly log-concave sampling.

Our proofs construct instance-dependent tests that target
samples typically pass but outputs of algorithms making
few queries typically fail.
For strongly log-concave targets, we embed a sequence of hidden Gaussian into high-degree tensor polynomial with smooth gates, forcing any algorithm to uncover the layers sequentially.
The log-Sobolev construction hides a constant fraction of the target mass in angular caps while controlling the log-Sobolev constant via delicate estimates.
\end{abstract}

\clearpage
\newpage

\setcounter{tocdepth}{2}
\setlength{\cftbeforesecskip}{3pt}
\setlength{\cftbeforesubsecskip}{0pt}
\tableofcontents
\clearpage
\newpage

\pagenumbering{arabic}
\section{Introduction}

We study the problem of sampling from a probability distribution $\pi_V$
on $\mathbb{R}^d$ with density proportional to $e^{-V}$, given oracle
access to the potential $V$ and its gradient $\nabla V$.
This problem is central to Bayesian inference,
statistical physics, machine
learning, differential
privacy, and randomized
algorithms~\cite{RobertCasella2004, FrenkelSmit2002,GoodfellowBengioCourville2016, DworkRoth2014, MotwaniRaghavan1995}.
We measure the cost of an algorithm by the number of oracle queries,
and the central question is the \emph{query complexity} of drawing a
sample whose distribution is within $\varepsilon$ of $\pi_V$ in total
variation distance, as a function of the dimension~$d$, the
accuracy~$\varepsilon$, and the regularity.
We consider two standard classes of target distributions $\pi_V$. 
In the first,
$\pi_V$ is $L$-log-smooth and $\alpha$-strongly log-concave, with
condition number $\kappa:=\frac L\alpha$. In the second, $\pi_V$ is
$L$-log-smooth and satisfies a log-Sobolev inequality ($\mathsf{LSI}$) with
constant~$\alpha$.
By the Bakry--\'Emery criterion, this second class
contains the first, and $\kappa:=\frac L\alpha$ plays the same role~\cite{BakryEmery1985}.

Recent progress on faster sampling algorithms  has been
driven largely by connections with optimization.
Viewing the Langevin diffusion as the Wasserstein gradient flow of relative
entropy \cite{JordanKinderlehrerOtto1998} links Langevin Monte Carlo to
gradient methods \cite{Wibisono2018,DurmusMajewskiMiasojedow2019} and
motivates the proximal sampler as a sampling analogue of the proximal point
method \cite{LeeShenTian2021,ChenChewiSalimWibisono2022}.
%
%
Recent advances include accelerated proximal and kinetic samplers \cite{chen2026accelerated,YuanGuoChen2026ReflectedProximal}, improved simulation and convergence analyses for piecewise-deterministic samplers \cite{LuLuo2026WindowedThinning,LuLuo2026EntropyHypocoercivity}, and Gaussian smoothing with high-order Picard approximations \cite{ChenChewiLuZhang2026SmoothedPicard,OAI:Subpolynomial-query-complexity-for-well-conditioned-log-concave-sampling-September-26-2026}.
In parallel, first-order rejection sampling enables high-accuracy sampling using exact or stochastic gradients \cite{ChenChewiDaskalakisRakhlin2026HighAccuracy,ChenChewiDaskalakisRakhlin2026Stochastic} and supports path-space and Gaussian cloud corrections that improve the dimension and condition-number dependence of sampling complexity \cite{ChenChewiRakhlinZhang2026,chen2026high}.

These algorithmic connections motivate comparing the first-order query complexities of sampling and optimization.
In smooth strongly convex optimization, achieving objective accuracy $\varepsilon$ takes $O\big(\sqrt\kappa\log\frac{1}{\varepsilon}\big)$ queries independent of the dimension \cite{Nesterov1983}, and the matching deterministic lower bound already holds on high-dimensional quadratic objectives \cite{NemirovskiYudin1983}.
These results thus show that quadratic functions capture the worst-case query complexity of strongly convex optimization.
For sampling, quadratic optimization correspond to Gaussian sampling, whose optimal query complexity is $\Theta(\log d)$ for constant $\kappa$ and accuracy
\cite{chewi2024query,zhou2026complexity}.
For general strongly log-concave targets, all state-of-the-art upper bounds have a super-polylogarithmic dependence on the dimension~$d$.
With a warm start, accelerated samplers achieve $\widetilde O\big(\sqrt\kappa\,d^{\frac{1}{4}}\operatorname{polylog}\frac{1}{\varepsilon}\big)$ query complexity \cite{chen2026accelerated,YuanGuoChen2026ReflectedProximal} while the fastest known sampler takes $d^{o(1)}$ queries only given the potential minimizer for $\kappa=2$ and constant accuracy \cite{OAI:Subpolynomial-query-complexity-for-well-conditioned-log-concave-sampling-September-26-2026}.
However, prior to this work, the strongest dimension-dependent lower bound
was only $\Omega(\log d)$, obtained from Gaussian hard instances
\cite{chewi2024query,zhou2026complexity}.
This gap leaves open whether general strongly log-concave sampling
also admits a polylogarithmic query bound, or whether non-quadratic
potentials force a larger dependence on the dimension.
We are thus led to the following question:

\begin{center}
\emph{Does $\mathrm{polylog}(d)$ complexity extend from Gaussian to strongly log-concave sampling?}
\end{center}

The second question concerns the dependence on $\kappa$. 
The log-Sobolev inequality is the standard relaxation of strong
log-concavity, as it is implied by it~\cite{BakryEmery1985}, it
survives bounded perturbations of the potential~\cite{HolleyStroock1987},
and it provides the exact functional framework under which the Langevin
diffusion converges exponentially in relative entropy~\cite{OttoVillani2000}. 
The best upper bounds under the two assumptions differ in their
dependence on $\kappa$, scaling as
$O(\sqrt\kappa\,d^{1/4}\operatorname{polylog}\frac1\varepsilon)$ under
strong log-concavity~\cite{chen2026accelerated,YuanGuoChen2026ReflectedProximal} and
$O(\kappa d^{1/5}\operatorname{polylog}\frac1\varepsilon)$ under a
log-Sobolev inequality~\cite{chen2026high}, thereby shifting from a
square-root to a linear dependence on $\kappa$.
In optimization, the analogous relaxation of strong convexity is the
Polyak--\L{}ojasiewicz condition~\cite{OttoVillani2000}, where
acceleration is impossible and $\Theta(\kappa\log\frac1\varepsilon)$
first-order queries are required~\cite{YueFangLin2023}. Whether the same holds for
sampling is open, and known lower bounds cannot separate the two
classes because they come from Gaussian targets, which are already
strongly log-concave. This leads to our second question.
\begin{center}
\emph{Does $\sqrt\kappa$ complexity extend from strongly log-concave to log-Sobolev sampling?}
\end{center}

\subsection{Our Results}

In this work, we make significant progress on both questions  by proving new
lower bounds for sampling from strongly log-concave and log-Sobolev
targets.
Combined with existing Gaussian lower bounds~\cite{zhou2026complexity} and
the upper bounds above, our results bracket the query complexity in
all three parameters. As a function of $d$, the  complexity of strongly log-concave
sampling is superpolylogarithmic yet subpolynomial, lying 
between $e^{\Omega(\log^{1/3} d)}$ and $d^{o(1)}$.
The dependence on the condition number~$\kappa$ is $\Theta(\kappa)$ for
log-Sobolev sampling versus $\Theta(\sqrt\kappa)$ under strong
log-concavity, while the accuracy dependence remains
$\operatorname{polylog}(\frac1\varepsilon)$ throughout.
Table~\ref{tab:complexity} summarizes the resulting picture.

\begin{table}[t]
\centering
\caption{First-order query complexity of Gaussian, strongly log-concave ($\mathsf{SLC}$), and log-Sobolev ($\mathsf{LSI}$) sampling.
Here a cold start supplies only the potential minimizer without warm start, and the side arrow indicates strict separations between classes.}
\label{tab:complexity}
\vspace{0.15cm}

\small
\setlength{\tabcolsep}{4pt}
\renewcommand{\arraystretch}{1.3}

\resizebox{\textwidth}{!}{%
\begin{tabular}{@{}c@{\quad}c@{}}
\rotatebox[origin=c]{90}{%
  \Large$\xleftarrow{\makebox[6cm]{\large strictly harder}}$%
}
&
\begin{tabular}{@{}lll>{\footnotesize}l>{\footnotesize}ll@{}}
\toprule
Setting & Bound & Query Complexity
  & \multicolumn{1}{l}{Initialization}
  & \multicolumn{1}{l}{Conditions}
  & Reference \\
\midrule

\multirow{2}{*}{Gaussian}
 & Upper
 & $O\bigl(\min\{d,
      \sqrt{\kappa}\log(\tfrac{d}{\varepsilon})\}\bigr)$
 & Cold start
 & -
 & \cite{chewi2024query} \\

 & Lower
 & $\Omega\bigl(\min\{d,
      \sqrt{\kappa}\log(1+
      \tfrac{d}{\sqrt{\kappa}\,\varepsilon})\}\bigr)$
 & Cold start
 & \makecell[l]{
     $\kappa\ge4,\ \varepsilon\le\tfrac{1}{16}$}
 & \cite{zhou2026complexity} \\

\midrule

\multirow{6}{*}{$\mathsf{SLC}$}
 & Upper
 & $O\bigl(\sqrt{\kappa}d^{1/4}\mathrm{polylog}(\frac{1}{\varepsilon})\bigr)$
 & Warm start
 & $D_2(\nu_0\|\pi)=O(1)$
 & \cite{chen2026accelerated,YuanGuoChen2026ReflectedProximal} \\

  & Upper
 & $\widetilde O\bigl(\sqrt{\kappa}\,d\,\,
      \mathrm{polylog}(\frac{1}{\varepsilon})\bigr)$
 & Cold start
 & -
 & \cite{LuLuo2026EntropyHypocoercivity} \\

 & Upper
 & $d^{o(1)}$
 & Cold start
 & \makecell[l]{
     $\kappa=2,\ \varepsilon=\tfrac{1}{10}$}
 & \cite{OAI:Subpolynomial-query-complexity-for-well-conditioned-log-concave-sampling-September-26-2026} \\

 & Lower
 & $\Omega\bigl(\min\{d,
      \sqrt{\kappa}\log(1+
      \tfrac{d}{\sqrt{\kappa}\,\varepsilon})\}\bigr)$
 & Cold start
 & \makecell[l]{
     $\kappa\ge4,\ \varepsilon\le\tfrac{1}{16}$}
 & \cite{zhou2026complexity} \\

 & Lower
 & $e^{\Omega(\log^{1/3}d)}$
 & Cold start
 & \makecell[l]{
     $\kappa=2,\ \varepsilon=\tfrac{1}{10}$}
 & Theorem~\ref{thm:main} \\

 & Lower
 & $e^{\Omega(\log^{1/3}d)}$
 & Warm start
 & \makecell[l]{
     $\kappa=2,\ \varepsilon=\tfrac{1}{10}$}
 & Theorem~\ref{thm:slc-warm} \\

\midrule

\multirow{4}{*}{$\mathsf{LSI}$}
 & Upper
 & $O\bigl(\kappa d^{1/5}\mathrm{polylog}(\frac{1}{\varepsilon})\bigr)$
 & Cold start
 & -
 & \cite{chen2026high} \\

 & Lower
 & $\Omega\bigl(\min\{d,
      \sqrt{\kappa}\log(1+
      \tfrac{d}{\sqrt{\kappa}\,\varepsilon})\}\bigr)$
 & Cold start
 & \makecell[l]{
     $\kappa\ge4,\ \varepsilon\le\tfrac{1}{16}$}
 & \cite{zhou2026complexity} \\

 & Lower
 & $e^{\Omega(\log^{1/3}d)}$
 & Cold start
 & \makecell[l]{
     $\kappa=2,\ \varepsilon=\tfrac{1}{10}$}
 & Theorem~\ref{thm:main} \\

 & Lower
 & $\Omega(\kappa)$
 & Warm start
 & \makecell[l]{
     $d\gtrsim\log^2\kappa,\ \varepsilon=\tfrac{1}{10}$}
 & Theorem~\ref{thm:lsi-sampling-lower-bound} \\

\bottomrule
\end{tabular}
\end{tabular}%
}
\end{table}

Our first result is the lower bound for strongly log-concave sampling in the dimension.
\begin{theorem}[\textbf{Informal, see Theorem~\ref{thm:main} and Theorem~\ref{thm:slc-warm}}]
\label{thm:main-informal}
For every sufficiently large~$d$, any sampler for $d$-dimensional
strongly log-concave distributions with condition number $\kappa=2$
to
accuracy~$\varepsilon = \frac{1}{10}$ requires
$e^{\Omega(\log^{1/3}d)}$ queries.
\end{theorem}
Combined with the $d^{o(1)}$ upper bound~\cite{OAI:Subpolynomial-query-complexity-for-well-conditioned-log-concave-sampling-September-26-2026}, this shows that strongly log-concave sampling in this regime is super-polylogarithmic yet subpolynomial in~$d$. In particular, it is strictly harder than Gaussian sampling, which requires only $\Theta(\log d)$ queries~\cite{chewi2024query,zhou2026complexity}, showing that quadratic potentials do not capture the dimensional hardness of the general problem.
This also establishes a sharp separation from smooth strongly convex optimization, which takes $O(1)$ queries independent of the dimension at fixed condition number and constant accuracy~\cite{Nesterov1983}.
Indeed, by Theorem~\ref{thm:slc-warm}, the $\exp\!\big(\Omega\big((\log d)^{\frac{1}{3}}\big)\big)$ sampling lower bound persists even when the algorithm receives a sample from an unknown $O(1)$-warm initial distribution and both the potential minimizer and the target mean are known to be the origin.


Our second result is the lower bound in the condition number.

\begin{theorem}[\textbf{Informal; see Theorem~\ref{thm:lsi-sampling-lower-bound}}]
\label{thm:lsi-informal}
For $d\gtrsim\log^2\kappa$, any sampler for $d$-dimensional
distributions satisfying a log-Sobolev inequality with condition
number~$\kappa$ to accuracy~$\frac1{10}$ requires $\Omega(\kappa)$
queries.
\end{theorem}

The lower bound holds for any dimension $d\gtrsim\log^2\kappa$,
matching the $\widetilde O(\kappa)$ upper bound for log-Sobolev
sampling at constant accuracy~\cite{ChenChewiSalimWibisono2022,pmlr-v195-fan23a,
AltschulerChewi2024WarmStarts,chen2026high} up to logarithmic
factors.
By contrast, strongly log-concave targets in this dimension
admit $\widetilde O(\sqrt\kappa)$ queries from a warm
start~\cite{chen2026accelerated,YuanGuoChen2026ReflectedProximal},
separating the two classes and ruling out $\sqrt\kappa$ acceleration
under a log-Sobolev inequality alone.
Moreover, the lower bound holds even when every potential has its
unique minimizer at the origin and the sampler is initialized from a
common known Gaussian $4$-warm start~$\nu$ satisfying
$\frac{d\nu}{d\pi_V}\le4$ for all targets in the hard family, showing
that neither a known mode nor a constant warm start removes the linear
cost in~$\kappa$.

\subsection{Technical Overview for Strongly Log-Concave Sampling}

\paragraph{Beyond Gaussian targets.}
Existing Gaussian lower bounds rely on the matrix--vector structure of
quadratic potentials, where a gradient query at~$x$ to a centered
Gaussian with precision matrix~$A$ simply returns~$Ax$.
Together with rotational symmetry of the hard instances, this
structure allows one to track the information revealed by adaptive
queries through growing linear subspaces and characterize the
conditional law of the remaining randomness
\cite{chewi2024query,zhou2026complexity}.
However, Gaussian targets admit $O(\log d)$-query samplers at
constant condition number and accuracy.
Stronger dimension-dependent lower bounds therefore require
non-quadratic potentials.
The challenge is to construct non-quadratic
potentials that hide information in richer nonlinear structures while
keeping the information revealed by adaptive queries quantitatively tractable.

\paragraph{An energy-test framework.}
To turn such hidden structure into a sampling lower bound, we need to identify a property of target samples that algorithms cannot reproduce from the information revealed by few queries.
We capture this property through an instance-dependent energy
test, reducing the distributional comparison to estimation of the probabilities
of passing a single test.
We draw a hidden Gaussian vector~$\Xi$ and use it to define both
a potential~$V^\Xi$ and a test set~$T^\Xi$.
%
%
For a suitably chosen query threshold
$\exp(\Theta((\log d)^{1/3}))$,
the output distribution~$\rho^\Xi$ of any sampler making fewer queries than this threshold satisfies
$\mathbb E_\Xi[\rho^\Xi(T^\Xi)]=o(1)$
(Theorem~\ref{hf:output-fails}).
Meanwhile, the target distribution~$\pi^\Xi\propto e^{-V^\Xi}$
assigns almost all its mass to~$T^\Xi$ on average over~$\Xi$,
giving $\mathbb E_\Xi[\pi^\Xi(T^\Xi)]=1-o(1)$
(Theorem~\ref{thm:target-good-mass}).
Since
$\operatorname{TV}(\rho^\Xi,\pi^\Xi)
\ge \pi^\Xi(T^\Xi)-\rho^\Xi(T^\Xi)$,
averaging over~$\Xi$ yields
$\mathbb E_\Xi[\operatorname{TV}(\rho^\Xi,\pi^\Xi)]\ge1-o(1)$.

\paragraph{Tractable nonlinear structure via tensor lifting.}
To realize this separation, we construct hidden Gaussian
fields using tensor features of the query directions,
following the Gaussian tensor representation of spherical
$p$-spin models
\cite{AuffingerBenArousCerny2013,AuffingerBenArous2013,HuangSellke2022}.
Tensor lifting offers two advantages.
First, at degree~$p$, the hidden Gaussian coefficients lie
in a $d^p$-dimensional tensor space, while each first-order
query reveals only $O(d)$ linear measurements.
Second, it allows us to control how much a query at one
location reveals about the field at other locations.

\begin{figure}[t]
  \centering
  \includegraphics[width=0.8\linewidth]{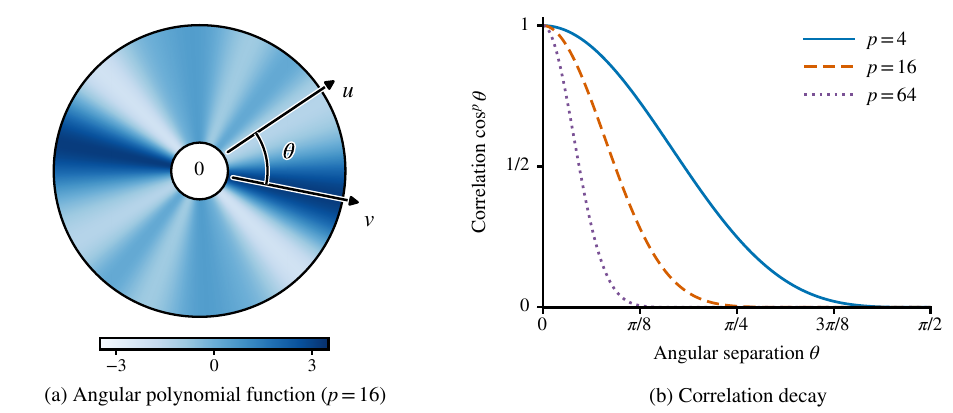}
\vspace{-0.3cm}
  \caption{
Angular polynomial and correlation decay.
\textbf{(a)} A realization of $x\mapsto F_p(\frac{x}{\|x\|})$ at $p=16$ on a
two-dimensional annulus before radial regularization, colored by
$\frac{F_p}{\sqrt d\,a_p}$, with directions $u$ and $v$ at an acute
angle~$\theta$.
\textbf{(b)} The correlation $\cos^p\theta$ between $F_p(u)$ and $F_p(v)$ over
the Gaussian coefficients, where $\theta=\arccos|\langle u,v\rangle|$.
Higher degrees turn a fixed angle into exponential correlation decay.
  }
  \label{fig:angular-function}
\end{figure}

For an even degree~$p$, we lift each direction
$u\in\mathbb S^{d-1}$ to the tensor feature
$\phi_p(u):=u^{\otimes p}\in(\mathbb R^d)^{\otimes p}$.
We define the random angular polynomial $F_p(u):=\sqrt d\,a_p\langle\Xi_p,\phi_p(u)\rangle$,
where $\Xi_p$ has independent standard Gaussian entries
and $a_p>0$ controls the amplitude.
Evaluating $F_p$ at $u=\frac{x}{\Vert{}x\Vert{}}$ for $x\ne0$ gives the angular function illustrated in Figure~\ref{fig:angular-function}(a), which we later combine with a smooth radial cutoff.

Since the field is linear in hidden Gaussian vector~$\Xi_p$, a function-value
and gradient query gives linear measurements of~$\Xi_p$
along $\phi_p(u)$ and its tangent derivatives.
These queries therefore fit the adaptive linear measurement
framework of Section~\ref{gauss:observation-rules} (Lemma~\ref{gauss:field-queries}).
In this angular setting, one value and $d-1$ tangential
derivatives increase the dimension of the observed subspace
by at most~$d$.
The Gaussian conditioning lemma of Section~\ref{gauss:conditioning} (Lemma~\ref{gauss:posterior})
then shows that, conditioned on the algorithm's random seed
and the full history of these linear measurements,
the unobserved component of~$\Xi_p$ remains a standard
Gaussian on the orthogonal complement.

The tensor lift introduces a nonlinear geometry on query
directions through the identity
$\langle\phi_p(u),\phi_p(v)\rangle=\langle u,v\rangle^p$
which gives
$\mathbb E[F_p(u)F_p(v)]=d\,a_p^2\langle u,v\rangle^p$.
Consequently, if $\vert{}\langle u,v\rangle\vert{}\le\rho<1$, the
correlation between $F_p(u)$ and $F_p(v)$ is at most~$\rho^p$, decaying
exponentially with the degree (Figure~\ref{fig:angular-function}(b)).
Thus, increasing the degree converts angular separation
between query directions into near orthogonality between
their feature vectors in the hidden coefficient space.

\paragraph{An $\Omega(\log^2 d)$ lower bound via mixing degrees.}
We next combine fields of different degrees to impose
simultaneous energy constraints on a common query geometry.
The potential depends on their sum, while the energy test
checks each degree component separately.
A successful output must therefore satisfy all component
tests using information obtained from the same collection
of queries.
Specifically, let $F(u)=\sum_{p\in\mathcal P}F_p(u)$,
where $\mathcal P$ is a finite set of even degrees and
the Gaussian tensors $(\Xi_p)_{p\in\mathcal P}$ are independent.
Fixing a smooth radial cutoff $C(x)=\chi(\frac{\|x\|}{\sqrt d})$ that
equals one on the typical shell $\{x\in\mathbb R^d:0.7\sqrt d\le\|x\|\le0.95\sqrt d\}$ and vanishes near the origin 
, we set
$H_p(x)=C(x)F_p(\frac{x}{\|x\|})$ for $x\ne0$ and $H_p(0)=0$.
Writing $H=\sum_{p\in\mathcal P}H_p$ for the total angular energy, we apply a smooth bounded
truncation~$U$ satisfying $U(s)=\beta s$ across the energy range
relevant to our test, where $\beta>0$ controls the perturbation
strength
.
We then define $V^\Xi(x)=\frac34\|x\|^2+U(H(x))$ and choose the
amplitudes~$a_p$ so that $I_d\preceq\nabla^2V^\Xi(x)\preceq2I_d$ holds
for all $x\in\mathbb R^d$ with high probability over~$\Xi$.

\begin{figure}[t]
    \centering
    \includegraphics[width=0.8\linewidth]{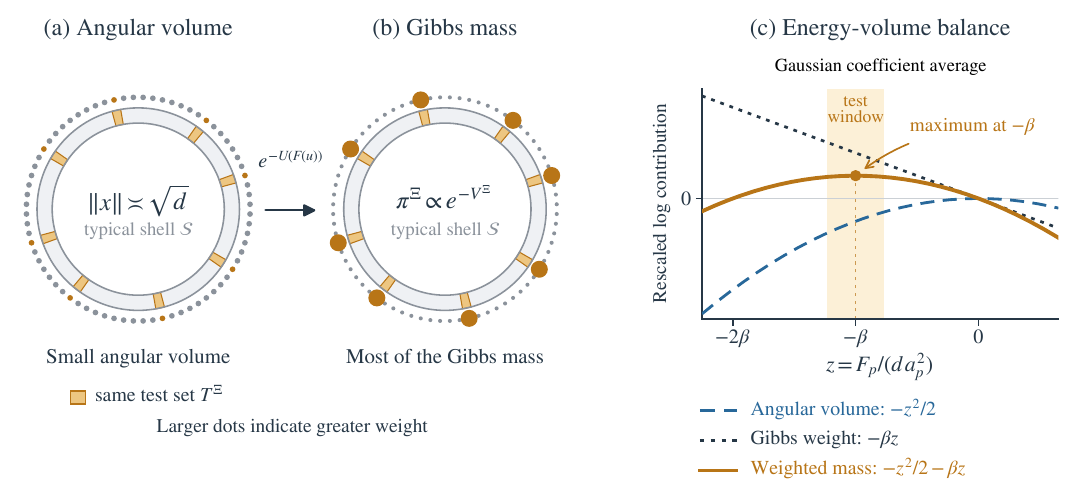}
    \vspace{-0.3cm}
    \caption{Geometry of the energy test.
\textbf{(a,b)} Sections of the typical shell showing that
$T^\Xi$ captures most Gibbs mass despite its small angular volume.
\textbf{(c)} Gaussian tilting shifts the normalized energy
$z_p = F_p/(d a_p^2)$ from $0$ to $-\beta$ at constant variance.
Fluctuations are marked by bars, with the test window
$|z_p+\beta|\le\delta\beta$ shaded for all $p\in\mathcal P$.}
    \label{fig:energy-test-geometry}
\end{figure}

We first explain why target samples satisfy these simultaneous energy constraints.
The energy test set $T^\Xi$ consists of all $x$ in the typical shell
whose degree-$p$ energy components $H_p(x)$ satisfy
$\left|\tfrac{H_p(x)}{d a_p^2}+\beta\right|\le\delta\beta$ for \emph{every}
$p\in\mathcal P$ 
, for a tolerance $\delta\in(0,1)$ chosen later.
The shift $-\beta$ arises from Gaussian exponential tilting, as tilting
the law of~$\Xi$ by $e^{-\beta F(u)}$ for a fixed direction~$u$ shifts
the mean of each $F_p(u)$ to $-\beta d a_p^2$.
At our chosen amplitudes, averaging over the many nearly orthogonal
directions keeps the normalizing constant close to its Gaussian
expectation, while concentration under the tilt makes the expected mass
outside~$T^\Xi$ negligible by comparison, yielding
$\pi^\Xi(T^\Xi)=1-o(1)$ with high probability over~$\Xi$.
Geometrically, the quadratic baseline concentrates the target
$\pi^\Xi\propto e^{-V^\Xi}$ on the typical shell, where directions with more negative values of~$F_p(u)$ receive a larger
factor~$e^{-\beta F_p(u)}$ in the density but occupy an exponentially
smaller fraction of the sphere, and balancing these two effects
concentrates the mass around $F_p(u)\approx-\beta d a_p^2$ for all
$p\in\mathcal P$, as shown in Figure~\ref{fig:energy-test-geometry}.


\begin{figure}[th]
  \centering
  \includegraphics[width=\linewidth]
    {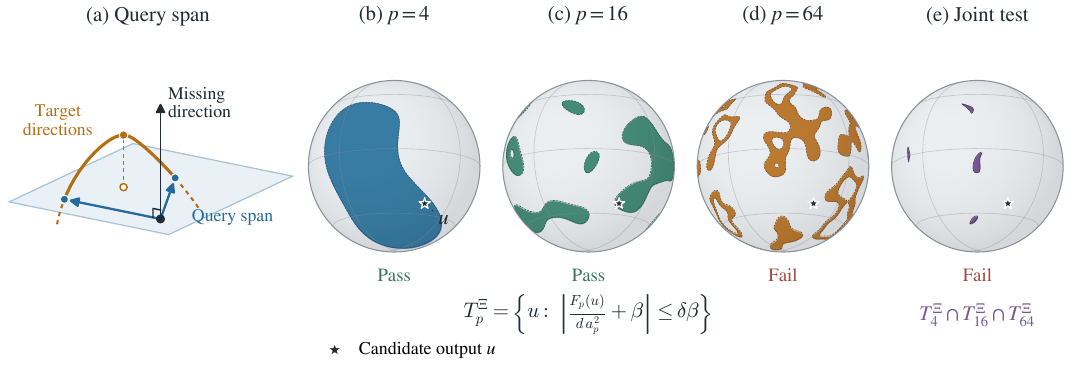}
    \vspace{-0.4cm}
\caption{Shared query constraints and the joint energy test.
\textbf{(a)} Each query contributes a vector shown as a blue arrow whose span
forms a low-dimensional subspace. Along any direction orthogonal to
this subspace, all query contributions vanish, whereas the target,
shown by the orange curve, has a projection exceeding the tolerance
whenever too few queries are made.
\textbf{(b)}--\textbf{(d)} Illustrative three-dimensional realizations of the
single-degree test regions
$T_p^\Xi=\{u:|\frac{F_p(u)}{d a_p^2}+\beta|\le\delta\beta\}$ for
$p=4,16,64$ with $\beta\sqrt d\,a_p=1$ and $\delta=\frac12$. The
candidate output~$u$ marked by a black star passes the first two tests
but fails the third.
\textbf{(e)} The joint test region $T_4^\Xi\cap T_{16}^\Xi\cap T_{64}^\Xi$ in
purple, which excludes~$u$.}
  \label{fig:algorithm-energy-geometry}
\end{figure}

It remains to show that few queries cannot produce an output satisfying all these energy constraints simultaneously.
Suppose a $q$-query algorithm returns~$X$ with a stronger oracle that
reveals the value and gradient of the mixed field
$H=\sum_{p\in\mathcal P}H_p$.
Once we condition on the algorithm's random seed and these
observations, the output~$X$ is fixed and each $H_p(X)$ splits into a
query-determined mean plus a centered Gaussian residual.
Crucially, across all degrees $p\in\mathcal P$, these query-determined
means are governed by the same angles between~$X$ and the query
points, so the algorithm must match the target energy at every degree
using a single shared query geometry, as illustrated in
Figure~\ref{fig:algorithm-energy-geometry}(b)--(e).
Once the Gaussian residuals and remaining projection errors are
bounded, the output energies have the shared profile
$p\sum_{t=1}^q B_tT_t^{p-1}$ up to the controlled errors,
where the signed coefficients~$B_t$ are independent of~$p$
and $T_t=\langle u,v_t\rangle$ is the overlap between the output
direction $u=\frac{X}{\Vert{}X\Vert{}}$ and the $t$-th query
direction $v_t=\frac{y_t}{\Vert{}y_t\Vert{}}$.
For even~$p$, dividing the target conditions by~$p$ and
absorbing the factors~$-1/(\beta T_t)$ into shared weights,
while omitting zero overlaps, reduces the task to the
exponential approximation problem
\[
  \Bigl|\sum_{t=1}^q c_t e^{-s_t p} - \frac1p\Bigr|
  \le O\Bigl(\frac\delta p\Bigr)
  \qquad\text{for all } p\in\mathcal P,
\]
where the weights~$c_t$ and rates $s_t=-\log|T_t|$ are \emph{shared
across all degrees}.
For $\mathcal P=\{4,6,\ldots,P\}$ with suitably chosen $P=d^{\Theta(1)}$
and $\delta=d^{-\Theta(1)}$, a quantitative Cauchy-matrix rank argument
shows that any such approximation requires
$\Omega(\log P\log\frac1\delta)=\Omega(\log^2 d)$ terms.

Geometrically, in the auxiliary approximation space,
too few queries leave a direction orthogonal to their
span where the target still exceeds the error tolerance, as shown in
Figure~\ref{fig:algorithm-energy-geometry}(a).
Comparing the target and output masses on~$T^\Xi$ then yields a
constant total variation gap on some admissible instance, establishing
the $\Omega(\log^2 d)$ query lower bound.

\paragraph{An
$e^{\Omega(\log^{1/3} d)}$ lower bound via multi-layer mixtures.}
Inspired by smooth zero-chain constructions in non-convex
optimization \cite{CarmonDuchiHinderSidford2020}
and recursive information hiding in submodular minimization
\cite{ChakrabartyGraurJiangSidford2022},
we organize the construction into $h+1$ layers.
With a suitable choice of parameters, the same approximation mechanism
yields the geometric separation needed for recursion, forcing any
algorithm that passes a layer to make two well-separated queries that
reveal information about that layer.
Smooth gates require each such query to pass the preceding layer's test.
To make the two costs add, we must also prevent information
obtained at one location from substantially reducing the
earlier-layer cost at the other.
We achieve this by making earlier-layer fields depend on
later-layer directions through nested tensor features.

\begin{figure}[t]
  \centering
  \includegraphics[width=0.78\linewidth]
    {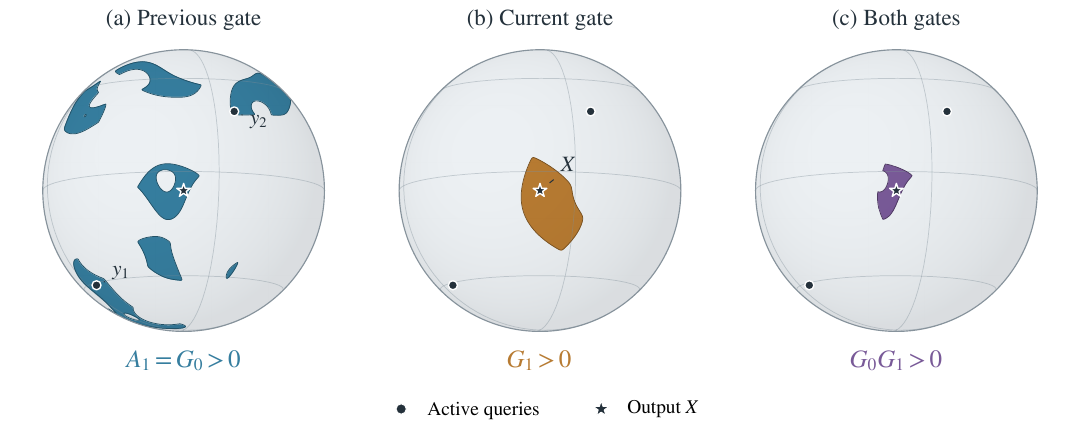}
    \vspace{-0.3cm}
\caption{Pointwise gating across two layers on a spherical slice.
\textbf{(a)} The active region $\{G_0>0\}$ for layer~$1$ in blue.
\textbf{(b)} The current gate region $\{G_1>0\}$ in orange.
\textbf{(c)} Their intersection $\{G_0G_1>0\}$ in purple, containing the
output~$X$ while excluding queries $y_1,y_2$ that pass only~$G_0$.}
  \label{fig:multilayer-query-geometry}
\end{figure}

We partition $x\in\mathbb R^d$ into blocks
$x_0,\ldots,x_h\in\mathbb R^n$ with $d=(h+1)n$.
From the block directions $u_j=\frac{x_j}{\|x_j\|}$, we define nested
compound directions $\psi_h=u_h$ and
$\psi_i=u_i\otimes\psi_{i+1}^{\otimes K}$ for a large integer~$K$.
At each layer~$i$, applying the above construction to~$\psi_i$ with
independent Gaussian coefficients yields degree-$p$ energies~$H_{i,p}(x)$,
total layer energy $H_i(x)=\sum_{p\in\mathcal P}H_{i,p}(x)$, and
smooth truncation~$U_i$ at tilt strength~$\beta_i$.
To couple the layers, we introduce a smooth \emph{gate} $G_i(x)\in[0,1]$ that
vanishes unless $\bigl|\frac{H_{i,p}(x)}{n a_{i,p}^2}+\beta_i\bigr|<\epsilon_g\beta_i$
for all $p\in\mathcal P$ (Lemma~\ref{hf:gate-prefix}), and multiply the layer-$i$ term in the
potential by the earlier gates $\prod_{j<i}G_j(x)$.
A query is \emph{active} at layer~$i$ when $\prod_{j<i}G_j(x)>0$,
whereas at any inactive query the layer-$i$ contribution to the value
and gradient vanishes, making layer~$i$ observable only at points that
pass all earlier gates (see Figure~\ref{fig:multilayer-query-geometry}).

\begin{figure}[t]
  \centering
  \includegraphics[width=0.9\linewidth]
    {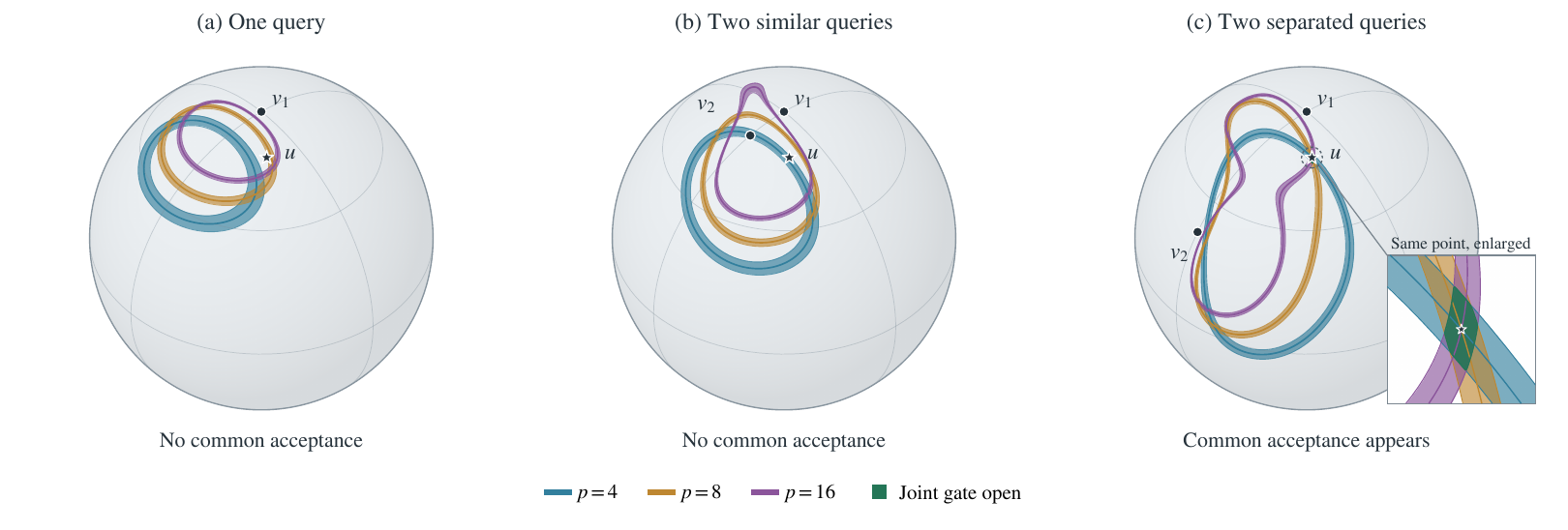}
    \vspace{-0.3cm}
\caption{Conditional-mean acceptance regions for $p=4,8,16$,
  where colored bands mark normalized conditional mean energies
  within $\frac1{10}\beta$ of~$-\beta$.
  \textbf{(a)} After one query at~$v_1$, the three bands have no common intersection.
  \textbf{(b)} A nearby second query at~$v_2$ deforms the bands without creating a common overlap.
  \textbf{(c)} A well-separated second query at~$v_2$ produces
  a common intersection near~$u$, opening the joint gate
  (dark green, inset).}
  \label{fig:gate-acceptance-obstruction}
\end{figure}

Suppose the algorithm produces a point~$Z$ that passes gate~$G_i$, so
that every layer-$i$ normalized degree energy at~$Z$ lies
near~$-\beta_i$.
Because inactive queries reveal nothing about layer~$i$, the same
conditioning decomposition as above reduces matching these
targets across $p\in\mathcal P$ to approximating~$\frac1p$ by a signed
exponential sum $\sum_t c_t e^{-s_t p}$, where the sum now runs only
over queries~$y_t$ active at layer~$i$ with overlaps
$T_t=\langle\psi_i(Z),\psi_i(y_t)\rangle$ and rates $s_t=-\log|T_t|$.
At any degree scale, terms with larger rates~$s_t$ have already decayed
away while terms with smaller rates vary too slowly to match the halving
of~$\frac1p$ as~$p$ doubles, so each scale requires an active query in
its rate window~(Lemma~\ref{out:rate-gap}).
Two separated degree scales therefore force two active queries at
distance at least~$P^{-\frac12}$ from each other up to sign (Lemma~\ref{out:two-directions}).
Geometrically, requiring the conditional mean energy at each
degree~$p$ to match its target value confines the candidate
direction~$u$ to a narrow spherical band.
Under a single active query~$v_1$, the bands for $p=4,8,16$ are
misaligned and have no common intersection
(Figure~\ref{fig:gate-acceptance-obstruction}(a)),
and adding a nearby second query~$v_2$ merely deforms them without
creating a common overlap
(Figure~\ref{fig:gate-acceptance-obstruction}(b)).
By contrast, two sufficiently separated queries align the bands
around~$u$, opening the joint acceptance gate
(Figure~\ref{fig:gate-acceptance-obstruction}(c)).

%

\begin{figure}[th]
  \centering
  \includegraphics[width=\linewidth]{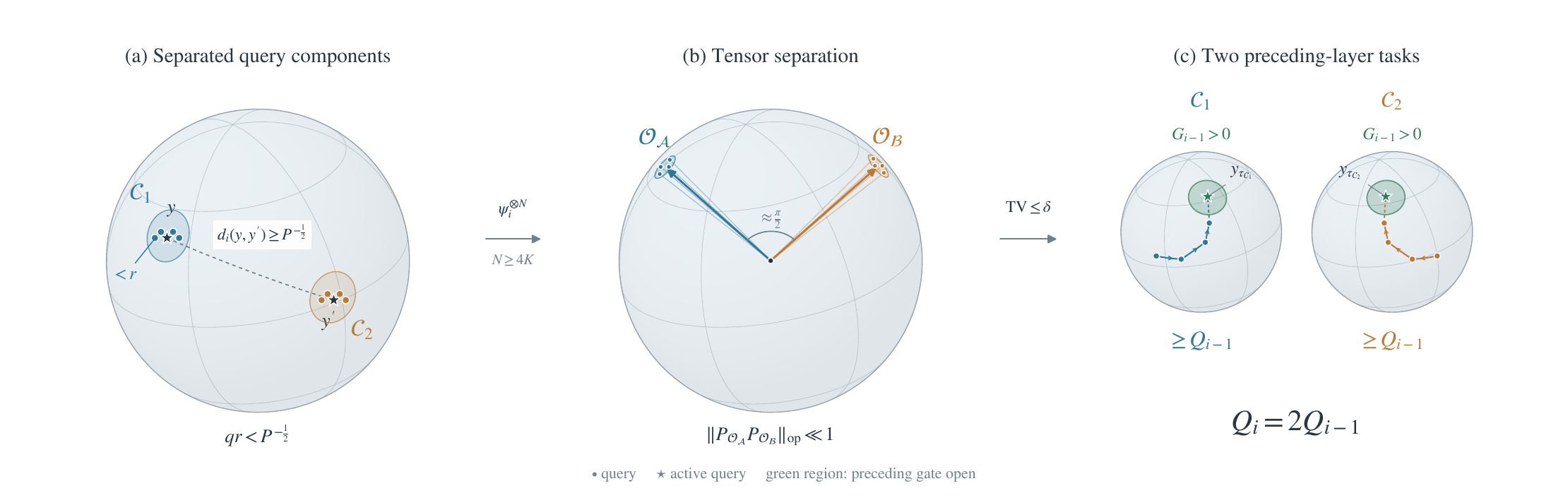}
  \vspace{-0.3cm}
\caption{Geometric mechanism behind the recursive query lower bound.
  \textbf{(a)} Linking nearby queries isolates the two separated
  active queries in distinct components.
  \textbf{(b)} High tensor powers render earlier-layer measurement
  spaces nearly orthogonal, enabling Gaussian decoupling.
  \textbf{(c)} Applying the comparison to each component
  charges at least~$Q_{i-1}$ queries before its first active query,
  giving~$Q_i=2Q_{i-1}$.}
  \label{fig:recursive-query-geometry}
\end{figure}
To convert this pairwise separation into a recursive cost for $i\ge 1$,
we link queries within distance~$r$ up to sign for $qr<P^{-\frac12}$,
placing the two active queries into disjoint connected components
separated by at least~$r$
(Lemma~\ref{out:two-components-lemma}; Figure~\ref{fig:recursive-query-geometry}(a)).
Because earlier-layer features contain high tensor powers of~$\psi_i$,
different components generate nearly orthogonal measurement spaces
(Lemma~\ref{out:ancestor-angle}).
Gaussian decoupling then permits a comparison with independent copies
of the earlier-layer coefficients inside and outside each component,
up to a negligible total variation error
(Theorem~\ref{gauss:decoupling} and Corollary~\ref{out:ancestor-comparison};
Figure~\ref{fig:recursive-query-geometry}(b)).
Before its first active query at layer~$i$, each component observes
only the earlier layers, while its first active point must pass
gate~$G_{i-1}$.
This charges at least $Q_{i-1}$ earlier queries to each of the two
disjoint components
(Lemma~\ref{out:component-charge}; Figure~\ref{fig:recursive-query-geometry}(c)),
yielding $Q_i=2Q_{i-1}$ from $Q_0=2$, so that $Q_h=2^{h+1}$.

For the target distribution, in the good region where all blocks lie in
their typical shells, all gates equal~$1$, and all energy truncations
are linear, the Gibbs weight relative to the quadratic baseline reduces
to $\exp(-\sum_{i=0}^h\beta_i H_i)$, while the gates and odd truncations
never increase the expected total mass.
By the same Gaussian tilting and directional-decorrelation argument as above, the target assigns probability $1-o(1)$ to this region
with high probability over the coefficients and hence passes the final
energy test.
Finally, preserving $I_d\preceq\nabla^2 V^\Xi\preceq 2I_d$ on admissible
realizations requires decreasing~$\beta_i$ across layers to offset the
growing tensor and gate derivatives, incurring a cumulative loss of
$\log\frac1{\beta_h}=O(h^3)$.
Choosing $h=\Theta(\log^{\frac13} n)$ keeps the final shift~$\beta_h$
above the error scale and yields
$Q_h=2^{h+1}=e^{\Theta(\log^{\frac13} d)}$, since
$\log d=\log n+O(\log h)$.
The same energy-test comparison as above then yields a constant
total variation gap on an admissible instance, establishing the
$e^{\Omega(\log^{\frac13} d)}$ query lower bound.

\paragraph{Extension to a warm start.}
The original energy test $T^\Xi$ cannot distinguish a target sample
from a warm initial sample, since both pass it with high probability.
We therefore introduce a hidden uniform direction $v\in S^{n-1}$ and
define the slabs $S_{v,t}=\{x:|v^\top x_h|\le t\}$.
A smooth bounded reweighting
$\nu_{\Xi,v}(dx)\propto f(v^\top x_h)\pi^\Xi(dx)$
yields a $3$-warm distribution satisfying
$\nu_{\Xi,v}(S_{v,a})\le\frac{3}{100}$ and
$\pi^\Xi(S_{v,b})>0.206$ for fixed $0<b<a$
by Lemma~\ref{lem:warm-initializer}.
The refined test set
$\mathcal T^{\mathrm{warm}}_{\Xi,v}=T^\Xi\cap S_{v,b}$
thus requires the algorithm to recover the missing slab mass while
passing the energy gates.
After separating out the event $x^\circ\in S_{v,a}$, which has probability
at most $\frac{3}{100}$, we apply Lemma~\ref{lem:warm-planting} to bound
the remaining event under a Gaussian reference model.
We then split outputs in $\mathcal T^{\mathrm{warm}}_{\Xi,v}$ according
to their distance from the initial direction.
Nearby outputs must cross from outside $S_{v,a}$ into $S_{v,b}$, which
is unlikely by Lemma~\ref{lem:warm-slab-geometry} because $v$ is
independent of the reference run.
For distant outputs, the reference mean shifts add
$-\beta_i\langle\psi_i(x),\psi_i(x^\circ)\rangle^p$
to the normalized degree-$p$ fields wherever the cutoffs equal $1$.
High degrees suppress these shifts, allowing the two-component recursion
to rule out distant outputs that pass the gate in few queries by
Lemma~\ref{lem:warm-localized-output}.
Consequently, the algorithm reaches $\mathcal T^{\mathrm{warm}}_{\Xi,v}$
with probability at most $\frac{3}{100}+o(1)$, whereas the target assigns
it average probability at least $0.206-o(1)$, giving the required total
variation gap.

\subsection{Technical Overview for $\mathsf{LSI}$ Sampling}

We prove our lower bound by constructing an instance-dependent cap test
that target samples pass with constant probability, whereas outputs of
algorithms making few queries rarely pass.
We draw a uniformly random direction $U\in\mathbb S^{d-1}$ and use it to
define a potential~$V_U$ and a test set~$T_U$ such that the target
satisfies $\pi_u(T_u)>0.55$ for every~$u\in\mathbb S^{d-1}$ (Proposition~\ref{prop:lsi-target-test}).
In contrast, the output law~$\rho_u$ of any randomized adaptive
algorithm making at most~$q$ queries satisfies
$\mathbb E_U[\rho_U(T_U)]\le C(q+1)p_d(t)$, where
$p_d(t) = \Theta(\frac1t e^{-\frac{t^2}2})$ is the spherical cap probability (Proposition~\ref{prop:lsi-output-test}).
Since $\operatorname{TV}(\rho_u,\pi_u)\ge\pi_u(T_u)-\rho_u(T_u)$, this
yields a constant total variation gap whenever~$q$ is a sufficiently
small multiple of~$p_d(t)^{-1}$.

\paragraph{The cap construction and cap test.}
Starting from the isotropic quadratic baseline $B(x)=\frac34\Vert{}x\Vert{}^2$
and the normalized latitude
$s_u(x)=\sqrt d\,\frac{\langle u,x\rangle}{\Vert{}x\Vert{}}$, we define the
cap test $T_u=\{x:\vert{}s_u(x)\vert{}\ge t\}$ and set
$V_u(x)=B(x)-\chi\bigl(\frac{\Vert{}x\Vert{}}{\sqrt d}\bigr)h_t(s_u(x))$,
where $\chi$ is a radial cutoff on the typical shell
$\Vert{}x\Vert{}=\Theta(\sqrt d)$ and $h_t$ is a smooth, even function that
vanishes below~$t$, grows with slope~$2t$ on $[t,3t]$, and saturates
above~$3t$.
For $d\gtrsim t^2$, the potential~$V_u$ is $O(t^2)$-smooth,
with the origin as its unique minimizer and the target mean.

\begin{figure}[th]
    \centering
    \includegraphics[width=\linewidth]{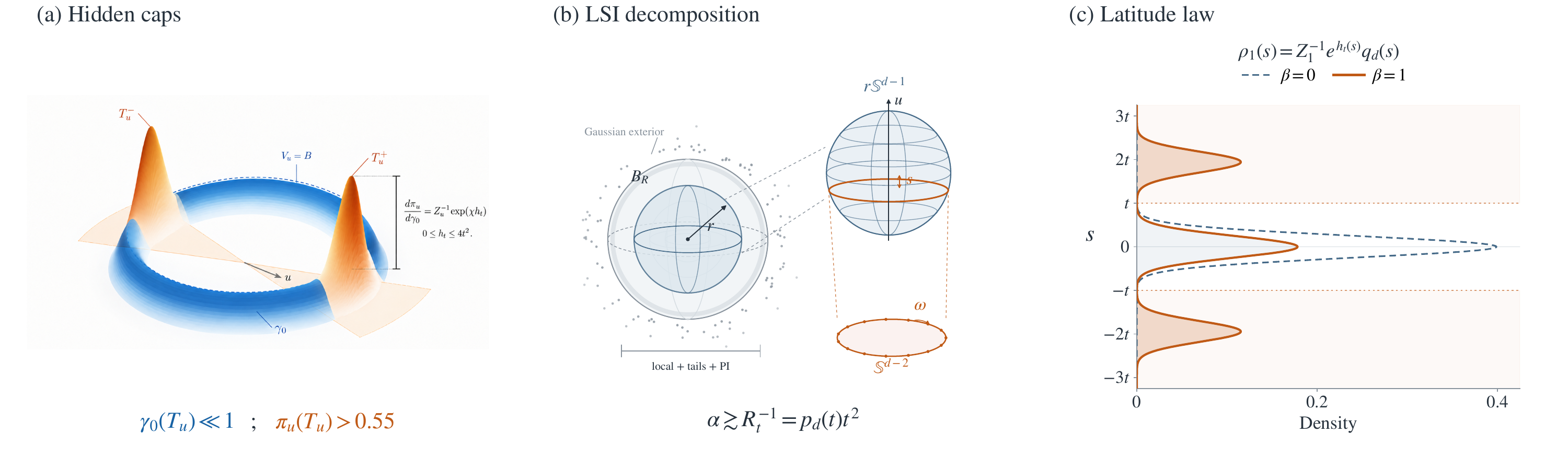}
    \vspace{-0.3cm}
\caption{Hidden-cap construction and $\mathsf{LSI}$ analysis.
  \textbf{(a)} The caps~$T_u$ have small reference probability under~$\gamma_0$
  but capture constant target mass.
  \textbf{(b)} Radial, latitude, and spherical-fiber decompositions
  yield a local $\mathsf{LSI}$, which combines with Poincar\'e
  bounds to establish the global $\mathsf{LSI}$.
  \textbf{(c)} Reference ($\beta=0$, dashed) and tilted ($\beta=1$, solid) latitude densities, with tilted modes near $s=\pm 2t$ separated by low-density barriers.
  }
  \vspace{-0.25cm}
    \label{fig:hidden-caps-lsi}
\end{figure}

Although the reference Gaussian $\gamma_0=\mathcal N(0,\frac23 I_d)$ places only probability~$p_d(t)$ on~$T_u$, the linear slope of~$h_t$ on the
typical shell shifts a constant fraction of the Gibbs mass to latitudes
near~$\pm 2t$ inside~$T_u$, while keeping the normalizing
constant~$O(1)$ so that~$\gamma_0$ serves as a common constant-warm
start for all~$u$.
Conversely, since $V_u$ and its gradient agree with~$B$ outside the
perturbed caps, coupling the algorithm on~$V_U$ with a reference run on~$B$
keeps the trajectory independent of~$U$ until a query hits the caps,
which occurs with probability $O(p_d(t))$ per query.
A union bound over the $q$ queries and the output therefore limits the
algorithm's acceptance probability to $O((q+1)p_d(t))$, forcing
$\Omega(p_d(t)^{-1})$ queries.
Figure~\ref{fig:hidden-caps-lsi}(a) illustrates this contrast
between the small reference mass of the hidden caps and their
constant target mass over the typical shell. Details are given in Sections~\ref{sec:lsi-family} and~\ref{sec:lsi-cap-test}.

\paragraph{Controlling the log-Sobolev rate.}
The main analytic challenge is to show that~$\pi_u$ satisfies a
log-Sobolev inequality with constant $\Omega(p_d(t)t^2)$ despite the
deep potential wells in the caps.
A direct Holley--Stroock perturbation argument gives only an
$\mathsf{LSI}$ rate of $\frac32 e^{-4t^2}$.
Combined with the $O(t^2)$ smoothness bound, this yields only an
$\widetilde\Omega(\kappa^{1/8})$ query lower bound
after calibrating the condition number.

Instead, writing $x=r\theta$ with $\theta\in\mathbb S^{d-1}$ and
$\beta=\chi(\frac r{\sqrt d})\in[0,1]$, we decompose the target into its
radial marginal and its conditional angular law at radius~$r$, which by
rotational symmetry around~$u$ further factors into the latitude
$s=\sqrt d\,\langle u,\theta\rangle$ and an independent uniform
transverse direction $\omega\in\mathbb S^{d-2}\subset u^\perp$
(Figure~\ref{fig:hidden-caps-lsi}(b)).
The resulting latitude density is
$\rho_\beta(s)=\frac{1}{Z_\beta}e^{\beta h_t(s)}q_d(s)$,
where $q_d$ is the latitude density of the uniform spherical measure
and $Z_\beta$ is the normalizing constant.
Among the three components, the radial marginal is a bounded
perturbation of the Gaussian radial law with a constant $\mathsf{LSI}$
rate on any interval $(0,R)$, while the transverse spherical fibers have
an intrinsic $\mathsf{LSI}$ rate of order~$\frac d{r^2}$ by spherical
curvature.
The main difficulty therefore lies in the one-dimensional tilted
latitude law.

To bound the latitude $\mathsf{LSI}$ rate, we apply the weighted
one-dimensional Hardy criterion (Proposition~\ref{prop:lsi-hardy}).
With the spherical gradient weight~$a_r(v)=\frac{d-v^2}{r^2}$,
the inverse $\mathsf{LSI}$ rate is controlled by
\[
  \sup_{0<s<\sqrt d}
  \left(\int_s^{\sqrt d}\rho_\beta(v)\,\mathrm dv\right)
  \log\frac{e}{\int_s^{\sqrt d}\rho_\beta(v)\,\mathrm dv}
  \int_0^s \frac{1}{a_r(v)\rho_\beta(v)}\,\mathrm dv.
\]
At full tilt~$\beta=1$, the solid curve in
Figure~\ref{fig:hidden-caps-lsi}(c) has two side modes near~$s=\pm2t$,
separated from the central mode by low-density barriers near~$s=\pm t$.
Each cap carries constant mass,
$\int_t^{\sqrt d}\rho_1(z)\,\mathrm dz=\Theta(1)$,
while $\rho_1(z)=\Theta(e^{-\frac{z^2}{2}})$ on the bulk~$[0,t]$.
The low density near~$s=t$ therefore makes the inverse-density
integral large, giving a Hardy product at~$s=t$ of order
$\frac{r^2}{d}\int_0^t e^{\frac{z^2}{2}}\,\mathrm dz
=\Theta\bigl(\frac{r^2}{d}R_t\bigr)$,
where $R_t=\frac{1}{p_d(t)t^2}
=\Theta\bigl(\frac1t e^{\frac{t^2}{2}}\bigr)$~(Fact~\ref{fact:lsi-angular-baseline}).
On the linear ramp~$[t,3t]$, completing the square in the
Gaussian approximation to the tilted density reveals a shifted
Gaussian centered at~$2\beta t$.
The resulting exponential prefactors cancel between the tail mass
and the inverse-density integral, while Gaussian tail estimates
control the remaining logarithmic factor
.
Together, these estimates bound the Hardy quantity by
$O\bigl(\frac{r^2}{d}R_t\bigr)$ uniformly over~$\beta\in[0,1]$
.
Extending the bound beyond the ramp and accounting for smoothing
yields a weighted latitude $\mathsf{LSI}$ rate of
$\Omega\bigl(\frac{d}{r^2R_t}\bigr)$.

We first lift the latitude bound to the full sphere of radius~$r$.
Combining it with the transverse fiber estimate via entropy
decomposition yields an $\mathsf{LSI}$ rate of
$\Omega\bigl(\frac{d}{r^2R_t}\bigr)$ for the conditional angular law~(Lemma~\ref{lem:lsi-angular}).
Within the perturbed shell~$r=\Theta(\sqrt d)$, this rate reduces to
$\Omega\bigl(\frac1{R_t}\bigr)$, whereas outside the shell the
angular law is uniform with the stronger rate
$\Omega\bigl(\frac d{r^2}\bigr)$.
Consequently, over all radii $0<r<R$, the inverse angular rate is
uniformly bounded by $O\bigl(R_t+\frac{R^2}{d}\bigr)$.

We next pass from fixed-radius spheres to the full ball by
incorporating the radial coordinate.
The radial marginal has a constant $\mathsf{LSI}$ rate,
but merging it with the angular bound requires controlling
how the conditional angular law varies with~$r$ through
$\beta(r)=\chi\bigl(\frac r{\sqrt d}\bigr)$.
Since $|\beta'(r)|=O\bigl(\frac1{\sqrt d}\bigr)$ and $h_t$ has oscillation
$O(t^2)$, the radial derivative of the angular log-density is
$O\bigl(\frac{t^2}{\sqrt d}\bigr)$.
This derivative enters quadratically in the radial--angular
entropy estimate (Lemma~\ref{lem:lsi-two-level}), contributing a coupling term of order
$O\bigl(\frac{t^4}{d}\bigr)$, which remains bounded for $d\gtrsim t^4$.
Combining the radial and angular inequalities therefore incurs
only a constant-factor loss, yielding a local $\mathsf{LSI}$
rate of $\Omega\bigl(\frac1{R_t+\frac{R^2}{d}}\bigr)$
for the target conditioned on the ball of radius~$R$ (Lemma~\ref{lem:lsi-local}).

To extend the local estimate to $\mathbb R^d$, we first establish
a global Poincar\'e inequality.
On a ball of radius $\Theta(\sqrt d)$, the local $\mathsf{LSI}$
implies a Poincar\'e inequality with rate
$\Omega\bigl(\frac1{R_t}\bigr)$.
A Lyapunov estimate for the quadratic tails controls the exterior
contribution, extending this Poincar\'e rate to the full space~(Lemma~\ref{lem:lsi-global-poincare}).
We next prove a global $\mathsf{LSI}$ with an $L^2$ remainder.
On a larger ball of radius $R=\Theta(\sqrt{dR_t})$,
the local $\mathsf{LSI}$ keeps rate
$\Omega\bigl(\frac1{R_t}\bigr)$ since
$R_t+\frac{R^2}{d}=\Theta(R_t)$.
We decompose a test function~$f$ into interior and exterior parts via
smooth cutoffs, applying the local $\mathsf{LSI}$ to the interior
part and the Gaussian reference $\mathsf{LSI}$ to the exterior part,
where the perturbation vanishes.
The cutoff gradients contribute an $L^2$ error of order
$O\bigl(\frac{R_t}{R^2}\bigr)=O\bigl(\frac1d\bigr)$, yielding
\[\operatorname{Ent}_{\pi_u}(f^2) \le CR_t\int\Vert{}\nabla f\Vert{}^2\,\mathrm d\pi_u +C\int f^2\,\mathrm d\pi_u.\]
To remove the $L^2$ term, we apply this bound to
$f-\mathbb E_{\pi_u}f$, converting $\int f^2\,\mathrm d\pi_u$ into
$\operatorname{Var}_{\pi_u}(f)$, while Rothaus' lemma (Lemma~\ref{lem:lsi-rothaus}) bounds
$\operatorname{Ent}_{\pi_u}(f^2)$ by the centered entropy plus
$2\operatorname{Var}_{\pi_u}(f)$.
The global Poincar\'e inequality then bounds all variance terms
by $O(R_t)\int\Vert{}\nabla f\Vert{}^2\,\mathrm d\pi_u$,
establishing the global $\mathsf{LSI}$ rate
$\Omega\bigl(\frac1{R_t}\bigr)=\Omega(p_d(t)t^2)$~(Theorem~\ref{thm:lsi-family}).

\paragraph{The lower bound.}
The $t^2$ factors in the smoothness and $\mathsf{LSI}$ bounds cancel,
giving $\kappa=\frac L\alpha=\Theta\bigl(\frac1{p_d(t)}\bigr)$
and hence an $\Omega(\kappa)$ query lower bound.
Since $t^2=\Theta(\log\kappa)$, the dimension requirement
$d\gtrsim t^4$ becomes $d\gtrsim\log^2\kappa$.

\subsection{Other Related Work}

The literature on continuous sampling is extensive, and we
focus here on results most relevant to our oracle model and
regularity assumptions.
For a broader treatment, see the monograph
\cite{Chewi2026LogConcaveSampling}.

\paragraph{Lower bounds for continuous sampling.}
Chewi et al.\ \cite{CGLGR22,chewi2024query} established
query lower bounds for log-concave sampling in one and
higher dimensions, including centered Gaussian targets.
Zhou and Suzuki \cite{zhou2026complexity} further sharpened
the Gaussian sampling lower bounds.
For uniform sampling from convex bodies, Vempala
\cite{Vempala2026MembershipSampling} proved nearly quadratic
membership-query lower bounds.
Beyond log-concavity, He and Zhang \cite{pmlr-v291-he25a}
established exponential-dimensional query lower bounds,
while Holzm\"uller and Bach \cite{HolzmuellerBach2025}
studied minimax sampling rates over smooth function classes
on compact domains.
For sampling under relative Fisher information guarantees,
Chewi et al.\ \cite{pmlr-v201-chewi23b} established query
lower bounds for smooth non-log-concave targets,
ruling out polylogarithmic dependence on inverse accuracy.

Chatterji, Bartlett, and Long \cite{ChatterjiBartlettLong2022}
proved lower bounds for stochastic-gradient sampling,
and Chen et al.\
\cite{ChenChewiDaskalakisRakhlin2026Stochastic}
studied the obstruction caused by heavy-tailed gradient noise.
Gopi, Lee, and Liu \cite{GopiLeeLiu2022} obtained lower bounds for composite targets with component-wise queries.
For diffusion sampling, Xun and Price \cite{XunPrice2026}
proved approximate-score query lower bounds.
Algorithm-specific lower bounds concern MALA
\cite{pmlr-v134-chewi21a,WuSchmidlerChen2022},
leapfrog Hamiltonian Monte Carlo
\cite{LeeShenTian2021LowerBounds},
and adaptive rejection sampling \cite{pmlr-v98-achddou19a}.
%

\paragraph{Upper bounds for log-concave sampling and
$\mathsf{LSI}$ sampling.}
Upper bounds for log-concave sampling have been developed
through several algorithmic approaches.
Geometric random walks include hit-and-run,
interior-point-inspired walks, and algorithmic diffusion
for convex bodies
\cite{LovaszVempala2006,
ChenDwivediWainwrightYu2018Polytopes,
KookVempalaZhang2024InAndOut}.
Non-asymptotic guarantees for overdamped Langevin algorithms
were developed in
\cite{DalalyanTsybakov2012,Dalalyan2017,DurmusMoulines2017},
with further analyses through optimization in Wasserstein space
\cite{Wibisono2018,DurmusMajewskiMiasojedow2019}.
Kinetic and Hamiltonian approaches include underdamped
Langevin algorithms, randomized midpoint discretization,
and Hamiltonian Monte Carlo
\cite{ChengChatterjiBartlettJordan2018,ShenLee2019,
ChenDwivediWainwrightYu2020HMC,ChenVempala2022}.
Metropolis-adjusted methods include MALA and its combination
with algorithmically constructed warm starts
\cite{DwivediChenWainwrightYu2018,pmlr-v134-chewi21a,
WuSchmidlerChen2022,AltschulerChewi2024WarmStarts}.
Moreau smoothing and mirror methods address non-smooth
potentials and constrained domains
\cite{DurmusMoulinesPereyra2018,pmlr-v125-zhang20a,
AhnChewi2021,pmlr-v247-srinivasan24a},
while proximal sampling methods use Gaussian-tilted distributions
\cite{LeeShenTian2021,ChenChewiSalimWibisono2022,
pmlr-v195-fan23a,LiuChewi2026Proximal}.
%

Recent high-accuracy approaches include first-order
rejection sampling
\cite{ChenChewiDaskalakisRakhlin2026HighAccuracy,
ChenChewiDaskalakisRakhlin2026Stochastic},
corrections on diffusion path space and Gaussian clouds
\cite{ChenChewiRakhlinZhang2026,chen2026high},
proximal kinetic and piecewise-deterministic samplers
\cite{chen2026accelerated,YuanGuoChen2026ReflectedProximal,
LuLuo2026WindowedThinning,LuLuo2026EntropyHypocoercivity},
and Gaussian smoothing with higher-order Picard approximation
\cite{ChenChewiLuZhang2026SmoothedPicard,
OAI:Subpolynomial-query-complexity-for-well-conditioned-log-concave-sampling-September-26-2026}.
Beyond log-concavity, Langevin, proximal, and parallel
sampling guarantees have also been established under
log-Sobolev and related functional inequalities
\cite{VempalaWibisono2019,ChewiErdogduLiShenZhang2025,
ChenChewiSalimWibisono2022,pmlr-v195-fan23a,
pmlr-v247-anari24a,chen2026high}.

\paragraph{Gaussian linear systems.}
Our lower bound hides information in Gaussian random fields
and analyzes what an adaptive algorithm learns about their
coefficients through exact first-order queries.
Oracle lower bounds for optimization have a long history
\cite{NemirovskiYudin1983}.
Randomized constructions include the random rotations of
\cite{CarmonDuchiHinderSidford2020}, the deformed Wigner
ensembles and truncated likelihood arguments of
\cite{SimchowitzElAlaouiRecht2018}, and the conditional
Wishart decompositions of
\cite{BravermanHazanSimchowitzWoodworth2020}.
Related models of adaptive linear information appear in
adaptive sensing \cite{PriceWoodruff2013,KachamWoodruff2023}
and information-based complexity
\cite{Novak1996,KonPlaskota2005}.
A closely related comparison is the Gaussian distinguishing
argument for Forrelation \cite{AaronsonAmbainis2018},
which compares adaptively selected measurements of
a shared Gaussian vector with independent Gaussian responses.
Our comparison instead replaces the coefficients shared
across two groups of measurements with independent copies
while preserving dependence within each group.
The total variation cost is controlled by the angle between
the full measurement spaces
(Theorem~\ref{gauss:decoupling}), allowing repeated or linearly
dependent measurements within either group.
For our tensor features, high tensor powers turn separation
of query locations into exponentially small overlap between
these spaces
(Lemma~\ref{out:ancestor-angle}).
Section~\ref{sec:linear-gaussian-system} contains precise
comparisons.

\subsection{Organization}

Section~\ref{sec:preliminaries} sets up the regularity assumptions, the oracle model, algorithm and the notions of query complexity.
Section~\ref{sec:linear-gaussian-system} develops our adaptive Gaussian
measurement framework, establishing sequential conditioning and the
decoupling theorem for separated measurement groups.
We then prove the strongly log-concave lower bound across
Sections~\ref{hf:section}--\ref{sec:target-good-mass}, constructing the
hard family and energy test in Section~\ref{hf:section}, showing that
few-query algorithms rarely pass the test in
Section~\ref{sec:output-fails}, and proving that target samples pass
with high probability in Section~\ref{sec:target-good-mass}.
Section~\ref{app:slc-warm} extends the strongly log-concave
lower bound to algorithms supplied with one sample from
an unknown constant-warm initialization law.
Section~\ref{sec:lsi-lower-bound} establishes the $\Omega(\kappa)$ lower
bound under a log-Sobolev inequality via the hidden-cap family, and
Section~\ref{sec:con} concludes with open problems.
\section{Preliminaries}
\label{sec:preliminaries}


Throughout, $x\cdot y=\langle x,y\rangle$ and $\|x\|$ denote
the Euclidean inner product and norm on the relevant ambient
space, and $S^{d-1}\subset\R^d$ denotes the unit sphere.
We write $\|\cdot\|_{\op}$ for the operator norm induced by
the Euclidean norms, $\|\cdot\|_\infty$ for the supremum norm,
and $P_W$ for the orthogonal projection onto a subspace~$W$.
The symbols $\partial_v$ and $D_v$ denote differentiation
in the direction~$v$, and $\log$ denotes the natural logarithm.
We write $\mathcal N(m,\Sigma)$ for the Gaussian law with
mean~$m$ and covariance~$\Sigma$, and $\mathbf 1_A$ for
the indicator of~$A$.
The notation $O,\Omega,\Theta,\lesssim,\gtrsim$ hides universal
multiplicative constants unless otherwise stated, while
$\widetilde O$ and $\widetilde\Omega$ additionally suppress
polylogarithmic factors.

\subsection{Assumptions}
\label{sec:prelim-assumptions}

A potential on $\R^d$ is a function $V\in C^1(\R^d)$ with
$\int_{\R^d}e^{-V(x)}\,dx<\infty$, and the associated target
distribution is the probability measure
$\pi_V(dx)\propto e^{-V(x)}\,dx$.

We describe the regularity and convexity of the target distribution
through properties of its potential, starting with smoothness.
\begin{definition}[\bf Smoothness]
\label{def:smooth}
A potential $V$ is $L$-smooth if
$\|\nabla V(x)-\nabla V(y)\|\le L\|x-y\|$ for all $x,y\in\R^d$, in
which case $\pi_V$ is called $L$-log-smooth. For $V\in C^2(\R^d)$,
$L$-smoothness is equivalent to $\|\nabla^2V(x)\|_{\op}\le L$ for all
$x\in\R^d$.
\end{definition}
We also consider a quantitative convexity condition on the potential,
which corresponds to strong log-concavity of the target distribution.
\begin{definition}[\bf Strong log-concavity]
\label{def:slc}
For $\alpha>0$, a potential $V$ is $\alpha$-strongly convex if
$V-\frac\alpha2\|\cdot\|^2$ is convex, in which case $\pi_V$ is called
$\alpha$-strongly log-concave. For $V\in C^2(\R^d)$, this is
equivalent to $\nabla^2V(x)\succeq\alpha I_d$ for all $x\in\R^d$. If
$V$ is $\alpha$-strongly convex and $L$-smooth, the ratio $\kappa = \frac{L}{\alpha}$
is its condition number.
\end{definition}

Beyond strong log-concavity, we also consider smooth potentials whose target distributions satisfy a functional inequality. To accommodate both the target $\pi_V$ on $\mathbb{R}^d$ and the auxiliary measures that arise in our later analysis, we state the next two definitions for a general probability measure $\mu$ supported on $\mathbb{R}^d$, a closed ball, a sphere, or an interval. We take the space of test functions to be $C_c^\infty(\mathbb{R}^d)$ when the domain is $\mathbb{R}^d$, and functions that are smooth up to the boundary in the remaining cases.
Given a nonnegative quadratic form $\mathcal E$ on test functions, which we refer to as an \emph{energy}, we default to the standard gradient energy $\mathcal E(f)=\int\Vert{}\nabla f\Vert{}^2\,d\mu$ defined with respect to the Euclidean, intrinsic, or one-dimensional gradient unless specified otherwise.

\begin{definition}[\bf Log-Sobolev inequality]
\label{def:lsi}
For $\alpha>0$, $\mu$ satisfies a \textbf{log-Sobolev inequality} ($\mathsf{LSI}(\alpha)$) with respect to
$\mathcal E$ if
$\operatorname{Ent}_\mu(f^2)\le\frac2\alpha\mathcal E(f)$ for every
test function $f$, where
$\operatorname{Ent}_\mu(g):=\int g\log g\,d\mu-\bigl(\int g\,d\mu\bigr)\log\bigl(\int g\,d\mu\bigr)$
with $0\log0:=0$.
\end{definition}
Replacing the entropy functional with the variance yields the Poincar\'e inequality.
\begin{definition}[\bf Poincar\'e inequality]
\label{def:pi}
For $\alpha>0$, $\mu$ satisfies a \textbf{Poincar\'e inequality} ($\mathsf{PI}(\alpha)$) with respect to
$\mathcal E$ if
$\operatorname{Var}_\mu(f)\le\frac1\alpha\mathcal E(f)$ for every test
function $f$, where
$\operatorname{Var}_\mu(f):=\int f^2\,d\mu-\bigl(\int f\,d\mu\bigr)^2$.
\end{definition}


This functional-inequality framework strictly generalizes the strongly log-concave regime. 
Specifically, if $\pi_V$ is $\alpha$-strongly log-concave, then $\pi_V$ satisfies $\mathsf{LSI}(\alpha)$ by the Bakry--\'Emery criterion (Lemma~\ref{prop:lsi-bakry-emery}). 
Consequently, for an $L$-smooth potential whose target distribution satisfies $\mathsf{LSI}(\alpha)$, the ratio $\kappa = \frac{L}{\alpha}$ serves as a natural generalization of the condition number.

By standard approximation arguments, both inequalities extend to the closure of the test functions under the norm $\Vert{}f\Vert{}_{L^2(\mu)}+\mathcal E(f)^{1/2}$. This closure contains all locally Lipschitz functions in $L^2(\mu)$ with finite energy, as well as all locally absolutely continuous functions with finite energy when the domain is an interval. Finally, in Section~\ref{sec:lsi-constant}, we also make use of weighted energies on an interval of the form $\int a\vert{}f'\vert{}^2\,d\mu$.

\subsection{Algorithm and Complexity}
\paragraph{Oracle.}
A potential $V$ is accessed through an exact first-order oracle, which
on a \emph{query} $y\in\R^d$ returns the pair
$(V(y),\nabla V(y))\in\R\times\R^d$.

\paragraph{Warm start.}
For $M\ge1$, a distribution $\nu$ is $M$-warm with respect to $\pi_V$ if $\nu\ll\pi_V$
and $\frac{d\nu}{d\pi_V}\le M$ holds $\pi_V$-almost everywhere.
In the warm-start model, used only when an initial sample is explicitly allowed,
each instance specifies a potential $V$ and a distribution $\nu$ that may depend on $\pi_V$.
The algorithm receives a single sample $X_{\mathrm{init}}\sim\nu$, without access to
the density of $\nu$ or to further samples from $\nu$, and its decisions may depend on
$X_{\mathrm{init}}$ and on an independent random seed.
%

\paragraph{Algorithms.}
For $t\ge0$, let $H_t:=\bigl(\R^d\times(\R\times\R^d)\bigr)^t$ be the
set of histories of $t$ queries and responses. A sampling algorithm
$\Alg$ is specified by Borel measurable query maps
$\mathsf A_t:[0,1]\times H_{t-1}\to\R^d$ for $t\ge1$, Borel stopping
sets $\mathsf S_t\subset[0,1]\times H_t$ for $t\ge0$, and Borel
output maps $\mathsf A_{\mathrm{out}}^{(t)}:[0,1]\times H_t\to\R^d$
for $t\ge0$. These maps are fixed independently of the unknown
potential $V$.

The algorithm draws a seed $\omega\sim\mathrm{Unif}[0,1]$,
independently of the oracle instance, and proceeds in rounds. After
$t$ rounds its history is $h_t:=((y_1,r_1),\dots,(y_t,r_t))$, with
$h_0$ empty. If $(\omega,h_t)\in\mathsf S_t$, it stops and outputs
$X:=\mathsf A_{\mathrm{out}}^{(t)}(\omega,h_t)$. Otherwise it queries
$y_{t+1}:=\mathsf A_{t+1}(\omega,h_t)$ and receives
$r_{t+1}:=(V(y_{t+1}),\nabla V(y_{t+1}))$. The number of rounds
before stopping is the number of queries $N$, and we require
$N<\infty$ almost surely for every potential under consideration. All
randomness, including any used at the output stage, is encoded in
$\omega$. The law of $X$ on $V$ is denoted by $\rho_V^{\Alg}$ and the
expected number of queries on $V$ by $\mathbb E_V[N]$. A $q$-query
algorithm is one with $N\le q$ almost surely on every potential.
\paragraph{Accuracy.}
Accuracy is measured in total variation distance, defined for
probability measures $\nu,\nu'$ on a common measurable space by
$\TV(\nu,\nu'):=\sup_B|\nu(B)-\nu'(B)|$ with $B$ ranging over all
measurable sets.

\paragraph{Query complexity.}
For a set $\mathcal V$ of potentials on $\R^d$ and
$\varepsilon\in(0,1)$, let
\[
  \QC(\mathcal V,\varepsilon):=\min\Bigl\{q\ge1:\
  \sup_{V\in\mathcal V}\TV(\rho_V^{\Alg},\pi_V)\le\varepsilon
  \text{ for some $q$-query algorithm }\Alg\Bigr\},
\]
with $\min\emptyset:=\infty$, and let
$\overline{\QC}(\mathcal V,\varepsilon)$ be the infimum of
$\sup_{V\in\mathcal V}\mathbb E_V[N]$ over all algorithms $\Alg$ with
$\sup_{V\in\mathcal V}\TV(\rho_V^{\Alg},\pi_V)\le\varepsilon$. A
$q$-query algorithm has $\mathbb E_V[N]\le q$, so
$\overline{\QC}\le\QC$.

\section{Linear Gaussian System}
\label{sec:linear-gaussian-system}
In this section, we study oracles whose replies are computed from inner products with the hidden Gaussian vector in a coefficient space whose dimension may far exceed
the query dimension.
Section~\ref{gauss:observation-rules} defines the measurement
process and explains how its measurements generate the
responses returned to the algorithm.
Section~\ref{gauss:conditioning} gives the conditional
distribution of $\Xi$ given the recorded measurements
and derives bounds on its projections.
Section~\ref{gauss:simulations} bounds the change in the
distribution of the complete record when shared Gaussian
coefficients are replaced by independent copies for two
groups of measurements whose spans are nearly orthogonal.
We specify how our construction fits this framework in
Section~\ref{hf:section}--\ref{sec:target-good-mass}.

\subsection{The Model}
\label{gauss:observation-rules}

Let $E$ be a finite-dimensional Euclidean space, and let
$\Xi\sim\mathcal N(0,I_E)$.
The vector $\Xi$ is sampled once and remains fixed throughout
the interaction. 
We note that the dimension of $E$ need not equal the query dimension $d$.
Let the seed $\omega$ be a random element of a standard Borel space
$\mathcal O$, independent of $\Xi$.
In Section~\ref{sec:preliminaries}, the random seed is uniform on $[0,1]$ and
we allow a general seed because it will later be enlarged by
coefficient vectors independent of $\Xi$, and every statement in
this section holds for any such enlargement.

\begin{definition}[\bf Adaptive linear measurement process]
\label{gauss:measurement-process}
Fix integers $q\ge1$ and $m_1,\ldots,m_q\ge0$.
A $q$-round \emph{adaptive linear measurement process} of $\Xi$ with
batch sizes $(m_k)$ is specified by Borel maps
\[
 \mathsf B_k:\mathcal O\times\mathbb R^{m_1}\times\cdots\times
 \mathbb R^{m_{k-1}}\to E^{m_k},
 \qquad 1\le k\le q.
\]
Its measurement vectors and measurements are defined recursively by
\[
 (a_{k,1},\ldots,a_{k,m_k})=\mathsf B_k(\omega,Z_1,\ldots,Z_{k-1}),
 \qquad
 Z_k=\bigl(\langle\Xi,a_{k,1}\rangle,\ldots,
 \langle\Xi,a_{k,m_k}\rangle\bigr),
\]
so that the whole batch of round $k$ is selected before any entry
of $Z_k$ is revealed.
Measurement vectors may be zero or linearly dependent.
The history and the measurement space after $k$ rounds are
\[
 \mathcal F_k=\sigma(\omega,Z_1,\ldots,Z_k),
 \qquad
 W_k=\operatorname{span}\{a_{j,b}:j\le k,\ b\le m_j\},
\]
with $\mathcal F_0=\sigma(\omega)$ and $W_0=\{0\}$.
The total number of scalar measurements is $M=\sum_{k=1}^qm_k$.
\end{definition}

Related batched measurement models with scalar measurement
budgets appear in adaptive sensing
\cite{PriceWoodruff2013,KachamWoodruff2023}.
Adaptive linear information is also classical in
information-based complexity \cite{Novak1996}, with Borel
measurable selection rules explicitly used in
\cite{KunschNovakWnuk2024}.
In our applications, each oracle query is simulated using
a batch of underlying linear measurements.
The algorithm receives only the resulting function value and gradient
reply, which need not reveal the individual measurements.

\paragraph{Oracle simulation.}
Let $\Alg$ be a $q$-query algorithm with seed $\omega$.
An oracle simulation for $\Alg$ consists of a $q$-round measurement
process $(\mathsf B_k)$ of $\Xi$ as in
Definition~\ref{gauss:measurement-process}, together with Borel
return maps
\[
 \mathsf R_k:\mathcal O\times\mathbb R^{m_1}\times\cdots\times
 \mathbb R^{m_k}\to\mathbb R\times\mathbb R^d,
 \qquad 1\le k\le q.
\]
At round $k$, the algorithm receives
\[
 r_k=\mathsf R_k(\omega,Z_1,\ldots,Z_k)
\]
and selects its next query from $\omega$ and the responses received
so far, as in Section~\ref{sec:preliminaries}, 
but the algorithm accesses these measurements only through
the returned responses.
The return maps may be nonlinear, as needed for hard examples in Section~\ref{hf:section}.
Consequently, each query $y_k$ is a Borel function of
$(\omega,Z_1,\ldots,Z_{k-1})$, so the batch $\mathsf B_k$ may depend
on the current query, and the transcript and the output of $\Alg$
are $\mathcal F_q$-measurable.


\begin{remark}[\bf Stopping and terminal measurements]
\label{gauss:terminal}
An algorithm that stops before round $q$ keeps its output, and the
remaining batches are zero vectors and this changes nothing in
Definition~\ref{gauss:measurement-process}.
After the output is fixed, the simulator may measure one terminal
batch $\mathsf B_{q+1}(\omega,Z_1,\ldots,Z_q)\in E^{m_{q+1}}$, which is
recorded but not returned. 
The internal measurements then form a
$(q+1)$-round process with $M=\sum_{k\le q+1}m_k$, while the
algorithm still makes $q$ queries.
A terminal batch lets a test that depends on the hidden coefficients,
such as a gate value at the output, be evaluated inside the process.
\end{remark}

\paragraph{Full transcript.}
Set $m_{q+1}=0$ if there is no terminal batch.
The full transcript of a simulation is
$\mathsf T=(\omega,Z_1,\ldots,Z_{q+1})$.
The algorithm's transcript $(\omega,(y_k,r_k)_{k\le q})$ and its
output are Borel functions of $(\omega,Z_1,\ldots,Z_q)$, and a
terminal test is a Borel function of $\mathsf T$. The algorithm sees
only the former.

\paragraph{Gaussian fields.}
Let $\Phi\in C^1(\mathbb R^d;E)$ be deterministic and let
$H^\Xi(x)=\langle\Xi,\Phi(x)\rangle$, $x\in\mathbb R^d$.
Then $H^\Xi$ is a centered Gaussian random field with covariance
kernel $k(x,y)=\langle\Phi(x),\Phi(y)\rangle$.

\begin{lemma}[\bf First-order queries to a Gaussian field]
\label{gauss:field-queries}
Let $\Alg$ be a $q$-query algorithm with first-order oracle access
to $H^\Xi$, and let $X$ be a Borel function of its transcript.
\begin{enumerate}[leftmargin=2em,label=(\roman*)]
\item For all $x,y\in\mathbb R^d$ and $v,w\in\mathbb R^d$,
 \[
  D_vH^\Xi(x)=\langle\Xi,D_v\Phi(x)\rangle,
  \qquad
  \mathbb E\bigl[D_vH^\Xi(x)\,D_wH^\Xi(y)\bigr]
  =\partial_s\partial_t\,k(x+sv,y+tw)\big|_{s=t=0}.
 \]
\item $\Alg$ is an oracle simulation with $m_k=1+d$, batch
 $\bigl(\Phi(y_k),D_{e_1}\Phi(y_k),\ldots,D_{e_d}\Phi(y_k)\bigr)$ at
 round $k$, and $r_k=Z_k$.
 Its measurement space after $k$ rounds is
 \[
  W_k=\operatorname{span}\{\Phi(y_j),\,D_v\Phi(y_j):
  j\le k,\ v\in\mathbb R^d\}.
 \]
\item $\Phi(X)$ and $D_v\Phi(X)$ are $\mathcal F_q$-measurable for
 every $v\in\mathbb R^d$.
\end{enumerate}
\end{lemma}

\begin{proof}
(i) The first identity holds because $\Xi$ does not depend on $x$.
Since $\mathbb E\langle\Xi,a\rangle\langle\Xi,b\rangle=\langle a,b\rangle$
for $a,b\in E$, the covariance equals
$\langle D_v\Phi(x),D_w\Phi(y)\rangle
=\partial_s\partial_t\langle\Phi(x+sv),\Phi(y+tw)\rangle|_{s=t=0}$
by bilinearity.
(ii) By (i), $\nabla H^\Xi(x)=(\langle\Xi,D_{e_i}\Phi(x)\rangle)_{i\le d}$,
and the span of $D_v\Phi(y_j)$ over all $v$ equals the span over the
basis directions.
(iii) The transcript is $\mathcal F_q$-measurable, and $\Phi$ and
$D_v\Phi$ are continuous.
\end{proof}
\subsection{Sequential Gaussian Conditioning}
\label{gauss:conditioning}

In this section, we describe what the measurements reveal about
$\Xi$.
After $k$ rounds, the seed and the measurements determine the
projection $P_{W_k}\Xi$, while the remainder $P_{W_k^\perp}\Xi$ is
still a centered Gaussian vector (see 
Lemma~\ref{gauss:posterior}).
Two consequences are used repeatedly later.
Corollary~\ref{gauss:linear-functional} describes the conditional
law of a linear functional whose vector may itself depend on the
history, and Lemma~\ref{gauss:predictable-projections} bounds the
total energy of projections onto subspaces chosen before each round.
A terminal batch, when present, counts as round $q+1$.
All conditioning in this section is on $\mathcal F_k$, which contains
the measurements themselves.
However, conditioning on nonlinear responses need not preserve Gaussianity
(see \cite[Theorem~3]{ChenHosseiniOwhadiStuart2025}).

We shall use conditional Gaussian laws in the following form.
Suppose that $\mathcal L(\Xi\mid\mathcal G)=\mathcal N(\mu,\Sigma)$
for a sub-$\sigma$-algebra $\mathcal G$ and $\mathcal G$-measurable
$\mu$ and $\Sigma$.
Then, for every $\mathcal G$-measurable random variable $Y$ with
values in a standard Borel space and every bounded Borel function
$f$,
\begin{equation}\label{gauss:freezing}
 \mathbb E[f(\Xi,Y)\mid\mathcal G]
 =\int_E f(x,Y)\,\mathcal N(\mu,\Sigma)(dx)
 \qquad\text{almost surely}.
\end{equation}
Indeed, for $f(x,y)=g(x)h(y)$ the factor $h(Y)$ is
$\mathcal G$-measurable and can be taken out of the conditional
expectation, and the general case follows by the monotone class
theorem.

The following lemma records the classical Gaussian
conditioning property for adaptive linear information
\cite{KonPlaskota2005}.
We include a proof for our batched model, allowing zero
and linearly dependent measurement vectors.
\begin{lemma}[\bf Gaussian conditioning]
\label{gauss:posterior}
For every $0\le k\le q$, the projection $P_{W_k}$ and the vector
$P_{W_k}\Xi$ are $\mathcal F_k$-measurable, and
\begin{equation}\label{gauss:posterior-law}
 \mathcal L(\Xi\mid\mathcal F_k)
 =\mathcal N\bigl(P_{W_k}\Xi,\,P_{W_k^\perp}\bigr)
 \qquad\text{almost surely}.
\end{equation}
\end{lemma}

\begin{proof}[Proof of Lemma~\ref{gauss:posterior}]
Let $A_{\le k}$ be the matrix whose columns are the measurement
vectors through round $k$, in orthonormal coordinates on $E$, and
let $\mathbf z_k=A_{\le k}^{\mathsf T}\Xi$ be the concatenation of
$Z_1,\ldots,Z_k$.
Then
\[
 P_{W_k}
 =A_{\le k}(A_{\le k}^{\mathsf T}A_{\le k})^\dagger A_{\le k}^{\mathsf T},
 \qquad
 P_{W_k}\Xi
 =A_{\le k}(A_{\le k}^{\mathsf T}A_{\le k})^\dagger\mathbf z_k,
\]
where $\dagger$ is the Moore--Penrose inverse, a Borel map by
$C^\dagger=\lim_{\delta\downarrow0}(C^{\mathsf T}C+\delta I)^{-1}C^{\mathsf T}$.
Both expressions are therefore $\mathcal F_k$-measurable.

We prove \eqref{gauss:posterior-law} by induction on $k$.
The case $k=0$ is the independence of $\Xi$ and $\omega$.
Assume the identity for $k-1$, and write $W=W_{k-1}$,
$\widehat\Xi=P_W\Xi$, and $R=\Xi-\widehat\Xi$.
By \eqref{gauss:freezing}, $R\sim\mathcal N(0,P_{W^\perp})$ given
$\mathcal F_{k-1}$.
Let $A_k$ be the matrix with columns $a_{k,1},\ldots,a_{k,m_k}$,
let $B=P_{W^\perp}A_k$, and let $\mathcal V=\operatorname{range}B$.
These objects are $\mathcal F_{k-1}$-measurable, and
$W_k=W\oplus\mathcal V$.
Since
$Z_k-A_k^{\mathsf T}\widehat\Xi=B^{\mathsf T}R=B^{\mathsf T}P_{\mathcal V}R$
and $B^{\mathsf T}$ is injective on
$\mathcal V=(\ker B^{\mathsf T})^\perp$, the batch $Z_k$ and the
projection $P_{\mathcal V}R$ determine each other given
$\mathcal F_{k-1}$.
Hence $\mathcal F_k=\mathcal F_{k-1}\vee\sigma(P_{\mathcal V}R)$.
Given $\mathcal F_{k-1}$, the projections $P_{\mathcal V}R$ and
$P_{W_k^\perp}R$ are jointly Gaussian with cross-covariance
$P_{\mathcal V}P_{W^\perp}P_{W_k^\perp}=0$, and are therefore
independent.
Consequently, the conditional law of $P_{W_k^\perp}R=P_{W_k^\perp}\Xi$
given $\mathcal F_k$ is still $\mathcal N(0,P_{W_k^\perp})$, while
$\Xi-P_{W_k^\perp}\Xi=\widehat\Xi+P_{\mathcal V}R=P_{W_k}\Xi$ is
$\mathcal F_k$-measurable.
This proves \eqref{gauss:posterior-law}.
\end{proof}

\begin{remark}[\bf Measurable subspaces]
\label{gauss:measurable-subspace}
A random subspace $S$ of $E$ is called $\mathcal G$-measurable if its
orthogonal projection $P_S$ is a $\mathcal G$-measurable random
matrix.
If $v_1,\ldots,v_r$ are $\mathcal G$-measurable vectors and $V$ is the
matrix with these columns, then
$P_{\operatorname{span}\{v_1,\ldots,v_r\}}=V(V^{\mathsf T}V)^\dagger V^{\mathsf T}$
is $\mathcal G$-measurable, since the Moore--Penrose inverse is a
Borel map.
In particular, $W_k$ is $\mathcal F_{k-1}$-measurable.
\end{remark}


We now record the two consequences of Lemma~\ref{gauss:posterior}
used later.
The first concerns a single linear functional of $\Xi$ whose vector
may depend on the history, for instance a feature vector evaluated
at the output of an algorithm.

\begin{corollary}[\bf Linear functionals]
\label{gauss:linear-functional}
Let $b$ be an $\mathcal F_k$-measurable vector in $E$.
Conditionally on $\mathcal F_k$,
$\langle\Xi,b\rangle\sim
\mathcal N\bigl(\langle P_{W_k}\Xi,b\rangle,\|P_{W_k^\perp}b\|^2\bigr)$ and
in particular, $\langle\Xi,b\rangle=\langle P_{W_k}\Xi,b\rangle$ is
determined by $\mathcal F_k$ when $b\in W_k$.
Moreover, if $b_1,\ldots,b_N$ are $\mathcal F_q$-measurable with
$\|b_j\|\le\sigma_j$ almost surely for constants $\sigma_j\ge0$, then
for every $t>0$,
\[
 \mathbb P\bigl\{\exists\,j\le N:
 |\langle P_{W_q^\perp}\Xi,b_j\rangle|>\sqrt{2t}\,\sigma_j\bigr\}
 \le2Ne^{-t}.
\]
\end{corollary}

\begin{proof}[Proof of Corollary~\ref{gauss:linear-functional}]
The first assertion is the image of \eqref{gauss:posterior-law}
under $x\mapsto\langle x,b\rangle$, by \eqref{gauss:freezing}.
For the second,
$\langle P_{W_q^\perp}\Xi,b_j\rangle=\langle\Xi,P_{W_q^\perp}b_j\rangle$
is conditionally centered Gaussian given $\mathcal F_q$ with variance
$\|P_{W_q^\perp}b_j\|^2\le\sigma_j^2$, so the Gaussian tail bound,
valid also for zero variance, gives conditional probability at most
$2e^{-t}$ for each $j$; take a union bound and expectations.
\end{proof}

The second consequence concerns the energy of $\Xi$ in subspaces
that are chosen before a round and revealed by it.
If $S_k$ is selected from the history before round $k$ and lies in
the part of $W_k$ not already contained in $W_{k-1}$, then
$P_{S_k}\Xi$ is a standard Gaussian vector in $S_k$ given that
history, and the total energy over all rounds has chi-squared tails
with degrees of freedom bounded by the total dimension.
\begin{lemma}[\bf Predictable projections]
\label{gauss:predictable-projections}
For $1\le k\le q$, let $S_k$ be an $\mathcal F_{k-1}$-measurable
subspace with $S_k\subseteq W_{k-1}^\perp\cap W_k$, and suppose that
$\sum_k\dim S_k\le D$ almost surely for a constant $D\ge0$.
Then the $S_k$ are mutually orthogonal and, for every $t>0$,
\begin{equation}\label{gauss:predictable-projection-tail}
 \mathbb P\Bigl\{\sum_{k=1}^q\|P_{S_k}\Xi\|^2>4(D+t)\Bigr\}\le e^{-t}.
\end{equation}
\end{lemma}

\begin{proof}[Proof of Lemma~\ref{gauss:predictable-projections}]
For $j<k$, $S_j\subseteq W_j\subseteq W_{k-1}$ and
$S_k\subseteq W_{k-1}^\perp$, so $S_j\perp S_k$.
Put $r_k=\dim S_k$ and $Y_k=\|P_{S_k}\Xi\|^2$; then $r_k$ is
$\mathcal F_{k-1}$-measurable, and $Y_k$ is $\mathcal F_k$-measurable
because $P_{S_k}\Xi=P_{S_k}P_{W_k}\Xi$.
Given $\mathcal F_{k-1}$, Lemma~\ref{gauss:posterior} and
$S_k\subseteq W_{k-1}^\perp$ give $P_{S_k}\Xi\sim\mathcal N(0,P_{S_k})$,
so $Y_k\sim\chi^2_{r_k}$ and
$\mathbb E[e^{Y_k/4}\mid\mathcal F_{k-1}]=2^{r_k/2}$.
Hence
$M_k=\exp\bigl(\tfrac14\sum_{j\le k}Y_j\bigr)\,2^{-\frac12\sum_{j\le k}r_j}$
is a martingale with $M_0=1$, and
$\mathbb E\exp(\tfrac14\sum_kY_k)\le2^{D/2}\,\mathbb EM_q=2^{D/2}$.
Markov's inequality gives
$\mathbb P\{\sum_kY_k>4(D+t)\}\le e^{-D-t}2^{D/2}\le e^{-t}$.
\end{proof}

\subsection{Decoupling Two Groups of Measurements}
\label{gauss:simulations}

In this section, each round of a measurement process is assigned to
one of two groups, and we ask how much the full transcript changes
when the two groups measure independent copies of $\Xi$ instead of
the same vector\footnote{We use ``decoupling'' in the sense of replacing shared
coefficients by independent copies and it is unrelated to decoupling
inequalities for Gaussian processes.}.
In the shared model, both groups measure the same Gaussian vector
$\Xi$.
In the independent model, each group measures its own independent
copy.
Theorem~\ref{gauss:decoupling} shows that the full transcripts of the
two models have nearly the same law whenever the measurement vectors
of the two groups span nearly orthogonal subspaces.
Later, the groups will consist of queries at separated locations,
and the theorem will allow us to treat them as if they had seen
independent coefficients.

\paragraph{Labels and the two coefficient models.}
Consider a measurement process as in
Definition~\ref{gauss:measurement-process}, with any terminal batch
counted among its $q$ rounds, together with the return maps
$\mathsf R_k$ of an oracle simulation.
In addition, fix Borel measurable \emph{labeling maps}
\[
 \mathsf L_k:\mathcal O\times\mathbb R^{m_1}\times\cdots\times
 \mathbb R^{m_{k-1}}\to\{\mathrm{local},\mathrm{far}\},
 \qquad 1\le k\le q.
\]
The label $\ell_k=\mathsf L_k(\omega,Z_1,\ldots,Z_{k-1})$ assigns the
batch of round $k$ to one of the two groups.
Like the measurement vectors, it is selected before the batch is
measured.

Let $\Xi^{\mathrm{local}}$ and $\Xi^{\mathrm{far}}$ be independent
$\mathcal N(0,I_E)$ vectors, independent of $\omega$.
The two models differ only in the vector that is measured:
\[
 Z_k=\bigl(\langle\Xi,a_{k,b}\rangle\bigr)_{b\le m_k}
 \quad\text{in the shared model},
 \qquad
 Z_k=\bigl(\langle\Xi^{\ell_k},a_{k,b}\rangle\bigr)_{b\le m_k}
 \quad\text{in the independent model}.
\]
Both models use the same maps $\mathsf B_k$, $\mathsf L_k$,
$\mathsf R_k$ and the same algorithm.
Since each model evaluates these maps on its own measurements, the
queries, labels, and measurement vectors may follow different
trajectories in the two models.

For fixed vectors $\xi^{\mathrm{local}},\xi^{\mathrm{far}}\in E$, let
$\mathsf T(\omega;\xi^{\mathrm{local}},\xi^{\mathrm{far}})$ be the full
transcript obtained when local batches are measured against
$\xi^{\mathrm{local}}$ and far batches against $\xi^{\mathrm{far}}$, and
put
\[
 P_{\mathrm{sh}}=\mathcal L\bigl(\mathsf T(\omega;\Xi,\Xi)\bigr),
 \qquad
 P_{\mathrm{ind}}=\mathcal L\bigl(
 \mathsf T(\omega;\Xi^{\mathrm{local}},\Xi^{\mathrm{far}})\bigr).
\]
By Section~\ref{gauss:observation-rules}, the algorithm's transcript and its
output are Borel functions of the full transcript, and these
functions are the same in both models.

\paragraph{The spans of the two groups.}
Fix a seed value $\omega$ and a point
$z=(z_1,\ldots,z_q)\in\mathbb R^{m_1}\times\cdots\times\mathbb R^{m_q}$.
Evaluating the maps along $z$, that is,
$(a_{k,b})_{b\le m_k}=\mathsf B_k(\omega,z_1,\ldots,z_{k-1})$ and
$\ell_k=\mathsf L_k(\omega,z_1,\ldots,z_{k-1})$, put
\[
 L(\omega,z)=\operatorname{span}\{a_{k,b}:\ell_k=\mathrm{local}\},
 \qquad
 F(\omega,z)=\operatorname{span}\{a_{k,b}:\ell_k=\mathrm{far}\}.
\]
These are the spans of the two groups along the prescribed sequence
of measurement values, whether or not that sequence can occur in
either model.

\begin{theorem}[\bf Gaussian decoupling]
\label{gauss:decoupling}
Let $0\le\rho<\tfrac12$, and suppose that
$\|P_{L(\omega,z)}P_{F(\omega,z)}\|_{\mathrm{op}}\le\rho$ for every seed
value $\omega$ and every
$z\in\mathbb R^{m_1}\times\cdots\times\mathbb R^{m_q}$.
Then, with $M=\sum_{k=1}^qm_k$,
\begin{equation}\label{gauss:decoupling-bound}
 \operatorname{TV}(P_{\mathrm{sh}},P_{\mathrm{ind}})\le\frac32\rho M.
\end{equation}
The same bound holds for the laws of the algorithm's transcript and
of its output.
\end{theorem}

The assumption concerns the spans of the two groups of
measurement vectors, because linear combinations from
the two groups may be strongly aligned even when all
pairwise inner products between the groups are small.
%
The quantity $\|P_LP_F\|_{\mathrm{op}}$ detects this.
On the other hand, the theorem imposes no condition within a group.
Measurement vectors with the same label may be zero, repeated, or
arbitrarily close to linearly dependent.

\begin{remark}[\bf Related comparisons]
The Gaussian distinguishing argument for Forrelation
\cite[Section~4]{AaronsonAmbainis2018} compares adaptive
measurements of a shared Gaussian vector along nearly
orthogonal test vectors with independent Gaussian responses.
Our comparison preserves dependence within each group
and controls the total variation cost through the angle
between the two full measurement spaces.
\end{remark}

\begin{proof}[Proof of Theorem~\ref{gauss:decoupling}]
We bound the total variation distance by the expected positive part
of the log-likelihood ratio under the shared model.
The two models differ only in the Gram matrices of the measurement
vectors, which near orthogonality makes comparable up to a factor
$1\pm\rho$ (Fact~\ref{gauss:gram-comparison}).
Expressing the likelihood ratio through these matrices
(Fact~\ref{gauss:likelihood}) bounds the log-likelihood ratio by
$\rho$ times the number of retained coordinates plus $\frac\rho2$
times the normalized energy of the measurements, and both have
expectation at most $M$ (Fact~\ref{gauss:innovations}).

\paragraph{Setup.}
Fix a seed value $\omega$, which will be suppressed from the notation.
List the entries of $Z_1,\ldots,Z_q$ in order as scalar measurements
$z_1,\ldots,z_M$, with measurement vectors $a_1,\ldots,a_M$ and labels
$\ell_1,\ldots,\ell_M$ inherited from their batches.
Each $a_t$ and $\ell_t$ is a Borel function of
$z_{<t}=(z_1,\ldots,z_{t-1})$, being determined by the batches that
precede its own.
This is a measurement process with batch sizes one, and its
measurement spaces agree with those of the original process at the
ends of batches.
Put $\Xi^{\mathrm{sh}}=\Xi$ and $a^{\mathrm{sh}}_t=a_t$.
Put $\Xi^{\mathrm{ind}}=(\Xi^{\mathrm{local}},\Xi^{\mathrm{far}})$ and
$a^{\mathrm{ind}}_t=(a_t,0)$ or $(0,a_t)$ according as $\ell_t$ is local
or far.
Then $\Xi^{\mathrm{ind}}$ is standard Gaussian in $E\oplus E$,
$\langle\Xi^{\mathrm{ind}},a^{\mathrm{ind}}_t\rangle
=\langle\Xi^{\ell_t},a_t\rangle$, and in model
$\alpha\in\{\mathrm{sh},\mathrm{ind}\}$ the scalar process measures
$\Xi^\alpha$ along the vectors $a^\alpha_t$.
The results of Section~\ref{gauss:conditioning} therefore apply to
both models.
Let $P^{\mathrm{sh}}$ and $P^{\mathrm{ind}}$ be the laws of
$z=(z_1,\ldots,z_M)$ in the two models, and let $W^\alpha_t$ be the
span of $a^\alpha_1,\ldots,a^\alpha_t$.
Along $z\in\mathbb R^M$, define the Gram matrices
\[
 \Gamma^\alpha_{st}=\langle a^\alpha_s,a^\alpha_t\rangle,
 \qquad 1\le s,t\le M,
\]
so that $\Gamma^{\mathrm{sh}}_{st}=\langle a_s,a_t\rangle$ and
$\Gamma^{\mathrm{ind}}_{st}=\mathbf 1_{\{\ell_s=\ell_t\}}\langle a_s,a_t\rangle$.
Their dependence on $z$ is suppressed.

\begin{fact}[\bf Gram comparison]
\label{gauss:gram-comparison}
For every $z\in\mathbb R^M$ and every index set
$J\subseteq\{1,\ldots,M\}$,
\begin{equation}\label{gauss:gram-bound}
 (1-\rho)\Gamma^{\mathrm{ind}}_{J,J}
 \preceq\Gamma^{\mathrm{sh}}_{J,J}
 \preceq(1+\rho)\Gamma^{\mathrm{ind}}_{J,J}.
\end{equation}
\end{fact}

\begin{proof}[Proof of Fact~\ref{gauss:gram-comparison}]
For $c\in\mathbb R^J$, put $u=\sum_{t\in J,\ \ell_t=\mathrm{local}}c_ta_t$
and $v=\sum_{t\in J,\ \ell_t=\mathrm{far}}c_ta_t$.
Then $c^{\mathsf T}\Gamma^{\mathrm{sh}}_{J,J}c=\|u+v\|^2$ and
$c^{\mathsf T}\Gamma^{\mathrm{ind}}_{J,J}c=\|u\|^2+\|v\|^2$.
Since $u\in L$ and $v\in F$,
\[
 \bigl|c^{\mathsf T}(\Gamma^{\mathrm{sh}}_{J,J}-\Gamma^{\mathrm{ind}}_{J,J})c\bigr|
 =2|\langle u,v\rangle|
 =2|\langle u,P_LP_Fv\rangle|
 \le2\rho\|u\|\|v\|
 \le\rho\,c^{\mathsf T}\Gamma^{\mathrm{ind}}_{J,J}c.
\]
\end{proof}

\paragraph{Removing redundant coordinates.}
The Gram matrices may be singular, so we first remove the
coordinates that are determined by earlier ones.
Within each label, scan the measurement vectors in order and retain
a vector if it lies outside the span of the previously retained
vectors with the same label.
Let $I$ be the set of retained indices and
$J_t=I\cap\{1,\ldots,t-1\}$.
Membership in $I$ is decided by ranks of Gram matrices, so it is a
Borel function of $z_{<t}$.

If $t\notin I$, then $a_t=\sum_{s\in J_t,\ \ell_s=\ell_t}c_{t,s}a_s$
for unique coefficients $c_{t,s}$, Borel in $z_{<t}$.
Since all vectors in this relation carry the same label, it also
holds for the vectors $a^{\mathrm{ind}}$, and in both models
\begin{equation}\label{gauss:omitted-value}
 z_t=\sum_{s\in J_t,\ \ell_s=\ell_t}c_{t,s}\,z_s
 \qquad(t\notin I).
\end{equation}
Thus an omitted measurement is the same deterministic function of
$z_{<t}$ in both models, and the retained vectors with index below
$t$ span $W^\alpha_{t-1}$.
They are also linearly independent, by construction in the
independent model, where different labels occupy orthogonal copies
of $E$, and by \eqref{gauss:gram-bound} in the shared model, which
gives $\Gamma^{\mathrm{sh}}_{J_t,J_t}\succeq(1-\rho)\Gamma^{\mathrm{ind}}_{J_t,J_t}\succ0$.
Hence they form a basis of $W^\alpha_{t-1}$.
In particular, writing $I=\{t_1<\cdots<t_r\}$,
$\widehat z=(z_{t_1},\ldots,z_{t_r})^{\mathsf T}$, and
$\widehat\Gamma^\alpha=\Gamma^\alpha_{I,I}$, the matrices
$\widehat\Gamma^\alpha$ and all their principal submatrices are
positive definite.

On the retained coordinates, both models have densities, and the
likelihood ratio takes the same form as for two centered Gaussian
vectors with covariances $\widehat\Gamma^{\mathrm{sh}}$ and
$\widehat\Gamma^{\mathrm{ind}}$, even though the measurements are not
jointly Gaussian when the vectors are chosen adaptively.

\begin{fact}[\bf Likelihood ratio]
\label{gauss:likelihood}
The laws $P^{\mathrm{sh}}$ and $P^{\mathrm{ind}}$ are mutually
absolutely continuous, and $P^{\mathrm{sh}}$-almost surely
\begin{equation}\label{gauss:likelihood-ratio}
 \log\frac{dP^{\mathrm{sh}}}{dP^{\mathrm{ind}}}(z)
 =\frac12\log
 \frac{\det\widehat\Gamma^{\mathrm{ind}}}{\det\widehat\Gamma^{\mathrm{sh}}}
 +\frac12\,\widehat z^{\mathsf T}\bigl[
 (\widehat\Gamma^{\mathrm{ind}})^{-1}-(\widehat\Gamma^{\mathrm{sh}})^{-1}
 \bigr]\widehat z,
\end{equation}
where for $I=\varnothing$ the determinant ratio is one and the
quadratic form is zero.
\end{fact}

\begin{proof}[Proof of Fact~\ref{gauss:likelihood}]
For $t\in I$ and $\alpha\in\{\mathrm{sh},\mathrm{ind}\}$, put
\[
 \mu^\alpha_t
 =\Gamma^\alpha_{t,J_t}(\Gamma^\alpha_{J_t,J_t})^{-1}z_{J_t},
 \qquad
 v^\alpha_t
 =\Gamma^\alpha_{tt}
 -\Gamma^\alpha_{t,J_t}(\Gamma^\alpha_{J_t,J_t})^{-1}\Gamma^\alpha_{J_t,t},
\]
with $\mu^\alpha_t=0$ and $v^\alpha_t=\Gamma^\alpha_{tt}$ if
$J_t=\varnothing$.
These are Borel functions of $z_{<t}$, and $v^\alpha_t>0$, being the
Schur complement of $\Gamma^\alpha_{J_t,J_t}$ in the positive definite
matrix $\Gamma^\alpha_{J_t\cup\{t\},J_t\cup\{t\}}$.
By Corollary~\ref{gauss:linear-functional}, applied to the scalar
process in model $\alpha$ after $t-1$ measurements with
$b=a^\alpha_t$, the conditional law of $z_t$ given $z_{<t}$ is
$\mathcal N(\langle P_{W^\alpha_{t-1}}\Xi^\alpha,a^\alpha_t\rangle,\,
\|P_{(W^\alpha_{t-1})^\perp}a^\alpha_t\|^2)$.
Expanding $P_{W^\alpha_{t-1}}\Xi^\alpha$ and $P_{W^\alpha_{t-1}}a^\alpha_t$
in the basis of retained vectors and taking inner products with that
basis shows that the mean is $\mu^\alpha_t$ and the variance is
$v^\alpha_t$.
If $t\notin I$, the same corollary shows that $z_t$ is the point mass
at the value \eqref{gauss:omitted-value}, which is common to both
models.

Let $p^\alpha_t(x\mid z_{<t})$ be the density of
$\mathcal N(\mu^\alpha_t,v^\alpha_t)$ for $t\in I$.
For $0\le t\le M$, let $P^\alpha_t$ be the law of $z_{\le t}$ under
model $\alpha$.
For $h\in\mathbb R^{t-1}$, let $K^\alpha_t(h,\cdot)$ be
$\mathcal N(\mu^\alpha_t(h),v^\alpha_t(h))$ if $t$ is retained along $h$,
and the point mass at the value \eqref{gauss:omitted-value}
otherwise.
By the preceding paragraph, $K^\alpha_t(z_{<t},\cdot)$ is a regular
conditional law of $z_t$ given $z_{<t}$ under $P^\alpha$.
Moreover, $K^{\mathrm{sh}}_t(h,dx)=R_t(h,x)\,K^{\mathrm{ind}}_t(h,dx)$,
where $R_t=p^{\mathrm{sh}}_t(x\mid h)/p^{\mathrm{ind}}_t(x\mid h)$ if $t$
is retained along $h$ and $R_t=1$ otherwise, and $R_t$ takes values
in $(0,\infty)$.
Suppose that $P^{\mathrm{sh}}_{t-1}$ and $P^{\mathrm{ind}}_{t-1}$ are
equivalent with density $D_{t-1}$.
Then for every bounded Borel $f$,
\begin{align}
 \int f\,dP^{\mathrm{sh}}_t
 &~=~\int\!\!\int f(h,x)\,K^{\mathrm{sh}}_t(h,dx)\,P^{\mathrm{sh}}_{t-1}(dh)\\
&~ =~\int\!\!\int f(h,x)\,D_{t-1}(h)R_t(h,x)\,
 K^{\mathrm{ind}}_t(h,dx)\,P^{\mathrm{ind}}_{t-1}(dh)\\
&~ =~\int f\,D_t\,dP^{\mathrm{ind}}_t,    
\end{align}
with $D_t=D_{t-1}R_t\in(0,\infty)$.
Starting from $D_0=1$, induction on $t$ shows that $P^{\mathrm{sh}}$
and $P^{\mathrm{ind}}$ are mutually absolutely continuous, with
\begin{equation}\label{gauss:density-product}
 \frac{dP^{\mathrm{sh}}}{dP^{\mathrm{ind}}}(z)
 =\prod_{t\in I(z)}
 \frac{p^{\mathrm{sh}}_t(z_t\mid z_{<t})}{p^{\mathrm{ind}}_t(z_t\mid z_{<t})}.
\end{equation}

Finally, fix $z$ and, for $0\le j\le r$, let
$G^\alpha_j=\Gamma^\alpha_{\{t_1,\ldots,t_j\},\{t_1,\ldots,t_j\}}$ and
$\widehat z^{(j)}=(z_{t_1},\ldots,z_{t_j})^{\mathsf T}$, with
$\det G^\alpha_0=1$ and the empty quadratic form equal to zero.
Since $J_{t_j}=\{t_1,\ldots,t_{j-1}\}$, the Schur complement of
$G^\alpha_{j-1}$ in $G^\alpha_j$ is $v^\alpha_{t_j}$, and the block
determinant and block inverse formulas give
\[
 \det G^\alpha_j=\det G^\alpha_{j-1}\,v^\alpha_{t_j},
 \qquad
 (\widehat z^{(j)})^{\mathsf T}(G^\alpha_j)^{-1}\widehat z^{(j)}
 =(\widehat z^{(j-1)})^{\mathsf T}(G^\alpha_{j-1})^{-1}\widehat z^{(j-1)}
 +\frac{(z_{t_j}-\mu^\alpha_{t_j})^2}{v^\alpha_{t_j}}.
\]
Induction on $j$ yields
\begin{equation}\label{gauss:schur-identities}
 \det\widehat\Gamma^\alpha=\prod_{t\in I}v^\alpha_t,
 \qquad
 \widehat z^{\mathsf T}(\widehat\Gamma^\alpha)^{-1}\widehat z
 =\sum_{t\in I}\frac{(z_t-\mu^\alpha_t)^2}{v^\alpha_t}.
\end{equation}
Taking logarithms in \eqref{gauss:density-product}, inserting the
Gaussian densities, and substituting \eqref{gauss:schur-identities}
proves \eqref{gauss:likelihood-ratio}.
\end{proof}

\begin{fact}[\bf Gaussian innovations]
\label{gauss:innovations}
Under the shared model,
\[
 \mathbb E_{\mathrm{sh}}\bigl[
 \widehat z^{\mathsf T}(\widehat\Gamma^{\mathrm{sh}})^{-1}\widehat z
 \bigr]
 =\mathbb E_{\mathrm{sh}}|I|\le M.
\]
\end{fact}

\begin{proof}[Proof of Fact~\ref{gauss:innovations}]
Put $\xi_t=(z_t-\mu^{\mathrm{sh}}_t)/\sqrt{v^{\mathrm{sh}}_t}$ for $t\in I$
and $\xi_t=0$ otherwise.
Since the event $\{t\in I\}$ is determined by $z_{<t}$ and the
conditional law of $z_t$ given $z_{<t}$ at a retained index is
$\mathcal N(\mu^{\mathrm{sh}}_t,v^{\mathrm{sh}}_t)$, we have
$\mathbb E_{\mathrm{sh}}[\xi_t^2\mid z_{<t}]=\mathbf 1_{\{t\in I\}}$.
By \eqref{gauss:schur-identities},
$\widehat z^{\mathsf T}(\widehat\Gamma^{\mathrm{sh}})^{-1}\widehat z
=\sum_{t=1}^M\xi_t^2$.
Taking expectations proves the assertion.
\end{proof}

\paragraph{Proof of Theorem~\ref{gauss:decoupling}.}
By \eqref{gauss:gram-bound} with $J=I$,
$\det\widehat\Gamma^{\mathrm{sh}}\ge(1-\rho)^{|I|}\det\widehat\Gamma^{\mathrm{ind}}$
and
$(\widehat\Gamma^{\mathrm{ind}})^{-1}\preceq(1+\rho)(\widehat\Gamma^{\mathrm{sh}})^{-1}$.
Substituting into \eqref{gauss:likelihood-ratio} and using
$-\log(1-\rho)\le2\rho$ for $\rho<\tfrac12$ gives,
$P^{\mathrm{sh}}$-almost surely,
\[
 \log\frac{dP^{\mathrm{sh}}}{dP^{\mathrm{ind}}}
 \le\rho|I|+\frac\rho2\,
 \widehat z^{\mathsf T}(\widehat\Gamma^{\mathrm{sh}})^{-1}\widehat z.
\]
The right-hand side is nonnegative.
Writing $x_+=\max\{x,0\}$ and using $(1-e^{-x})_+\le x_+$, we obtain
\[
\begin{aligned}
 \operatorname{TV}(P^{\mathrm{sh}},P^{\mathrm{ind}})
 &~=~\mathbb E_{\mathrm{sh}}\Bigl[
 \Bigl(1-\frac{dP^{\mathrm{ind}}}{dP^{\mathrm{sh}}}\Bigr)_+\Bigr]
 \le\mathbb E_{\mathrm{sh}}\Bigl[
 \Bigl(\log\frac{dP^{\mathrm{sh}}}{dP^{\mathrm{ind}}}\Bigr)_+\Bigr]\\
 &~\le~\rho\,\mathbb E_{\mathrm{sh}}|I|
 +\frac\rho2\,\mathbb E_{\mathrm{sh}}\bigl[
 \widehat z^{\mathsf T}(\widehat\Gamma^{\mathrm{sh}})^{-1}\widehat z\bigr]
 =\frac32\rho\,\mathbb E_{\mathrm{sh}}|I|
 \le\frac32\rho M,    
\end{aligned}
\]
where the equality is Fact~\ref{gauss:innovations}.

This bound holds for every seed value.
Since the seed has the same law in both models, averaging over it
gives the same bound for the joint law of the seed and the scalar
measurements.
The full transcript, the algorithm's transcript, and its output are
Borel functions of these data, the same in both models, and total
variation does not increase under such maps.
This proves the theorem.
\end{proof}

\section{The Hard Family and Energy Test}\label{hf:section}
We now construct strongly log-concave potentials whose
first-order replies can be simulated by the linear Gaussian
measurements of Section~\ref{sec:linear-gaussian-system}.
The construction uses nested tensor features to make
earlier layer measurement spaces nearly orthogonal across
query groups separated in the compound direction space.
This geometry allows us to apply
Theorem~\ref{gauss:decoupling} to control information shared
between these groups.
Combined with mixed degrees and smooth gates that restrict
access to successive layers, it yields the following
query lower bound.

\begin{theorem}[\bf Lower bound for Strongly log-concave sampling]
\label{thm:main}
There exist universal constants $c>0$ and $d_0$ such that for every
$d\ge d_0$ there is a family $\mathcal V_d$ of $C^\infty$ potentials on
$\R^d$ with the following properties:
\begin{enumerate}
\item $I_d\preceq\nabla^2 V(x)\preceq2I_d$ for every $V\in\mathcal V_d$
and every $x\in\R^d$;
\item $V(-x)=V(x)$, the unique minimizer of $V$ is $0$, and
$\pi_V(dx)\propto e^{-V(x)}dx$ has mean zero;
\item Sampling from the class $\{\pi_V:V\in\mathcal V_d\}$ with
accuracy $1/10$ in total variation distance requires at least
$\Omega(\exp(c\log^{1/3}d))$ exact value-and-gradient queries.
\end{enumerate}
\end{theorem}

This section constructs the hard family and the test that drives the
lower bound. Section~\ref{hf:hard-functions} defines a random family
of potentials of the required class, Section~\ref{hf:basic-properties}
records their basic properties, and
Section~\ref{hf:energy-test-section} defines the energy test, which
separates the output of an algorithm from a sample of the target.

To complete the proof of Theorem~\ref{thm:main}, Section~\ref{sec:output-fails} shows that an algorithm making few queries rarely passes the test, Section~\ref{sec:target-proof} shows that a target sample passes it with high probability, and Section~\ref{sec:parameters-and-lower-bound} combines these bounds to establish the theorem.
Building on the same hard potentials, Section~\ref{app:slc-warm} further establishes the warm-start extension.

\subsection{Hard Functions}\label{hf:hard-functions}

This subsection defines the random potentials behind the lower
bound. Section~\ref{hf:construction} defines the random potentials, a
quadratic potential plus one gated Gaussian field per level.
Section~\ref{hf:admissible} shows that they are smooth with
well-defined targets and singles out the admissible realizations.

\subsubsection{The Construction}
\label{hf:construction}
Let $n$ be a sufficiently large integer, put
$h:=\lfloor\frac1{32}(\log n)^{1/3}\rfloor$ and $d_n:=(h+1)n$, and
write $x=(x_0,\dots,x_h)\in\R^{d_n}$ with blocks $x_j\in\R^n$. The
hard potential has the form
\vspace{2.6em}
\begin{equation}
  V^\Xi(x)=\frac34\|x\|^2+\sum_{i=0}^h
  \eqnmarkbox[cGate]{gate}{\Bigl(\prod_{j<i}G_j^\Xi(x)\Bigr)}\,
  \eqnmarkbox[cTrunc]{trunc}{U_i}\Bigl(
  \eqnmarkbox[cCut]{cut}{C_i(x)}\,
  \eqnmarkbox[cField]{field}{F_i^\Xi\Bigl(\frac{x_i}{\|x_i\|},\dots,
  \frac{x_h}{\|x_h\|}\Bigr)}\Bigr).
  \label{hf:potential}
\end{equation}
\annotate[yshift=1.2em]{above,left}{gate}{gates of the lower levels}
\annotate[yshift=-1.4em]{below,left}{trunc}{truncation of $s\mapsto\beta_is$}
\annotate[yshift=1.2em]{above,right}{cut}{radial cutoff}
\annotate[yshift=-1em]{below,right}{field}{Gaussian angular function\\ of high levels}
\vspace{1.8em}

\noindent
Here $\Xi$ is a Gaussian coefficient vector. For each level
$0\le i\le h$, $\textcolor{cField}{F_i^\Xi}$ is a Gaussian random
function of the directions of the blocks $x_i,\dots,x_h$, defined in
\eqref{hf:angular-fields}, $\textcolor{cCut}{C_i}$ is the radial
cutoff \eqref{hf:radial-cutoff}, which switches the level off unless
these blocks have typical norms, $\textcolor{cTrunc}{U_i}$ is the
smooth truncation \eqref{hf:energy-truncation} of the linear map
$s\mapsto\beta_is$, and $\textcolor{cGate}{G_j^\Xi}\in[0,1]$ is the
gate of level $j$, defined in \eqref{hf:gates}. 
The summand of index
$i$ is defined to be zero whenever $x_j=0$ for some $j\ge i$, and the
empty product equals one. 
We call
$x\mapsto C_i(x)F_i^\Xi\bigl(\frac{x_i}{\|x_i\|},\ldots,\frac{x_h}{\|x_h\|}\bigr)$
the field of level $i$, $G_i^\Xi$ its gate,
$A_i^\Xi:=\prod_{j<i}G_j^\Xi$ its gate product, and the summand of
index $i$ its load. The load of level $i$ contains the gates of the
lower levels but not its own gate, and the gate of level $h$ appears
in no load.


\paragraph{Angular functions $F_i^\Xi$.}
Our construction hides the angular configurations by Gaussian random
functions. Put $K:=2^{74(h+1)}$, $P:=2^{64(h+1)}$, and
$S_i:=\sum_{k=0}^{h-i}K^k$ for $0\le i\le h$. For unit directions
$u_0,\ldots,u_h\in\mathbb S^{n-1}$ and $0\le i\le h$, define the unit
tensor of order $S_i$
\[\psi_i(u_i,\ldots,u_h)
=u_i\otimes u_{i+1}^{\otimes K}\otimes\cdots
\otimes u_h^{\otimes K^{h-i}}.\]
For each $i$ and each even degree $p\in\mathcal P:=\{4,6,\ldots,P\}$,
let $\Xi_{i,p}$ be a standard Gaussian tensor in the Euclidean tensor
space $\mathbb R^{\,n^{pS_i}}$, independently across $i,p$, and set
\begin{equation}\label{hf:field-normalization}
  a_{i,p}:=\frac{1}{p^4S_i^2\sqrt{(h-i+1)\log(2PS_i)}} .
\end{equation}
Define the angular polynomial
$F_i^\Xi:(\mathbb S^{n-1})^{h-i+1}\mapsto\mathbb R$ and its degree-$p$
components $F_{i,p}^\Xi:(\mathbb S^{n-1})^{h-i+1}\allowbreak\mapsto\mathbb R$ by
\begin{defband}{cField}
\begin{equation}\label{hf:angular-fields}
 F_i^\Xi=\sum_{p\in\mathcal P}F_{i,p}^\Xi,\qquad
 F_{i,p}^\Xi(u_i,\ldots,u_h)
 =\sqrt n\,a_{i,p}
   \langle\Xi_{i,p},\psi_i(u_i,\ldots,u_h)^{\otimes p}\rangle,
\end{equation}
\end{defband}
\noindent where $\langle\cdot,\cdot\rangle$ is the entry-wise inner
product of tensors. Since $\psi_i$ is a unit tensor,
$F_{i,p}^\Xi(u_i,\ldots,u_h)\sim\mathcal N(0,na_{i,p}^2)$ for each
fixed $(u_i,\ldots,u_h)$. Collect the Gaussian vectors into
$\Xi=(\Xi_{i,p})_{i,p}$. Their joint Euclidean coefficient space is
\[
 \mathcal G_n
 =\bigoplus_{i=0}^h\bigoplus_{p\in\mathcal P}
   \mathbb R^{\,n^{pS_i}}.
\]
Since the coefficient vectors are independent standard Gaussians,
their joint law is $\mu_n=\mathcal N(0,I_{\mathcal G_n})$.

\paragraph{Two smooth truncations.}
We use two smooth truncations, acting on the location and on the
field value, respectively.
\begin{enumerate}
\item \textbf{Radial cutoff $C_i$.}
We use a fixed smooth radial cutoff
$\chi:[0,\infty)\to[0,1]$.\footnote{
Define $\rho(t)=e^{-1/t}$ for $t>0$ and $\rho(t)=0$ otherwise, and set
$\chi(r)=
 \frac{\rho(r-0.65)}{\rho(r-0.65)+\rho(0.7-r)}
 \frac{\rho(1-r)}{\rho(1-r)+\rho(r-0.95)}$.
The flatness of $\rho$ at zero makes $\chi$ smooth.}
It satisfies $\chi=1$ on $[0.7,0.95]$,
$\operatorname{supp}(\chi)=[0.65,1]\subset(0.6,1.05)$, and
$\|\chi'\|_\infty+\|\chi''\|_\infty\le C$ for an absolute constant
$C$. We define the radial cutoff function $C_i:\R^{d_n}\to[0,1]$ as
\begin{defband}{cCut}
\begin{equation}\label{hf:radial-cutoff}
C_i(x)=\prod_{j=i}^h\chi\Bigl(\frac{\|x_j\|}{\sqrt n}\Bigr).
\end{equation}
\end{defband}
\noindent Then $C_i(x)=1$ if $\frac{\|x_j\|}{\sqrt n}\in[0.7,0.95]$
for every $j=i,\ldots,h$, and $C_i(x)=0$ if
$\frac{\|x_j\|}{\sqrt n}\notin(0.6,1.05)$ for at least one such $j$.

\item \textbf{Energy truncation $U_i$.}
We use a fixed smooth odd function
$\varphi:\mathbb R\to\mathbb R$.\footnote{
With $\rho(t)=e^{-1/t}$ for $t>0$ and $\rho(t)=0$ otherwise, set
$\varphi(t)=\int_0^t
 \frac{\rho(2-|s|)}{\rho(2-|s|)+\rho(|s|-1)}\,ds$.
The integrand is smooth and even, takes values in $[0,1]$, equals one
for $|s|\le1$, and vanishes for $|s|\ge2$.}
It satisfies $\varphi(t)=t$ for $|t|\le1$,
$|\varphi(t)|\le\min\{|t|,2\}$, $|\varphi'|\le1$, and
$\|\varphi''\|_\infty\le C$ for an absolute constant $C$. We define
the energy truncation function $U_i:\R\to\R$ as
\begin{defband}{cTrunc}
\begin{equation}\label{hf:energy-truncation}
  U_i(s)=\frac{n\beta_i^2}{\lambda}\,
  \varphi\Bigl(\frac{\lambda s}{n\beta_i}\Bigr),
\end{equation}
\end{defband}
\noindent where $\lambda:=\frac{\lambda_*}{(h+1)^2}$ for a
sufficiently small absolute constant $\lambda_*>0$, and the
amplitudes $\beta_i>0$ are chosen successively by
\begin{equation}\label{hf:amplitudes}
\begin{aligned}
  \beta_0&:=\eta_*\lambda,\qquad
  \beta_i:=\eta_*\lambda\Bigl(\sum_{j=0}^{i-1}\frac{D_j}{\beta_j}\Bigr)^{-1}
  \quad(1\le i\le h),\\
  D_i&:=\frac{C_gP^5(h-i+1)\sqrt{\log(2P)}}{\epsilon_g}\,
  S_i^3\log(2PS_i),
\end{aligned}
\end{equation}
where $\eta_*>0$ is a sufficiently small and $C_g$ a sufficiently
large absolute constant, and $\epsilon_g:=\frac1{100}$ is the energy
tolerance that also enters the gates below. 
Then $U_i(s)=\beta_is$
for $|s|\le\frac{n\beta_i}{\lambda}$, while
$|U_i(s)|\le\frac{2n\beta_i^2}{\lambda}$ globally.
\end{enumerate}

\paragraph{Gate functions $ G_i^\Xi$.}
We use two fixed smooth cutoff functions:
\begin{enumerate}
\item A radial cutoff $J:[0,\infty)\to[0,1]$, with
$\operatorname{supp}J\subset(0.7,0.95)$ and $J=1$ on
$[0.75,0.9]$.\footnote{
Using $\rho$ from the definition of $\chi$, one may take
$J(r)=
 \frac{\rho(r-0.725)}{\rho(r-0.725)+\rho(0.75-r)}
 \frac{\rho(0.925-r)}{\rho(0.925-r)+\rho(r-0.9)}$.}
In particular, $\chi=1$ wherever $J>0$.
\item An energy-error cutoff $g:\mathbb R\to[0,1]$, with $g=1$ on
$(-\infty,\frac12]$ and $g=0$ on $[1,\infty)$.\footnote{
One may take
$g(t)=\rho(1-t)/\bigl(\rho(1-t)+\rho(t-\frac12)\bigr)$.}
\end{enumerate}
We set $\tau:=8\log(2P)$ and, with the energy tolerance
$\epsilon_g=\frac1{100}$, define the gate function
$G_i^\Xi:\R^{d_n}\to[0,1]$ as
\begin{defband}{cGate}
\begin{equation}\label{hf:gates}
 G_i^\Xi(x)
 =\left(\prod_{j=i}^h
 J\!\left(\frac{\|x_j\|}{\sqrt n}\right)\right) g\!\left(
 \frac1\tau\log\sum_{p\in\mathcal P}
 2\cosh\!\left(
 \frac{\tau}{\epsilon_g\beta_i}
 \left[
 \frac{
 C_i(x)F_{i,p}^\Xi\!\left(
 \tfrac{x_i}{\|x_i\|},\ldots,\tfrac{x_h}{\|x_h\|}
 \right)}
 {na_{i,p}^2}
 +\beta_i
 \right]\right)\right).
\end{equation}
\end{defband}
\noindent Each gate depends only on the coefficients
$(\Xi_{i,p})_{p\in\mathcal P}$ of its own level and on the location
$x$.

\subsubsection{Admissible Instances}
\label{hf:admissible}
We first check that every coefficient realization defines a smooth
potential with a well-defined target, and then single out the
realizations that we retain as hard instances.

\paragraph{Well-definedness.}
Write $B(x):=\frac34\|x\|^2$. For $-1\le i\le h$, the prefix
$V_i^\Xi$ is the sum of $B$ and the loads of the levels $k\le i$, so
that $V_{-1}^\Xi=B$ and $V_h^\Xi=V^\Xi$. Let
$\gamma_n:=\mathcal N(0,\frac23I_{d_n})$ be the target distribution
of $B$, and put
\begin{equation}\label{hf:targets}
  w_i^\Xi:=e^{-(V_i^\Xi-B)},\qquad
  Z_i^\Xi:=\int w_i^\Xi\,d\gamma_n,\qquad
  \pi_i^\Xi:=\frac{w_i^\Xi}{Z_i^\Xi}\,\gamma_n
  \qquad(-1\le i\le h).
\end{equation}
Since $\gamma_n$ has density proportional to $e^{-B}$, $\pi_i^\Xi$ is
the target distribution of $V_i^\Xi$ whenever $0<Z_i^\Xi<\infty$,
which the next lemma verifies. We write $w^\Xi$, $Z^\Xi$, and
$\pi^\Xi$ for $w_h^\Xi$, $Z_h^\Xi$, and $\pi_h^\Xi$.

\begin{lemma}[\bf Smoothness, symmetry, and normalization]
\label{hf:smoothness}
For every coefficient realization $\Xi$ and every $-1\le i\le h$:
\begin{enumerate}[leftmargin=*]
\item \textbf{Smoothness.} $V_i^\Xi\in C^\infty(\R^{d_n})$.
\item \textbf{Symmetry.} $V_i^\Xi$ is even in each block. Wherever
$C_0=0$, in particular at $0$, $V_i^\Xi$ and $B$ have the same value
and gradient.
\item \textbf{Normalization.}
$\bigl\|V_i^\Xi-B\bigr\|_\infty\le\frac{2n}{\lambda}\sum_{j=0}^i\beta_j^2$.
Hence $0<Z_i^\Xi<\infty$, so $\pi_i^\Xi$ is well defined for every
$\Xi$, and $\Xi\mapsto Z_i^\Xi$ is continuous.
\end{enumerate}
\end{lemma}

\begin{proof}[Proof of Lemma~\ref{hf:smoothness}]
Throughout the proof write $u_j:=\frac{x_j}{\|x_j\|}$ for $x_j\ne0$,
so that $H_k^\Xi(x)=C_k(x)F_k^\Xi(u_k,\ldots,u_h)$ and
\[
  V_i^\Xi=B+\sum_{k=0}^iA_k^\Xi\,U_k(H_k^\Xi),\qquad
  A_k^\Xi=\prod_{j<k}G_j^\Xi .
\]

\smallskip\noindent\textit{Proof of (1).}
The formula for $H_k^\Xi$ is smooth wherever $x_j\ne0$ for all
$j\ge k$, and $H_k^\Xi$ vanishes on the open set
$\{x:\|x_j\|<0.65\sqrt n\text{ for some }j\ge k\}$, which contains
every point where the formula is undefined, because $\chi=0$ on
$[0,0.65)$. Hence $H_k^\Xi\in C^\infty(\R^{d_n})$, and likewise
$G_k^\Xi$, with $J$ in place of $\chi$. Therefore $A_k^\Xi$,
$A_k^\Xi U_k(H_k^\Xi)$, and $V_i^\Xi$ are smooth.

\smallskip\noindent\textit{Proof of (2).}
Let $\sigma_j$ change the sign of the block $x_j$, which replaces
$u_j$ by $-u_j$. For $k\le j$,
\[
  \psi_k(u_k,\ldots,-u_j,\ldots,u_h)
  =(-1)^{K^{j-k}}\psi_k(u_k,\ldots,u_j,\ldots,u_h),
\]
so $\psi_k^{\otimes p}$, and hence $F_{k,p}^\Xi$, is invariant under
$u_j\mapsto-u_j$ because $p$ is even, while $C_k$ and
$J(\frac{\|x_j\|}{\sqrt n})$ are invariant under $\sigma_j$. Thus
$H_k^\Xi\circ\sigma_j=H_k^\Xi$ and $G_k^\Xi\circ\sigma_j=G_k^\Xi$.
For $k>j$ these functions do not depend on $x_j$. Hence
$V_i^\Xi\circ\sigma_j=V_i^\Xi$ for every $j$.

Now let $C_0(x)=0$, so that $\chi(\frac{\|x_j\|}{\sqrt n})=0$ for
some $j$. Since $\{J>0\}\subseteq\{\chi=1\}$,
\[
  J\Bigl(\frac{\|x_j\|}{\sqrt n}\Bigr)=0,\qquad
  G_0^\Xi(x)=0,\qquad
  A_k^\Xi(x)=0\quad(k\ge1).
\]
The functions $C_0$ and $A_k^\Xi$ are smooth, nonnegative, and
vanish at $x$, hence $\nabla C_0(x)=0$ and $\nabla A_k^\Xi(x)=0$ for
$k\ge1$. By the product rule, $H_0^\Xi(x)=0$ and
$\nabla H_0^\Xi(x)=0$ (if a block of $x$ vanishes, $H_0^\Xi\equiv0$
near $x$), so
\[
  U_0(H_0^\Xi(x))=U_0(0)=0,\qquad
  \nabla\bigl[U_0(H_0^\Xi)\bigr](x)=U_0'(0)\,\nabla H_0^\Xi(x)=0,
\]
and for $k\ge1$
\[
  A_k^\Xi(x)\,U_k(H_k^\Xi(x))=0,
\]
and
\[  \nabla\bigl[A_k^\Xi\,U_k(H_k^\Xi)\bigr](x)
  =U_k(H_k^\Xi(x))\,\nabla A_k^\Xi(x)
  +A_k^\Xi(x)\,U_k'(H_k^\Xi(x))\,\nabla H_k^\Xi(x)=0 .\]
Summing, $V_i^\Xi(x)=B(x)$ and $\nabla V_i^\Xi(x)=\nabla B(x)$.
Finally $C_0(0)=\chi(0)^{h+1}=0$.

\smallskip\noindent\textit{Proof of (3).}
Since $0\le A_k^\Xi\le1$ and $|\varphi|\le2$,
\[
  \bigl|A_k^\Xi\,U_k(H_k^\Xi)\bigr|\le\bigl|U_k(H_k^\Xi)\bigr|
  =\frac{n\beta_k^2}{\lambda}\Bigl|\varphi\Bigl(\frac{\lambda H_k^\Xi}{n\beta_k}\Bigr)\Bigr|
  \le\frac{2n\beta_k^2}{\lambda},
  \qquad
  |V_i^\Xi-B|\le\frac{2n}{\lambda}\sum_{k=0}^i\beta_k^2=:L_i.
\] Hence
$e^{-L_i}\le w_i^\Xi\le e^{L_i}$, and integrating against the
probability measure $\gamma_n$ gives $e^{-L_i}\le Z_i^\Xi\le e^{L_i}$.
For each fixed $x$, $\Xi\mapsto w_i^\Xi(x)$ is continuous and bounded
by $e^{L_i}$, so $\Xi\mapsto Z_i^\Xi=\int w_i^\Xi\,d\gamma_n$ is
continuous by dominated convergence.
\end{proof}

\paragraph{Admissible event.}
The main theorem requires $I\preceq\nabla^2V\preceq2I$. Because
Gaussian fields are unbounded, this condition can fail for rare
realizations of $\Xi$. 
We therefore restrict to the realizations on
which every degree component at each level stays within its typical
magnitude.
Fix an absolute constant $C_f>0$,
to be chosen sufficiently large, and define
\begin{equation}\label{hf:admissible-event}
 \mathcal E_n
 :=\bigcap_{i=0}^h\bigcap_{p\in\mathcal P}
 \Bigl\{\Xi\in\mathcal G_n:
 \sup_{u_i,\ldots,u_h\in\mathbb S^{n-1}}
 \bigl|F_{i,p}^\Xi(u_i,\ldots,u_h)\bigr|
 \le C_fna_{i,p}\sqrt{(h-i+1)\log(2PS_i)}\Bigr\}.
\end{equation}
The threshold matches the typical order of the supremum, since
$F_{i,p}^\Xi$ has pointwise standard deviation $\sqrt n\,a_{i,p}$ and
its supremum over the $n(h-i+1)$-dimensional product of spheres
exceeds this scale by a factor of order $\sqrt{n(h-i+1)\log(2PS_i)}$.

We call
$\Xi$ admissible if $\Xi\in\mathcal E_n$. The probabilistic estimates
in Sections~\ref{sec:output-fails} and~\ref{sec:target-good-mass} use
the Gaussian law $\mu_n$ itself, and conditioning on $\mathcal E_n$
is deferred to Section~\ref{sec:parameters-and-lower-bound}.
\begin{proposition}[\bf Admissibility and parameter bounds]
\label{hf:parameter-bounds}
The absolute constants $C_f,C_g,\eta_*,\lambda_*$ can be chosen so that, for all sufficiently large $n$, the following hold.
\begin{enumerate}[leftmargin=*]
\item $\mu_n(\mathcal E_n^c)\le2(h+1)Pe^{-n}$.
\item For every $\Xi\in\mathcal E_n$, $0\le i\le h$, and $x\in\R^{d_n}$,
$I_{d_n}\preceq\nabla^2V_i^\Xi(x)\preceq2I_{d_n}$. Consequently, $0$ is
the unique minimizer of $V_i^\Xi$, and its target $\pi_i^\Xi$ has mean
zero.
\item The amplitudes decrease and satisfy $\beta_h\ge n^{-\frac1{16}}$.
Moreover, $\log S_0=O((\log n)^{\frac23})$ and
$\min_{i,p}a_{i,p}=n^{-o(1)}$.
\end{enumerate}
\end{proposition}

\begin{proof}[Proof of Proposition~\ref{hf:parameter-bounds}]
We choose $C_f$ first, then $C_g$, and finally $\eta_*,\lambda_*>0$,
with $C_g\ge1$, $\eta_*\le1$, $\lambda_*\le\frac12$, and
$\eta_*\lambda_*\le\frac14$.
Constants $C$ in deterministic derivative bounds may depend on the
fixed smooth functions and on $C_f$, but not on $n,h,i,p$, or $\Xi$.
Write $b_i:=h-i+1$, $\ell_i:=\log(2PS_i)$,
$H_{i,p}^\Xi(x):=C_i(x)F_{i,p}^\Xi
\bigl(\frac{x_i}{\|x_i\|},\ldots,\frac{x_h}{\|x_h\|}\bigr)$, and
$H_i^\Xi:=\sum_{p\in\mathcal P}H_{i,p}^\Xi$.

By~\eqref{hf:amplitudes}, we have $D_i\ge1$,
$\beta_0=\eta_*\lambda\le\frac14$, and
$\sum_{j<i}\frac{D_j}{\beta_j}=\frac{\beta_0}{\beta_i}$ for $i\ge1$,
which implies
$\frac{\beta_0}{\beta_{i+1}}=\frac{\beta_0}{\beta_i}+\frac{D_i}{\beta_i}$.
Rearranging gives
\begin{equation}\label{hf:ver-amplitude-step}
 \beta_1=\frac{\beta_0^2}{D_0}\le\beta_0,\qquad
 \beta_{i+1}=\frac{\beta_0\beta_i}{\beta_0+D_i}\le\beta_i\quad(1\le i<h),
\end{equation}
so $\beta_i\le\beta_0\le\lambda$. Moreover, since $\eta_*\le1$, we
obtain $\sum_{j<i}\frac{D_j}{\beta_j}\le\frac{\lambda}{\beta_i}$ for
all $0\le i\le h$, where the sum is empty for $i=0$.

\smallskip\noindent\emph{Proof of (1).}
We bound $\sup|F_{i,p}^\Xi|$ over the product of spheres by its maximum over a finite net and apply Gaussian tail bounds at the net points. The net step requires derivative bounds for $F_{i,p}^\Xi$, which we obtain by applying the following estimate along great circles. The same derivative bounds are used again in the proof of (2).
\begin{fact}
\label{fact:ver-bernstein}
If $T(t)=\sum_{k=-m}^mc_ke^{ikt}$ with $m\ge1$, then
$\|T^{(r)}\|_\infty\le r!\,e\,m^r\|T\|_\infty$ on $\R$ for $r=1,2$.
\end{fact}
\begin{proof}
Let $M:=\|T\|_\infty$ and $P(w):=\sum_{k=-m}^mc_kw^{k+m}$, so that
$T(z)=e^{-imz}P(e^{iz})$ and $|P|\le M$ on $|w|=1$. Applying the
maximum modulus principle to the polynomials $P(w)$ and
$w^{2m}P(\frac1w)$ on $|w|\le1$ gives
\[
 |P(w)|\le M\max\{1,|w|^{2m}\},
 \qquad\text{and hence}\qquad
 |T(t+is)|=e^{ms}\,|P(e^{it-s})|\le Me^{m|s|}.
\]
Cauchy's estimate on the disk of radius $\frac1m$ centered at $t\in\R$
then yields $|T^{(r)}(t)|\le r!\,e\,m^rM$.
\end{proof}
Fix $i,p,\Xi$, and set
$M:=\sup_{(\mathbb S^{n-1})^{b_i}}|F_{i,p}^\Xi|$ and
$m_j:=pK^{j-i}$ for $i\le j\le h$, so that $\sum_{j=i}^hm_j=pS_i$.
Since $\psi_i^{\otimes p}$ is homogeneous of degree $m_j$ in $u_j$,
along any great circle of the $j$-th sphere $F_{i,p}^\Xi$ is a
trigonometric polynomial of degree at most $m_j$. Writing $\partial_j$
for a unit tangential derivative on the $j$-th sphere and applying
Fact~\ref{fact:ver-bernstein}, successively across two spheres for the
mixed derivatives, we obtain
\[
 |\partial_jF_{i,p}^\Xi|\le em_jM,\qquad
 |\partial_j^2F_{i,p}^\Xi|\le2em_j^2M,\qquad
 |\partial_j\partial_kF_{i,p}^\Xi|\le e^2m_jm_kM\quad\text{for }j\ne k.
\]
Summing over the blocks yields
\begin{equation}\label{hf:ver-angular-derivatives}
 \|\nabla_{\!S}F_{i,p}^\Xi\|\le e\,pS_iM,\qquad
 \|\operatorname{Hess}_SF_{i,p}^\Xi\|_{\op}\le e^2(pS_i)^2M,
\end{equation}
where $\nabla_{\!S}$ and $\operatorname{Hess}_S$ denote the intrinsic
gradient and Hessian on the product of spheres.

In each of the $b_i$ spheres, fix a geodesic $\delta$-net with
$\delta:=\frac1{4epS_i}$ and at most $(\frac C\delta)^n$ points, which
exists by volume comparison since geodesic distance $\delta$ implies
chordal distance at least $\frac{2\delta}\pi$. Their product
$\mathcal N$ satisfies
\[
 \log|\mathcal N|\le b_in\log\frac C\delta\le C_0\,b_in\ell_i ,
\]
with $C_0$ independent of $C_f$. Moving to a net point one sphere at a
time and using the first-derivative bounds yields
\[
 M\le\max_{\mathcal N}|F_{i,p}^\Xi|+\delta\sum_{j\ge i}em_jM
 =\max_{\mathcal N}|F_{i,p}^\Xi|+\frac M4,
\]
and hence $M\le2\max_{\mathcal N}|F_{i,p}^\Xi|$ .
By~\eqref{hf:angular-fields}, each $F_{i,p}^\Xi(u)$ for
$u\in\mathcal N$ is a centered Gaussian with variance $na_{i,p}^2$.
Consequently,
\begin{align*}
 \mu_n\bigl\{M>C_fna_{i,p}\sqrt{b_i\ell_i}\bigr\}
 &\le|\mathcal N|\max_{u\in\mathcal N}\mu_n\Bigl\{|F_{i,p}^\Xi(u)|>\frac{C_f}{2}na_{i,p}\sqrt{b_i\ell_i}\Bigr\}\\
 &\le2\exp\Bigl[\Bigl(C_0-\frac{C_f^2}8\Bigr)b_in\ell_i\Bigr]\le2e^{-n}
\end{align*}
once $C_f^2\ge8(C_0+1)$, as $b_i\ell_i\ge1$. A union bound over the
$(h+1)|\mathcal P|\le(h+1)P$ pairs $(i,p)$ gives
$\mu_n(\mathcal E_n^c)\le2(h+1)Pe^{-n}$, establishing~(1).

\smallskip\noindent\emph{Proof of (2).}
Fix $\Xi\in\mathcal E_n$. We bound the gradient and Hessian of the fields, then of the gates, and finally of the loads, using the following product rule.

\begin{fact}[\bf Products of bounded functions]\label{fact:ver-products}
Let $a_j\ge0$ and let $f_j$ be smooth functions satisfying $|f_j|\le1$, $\sqrt n\|\nabla f_j\|\le a_j$, and $n\|\nabla^2f_j\|_{\op}\le a_j^2$. Then $\prod_jf_j$ satisfies the same three bounds with parameter $a:=\sum_ja_j$.
\end{fact}
\begin{proof}
By the product rule, $\sqrt n\bigl\|\nabla\prod_jf_j\bigr\|\le\sum_j\sqrt n\|\nabla f_j\|\le a$ and $n\bigl\|\nabla^2\prod_jf_j\bigr\|_{\op}\le\sum_ja_j^2+\sum_{j\ne k}a_ja_k=a^2$.
\end{proof}

By the proof of Lemma~\ref{hf:smoothness}(1), the fields and gates at level $i$ vanish identically unless $\|x_j\|>0.6\sqrt n$ for all $j\ge i$, so we work on this set.
For $v\in\R^{d_n}$ and $u_j=\frac{x_j}{\|x_j\|}$, the curve $t\mapsto u_j(x+tv)$ has at $t=0$ velocity $\frac{P_{u_j^\perp}v_j}{\|x_j\|}$ of norm at most $\frac{C\|v_j\|}{\sqrt n}$ and covariant acceleration $-\frac{2\langle u_j,v_j\rangle}{\|x_j\|^2}P_{u_j^\perp}v_j$ of norm at most $\frac{C\|v_j\|^2}{n}$.
By the chain rule and \eqref{hf:ver-angular-derivatives}, the pullback $\widetilde F:=F_{i,p}^\Xi(u_i,\ldots,u_h)$ satisfies
\[
 \sqrt n\|\nabla\widetilde F\|\le CpS_iM,\qquad \mbox{and}\qquad n\|\nabla^2\widetilde F\|_{\rm op}\le C(pS_i)^2M .
\]
Each radial factor satisfies Fact~\ref{fact:ver-products} with an absolute constant $a_j=C$, since
\[
 \nabla\chi\Bigl(\frac{\|x_j\|}{\sqrt n}\Bigr)=\frac{\chi'}{\sqrt n}\,u_j,\qquad
 \nabla^2\chi\Bigl(\frac{\|x_j\|}{\sqrt n}\Bigr)=\frac{\chi''}{n}\,u_ju_j^{\top}+\frac{\chi'}{\sqrt n\,\|x_j\|}P_{u_j^\perp},
\]
and likewise for $J$.
If $M=0$, then $H_{i,p}^\Xi=0$. Otherwise, applying Fact~\ref{fact:ver-products} to the $b_i$ factors of $C_i$ and to $\frac{\widetilde F}{M}$ shows that $\frac{H_{i,p}^\Xi}{M}=C_i\frac{\widetilde F}{M}$ has parameter $C(b_i+pS_i)\le CpS_i$. Since $\Xi\in\mathcal E_n$, \eqref{hf:field-normalization} gives $M\le C_fna_{i,p}\sqrt{b_i\ell_i}=\frac{C_fn}{p^4S_i^2}$, and hence
\begin{equation}\label{hf:ver-degree-bounds}
 \|\nabla H_{i,p}^\Xi\|\le\frac{CpS_iM}{\sqrt n}\le\frac{C\sqrt n}{p^3S_i},\qquad
 \|\nabla^2H_{i,p}^\Xi\|_{\op}\le\frac{C(pS_i)^2M}{n}\le\frac C{p^2}.
\end{equation}
Summing over $p\in\mathcal P$ with $\sum_pp^{-3}\le\sum_pp^{-2}\le1$ then yields $\|\nabla H_i^\Xi\|\le C\sqrt n$ and $\|\nabla^2H_i^\Xi\|_{\op}\le C$.

Next, set $z_{i,p}:=\frac{1}{\epsilon_g\beta_i}\bigl(\frac{H_{i,p}^\Xi}{na_{i,p}^2}+\beta_i\bigr)$, $M_i:=\frac1\tau\log\sum_{p\in\mathcal P}2\cosh(\tau z_{i,p})$, and $R_i:=\prod_{j\ge i}J\bigl(\frac{\|x_j\|}{\sqrt n}\bigr)$, so that $G_i^\Xi=R_i\,g(M_i)$ by~\eqref{hf:gates}. By~\eqref{hf:field-normalization}, \eqref{hf:amplitudes}, and $\tau=8\log(2P)$, we have $\frac1{na_{i,p}^2}=\frac{p^8S_i^4b_i\ell_i}{n}$ and $D_i=\frac{C_g\sqrt{\frac\tau8}}{\epsilon_g}\,P^5S_i^3b_i\ell_i$. Combining~\eqref{hf:ver-degree-bounds} with $p\le P$, $\epsilon_g\beta_i\le1$, and $S_i,b_i,\ell_i\ge1$ gives
\begin{align*}
 \sqrt n\|\nabla z_{i,p}\|&=\frac{p^8S_i^4b_i\ell_i}{\sqrt n\,\epsilon_g\beta_i}\|\nabla H_{i,p}^\Xi\|
 \le\frac{CP^5S_i^3b_i\ell_i}{\epsilon_g\beta_i}=\frac{CD_i}{C_g\beta_i\sqrt\tau},\\
n\|\nabla^2z_{i,p}\|_{\op}&=\frac{p^8S_i^4b_i\ell_i}{\epsilon_g\beta_i}\|\nabla^2H_{i,p}^\Xi\|_{\op}
 \le\frac{CP^6S_i^4b_i\ell_i}{\epsilon_g\beta_i}\le\frac{CD_i^2}{C_g^2\beta_i^2\tau}.
\end{align*}
Let $\sigma$ be the probability distribution on $\mathcal P\times\{\pm1\}$ with $\sigma_{p,s}\propto e^{s\tau z_{i,p}}$. Differentiating $M_i$ gives $\nabla M_i=\mathbb E_\sigma[s\nabla z_{i,p}]$ and $\nabla^2M_i=\mathbb E_\sigma[s\nabla^2z_{i,p}]+\tau\operatorname{Cov}_\sigma(s\nabla z_{i,p})$. Since $\|\operatorname{Cov}_\sigma(s\nabla z_{i,p})\|_{\op}\le\max_p\|\nabla z_{i,p}\|^2$ and $\tau\ge1$, we obtain
\[
 \sqrt n\|\nabla M_i\|\le\frac{CD_i}{C_g\beta_i},\qquad
 n\|\nabla^2M_i\|_{\op}\le\frac{CD_i^2}{C_g^2\beta_i^2\tau}+\tau\cdot\frac{CD_i^2}{C_g^2\beta_i^2\tau}\le\frac{CD_i^2}{C_g^2\beta_i^2}.
\]
Since $0\le g\le1$ with $g',g''$ bounded, the chain rule shows that $g(M_i)$ satisfies Fact~\ref{fact:ver-products} with $a=\frac{CD_i}{C_g\beta_i}$, and $R_i$ satisfies it with $a=Cb_i\le\frac{CD_i}{C_g\beta_i}$. Hence $G_i^\Xi=R_i\,g(M_i)$ has parameter $\frac{CD_i}{C_g\beta_i}\le\frac{D_i}{\beta_i}$ once $C_g\ge C$, that is,
\begin{equation}\label{hf:ver-gate-derivatives}
 \sqrt n\|\nabla G_i^\Xi\|\le\frac{D_i}{\beta_i},\qquad
 n\|\nabla^2G_i^\Xi\|_{\op}\le\Bigl(\frac{D_i}{\beta_i}\Bigr)^2 .
\end{equation}

Set $W_i:=\frac12\varphi\bigl(\frac{\lambda H_i^\Xi}{n\beta_i}\bigr)$, so that $A_i^\Xi\,U_i(H_i^\Xi)=\frac{2n\beta_i^2}{\lambda}\,A_i^\Xi W_i$ and $|W_i|\le1$ by~\eqref{hf:energy-truncation}.
By the chain rule, the bounds on $H_i^\Xi$, and $\frac\lambda{\beta_i}\ge1$, we obtain
\[
 \sqrt n\|\nabla W_i\|\le\frac{C\lambda}{n\beta_i}\sqrt n\|\nabla H_i^\Xi\|\le\frac{C\lambda}{\beta_i},\qquad
 n\|\nabla^2W_i\|_{\op}\le\frac{C\lambda^2}{n\beta_i^2}\|\nabla H_i^\Xi\|^2+\frac{C\lambda}{\beta_i}\|\nabla^2H_i^\Xi\|_{\op}
 \le C\Bigl(\frac\lambda{\beta_i}\Bigr)^2 .
\]
Applying Fact~\ref{fact:ver-products} to $W_i$ and $G_j^\Xi$ for $j<i$, together with~\eqref{hf:ver-gate-derivatives} and the bound $\sum_{j<i}\frac{D_j}{\beta_j}\le\frac{\lambda}{\beta_i}$, shows that $A_i^\Xi W_i$ has parameter $\frac{C\lambda}{\beta_i}+\sum_{j<i}\frac{D_j}{\beta_j}\le\frac{C\lambda}{\beta_i}$, and hence
\begin{equation}\label{hf:ver-load-hessian}
 \bigl\|\nabla^2[A_i^\Xi U_i(H_i^\Xi)]\bigr\|_{\op}
 \le\frac{2n\beta_i^2}{\lambda}\cdot\frac1n\Bigl(\frac{C\lambda}{\beta_i}\Bigr)^2=C\lambda .
\end{equation}
Since $V_i^\Xi=B+\sum_{k\le i}A_k^\Xi U_k(H_k^\Xi)$ and $\nabla^2B=\frac32I_{d_n}$, we have
\[
 \Bigl\|\nabla^2V_i^\Xi-\frac32I_{d_n}\Bigr\|_{\op}\le(h+1)C\lambda=\frac{C\lambda_*}{h+1}\le\frac14
 \qquad\text{once }\lambda_*\le\frac1{4C},
\]
so that $I_{d_n}\preceq\nabla^2V_i^\Xi\preceq2I_{d_n}$. By Lemma~\ref{hf:smoothness}(2), $\nabla V_i^\Xi(0)=0$ and $V_i^\Xi$ is even, which implies that $V_i^\Xi(x)\ge V_i^\Xi(0)+\frac12\|x\|^2$, $0$ is the unique minimizer of $V_i^\Xi$, $\int\|x\|\,d\pi_i^\Xi(x)<\infty$, and $\int x\,d\pi_i^\Xi(x)=0$. This proves~(2).

\smallskip\noindent\emph{Proof of (3).}
The amplitudes decrease by \eqref{hf:ver-amplitude-step}. Increase the threshold on $n$ so that $h\ge1$. By \eqref{hf:amplitudes}, $D_i\le D_0$,  so \eqref{hf:ver-amplitude-step} and $\beta_0\le1\le D_i$ give
\[
 \beta_{i+1}\ge\frac{\beta_0\beta_i}{2D_0}\quad(0\le i<h),\qquad
 \beta_h\ge\beta_0\Bigl(\frac{\beta_0}{2D_0}\Bigr)^h .
\]
With $S_0\le2K^h$, $\log K=74(h+1)\log2$, and $\log P=64(h+1)\log2$,
\[
 \log D_0=3\log S_0+5\log P+O(\log(h+1))\le222(\log2)h^2+O(h),
\]
and
\[ \log\frac1{\beta_0}=\log\frac{(h+1)^2}{\eta_*\lambda_*}=O(\log(h+1)),\]
hence, for all sufficiently large $n$,
\[
 \log\frac1{\beta_h}\le(h+1)\log\frac1{\beta_0}+h\log(2D_0)
 \le222(\log2)h^3+O(h^2)\le1024h^3\le\frac{\log n}{32},
\]
the last step by $h\le\frac1{32}(\log n)^{\frac13}$. Thus $\beta_h\ge n^{-\frac1{32}}\ge n^{-\frac1{16}}$, and $\log S_0\le\log2+74h(h+1)\log2=O((\log n)^{\frac23})$. Finally, the denominator in \eqref{hf:field-normalization} is largest at $(i,p)=(0,P)$, so
\[
 \log\frac1{\min_{i,p}a_{i,p}}=4\log P+2\log S_0+\frac12\log((h+1)\ell_0)=O(h^2)=o(\log n).
\]
This proves (3).
\end{proof}

\subsection{Basic Properties}\label{hf:basic-properties}

This subsection records three properties of the construction that
hold for every realization of $\Xi$ and are used throughout
Sections~\ref{sec:output-fails} and~\ref{sec:target-proof}: the
fields are linear in the Gaussian coefficients
(Lemma~\ref{hf:representers}), which determines their covariance
(Lemma~\ref{hf:covariance}), and the gates certify the normalized
energies of their level and decide which levels enter an oracle reply
(Lemma~\ref{hf:gate-prefix}).

\paragraph{Linear representation.}
For each $0\le i\le h$, write $\Xi_i=(\Xi_{i,p})_{p\in\mathcal P}$.
We call $x\mapsto\langle\Xi_i,\Phi_i(x)\rangle$, defined below,
the field of level~$i$.
The corresponding gate is $G_i^\Xi$, and the load of level~$i$
is the summand of index~$i$ in \eqref{hf:potential}.
Define the vector-valued features
\[
 \Phi_{i,p}(x)
 =\sqrt n\,a_{i,p}C_i(x)
 \psi_i\!\left(
 \frac{x_i}{\|x_i\|},\ldots,\frac{x_h}{\|x_h\|}
 \right)^{\otimes p},
 \qquad
 \Phi_i(x)=\bigoplus_{p\in\mathcal P}\Phi_{i,p}(x).
\]
Set $\Phi_{i,p}(x)=0$ whenever a participating block vanishes.
The same zero-extension convention applies to all expressions
$C_i(x)F_{i,p}^\Xi$ and $C_i(x)F_i^\Xi$ below, where the
angular functions are evaluated at the normalized block directions.

\begin{lemma}[\bf Linear representation]
\label{hf:representers}
The features $\Phi_{i,p}$ and $\Phi_i$ are globally smooth.
For every coefficient realization $\Xi$,
\begin{equation}\label{hf:linear-representers}
 C_i(x)F_{i,p}^\Xi\!\left(
 \tfrac{x_i}{\|x_i\|},\ldots,\tfrac{x_h}{\|x_h\|}
 \right)=\langle\Xi_{i,p},\Phi_{i,p}(x)\rangle,
 \qquad
 C_i(x)F_i^\Xi\!\left(
 \tfrac{x_i}{\|x_i\|},\ldots,\tfrac{x_h}{\|x_h\|}
 \right)=\langle\Xi_i,\Phi_i(x)\rangle.
\end{equation}
The derivatives are also linear in the Gaussian coefficients.
For every direction $v\in\mathbb R^{(h+1)n}$,
\[
 D_v\!\left[
 C_i(x)F_i^\Xi\!\left(
 \tfrac{x_i}{\|x_i\|},\ldots,\tfrac{x_h}{\|x_h\|}
 \right)\right]
 =\langle\Xi_i,D_v\Phi_i(x)\rangle.
\]
\end{lemma}

\begin{proof}
On the open set where all participating blocks are nonzero,
$\Phi_{i,p}$ is a composition of smooth maps and near every other point
$C_i$ vanishes, so $\Phi_{i,p}$ is $0$ there. Hence $\Phi_{i,p}$ and
$\Phi_i$ are $C^\infty$. The first identity is the definition of
$F^\Xi_{i,p}$, the second is its sum over $p$, and the third follows
by differentiating the second, as $\Xi_i$ does not depend on $x$.
\end{proof}

\paragraph{Covariance.}
For $x,y$ whose blocks $i,\ldots,h$ are nonzero, define
\begin{equation}\label{hf:compound-correlation}
 T_i(x,y)
 =\prod_{k=i}^h
 \left(
 \frac{\langle x_k,y_k\rangle}{\|x_k\|\|y_k\|}
 \right)^{K^{k-i}}.
\end{equation}

\begin{lemma}[\bf Covariance]
\label{hf:covariance}
Under $\mu_n$, the random fields
$x\mapsto C_i(x)F_{i,p}^\Xi\!\left(
\tfrac{x_i}{\|x_i\|},\ldots,\tfrac{x_h}{\|x_h\|}
\right)$
are centered Gaussian and independent for distinct source--degree
pairs $(i,p)$.
\begin{enumerate}
\item \textbf{Gaussian structure and covariance.}
Their covariance kernels are
\begin{equation}\label{hf:degree-covariance}
 \langle\Phi_{i,p}(x),\Phi_{i,p}(y)\rangle
 =na_{i,p}^2C_i(x)C_i(y)T_i(x,y)^p.
\end{equation}
The kernel is zero whenever a participating block vanishes.

\item \textbf{Derivative covariance.}
For any directions $v,w$,
\[
 \langle D_v\Phi_{i,p}(x),D_w\Phi_{i,p}(y)\rangle
 =na_{i,p}^2
 \left.\frac{\partial^2}{\partial s\,\partial t}
 \left[
 C_i(x+sv)C_i(y+tw)T_i(x+sv,y+tw)^p
 \right]\right|_{s=t=0}.
\]

\item \textbf{Ancestor correlation.}
For $0\le j<i\le h$,
\begin{equation}\label{hf:ancestor-correlation}
 |T_j(x,y)|\le |T_i(x,y)|^{K^{i-j}},
\end{equation}
whenever all blocks $j,\ldots,h$ of both points are nonzero.
\end{enumerate}
\end{lemma}

\begin{proof}
By Lemma~\ref{hf:representers}, the values of the field $(i,p)$ at
finitely many points are linear functions of the standard Gaussian
vector $\Xi_{i,p}$, so they are jointly centered Gaussian; the fields
of distinct pairs $(i,p)$ are independent because the vectors
$\Xi_{i,p}$ are.

(1) For a standard Gaussian vector $\zeta$ and fixed vectors $a,b$,
$\E[\langle\zeta,a\rangle\langle\zeta,b\rangle]=\langle a,b\rangle$.
Hence the covariance kernel of the field $(i,p)$ is
$\langle\Phi_{i,p}(x),\Phi_{i,p}(y)\rangle$, and by the definition of
$\Phi_{i,p}$ and $\langle a^{\otimes p},b^{\otimes p}\rangle
=\langle a,b\rangle^p$,
\[
 \langle\Phi_{i,p}(x),\Phi_{i,p}(y)\rangle
 =na_{i,p}^2\,C_i(x)C_i(y)\,
 \bigl\langle\psi_i(u_{\ge i}(x)),\psi_i(u_{\ge i}(y))\bigr\rangle^p
 =na_{i,p}^2\,C_i(x)C_i(y)\,T_i(x,y)^p ,
\]
the last step by the definition of $T_i$. If a participating block of
$x$ or $y$ vanishes, both sides are $0$ by the zero-extension
convention.

(2) Since $\Phi_{i,p}$ is $C^\infty$ and the inner product is
bilinear, $\langle D_v\Phi_{i,p}(x),D_w\Phi_{i,p}(y)\rangle
=\partial_s\partial_t\langle\Phi_{i,p}(x+sv),\Phi_{i,p}(y+tw)\rangle
\big|_{s=t=0}$, and the inner product on the right is given by (1).

(3) By the definition of $T_j$ and $T_i$,
\[
 T_j(x,y)=\prod_{k=j}^{i-1}
 \Bigl(\frac{\langle x_k,y_k\rangle}{\|x_k\|\,\|y_k\|}\Bigr)^{K^{k-j}}
 \cdot T_i(x,y)^{K^{i-j}} ,
\]
and each factor of the product has absolute value at most $1$ by
Cauchy--Schwarz.
\end{proof}

\paragraph{The role of the gates.}
The gates control which sources can affect the oracle reply.
For $0\le i\le h$, write
$A_i^\Xi(x)=\prod_{j=0}^{i-1}G_j^\Xi(x)$, with $A_0^\Xi=1$.
By Lemma~\ref{hf:representers},
\[
 V_i^\Xi(x)-V_{i-1}^\Xi(x)
 =A_i^\Xi(x)U_i\!\left(\langle\Xi_i,\Phi_i(x)\rangle\right).
\]

\begin{lemma}[\bf Gate certificates and prefix oracle replies]
\label{hf:gate-prefix}
For every level $i$ and point $x$, the following hold.
\begin{enumerate}[leftmargin =* ]
\item \textbf{Energy certificate.}
If $G_i^\Xi(x)>0$, then $C_i(x)=1$ and
\[\left|\frac{\langle\Xi_{i,p},\Phi_{i,p}(x)\rangle}
{na_{i,p}^2}+\beta_i\right|<\epsilon_g\beta_i\]
for every $p\in\mathcal P$.

\item \textbf{Gate activation.}
If $J(\tfrac{\|x_j\|}{\sqrt n})=1$ for every $j\ge i$ and
$\left|\tfrac{\langle\Xi_{i,p},\Phi_{i,p}(x)\rangle}
{na_{i,p}^2}+\beta_i\right|
\le\tfrac14\epsilon_g\beta_i$
for every $p\in\mathcal P$, then $G_i^\Xi(x)=1$.

\item \textbf{Inactive level.}
If $A_i^\Xi(x)=0$, then $\nabla A_i^\Xi(x)=0$, and
$V_i^\Xi$ and $V_{i-1}^\Xi$ have the same value and gradient at $x$.

\item \textbf{Prefix simulation.}
Given $\Xi_0,\ldots,\Xi_{i-1}$, a joint value--gradient query
to $V_i^\Xi$ at $x$ can be simulated using one joint query to
the smooth function
$x\mapsto\langle\Xi_i,\Phi_i(x)\rangle$
when $A_i^\Xi(x)>0$.
When $A_i^\Xi(x)=0$, no query to this function is needed.
\end{enumerate}
\end{lemma}

\begin{proof}
Let $m:=\max_{p\in\mathcal P}|z_p|$. The sum
$\sum_p2\cosh(\tau z_p)$ has $2|\mathcal P|$ positive terms, the
largest of which is $e^{\tau m}$, so
\begin{equation}\label{hf:softmax-bounds}
 m\le\frac1\tau\log\sum_{p\in\mathcal P}2\cosh(\tau z_p)
 \le m+\frac{\log(2|\mathcal P|)}{\tau}\le m+\frac18 ,
\end{equation}
the last step by $|\mathcal P|\le P$ and $\tau=8\log(2P)$.

(1) If $G_i^\Xi(x)>0$, all its factors are positive. Positivity of
the radial factors gives $C_i(x)=1$, since $\chi=1$ wherever $J>0$.
Positivity of the outer factor gives that the argument of $g$ is less
than $1$, hence $m<1$ by \eqref{hf:softmax-bounds}, which is the
claim.

(2) Under the hypotheses, the radial factors equal $1$ and
$m\le\tfrac14$, so the argument of $g$ is at most $\tfrac38$ by
\eqref{hf:softmax-bounds}, and $g=1$ there. Hence $G_i^\Xi(x)=1$.

(3) $A_i^\Xi$ is smooth and nonnegative, so each of its zeros is a
minimum and $\nabla A_i^\Xi$ vanishes there. The product rule applied
to $V_i^\Xi-V_{i-1}^\Xi=A_i^\Xi\,U_i(\langle\Xi_i,\Phi_i\rangle)$
then shows that this difference and its gradient vanish at $x$.

(4) The vectors $\Xi_0,\ldots,\Xi_{i-1}$ determine $V_{i-1}^\Xi$,
$A_i^\Xi$ and their gradients at $x$. If $A_i^\Xi(x)=0$, (3) gives
the reply. Otherwise let $(s,v)$ be the value and gradient of
$\langle\Xi_i,\Phi_i(\cdot)\rangle$ at $x$; then
$V_i^\Xi(x)=V_{i-1}^\Xi(x)+A_i^\Xi(x)U_i(s)$ and
\[
 \nabla V_i^\Xi(x)=\nabla V_{i-1}^\Xi(x)+U_i(s)\,\nabla A_i^\Xi(x)
 +A_i^\Xi(x)\,U_i'(s)\,v .
\]
\end{proof}

\subsection{The Energy Test}\label{hf:energy-test-section}
\label{sec:energy-test}
We compare the algorithm's output with a target sample through
the degree energies of the final source.
For fixed $u\in\mathbb S^{n-1}$ and $\Xi\sim\mu_n$,
$F_{h,p}^\Xi(u)\sim\mathcal{N}(0,na_{h,p}^2)$.
We therefore call
$\tfrac{1}{na_{h,p}^2}
F_{h,p}^\Xi(\tfrac{x_h}{\|x_h\|})$
the normalized energy of $x$ in degree $p$.

\begin{definition}[\bf Energy test]\label{hf:energy-tests}
Let
\[
 \mathcal A_n
 =\left\{
 x\in\mathbb R^{(h+1)n}:
 0.75\sqrt n\le\|x_h\|\le0.9\sqrt n
 \right\}.
\]
Define
\[
 \mathcal T_n
 =\left\{
 (\Xi,x)\in\mathcal G_n\times\mathcal A_n:
 \max_{p\in\mathcal P}
 \left|
 \frac{F_{h,p}^\Xi(\tfrac{x_h}{\|x_h\|})}
      {na_{h,p}^2}
 +\beta_h
 \right|
 \le\frac{\epsilon_g\beta_h}{4}
 \right\}.
\]
We say that $(\Xi,x)$ passes the energy test if
$(\Xi,x)\in\mathcal T_n$.
\end{definition}

The set $\mathcal T_n$ is closed: $\mathcal A_n$ is closed,
and the energy maps are continuous on
$\mathcal G_n\times\mathcal A_n$.
By Lemma~\ref{hf:gate-prefix}, passing the test implies
$G_h^\Xi(x)=1$.
The test involves only the final source; the earlier gates
control how this source enters the oracle replies.

\paragraph{The reference energy.}
Reweighting $\mathcal{N}(0,na_{h,p}^2)$ by $e^{-\beta_h y}$
and normalizing gives
$\mathcal{N}(-\beta_h na_{h,p}^2,na_{h,p}^2)$.
Thus the normalized energy under this tilt has mean $-\beta_h$.
Section~\ref{sec:target-proof} justifies this reference level for a target sample,
accounting for the gates, energy truncations, and
instance-dependent normalizing constant.

\paragraph{The role of the test.}
For all sufficiently large $n$,
Theorem~\ref{hf:output-fails} shows that any randomized adaptive
algorithm with a worst-case budget of fewer than $Q$ joint
value--gradient queries passes the test with probability at most
$n^{-10}$.
Its output need not be a queried point.
Theorem~\ref{thm:target-good-mass} shows that a target sample passes
with probability at least $1-Cn^{-1/2}$, for an absolute $C>0$.
Both probabilities are averaged over $\Xi\sim\mu_n$ and,
for the algorithm, its independent random seed.

For fixed $\Xi$, let
$\mathcal T_n(\Xi)=\{x:(\Xi,x)\in\mathcal T_n\}$,
and let $\rho_\Xi^{\mathsf A}$ be the algorithm's output law.
By the definition of total variation,
\begin{equation}\label{hf:tv-test-interface}
 \operatorname{TV}(\rho_\Xi^{\mathsf A},\pi^\Xi)
 \ge
 \pi^\Xi(\mathcal T_n(\Xi))
 -\rho_\Xi^{\mathsf A}(\mathcal T_n(\Xi)).
\end{equation}
Section~\ref{sec:parameters-and-lower-bound} averages this inequality over admissible instances
and uses Proposition~\ref{hf:parameter-bounds} to obtain
the sampling lower bound.

\section{The Output Fails the Energy Test}\label{sec:output-fails}

We prove Theorem~\ref{hf:output-fails} under the original independent
Gaussian law $\mu_n$. 

\begin{theorem}[\bf The output fails the test]
\label{hf:output-fails}
For all sufficiently large $n$, and  any randomized
adaptive algorithm $\mathsf A$ making at most $q<Q$ 
queries, its output $X$ satisfies
\begin{equation}\label{hf:output-test-bound}
 \mathbb P_{\Xi\sim\mu_n,\,\mathsf A}
 \{(\Xi,X)\in\mathcal T_n\}
 \le
 \mathbb P_{\Xi\sim\mu_n,\,\mathsf A}
 \{G_h^\Xi(X)>0\}
 \le n^{-10}.
\end{equation}
\end{theorem}


Section~\ref{out:single-source} shows that, for an algorithm querying the field of one level, the normalized energies at the output share a common degree profile $p\sum_tB_tT_t^{p-1}$, up to an error controlled by the Gaussian conditioning of Section~\ref{gauss:conditioning}. Section~\ref{out:approximation} shows that passing a gate forces this profile to approximate a constant, which requires two queries with separated directions $\psi_i$. Section~\ref{out:locality} applies the decoupling theorem of Section~\ref{gauss:simulations}: queries with separated directions span nearly orthogonal measurement spaces for every field below the level, so the two groups may be treated as seeing independent coefficients. Section~\ref{out:recursive-charge} combines these facts into an induction over the levels in which the query threshold doubles at every level.

\subsection{Energy on $i$-th Level}
\label{out:single-source}

In this subsection, for fixed $0\le i\le h$, we derive a simultaneous representation
of the normalized degree energies at the algorithm's output,
with coefficients that are independent of the degree $p$ (see~\eqref{out:profile-formula}). Consider a randomized algorithm making at most $q$ queries to the first-order oracle of the field of level $i$ without gate,
\[
 x\longmapsto\langle\Xi_i,\Phi_i(x)\rangle=\sum_{p\in\mathcal P}\langle\Xi_{i,p},\Phi_{i,p}(x)\rangle .
\]
By Lemma~\ref{gauss:field-queries}, it is an oracle simulation of $\Xi_i$ in the sense of Section~\ref{gauss:observation-rules}. Write $y_1,\ldots,y_q$ for its queries, $(s_t,g_t)$ for the value and gradient returned at $y_t$, $g_{t,k}\in\R^n$ for block $k$ of $g_t$, and $\partial_{k,v}$ for differentiation along $v\in\R^n$ in block $k$. By Lemma~\ref{hf:representers},
\[
 s_t=\langle\Xi_i,\Phi_i(y_t)\rangle,\qquad
 \langle g_{t,k},v\rangle=\langle\Xi_i,\partial_{k,v}\Phi_i(y_t)\rangle\qquad(i\le k\le h),
\]
so the histories and the measurement spaces of the simulation are
\[
 \mathcal F_t=\sigma\bigl(\omega,(s_a,g_a)_{a\le t}\bigr),\qquad
 W_t=\operatorname{span}\{\Phi_i(y_a),\partial_{k,v}\Phi_i(y_a):a\le t,\ i\le k\le h,\ v\in\R^n\},\qquad W_0=\{0\},
\]
and $y_t$ is $\mathcal F_{t-1}$-measurable. The output $X$ is any $\mathcal F_q$-measurable point, whether or not it has been queried, and a run that stops early is padded by zero batches (Remark~\ref{gauss:terminal}).
The following theorem gives a common representation of the output
energies across all degrees $p \in \mathcal{P}$.

\begin{theorem}
\label{out:profile}
For each query $y_t$, set
$T_t=\begin{cases}
T_i(X,y_t),& C_i(X)C_i(y_t)>0,\\
0,&\text{otherwise},
\end{cases}$
where $T_i$ is defined in Eq.~\eqref{hf:compound-correlation}.
There exist $\mathcal F_q$-measurable real coefficients
$B_1,\ldots,B_q$ and random errors $(E_p)_{p\in\mathcal P}$
such that
\begin{equation}\label{out:profile-formula}
 \frac{\langle\Xi_{i,p},\Phi_{i,p}(X)\rangle}{na_{i,p}^2}
 =p\sum_{t=1}^q B_tT_t^{p-1}+E_p
 \qquad\text{for every }p\in\mathcal P.
\end{equation}
The coefficients $B_t$ are common to all degrees.
Moreover, for every $0<\zeta<\tfrac12$, with probability
at least $1-\zeta$ under $\mu_n$ and the algorithm's seed,
\[
 \max_{p\in\mathcal P}|E_p|
 \le
 \frac{C\left(\sqrt{h-i+1}\,q+
 \sqrt{\log\left(\tfrac{2P}{\zeta}\right)}\right)}
 {\sqrt n\,\min_{p\in\mathcal P}a_{i,p}}
 =o(\beta_i),
\]
provided $q<Q$ and $\log(1/\zeta)=n^{o(1)}$, where $C>0$ is an absolute constant.

\end{theorem}

Section~\ref{out:decomposition-section} decomposes each output energy into fresh, unobserved, and seen contributions, and Section~\ref{out:fresh-section} expresses the fresh contribution through the common coefficients $B_t$ and nodes $T_t$. The two error terms are then bounded by Corollary~\ref{gauss:linear-functional} and Lemma~\ref{gauss:predictable-projections} in the proof of the theorem.

\subsubsection{Decomposition of the Output Energies}
\label{out:decomposition-section}

For each block $k\in\{i,\ldots,h\}$, define
\[
 K_{t,k}=\operatorname{span}\{y_{1,k},\ldots,y_{t,k},
 g_{1,k},\ldots,g_{t-1,k}\},
 \qquad \dim K_{t,k}\le2t-1.
\]
Within the new coefficient observation space $W_t\cap W_{t-1}^{\perp}$,
define the fresh subspace $\widetilde{\mathcal N}_t$
and the seen subspace $\widetilde{\mathcal S}_t$ by
\[
 \begin{aligned}
 \widetilde{\mathcal N}_t
 &=P_{W_{t-1}^\perp}\operatorname{span}
 \{\partial_{k,v}\Phi_i(y_t):
 i\le k\le h,\ v\in K_{t,k}^\perp\},\\
 \widetilde{\mathcal S}_t
 &=W_t\cap(W_{t-1}\oplus\widetilde{\mathcal N}_t)^\perp.
 \end{aligned}
\]
Define the fresh and seen subspaces over all $q$ queries by
\[
 \mathcal N=\sum_{t=1}^q\widetilde{\mathcal N}_t,
 \qquad
 \mathcal S=\sum_{t=1}^q\widetilde{\mathcal S}_t.
\]
The unobserved subspace is $\mathcal U=W_q^\perp$,
consisting of directions orthogonal to every measurement representer.

\begin{lemma}[\bf Decomposition of the output energies]
\label{out:decomposition}
For every coefficient realization and seed, the source coefficient
space decomposes orthogonally as
$\mathcal N\oplus\mathcal S\oplus\mathcal U$.
Consequently, for every $p\in\mathcal P$,
\begin{equation}\label{out:energy-decomposition}
 \langle\Xi_{i,p},\Phi_{i,p}(X)\rangle
 =
 \underbrace{\langle\Phi_{i,p}(X),P_{\mathcal N}\Xi_i\rangle}
 _{M_p\;\text{(fresh)}}
 +
 \underbrace{\langle\Phi_{i,p}(X),P_{\mathcal U}\Xi_i\rangle}
 _{U_p\;\text{(unobserved)}}
 +
 \underbrace{\langle\Phi_{i,p}(X),P_{\mathcal S}\Xi_i\rangle}
 _{L_p\;\text{(seen)}}.
\end{equation}
Moreover, $\widetilde{\mathcal N}_t$ and
$\widetilde{\mathcal S}_t$ are $\mathcal F_{t-1}$-measurable, and
\[
 \dim\widetilde{\mathcal S}_t\le1+(h-i+1)(2t-1),
 \qquad
 \dim\mathcal S\le q+(h-i+1)q^2.
\]
\end{lemma}

\begin{proof}
Fix $t$ and put
\[
 S_t:=\operatorname{span}\{\Phi_i(y_t),\ \partial_{k,e}\Phi_i(y_t):\ i\le k\le h,\ e\in K_{t,k}\},
\]
and
\[ N_t:=\operatorname{span}\{\partial_{k,v}\Phi_i(y_t):\ i\le k\le h,\ v\in K_{t,k}^\perp\},\]
so that $\widetilde{\mathcal N}_t=P_{W_{t-1}^\perp}N_t$. Since
$\R^n=K_{t,k}\oplus K_{t,k}^\perp$ for each $k$ and
$\partial_{k,v}\Phi_i(y_t)$ is linear in $v$,
\begin{equation}\label{out:Wt-split}
 W_t=W_{t-1}+S_t+N_t .
\end{equation}

As $N_t\subset W_t$ and $W_{t-1}\subset W_t$, the projection
$P_{W_{t-1}^\perp}$ maps $N_t$ into $W_t$, so
$\widetilde{\mathcal N}_t\subset W_t\cap W_{t-1}^\perp$. By the
definition of $\widetilde{\mathcal S}_t$,
$W_t\cap W_{t-1}^\perp=\widetilde{\mathcal N}_t\oplus\widetilde{\mathcal S}_t$,
and $W_t=W_{t-1}\oplus(W_t\cap W_{t-1}^\perp)$ gives the first claim.
Iterating from $W_0=\{0\}$,
$W_q=\bigoplus_t\widetilde{\mathcal N}_t\oplus\bigoplus_t\widetilde{\mathcal S}_t
=\mathcal N\oplus\mathcal S$, and $\mathcal U=W_q^\perp$ completes the
decomposition. Then
$\Xi_i=P_{\mathcal N}\Xi_i+P_{\mathcal S}\Xi_i+P_{\mathcal U}\Xi_i$,
and pairing with $\Phi_{i,p}(X)$ gives
Eq.~\eqref{out:energy-decomposition}.

By Eq.~\eqref{out:Wt-split},
$W_t\cap W_{t-1}^\perp=P_{W_{t-1}^\perp}S_t+\widetilde{\mathcal N}_t$,
so 
\[
 \dim\widetilde{\mathcal S}_t
 =\dim(W_t\cap W_{t-1}^\perp)-\dim\widetilde{\mathcal N}_t
 \le\dim S_t
 \le1+\sum_{k=i}^h\dim K_{t,k}
 \le1+(h-i+1)(2t-1),
\]
and summing over $t\le q$ gives $\dim\mathcal S\le q+(h-i+1)q^2$.

Finally, $y_t$, the spaces $\mathcal K_{t,k}$, and $W_{t-1}$ are $\mathcal F_{t-1}$-measurable, so $N_t$ and $S_t$ are spanned by $\mathcal F_{t-1}$-measurable vectors, and $\widetilde N_t$, $\widetilde S_t$ are obtained from $N_t$, $S_t$, $W_{t-1}$ by sums, projections, and orthogonal complements, which are Borel operations on the projections. Hence they are $\mathcal F_{t-1}$-measurable by Remark~\ref{gauss:measurable-subspace}.
\end{proof}


\subsubsection{Formulation of the Fresh Part}
\label{out:fresh-section}

Recall for each query $y_t$, set
$T_t=\begin{cases}
T_i(X,y_t),& C_i(X)C_i(y_t)>0,\\
0,&\text{otherwise},
\end{cases}$
where $T_i$ is defined in Eq.~\eqref{hf:compound-correlation}.

\begin{proposition}[\bf Formulation of the fresh part]
\label{out:fresh-profile}
There are real coefficients $B_1,\ldots,B_q$, 
such that
\begin{equation}\label{out:fresh-formula}
 M_p = \langle\Phi_{i,p}(X),P_{\mathcal N}\Xi_i\rangle=na_{i,p}^2p\sum_{t=1}^qB_tT_t^{p-1}
 \qquad\text{for every }p\in\mathcal P.
\end{equation}
The $B_t$ and $T_t$ serve every degree $p$.
Queries with $C_i(y_t)=0$ have $B_t=0$.
\end{proposition}

\begin{proof}
If $C_i(X)=0$, then every $M_p$ vanishes, and we take
$B_a=0$ for every $a$. Hence assume $C_i(X)>0$.

We first identify the derivative representers that give the
desired dependence on $p$.
For a query $y_a$ with $C_i(y_a)>0$ and a direction
$v\perp y_{a,k}$, differentiating the identity
$\langle\Phi_{i,p}(X),\Phi_{i,p}(y)\rangle
=na_{i,p}^2C_i(X)C_i(y)T_i(X,y)^p$
from Lemma~\ref{hf:covariance}
in block $k$ of $y$ along $v$, and evaluating at $y=y_a$, gives
\begin{equation}\label{out:output-normal}
 \begin{aligned}
 \frac{\langle\Phi_{i,p}(X),
 \partial_{k,v}\Phi_i(y_a)\rangle}{na_{i,p}^2}
 &=\partial_{k,v}\!\left[
 C_i(X)C_i(\cdot)T_i(X,\cdot)^p
 \right](y_a)\\
 &=C_i(X)C_i(y_a)
 \partial_{k,v}[T_i(X,\cdot)^p](y_a)\\
 &=pT_a^{p-1}C_i(X)C_i(y_a)
 \partial_{k,v}[T_i(X,\cdot)](y_a).
 \end{aligned}
\end{equation}
The second equality uses $\partial_{k,v}C_i(y_a)=0$,
since $v\perp y_{a,k}$. In the last line, the derivative acts on the second argument of $T_i$,
and the factor multiplying $pT_a^{p-1}$ is independent of $p$.

It therefore suffices to express $P_{\mathcal N}\Xi_i$
as a linear combination of vectors
$\partial_{k,v}\Phi_i(y_a)$ with
$1\le a\le q$, $i\le k\le h$, and $v\perp y_{a,k}$.
By definition, $\mathcal N$ is spanned by
$P_{W_{t-1}^{\perp}}\partial_{k,v}\Phi_i(y_t)$,
where $1\le t\le q$, $i\le k\le h$, and
$v\in K_{t,k}^{\perp}$.
The following fact shows that projecting
$\partial_{k,v}\Phi_i(y_t)$ onto $W_{t-1}$
gives a linear combination of
$\partial_{k,v}\Phi_i(y_1),\ldots,
\partial_{k,v}\Phi_i(y_{t-1})$.

\begin{fact}[\bf Projection preserves the derivative direction]
\label{out:fresh-projection}
For $1\le t\le q$, $i\le k\le h$, and
$v\in K_{t,k}^{\perp}$,
\[
 P_{W_{t-1}}\partial_{k,v}\Phi_i(y_t)
 \in
 \operatorname{span}
 \{\partial_{k,v}\Phi_i(y_a):a<t\}.
\]
\end{fact}

\begin{proof}[Proof of Fact~\ref{out:fresh-projection}]
The case $v=0$ is immediate, so assume $v\ne0$.
Let
\[
\begin{aligned}
 L
 &=\operatorname{span}
 \{\partial_{k,v}\Phi_i(y_a):1\le a<t\},\\
 R
 &=\operatorname{span}\Bigl(
 \{\Phi_i(y_a):1\le a<t\}\\
 &\qquad{}\cup
 \{\partial_{\ell,w}\Phi_i(y_a):
 1\le a<t,\ i\le\ell\le h,\ \ell\ne k,\ w\in\mathbb R^n\}\\
 &\qquad{}\cup
 \{\partial_{k,w}\Phi_i(y_a):
 1\le a<t,\ w\in\mathbb R^n,\ w\perp v\}
 \Bigr).
\end{aligned}
\]
By definition, we have $W_{t-1}=L+R$.

To show orthogonality to $R$, we first show that
$\partial_{k,v}\Phi_i(y_a)$ is orthogonal to $\Phi_i(z)$
whenever $a\le t$ and $z_k\perp v$.
If either cutoff vanishes, the corresponding value or derivative
feature is zero.
Otherwise, let $e$ be zero outside block $k$, with $e_k=v$.

Since $v\perp y_{a,k}$ and $v\perp z_k$,
\[
 \begin{aligned}
 \|y_{a,k}\pm sv\|^2
 &=\|y_{a,k}\|^2+s^2\|v\|^2,\\
 \langle y_{a,k}\pm sv,z_k\rangle
 &=\langle y_{a,k},z_k\rangle.
 \end{aligned}
\]
By Lemma~\ref{hf:covariance}, these identities give
\[
 \begin{aligned}
 \langle\Phi_i(y_a+se),\Phi_i(z)\rangle
 &=nC_i(y_a+se)C_i(z)
   \sum_{p\in\mathcal P}a_{i,p}^2T_i(y_a+se,z)^p\\
 &=nC_i(y_a-se)C_i(z)
   \sum_{p\in\mathcal P}a_{i,p}^2T_i(y_a-se,z)^p\\
 &=\langle\Phi_i(y_a-se),\Phi_i(z)\rangle.
 \end{aligned}
\]
Differentiating this at $s=0$ yields
\[
 \begin{aligned}
 \langle\partial_{k,v}\Phi_i(y_a),\Phi_i(z)\rangle
 &=\left.\frac{d}{ds}
   \langle\Phi_i(y_a+se),\Phi_i(z)\rangle\right|_{s=0}\\
 &=\left.\frac{d}{ds}
   \langle\Phi_i(y_a-se),\Phi_i(z)\rangle\right|_{s=0}\\
 &=-\langle\partial_{k,v}\Phi_i(y_a),\Phi_i(z)\rangle.
 \end{aligned}
\]
Hence $\langle\partial_{k,v}\Phi_i(y_a),\Phi_i(z)\rangle=0$.

We apply the preceding identity to the vectors spanning $R$.
For $a\le t$ and $b<t$, taking $z=y_b$ gives
$\langle\partial_{k,v}\Phi_i(y_a),\Phi_i(y_b)\rangle=0$,
since $y_{b,k}\perp v$.
For the derivative vectors, let $f$ be zero outside block
$\ell$, with $f_\ell=w$.
If $\ell\ne k$ or $w\perp v$, then
\[
 \langle v,(y_b+rf)_k\rangle
 =
 \begin{cases}
 \langle v,y_{b,k}\rangle,
 &\ell\ne k,\\
 \langle v,y_{b,k}\rangle+r\langle v,w\rangle,
 &\ell=k,\ w\perp v
 \end{cases}
 =0.
\]
Applying the identity along this line and differentiating gives
\[
 \begin{aligned}
 \langle\partial_{k,v}\Phi_i(y_a),
         \partial_{\ell,w}\Phi_i(y_b)\rangle
 &=
 \frac{d}{dr}
 \underbrace{
 \langle\partial_{k,v}\Phi_i(y_a),\Phi_i(y_b+rf)\rangle
 }_{=\,0\text{ for every }r}
 \bigg|_{r=0}=0.
 \end{aligned}
\]
Thus $\partial_{k,v}\Phi_i(y_a)\perp R$ for every $a\le t$.
Taking $a<t$ gives $L\perp R$, while taking $a=t$ gives
$\partial_{k,v}\Phi_i(y_t)\perp R$.
Since $W_{t-1}=L\oplus R$, we conclude that
\[
 \begin{aligned}
 P_{W_{t-1}}\partial_{k,v}\Phi_i(y_t)
 &=P_L\partial_{k,v}\Phi_i(y_t)
   +\underbrace{P_R\partial_{k,v}\Phi_i(y_t)}_{=\,0}\\
 &=P_L\partial_{k,v}\Phi_i(y_t)\in L.
 \end{aligned}
\]
\end{proof}

By the fact,
\[
 P_{W_{t-1}^{\perp}}\partial_{k,v}\Phi_i(y_t)
 \in
 \operatorname{span}
 \{\partial_{k,v}\Phi_i(y_a):a\le t\}
 \qquad(v\in K_{t,k}^{\perp}).
\]
Every derivative on the right is tangential at its query,
since $v\perp y_{a,k}$ for $a\le t$.
The definitions of $\widetilde{\mathcal N}_t$ and
$\mathcal N$ therefore imply
\[
 \mathcal N
 \subseteq
 \operatorname{span}\{
 \partial_{k,v}\Phi_i(y_a):
 1\le a\le q,\ i\le k\le h,\ v\perp y_{a,k}\}.
\]
Since differentiation is linear in its direction,
terms at the same query and block can be combined.
Thus there are vectors $v_{a,k}\perp y_{a,k}$ such that
\begin{equation}\label{out:fresh-tangent-representation}
 P_{\mathcal N}\Xi_i
 =
 \sum_{a=1}^q\sum_{k=i}^h
 \partial_{k,v_{a,k}}\Phi_i(y_a).
\end{equation}

These vectors can be chosen $\mathcal F_q$-measurably: $P_{W_q}\Xi_i$ is $\mathcal F_q$-measurable by Lemma~\ref{gauss:posterior}, hence so is $P_N\Xi_i=P_NP_{W_q}\Xi_i$, and we take the minimum-norm solution of the linear system \eqref{out:fresh-tangent-representation} subject to $v_{a,k}\perp y_{a,k}$. For queries with $C_i(y_a)=0$ the derivative representers vanish, and we take $v_{a,k}=0$.

Finally, pair \eqref{out:fresh-tangent-representation}
with $\Phi_{i,p}(X)$ and apply \eqref{out:output-normal}.
This gives
\[
 M_p
 =na_{i,p}^2p
 \sum_{a=1}^q
 \left(
 C_i(X)C_i(y_a)
 \sum_{k=i}^h
 \partial_{k,v_{a,k}}[T_i(X,\cdot)](y_a)
 \right)T_a^{p-1},
\]
where terms with $C_i(y_a)=0$ are omitted.
Thus \eqref{out:fresh-formula} holds with
\[B_a=C_i(X)C_i(y_a)\sum_{k=i}^h
\partial_{k,v_{a,k}}[T_i(X,\cdot)](y_a)\]
for $C_i(y_a)>0$.
Queries with zero cutoff contribute nothing, so we take
$B_a=0$ for those queries.
\end{proof}

\subsubsection{Proof of Theorem~\ref{out:profile}}
\begin{proof}[Proof of Theorem~\ref{out:profile}]
By Lemma~\ref{gauss:field-queries}(ii), the algorithm is an oracle simulation of $\Xi_i$ whose measurement space after $t$ rounds is $W_t$, and $\Phi_{i,p}(X)$ is $\mathcal F_q$-measurable by (iii). Lemma~\ref{out:decomposition} and Proposition~\ref{out:fresh-profile} give \eqref{out:profile-formula} with
\[
 E_p=\frac{U_p+L_p}{na_{i,p}^2},\qquad
 U_p=\bigl\langle\Phi_{i,p}(X),P_{W_q^\perp}\Xi_i\bigr\rangle,\qquad
 L_p=\bigl\langle\Phi_{i,p}(X),P_{\mathcal S}\Xi_i\bigr\rangle ,
\]
as $\mathcal U=W_q^\perp$. Since $\|\Phi_{i,p}(X)\|\le\sqrt n\,a_{i,p}$, Corollary~\ref{gauss:linear-functional} with $b_p=\Phi_{i,p}(X)$ and $\sigma_p=\sqrt n\,a_{i,p}$ gives
\[
 \mathbb P\bigl\{\exists p\in\mathcal P:|U_p|>\sqrt{2nt}\,a_{i,p}\bigr\}\le2|\mathcal P|e^{-t}\qquad(t>0).
\]
By Lemma~\ref{out:decomposition}, the spaces $\widetilde{\mathcal S}_t$ are $\mathcal F_{t-1}$-measurable, lie in $W_{t-1}^\perp\cap W_t$, and have total dimension at most $D:=q+(h-i+1)q^2$, so Lemma~\ref{gauss:predictable-projections} and Cauchy--Schwarz give
\[
 \mathbb P\bigl\{\|P_{\mathcal S}\Xi_i\|^2>4(D+t)\bigr\}\le e^{-t}\qquad(t>0),\qquad
 |L_p|\le\sqrt n\,a_{i,p}\,\|P_{\mathcal S}\Xi_i\| .
\]
Fix $0<\zeta<\frac12$ and take $t=\log\frac{4|\mathcal P|}{\zeta}$ in the first bound and $t=\log\frac2\zeta$ in the second. Outside an event of probability at most $\zeta$, for every $p\in\mathcal P$,
\[
 |E_p|\le\frac{\sqrt{2\log\frac{4|\mathcal P|}{\zeta}}+2\sqrt{D+\log\frac2\zeta}}{\sqrt n\,a_{i,p}}
 \le\frac{C\bigl(\sqrt{h-i+1}\,q+\sqrt{\log\frac{2P}{\zeta}}\bigr)}{\sqrt n\,\min_{p\in\mathcal P}a_{i,p}},
\]
using $\sqrt{a+b}\le\sqrt a+\sqrt b$, $D\le2(h-i+1)q^2$, and $4|\mathcal P|\le2P$.
\end{proof}

\subsection{Passing a Gate Requires Separated Queries}
\label{out:approximation}

Theorem~\ref{out:profile} shows that the normalized degree-$p$
energy at the algorithm's output is approximated by
$p\sum_{t=1}^q B_tT_t^{p-1}$, where both the coefficients $B_t$
and the query correlations $T_t$ are independent of the degree $p$.
If the output $X$ passes gate $i$, that is, if $G_i^\Xi(X)>0$,
then Lemma~\ref{hf:gate-prefix} implies that the normalized energy
in every degree differs from $-\beta_i$ by less than
$\epsilon_g\beta_i$.

In this section, we show that passing gate $i$ requires
two queried points $y,y'$ satisfy the condition that both
$\|\psi_i(y)-\psi_i(y')\|\ge P^{-1/2}$ and
$\|\psi_i(y)+\psi_i(y')\|\ge P^{-1/2}$ with high probability (see Proposition~\ref{out:source-separation}).
Recall here
$\psi_i(u_i,\ldots,u_h)
=u_i\otimes u_{i+1}^{\otimes K}\otimes\cdots
\otimes u_h^{\otimes K^{h-i}}$
for unit directions $u_i,\ldots,u_h$.
Whenever $C_i(x)>0$, we write
$\psi_i(x)=\psi_i(\tfrac{x_i}{\|x_i\|},\ldots,
\tfrac{x_h}{\|x_h\|})$.
Thus $\psi_i(x)$ is a unit vector and
$\langle\psi_i(x),\psi_i(y)\rangle=T_i(x,y)$
whenever both cutoffs are positive.

For even $p$ and $0<|t|\le1$, we have
$t^{p-1}=t^{-1}e^{-sp}$, where $s=-\log|t|\ge0$.
Thus, after rescaling the coefficients, approximating a nonzero
constant by $p\sum_t B_tT_t^{p-1}$ is equivalent to approximating
$1/p$ by an exponential sum.
The following lemma shows that such an approximation forces
a rate $s$ in a prescribed interval, even when the coefficients
have arbitrary signs.
We will apply it at two scales to obtain the separated directions
in Lemma~\ref{out:two-directions}. The proof can be found in appendix~\ref{app:rate-window}.
\begin{lemma}[\bf Rate window]
\label{out:rate-gap}
There exist universal constants $c,C>0$ such that the following
holds. Let $P\ge 8$ be even, and let $1\le q\le Q$ be integers.
Suppose that
\[
    f(p)=\sum_{j=1}^{q} c_j e^{-s_jp},
    \qquad c_j\in\mathbb R,\quad s_j\ge 0,
\]
satisfies
\[
    |pf(p)-1|\le \frac{1}{10}
    \qquad\text{for }p=4,6,\ldots,P.
\]
Then, for every $p_0\in\mathbb R$ satisfying
\[
    CQ\le p_0\le \frac{cP}{Q^2},
\]
there exists $j\in\{1,\ldots,q\}$ with $c_j\ne 0$ such that
\begin{equation}
    \frac{c}{Q^2p_0}<s_j<\frac{CQ}{p_0}.
\end{equation}
One may take $c=2^{-20}$ and $C=10240$.
\end{lemma}

We next use Lemma~\ref{out:rate-gap} to derive a geometric
separation statement.
Because its rate interval scales inversely with $p_0$,
we can choose two values of $p_0$ that yield rates $s,s'$
satisfying $P^{-1}<s<s'/16$ and $s'<1/10$.
We show that this gap forces the corresponding query vectors
to be separated both from each other and from each other's
negatives.

\begin{lemma}[\bf Two separated query directions]
\label{out:two-directions}
Let $Q\ge4$ be an even integer, let $1\le q\le Q$ be an integer,
and set $P=Q^{64}$.
For arbitrary unit vectors $w_*,w_1,\ldots,w_q$ in a real
Hilbert space, define $t_j=\langle w_*,w_j\rangle$.
Suppose that $c\neq 0$ and
$B_1,\ldots,B_q\in\mathbb R$ satisfy
\[
 \left|p\sum_{j=1}^q B_jt_j^{p-1}-c\right|
 \le\frac{|c|}{10}
 \qquad\text{for all }p=4,6,\ldots,P.
\]
Then there exist indices $1\le j<k\le q$ such that both
$\|w_j-w_k\|\ge P^{-1/2}$ and
$\|w_j+w_k\|\ge P^{-1/2}$.
\end{lemma}



\begin{proof}[Proof of Lemma~\ref{out:two-directions}]
Terms with $t_j=0$ contribute nothing to the approximation,
so we omit them, retaining the original indices.
At least one term remains, since otherwise the approximation
condition cannot hold.
For each remaining term, set
$s_j=-\log|t_j|$ and $d_j=B_j/(ct_j)$.
Then $s_j\ge0$, and, since $p$ is even,
\[
 \begin{aligned}
 f(p)
 :=\frac1c\sum_j B_jt_j^{p-1}
 &=\sum_j\frac{B_j}{ct_j}|t_j|^p=\sum_j d_je^{-s_jp}.
 \end{aligned}
\]
The hypothesis gives $|pf(p)-1|\le1/10$ for
$p=4,6,\ldots,P$.

We use the constants
$c_{\mathrm{rg}}=2^{-20}$ and $C_{\mathrm{rg}}=10240$
established in the proof of Lemma~\ref{out:rate-gap} (see \eqref{app:rate-window-constants}).
Because the number of remaining terms is at most $Q$,
that lemma applies at every scale $u$ satisfying
$C_{\mathrm{rg}}Q\le u\le c_{\mathrm{rg}}P/Q^2$,
and supplies a rate with nonzero coefficient in
$(c_{\mathrm{rg}}/(Q^2u),C_{\mathrm{rg}}Q/u)$.
Indeed, using $Q$ in place of the actual number of terms
restricts the admissible scales and enlarges the rate interval.

Choose the two scales
\[
 p_0=10C_{\mathrm{rg}}Q,
 \qquad
 p_1=\frac{c_{\mathrm{rg}}P}{Q^2}.
\]
We first verify that both are admissible.
Since $C_{\mathrm{rg}}<2^{14}$,
$c_{\mathrm{rg}}^{-1}=2^{20}$, and $Q\ge4$,
\[
 160\left(\frac{C_{\mathrm{rg}}}{c_{\mathrm{rg}}}\right)^2Q^6
 <2^{76}Q^6
 <Q^{64}=P,
\]
where the second inequality follows from
$Q^{58}\ge4^{58}=2^{116}>2^{76}$.
Consequently,
$p_1/p_0=c_{\mathrm{rg}}P/(10C_{\mathrm{rg}}Q^3)
>16C_{\mathrm{rg}}Q^3/c_{\mathrm{rg}}>1$.
Thus
\begin{equation}\label{eq:admissible}
 C_{\mathrm{rg}}Q
 \le p_0<p_1
 =\frac{c_{\mathrm{rg}}P}{Q^2}.
\end{equation}

Applying Lemma~\ref{out:rate-gap} at $p_1$ and $p_0$,
respectively, gives rates $s$ and $s'$ such that
\[
 \begin{aligned}
 \frac1P
 &<s<\frac{C_{\mathrm{rg}}Q^3}{c_{\mathrm{rg}}P},\\
 \frac{c_{\mathrm{rg}}}{10C_{\mathrm{rg}}Q^3}
 &<s'<\frac1{10}.
 \end{aligned}
\]
The preceding lower bound on $P$ gives
$16C_{\mathrm{rg}}Q^3/(c_{\mathrm{rg}}P)
<c_{\mathrm{rg}}/(10C_{\mathrm{rg}}Q^3)$.
Hence $s'>16s$, so the selected rates correspond to distinct
original indices $j,k$.

Set $v=\operatorname{sign}(t_j)w_j$ and
$v'=\operatorname{sign}(t_k)w_k$, with $j,k$ corresponding
to $s,s'$, respectively.
Then $\langle w_*,v\rangle=e^{-s}$ and
$\langle w_*,v'\rangle=e^{-s'}$.
Since these vectors are unit vectors,
$\|v-w_*\|^2=2(1-e^{-s})$ and
$\|v'-w_*\|^2=2(1-e^{-s'})$.

For $0\le r\le1/10$, integrating
$1/2\le e^{-u}\le1$ over $u\in[0,r]$ gives
$r/2\le1-e^{-r}\le r$.
As $0<s<s'<1/10$, the reverse triangle inequality yields
\[
 \begin{aligned}
 \|v-v'\|
 &\ge\|v'-w_*\|-\|v-w_*\|\ge\sqrt{s'}-\sqrt{2s}>(4-\sqrt2)\sqrt{s}>P^{-1/2}.
 \end{aligned}
\]
Here the third line uses $s'>16s$, and the last uses
$s>1/P$ and $4-\sqrt2>1$.
For the sum, Cauchy--Schwarz gives
\[
 \begin{aligned}
 \|v+v'\|
 &\ge\langle w_*,v+v'\rangle=e^{-s}+e^{-s'}>2e^{-1/10}>1>P^{-1/2}.
 \end{aligned}
\]
Changing the sign of either vector only interchanges
the two distances $\|w_j-w_k\|$ and $\|w_j+w_k\|$.
Both bounds therefore hold for the original vectors.
Reordering the selected indices if necessary gives $j<k$.
\end{proof}

We now combine Theorem~\ref{out:profile}, the gate certificate of
Lemma~\ref{hf:gate-prefix}(1), and Lemma~\ref{out:two-directions}
into the statement used in Section~\ref{out:recursive-charge}.

\begin{proposition}[\bf Passing a gate requires separated queries]
\label{out:source-separation}
Fix $0\le i\le h$, and put $\zeta=e^{-Q^2}n^{-300}$.
Consider an algorithm making at most $q<Q$ queries to the value
and gradient of $x\mapsto\langle\Xi_i,\Phi_i(x)\rangle$, with a
seed independent of $\Xi_i$ and an arbitrary output $Z$.
For all sufficiently large $n$, with probability at least
$1-\zeta$, the following implication holds: if $G_i(Z)>0$, then
two queried points $y,y'$ satisfy $C_i(y)C_i(y')>0$ and
\[
 \|\psi_i(y)-\psi_i(y')\|\ge P^{-1/2},
 \qquad
 \|\psi_i(y)+\psi_i(y')\|\ge P^{-1/2}.
\]
\end{proposition}

\begin{proof}
The deterministic parameter bounds in
Proposition~\ref{hf:parameter-bounds} give $Q=n^{o(1)}$,
$h+1=n^{o(1)}$, $\min_{j,p}a_{j,p}=n^{-o(1)}$, and
$\beta_i\ge n^{-1/16}$.
Moreover, $\log(2P/\zeta)=Q^2+300\log n+\log(2P)$.
Thus the error bound in Theorem~\ref{out:profile} satisfies
\[
 \max_{p\in\mathcal P}|E_p|
 \le\frac{C\bigl(\sqrt{h+1}\,Q
   +\sqrt{Q^2+300\log n+\log(2P)}\bigr)}
   {\sqrt n\,\min_{j,p}a_{j,p}}
 =n^{-1/2+o(1)}
 \le\frac{n^{-1/16}}{20}
 \le\frac{\beta_i}{20}
\]
outside an event of probability at most $\zeta$.
The theorem applies because $q<Q$ and $\log(1/\zeta)=n^{o(1)}$.

Work on this event and suppose that $G_i(Z)>0$.
Lemma~\ref{hf:gate-prefix}(1) gives $C_i(Z)=1$ and a degreewise
normalized energy error less than $\epsilon_g\beta_i$.
Combining this with the profile formula
\eqref{out:profile-formula} yields
\[
 \Bigl|p\sum_{t=1}^qB_tT_t^{p-1}+\beta_i\Bigr|
 \le\Bigl(\epsilon_g+\frac1{20}\Bigr)\beta_i
 <\frac{\beta_i}{10}
 \qquad(p\in\mathcal P).
\]
Terms with $C_i(y_t)=0$ have $T_t=0$ and contribute zero.
After removing them, every remaining node is
$T_t=\langle\psi_i(Z),\psi_i(y_t)\rangle$.
If no terms remain, the displayed inequality is impossible.
Otherwise, apply Lemma~\ref{out:two-directions} with
$w_*=\psi_i(Z)$, the remaining $w_j=\psi_i(y_t)$, and
$c=-\beta_i$.
It supplies the required two queried points.
\end{proof}

\subsection{Separated Queries See Independent Coefficients}
\label{out:locality}

Fix $1\le i\le h$.
We compare two simulations that measure the fields indexed by
$j<i$ at two groups of query points.
One simulation uses the same Gaussian coefficients for both groups,
the other an independent copy for each group.
Corollary~\ref{out:ancestor-comparison} shows that their full
transcripts have close laws when the compound directions $\psi_i$ of
points in different groups are separated.
The probabilistic part is Theorem~\ref{gauss:decoupling}.
What remains is geometry: separation of the directions $\psi_i$
makes the two measurement spaces nearly orthogonal, because every
feature of a field below level $i$ contains a high tensor power of
$\psi_i$ (Lemma~\ref{out:ancestor-angle}).

\subsubsection{The Two Simulations}
\label{out:simulation-setup}

Write
\[
 \mathcal G_{<i}
 =\bigoplus_{j<i}\bigoplus_{p\in\mathcal P}
 \mathbb R^{\,n^{p\sum_{k=0}^{h-j}K^k}}
\]
for the coefficient space of the fields below level $i$, so that
$\Xi_{<i}=(\Xi_{j,p})_{j<i,\,p\in\mathcal P}\sim\mathcal N(0,I_{\mathcal G_{<i}})$
under $\mu_n$.
Each $\Phi_{j,p}(x)$ and $D_v\Phi_{j,p}(x)$ with $j<i$ is embedded in
$\mathcal G_{<i}$ by setting the other components to zero.

Consider an oracle simulation for an algorithm $\Alg$, in the sense
of Sections~\ref{gauss:observation-rules} and~\ref{gauss:simulations}, whose
hidden vector is $\Xi_{<i}$ and whose seed is independent of
$\Xi_{<i}$.
The seed may contain the coefficient vectors $\Xi_i,\ldots,\Xi_h$.
The simulation has $q$ rounds with batch sizes $m_1,\ldots,m_q$ and
$M=\sum_km_k$, a terminal batch being counted as a round as in
Remark~\ref{gauss:terminal}.
At round $k$, a point $y_k\in\mathbb R^{(h+1)n}$, which is the
current query or the output for a terminal batch, and a label
$\ell_k\in\{\mathrm{local},\mathrm{far}\}$ are selected as Borel
functions of the seed and the preceding measurements, and the batch
consists of vectors of the form $\Phi_{j,p}(y_k)$ or
$D_v\Phi_{j,p}(y_k)$ with $j<i$, together with zero vectors.
Only the response computed by the return map is supplied to $\Alg$.
In the shared model, all batches measure $\Xi_{<i}$.
In the independent model, local and far batches measure independent
copies $\Xi_{<i}^{\mathrm{local}}$ and $\Xi_{<i}^{\mathrm{far}}$,
both independent of the seed.
Both models use the same maps, evaluated on their own histories, and
$P_{\mathrm{sh}}$ and $P_{\mathrm{ind}}$ denote the laws of the full
transcript, as in Section~\ref{gauss:simulations}.

\begin{corollary}[\bf Decoupling of separated query groups]
\label{out:ancestor-comparison}
Let $s,t\ge1$ be integers with $K\ge2st$, let $r>0$, and put
$\rho=2\sqrt{st}\exp(-Kr^2/(4st))$.
Suppose that, for every seed value and every
$z\in\mathbb R^{m_1}\times\cdots\times\mathbb R^{m_q}$, the points
and labels evaluated along $z$ satisfy the following two conditions.
\begin{enumerate}[leftmargin=2em,label=(\roman*)]
\item At most $s$ rounds are labelled local and at most $t$ rounds
 are labelled far.
\item For every local round $k$ and far round $\ell$ with
 $C_i(y_k)C_i(y_\ell)>0$,
 \[
  \|\psi_i(y_k)-\psi_i(y_\ell)\|\ge r
  \qquad\text{and}\qquad
  \|\psi_i(y_k)+\psi_i(y_\ell)\|\ge r.
 \]
\end{enumerate}
If $\rho<\tfrac12$, then
\begin{equation}\label{out:ancestor-transcript-bound}
 \operatorname{TV}(P_{\mathrm{sh}},P_{\mathrm{ind}})\le\frac32\rho M,
\end{equation}
and the same bound holds for the laws of the algorithm's transcript
and of its output.
\end{corollary}

Only separation between the two groups is required.
Points within a group may coincide, and the queries may depend
adaptively on responses from both groups.
A test that depends on the hidden coefficients at the output, such
as the value of a gate, is included in the comparison by evaluating
the required measurements in a terminal batch, which is how the
corollary is used in Section~\ref{out:recursive-charge}.

\subsubsection{A Tensor Estimate}
\label{out:gaussian-comparison-section}

Let $E$ be a finite-dimensional Euclidean space.
For a nonempty finite set $A$ of unit vectors in $E$ and an integer
$N\ge1$, define
\[
 W_A^N
 =\operatorname{span}\Bigl\{
 v^{\otimes N},\
 \frac{d}{d\epsilon}(v+\epsilon e)^{\otimes N}\Big|_{\epsilon=0}:
 v\in A,\ e\in E\Bigr\},
\]
the span of the values and first derivatives of
$v\mapsto v^{\otimes N}$ at the points of $A$.

\begin{lemma}[\bf Angle between two tensor subspaces]
\label{out:two-jet}
Let $A,B$ be nonempty sets of $s,t$ unit vectors in $E$,
respectively, and let $r>0$.
Suppose that $\|a-b\|\ge r$ and $\|a+b\|\ge r$ for every $a\in A$
and $b\in B$.
Then, for every integer $N\ge8st$,
\begin{equation}\label{out:two-jet-bound}
 \|P_{W_A^N}P_{W_B^N}\|_{\mathrm{op}}
 \le2\sqrt{st}\,\exp\!\Bigl(-\frac{Nr^2}{16st}\Bigr).
\end{equation}
\end{lemma}

The proof is given in Appendix~\ref{out:tensor-spaces}.
For individual tensor powers, the decay in $N$ follows from
$\langle a^{\otimes N},b^{\otimes N}\rangle=\langle a,b\rangle^N$.
The lemma controls arbitrary linear combinations of values and
first derivatives, even when points within a group approach one
another and their measurement vectors become nearly linearly
dependent.
Both signs enter the separation condition because
$W^N_{\{a\}}=W^N_{\{-a\}}$.

\subsubsection{Locality and Proof of the Comparison}
\label{out:ancestor-comparison-proof}

The features of the fields below level $i$ contain
$\psi_i(x)^{\otimes N}$ with $N\ge4K$, so Lemma~\ref{out:two-jet}
applies to them with an exponent of order $K$.

\begin{lemma}[\bf Locality of the measurements of earlier fields]
\label{out:ancestor-angle}
Let $s,t\ge1$ be integers with $K\ge2st$, and let $r>0$.
Let $\mathcal A,\mathcal B\subset\mathbb R^{(h+1)n}$ be finite sets
with $|\mathcal A|\le s$ and $|\mathcal B|\le t$ such that
\[
 \|\psi_i(x)-\psi_i(y)\|\ge r
 \quad\text{and}\quad
 \|\psi_i(x)+\psi_i(y)\|\ge r
 \qquad\text{whenever }x\in\mathcal A,\ y\in\mathcal B,\
 C_i(x)C_i(y)>0.
\]
Put
\[
 \mathcal O_{\mathcal A}
 =\operatorname{span}\bigl\{\Phi_{j,p}(x),\ D_v\Phi_{j,p}(x):
 j<i,\ p\in\mathcal P,\ x\in\mathcal A,\ v\in\mathbb R^{(h+1)n}\bigr\}
 \subseteq\mathcal G_{<i},
\]
and define $\mathcal O_{\mathcal B}$ in the same way.
Then
\begin{equation}\label{out:ancestor-angle-bound}
 \|P_{\mathcal O_{\mathcal A}}P_{\mathcal O_{\mathcal B}}\|_{\mathrm{op}}
 \le2\sqrt{st}\,\exp\Bigl(-\frac{Kr^2}{4st}\Bigr).
\end{equation}
\end{lemma}

\begin{proof}[Proof of Lemma~\ref{out:ancestor-angle}]
Points with vanishing cutoff contribute nothing.
If $C_j(x)=0$ for some $j<i$, then $\nabla C_j(x)=0$ because $C_j$
is smooth and nonnegative, so $C_j(x+uv)=o(|u|)$ as $u\to0$ for
every direction $v$, and the bound
$\|\Phi_{j,p}(x+uv)\|\le\sqrt n\,a_{j,p}C_j(x+uv)$ gives
$\Phi_{j,p}(x)=0$ and $D_v\Phi_{j,p}(x)=0$.
Since the product defining $C_j$ contains every factor of $C_i$,
this applies to every point with $C_i(x)=0$.
We may therefore remove such points from $\mathcal A$ and
$\mathcal B$.
If either set becomes empty, the corresponding space is $\{0\}$ and
there is nothing to prove.
For the remaining points, put $A=\{\psi_i(x):x\in\mathcal A\}$ and
$B=\{\psi_i(y):y\in\mathcal B\}$, so that $|A|\le s$, $|B|\le t$, and
$\|a-b\|\ge r$, $\|a+b\|\ge r$ for all $a\in A$ and $b\in B$.

Fix $j<i$ and $p\in\mathcal P$, and put $N=pK^{i-j}$.
At a point $x$ with $C_j(x)>0$, all blocks $x_j,\ldots,x_h$ are
nonzero, and after a permutation of the tensor factors that does not
depend on $x$,
\[
 \Phi_{j,p}(x)
 =\sqrt n\,a_{j,p}\,C_j(x)
 \Bigl(\bigotimes_{m=j}^{i-1}
 \Bigl(\frac{x_m}{\|x_m\|}\Bigr)^{\otimes pK^{m-j}}\Bigr)
 \otimes\psi_i(x)^{\otimes N},
\]
because $\psi_j^{\otimes p}$ contains $pK^{m-j}=NK^{m-i}$ copies of
$x_m/\|x_m\|$ for every $m\ge i$.
Let $\mathcal R_{j,p}$ be the tensor space containing the factor in
parentheses.
Differentiating the cutoff or that factor leaves
$\psi_i(x)^{\otimes N}$ unchanged, and differentiating the last
factor gives, by the chain rule, a first derivative of
$u\mapsto u^{\otimes N}$ at $u=\psi_i(x)$ in the direction
$D_v\psi_i(x)$.
Hence
$\Phi_{j,p}(x),D_v\Phi_{j,p}(x)\in\mathcal R_{j,p}\otimes W^N_{\{\psi_i(x)\}}$
for every $v$.
Every measurement vector lies in a single coordinate block of
$\mathcal G_{<i}$, so $\mathcal O_{\mathcal A}$ is the orthogonal
direct sum of its components $\mathcal O^{(j,p)}_{\mathcal A}$ over
the blocks $(j,p)$, and the inclusion above gives
$\mathcal O^{(j,p)}_{\mathcal A}\subseteq\mathcal R_{j,p}\otimes W^N_A$
and
$\mathcal O^{(j,p)}_{\mathcal B}\subseteq\mathcal R_{j,p}\otimes W^N_B$.

Since $p\ge4$, $j<i$, and $K\ge2st$, we have $N\ge4K\ge8st$, and
Lemma~\ref{out:two-jet} gives
\[
 \|P_{W^N_A}P_{W^N_B}\|_{\mathrm{op}}
 \le2\sqrt{|A||B|}\exp\!\Bigl(-\frac{Nr^2}{16|A||B|}\Bigr)
 \le2\sqrt{st}\exp\!\Bigl(-\frac{Kr^2}{4st}\Bigr),
\]
where the second inequality uses $|A|\le s$, $|B|\le t$, and
$N\ge4K$.
The projection onto $\mathcal R_{j,p}\otimes W^N_A$ is
$I_{\mathcal R_{j,p}}\otimes P_{W^N_A}$, and similarly for $B$, so
tensoring with this identity preserves the norm of the product of
the two projections, and passing to subspaces does not increase it.
The same bound therefore holds for
$\|P_{\mathcal O^{(j,p)}_{\mathcal A}}
P_{\mathcal O^{(j,p)}_{\mathcal B}}\|_{\mathrm{op}}$.
Finally, the norm of the product of the projections onto two
orthogonal direct sums is the maximum of the norms over the blocks.
This proves \eqref{out:ancestor-angle-bound}.
\end{proof}

\begin{proof}[Proof of Corollary~\ref{out:ancestor-comparison}]
Fix a seed value and $z$, and let $L$ and $F$ be the spans of the
local and far measurement vectors along $z$, as in
Section~\ref{gauss:simulations}.
Let $\mathcal A$ and $\mathcal B$ be the sets of points of the local
and far rounds.
Then $|\mathcal A|\le s$ and $|\mathcal B|\le t$ by (i), and (ii) is
the separation hypothesis of Lemma~\ref{out:ancestor-angle}, so
$\|P_{\mathcal O_{\mathcal A}}P_{\mathcal O_{\mathcal B}}\|_{\mathrm{op}}\le\rho$.
Since $L\subseteq\mathcal O_{\mathcal A}$ and
$F\subseteq\mathcal O_{\mathcal B}$,
\[
 \|P_LP_F\|_{\mathrm{op}}
 =\|P_LP_{\mathcal O_{\mathcal A}}P_{\mathcal O_{\mathcal B}}P_F\|_{\mathrm{op}}
 \le\|P_{\mathcal O_{\mathcal A}}P_{\mathcal O_{\mathcal B}}\|_{\mathrm{op}}
 \le\rho.
\]
This holds for every seed value and every $z$.
The simulation is a measurement process of
Definition~\ref{gauss:measurement-process} with batch sizes
$m_1,\ldots,m_q$, so Theorem~\ref{gauss:decoupling} applies and
gives all three bounds.
\end{proof}
\subsection{Recursive Query Lower Bound}
\label{out:recursive-charge}

We prove Theorem~\ref{hf:output-fails} by induction on $0\le i\le h$
over the prefixes~$V_i^\Xi$.
Specifically, for $Q_i=2^{i+1}$, any algorithm making fewer than~$Q_i$
queries to~$V_i^\Xi$ passes gate~$i$ with probability at most~$e_i$,
where $e_i$ is defined below and satisfies $e_h\le n^{-10}$.
Since $Q_h=Q$ and $V_h^\Xi=V^\Xi$, the case $i=h$ yields the theorem.
Throughout the proof, the coefficients retain their unconditioned
Gaussian law.

The induction step has two ingredients.
Proposition~\ref{out:source-separation}, applied to a simulation
that queries the field of level $i$ only at the active points of
the run, shows that passing gate $i$ requires two active queries
whose directions $\psi_i$ are separated.
A graph on the query indices, with edges between queries of nearby
directions, places these two queries in distinct components.
Corollary~\ref{out:ancestor-comparison} then shows that each
component is, up to a small decoupling error, an algorithm
querying the preceding prefix $V_{i-1}^\Xi$, so by the induction
hypothesis each active component contains at least $Q_{i-1}$
queries before its first active one.
Two disjoint components therefore require $2Q_{i-1}=Q_i$ queries.

Section~\ref{out:query-components} defines the parameters, the query
graph, and the bad event~$\mathsf B_i(T)$.
Section~\ref{out:two-components} shows that passing~$G_i$ requires two
distinct active components with high probability.
Section~\ref{out:component-bound-proof} applies the preceding-level
bound to show that each active component must make at least~$T$
queries before its first active point, up to a small failure
probability.
Section~\ref{out:recursive-query-bound} combines these bounds to obtain
$Q_i=2Q_{i-1}$ and completes the induction.

\subsubsection{Setup}
\label{out:query-components}

Throughout this section, $i$ denotes a fixed level,
$0\le i\le h$, and $\Alg$ a randomized algorithm making at most
$q\le Q$ first-order queries to $V_i$, with a seed $\omega$
independent of $\Xi_{\le i}=(\Xi_0,\ldots,\Xi_i)$ and with
output $Z$.
We may assume that every run makes exactly $q$ queries: after the
output has been fixed, pad a shorter run with queries at the
origin.
Since $C_j(0)=0$ for every $j$, all fields, gates, and loads
vanish at the origin together with their gradients, so every
padding response is $(0,0)$, whatever $\Xi$ is, and the output is
unchanged.
Probabilities are taken under $\mu_n$ and the law of the seed.

\paragraph{Parameters.}
Recall that $Q=2^{h+1}$, $P=Q^{64}$, and $K=Q^{74}$, so that
$\log Q=\Theta(\log^{1/3} n)$ and $P,Q,h=n^{o(1)}$.
Let $\zeta=e^{-Q^2}n^{-300}$ be the exceptional probability in
Proposition~\ref{out:source-separation}, and put
\[
 Q_i=2^{i+1}\quad(0\le i\le h),
 \qquad
 r=\Bigl(\frac{1204\,(Q+1)^2(Q^2+\log n)}{K}\Bigr)^{1/2},
\]
\[
 M=|\mathcal P|\bigl[Q(h+1)(d_n+1)+1\bigr],
 \qquad
 \delta=\frac32\zeta M,
 \qquad
 e_0=\zeta,
 \quad
 e_i=\zeta+2^Q(e_{i-1}+\delta)\quad(1\le i\le h).
\]
Here $Q_i$ is the level-$i$ query threshold, and $e_i$ bounds
the probability of passing~$G_i$ below this threshold.
The edge threshold~$r$ keeps query components small while
ensuring sufficient separation for decoupling.
Each comparison uses at most~$M$ scalar measurements and
incurs total variation error at most~$\delta$.
The recursion adds the separation failure probability~$\zeta$
to a union bound over at most~$2^Q$ candidate simulations,
each contributing~$e_{i-1}+\delta$.

\begin{lemma}[\bf Parameter bounds]
\label{out:parameter-bounds}
For all sufficiently large $n$:
\begin{enumerate}[leftmargin=2em,label=(\roman*)]
\item $K\ge2(Q+1)^2$ and
 $2(Q+1)e^{-\frac{Kr^2}{4(Q+1)^2}}\le\zeta<\tfrac12$;
\item $Qr<P^{-1/2}$;
\item $M\le n^2$ and $e_h\le n^{-10}$.
\end{enumerate}
\end{lemma}

\begin{proof}[Proof of Lemma~\ref{out:parameter-bounds}]
(i) $K=Q^{74}\ge2(Q+1)^2$ because $Q\ge2$.
By the definition of $r$,
\[
 2(Q+1)\exp\Bigl(-\frac{Kr^2}{4(Q+1)^2}\Bigr)
 =\exp\bigl(\log(2(Q+1))-301(Q^2+\log n)\bigr)
 \le e^{-Q^2}n^{-300}=\zeta,
\]
since $\log(2(Q+1))\le300Q^2$, and $\zeta<\tfrac12$.

(ii) Using $P=Q^{64}$, $K=Q^{74}$, and $(Q+1)^2\le4Q^2$,
\[
 P(Qr)^2
 =\frac{1204\,PQ^2(Q+1)^2(Q^2+\log n)}{K}
 \le4816\bigl(Q^{-4}+(\log n)Q^{-6}\bigr)
 \longrightarrow0,
\]
because $\log\log n-6\log Q\to-\infty$.
Hence $Qr<P^{-1/2}$ for all sufficiently large $n$.

(iii) Since $|\mathcal P|\le P$ and $d_n+1\le2(h+1)n$,
$M\le P\bigl[2Q(h+1)^2n+1\bigr]=n^{1+o(1)}\le n^2$.
Solving the recursion gives, for $h\ge1$,
\[
 e_h
 =2^{Qh}\zeta+(\zeta+2^Q\delta)\sum_{k=0}^{h-1}2^{Qk}
 \le2^{Qh}\zeta+2\cdot2^{Q(h-1)}
   \Bigl(\zeta+\frac32\,2^QM\zeta\Bigr)
 \le3(1+M)2^{Qh}\zeta,
\]
and the final bound also holds for $h=0$.
Since $h/Q\to0$, we have $2^{Qh}=e^{o(Q^2)}\le e^{Q^2/2}$ for
large $n$, and therefore
$e_h\le6n^2e^{Q^2/2}e^{-Q^2}n^{-300}\le n^{-10}$.
\end{proof}

\paragraph{Active points and the query graph.}
Recall that $A_0=1$ and $A_i=\prod_{j<i}G_j$.
A point $y$ is \emph{active} if $A_i(y)>0$ and \emph{inactive} otherwise. 
For
$i\ge1$, an active point passes gate $i-1$, since $A_i(y)>0$
forces $G_{i-1}(y)>0$.
By Lemma~\ref{hf:gate-prefix}(3), $V_i$ and $V_{i-1}$ have the
same value and gradient at every inactive point, and by
Lemma~\ref{hf:gate-prefix}(4), at an active point the response of
$V_i$ is determined by $\Xi_{<i}$ and one first-order query to the
field of level $i$.
For points $x,y$ with $C_i(x)C_i(y)>0$, put
\[
 d_i(x,y)=\min\{\|\psi_i(x)-\psi_i(y)\|,\|\psi_i(x)+\psi_i(y)\|\}.
\]
This is a pseudometric on $\{x:C_i(x)>0\}$.
For the triangle inequality, multiply the signs attaining the
minima for $(x,y)$ and $(y,z)$.
The condition $d_i(x,y)\ge r$ is exactly the separation condition
of Corollary~\ref{out:ancestor-comparison}.
For a run of $\Alg$ with queries $y_1,\ldots,y_q$, each
level~$i$ induces a graph
on $\{1,\ldots,q\}$ in which distinct indices $s,s'$ are joined by
an edge whenever $C_i(y_s)C_i(y_{s'})>0$ and $d_i(y_s,y_{s'})<r$,
and call its connected components simply components. 
Queries with
$C_i=0$ are isolated vertices, and the graph depends on the whole
run.
A component is active if it contains an active query, and for an
active component $\mathcal C$ we write
$\tau_{\mathcal C}=\min\{s\in\mathcal C:A_i(y_s)>0\}$ for its
first active index.

For an integer $T\ge1$, let $\mathsf B_i(T)$ be the bad event
that some active component contains fewer than~$T$ queries
before its first active query, i.e.,
\[
 \mathsf B_i(T)
 :=
 \left\{
 \exists\text{ active component }\mathcal C:
 \#\{s\in\mathcal C:s<\tau_{\mathcal C}\}<T
 \right\}.
\]
Two indices in a component are joined by a path of at most $q-1$
edges, so two queries with positive cutoffs in the same component
satisfy $d_i<(q-1)r<P^{-1/2}$ by
Lemma~\ref{out:parameter-bounds}(ii).

\subsubsection{Two Active Queries in Distinct Components}
\label{out:two-components}

Proposition~\ref{out:source-separation} concerns algorithms that
query the field of level $i$ directly.
The next lemma transfers it to $\Alg$, which queries $V_i$, and
states its conclusion in terms of the query graph.
\begin{lemma}[\bf Two active queries in distinct components]
\label{out:two-components-lemma}
Suppose $q<Q$.
For all sufficiently large $n$, with probability at least
$1-\zeta$, the following holds: if $G_i(Z)>0$, then two active
queries of $\Alg$ lie in distinct components.
\end{lemma}

\begin{proof}[Proof of Lemma~\ref{out:two-components-lemma}]
Let $\Alg^{(i)}$ be the algorithm with seed $(\omega,\Xi_{<i})$
and first-order oracle access to the field of level $i$ that runs
$\Alg$, answers each inactive query from its seed, which
determines the response of $V_{i-1}$, makes one oracle query at
each active query and answers with the response of $V_i$ by
Lemma~\ref{hf:gate-prefix}(4), and outputs $Z$.
Its seed is independent of $\Xi_i$, it makes at most $q<Q$ oracle
queries, and its queries are the active queries of $\Alg$.
By Proposition~\ref{out:source-separation}, with probability at
least $1-\zeta$, if $G_i(Z)>0$ then two of its queries $y,y'$
satisfy $C_i(y)C_i(y')>0$ and $d_i(y,y')\ge P^{-1/2}$.
By the diameter bound of Section~\ref{out:query-components}, such
queries lie in distinct components, and both components are
active.
\end{proof}

In the induction step, the lemma supplies two active components
with disjoint index sets, each of which will be charged
$Q_{i-1}$ queries by Lemma~\ref{out:component-charge}.
For $i=0$, every point is active, and the lemma says that passing
gate $0$ requires at least two queries.

\subsubsection{Charging a Component to the Preceding Prefix}
\label{out:component-bound-proof}


To turn the two-component requirement into a recursive query
bound, we must control the number of queries in each component
before its first active point.
The next lemma uses the preceding-level hypothesis to bound
the probability that any active component has fewer than~$T$
such queries.

\begin{lemma}[\bf Charging a component to the preceding prefix]
\label{out:component-charge}
Let $i\ge1$, let $T\ge1$ be an integer, and let $e\ge0$.
Suppose that 
 every randomized 
algorithm making fewer than $T$
first-order queries to $V_{i-1}$
with 
a seed independent of $\Xi_{<i}$ and 
an arbitrary
output $Z'$,
satisfies
$\mathbb P\{G_{i-1}(Z')>0\}\le e$.
Then $\mathbb P(\mathsf B_i(T))\le2^Q(e+\delta)$.
\end{lemma}

\begin{proof}[Proof of Lemma~\ref{out:component-charge}]
On~$\mathsf B_i(T)$, there is an active component~$\mathcal C$
such that, writing $\tau=\tau_{\mathcal C}$ and
$S=\{s\in\mathcal C:s<\tau\}$, we have $|S|<T$.
Because component membership may depend on future queries,
it cannot be used directly as an online labeling rule.
We therefore construct a simulation for each fixed candidate
pair~$(\tau,S)$ and prove two facts about these simulations.

\paragraph{The candidate simulation.}
Fix $\tau\in\{1,\dots,q\}$ and $S\subseteq\{1,\dots,\tau-1\}$ with
$\vert{}S\vert{}<T$.
Let $\mathrm{Alg}_\tau$ execute the first $\tau-1$ queries of
$\mathrm{Alg}$ and output the $\tau$-th point it would query next,
without receiving the $\tau$-th reply.
We simulate $\mathrm{Alg}_\tau$ with hidden vector~$\Xi_{<i}$ and
independent seed $(\omega,\Xi_i)$.
The indices in $S\cup\{\tau\}$ are labeled $\mathsf{local}$, and the
remaining indices in $\{1,\dots,\tau-1\}\setminus S$ are labeled
$\mathsf{far}$.

At each round $k=1,\dots,\tau$, if the simulation has not
failed, we check whether the proposed point and any earlier
admitted point of the opposite label both have positive
cutoffs~$C_i$ and lie at $d_i$-distance less than~$r$.
Here the proposed point at $k=\tau$ is the output
of~$\mathrm{Alg}_\tau$.
If no such pair exists, the proposed point is admitted as~$x_k$.
Otherwise, the simulation enters an absorbing failed state:
the current and all subsequent rounds use $x_k=0$ and
zero measurement batches, with reply~$(0,0)$ at
nonterminal rounds.

At an admitted round $k<\tau$, we record the measurements
of~$\Xi_{<i}$ along $\Phi_{j,p}(x_k)$ and
$D_{e_a}\Phi_{j,p}(x_k)$ for $j<i$, $p\in\mathcal P$,
and $1\le a\le d_n$.
By Lemma~\ref{hf:representers}, these determine the values
and gradients of~$V_{i-1}$ and~$A_i$ at~$x_k$.
At a $\mathsf{local}$ round ($k\in S$), we return the value
and gradient of~$V_{i-1}$ at~$x_k$.
At a $\mathsf{far}$ round ($k\notin S$), we combine the
recorded batch with the seed vector~$\Xi_i$ to return
the value and gradient of~$V_i$ at~$x_k$, using
Lemma~\ref{hf:gate-prefix}(4).

At the terminal round~$\tau$, if the proposed point is admitted
as~$x_\tau$, we record the measurements along
$\Phi_{i-1,p}(x_\tau)$ for $p\in\mathcal P$.
This terminal batch determines $G_{i-1}(x_\tau)$ but is
not returned to~$\mathrm{Alg}_\tau$, as in
Remark~\ref{gauss:terminal}.
Let $E_{\tau,S}:=\{G_{i-1}(x_\tau)>0\}$.
Since $G_{i-1}(0)=0$, this event can occur only if
all~$\tau$ checks pass.

For each candidate~$(\tau,S)$, let $\mathbb P_{\mathrm{sh}}$
and $\mathbb P_{\mathrm{ind}}$ denote the full-transcript
laws under the shared and independent coefficient models
of Section~\ref{gauss:simulations}.
The shared model measures every batch against the same
vector~$\Xi_{<i}$.
The independent model uses two independent copies:
$\mathsf{local}$ batches, including the terminal batch,
are measured against~$\Xi_{<i}^{\mathrm{local}}$, and
$\mathsf{far}$ batches against~$\Xi_{<i}^{\mathrm{far}}$.
The seed remains~$(\omega,\Xi_i)$ in both models.
In particular, in the independent model the terminal gate
$G_{i-1}(x_\tau)$ is evaluated using
$\Xi_{<i}^{\mathrm{local}}$.

\begin{fact}[\bf Bound for a fixed candidate]
\label{fact7}
For every candidate~$(\tau,S)$,
\begin{equation}\label{out:fixed-reference-bound}
 \mathbb P_{\mathrm{sh}}(\mathsf E_{\tau,S})\le e+\delta.
\end{equation}
\end{fact}

\begin{proof}[Proof of Fact~\ref{fact7}]
The batch sizes are
$m_k=i|\mathcal P|(d_n+1)$ for $k<\tau$ and
$m_\tau=|\mathcal P|$, including after failure.
Thus the total number of scalar measurements is at most~$M$.
The checks and measurement vectors are Borel functions of
the seed and preceding measurements, while the returned
replies are Borel functions of the seed and the measurements
recorded so far.
The terminal batch makes~$\mathsf E_{\tau,S}$ measurable
from the full transcript.
Hence the construction is an oracle simulation of the form
in Section~\ref{out:simulation-setup}.

There are $\vert{}S\vert{}+1\le Q$ rounds labeled
$\mathsf{local}$ and $\tau-1-\vert{}S\vert{}\le Q$ rounds
labeled $\mathsf{far}$.
%
%
Every round at or after the first failed check uses $x_k=0$,
with $C_i(0)=0$.
Thus, any pair of opposite-label points with
$C_i(x_k)C_i(x_\ell)>0$ consists of admitted points and
satisfies $d_i(x_k,x_\ell)\ge r$.
This separation holds on every history, without conditioning
on passing the checks.
By Lemma~\ref{out:parameter-bounds}(i),
$K\ge2(Q+1)^2$ and
$\rho=2(Q+1)e^{-\frac{Kr^2}{4(Q+1)^2}}
\le\zeta<\frac12$.
Corollary~\ref{out:ancestor-comparison} therefore applies
with both group-size bounds equal to~$Q+1$, yielding
$\mathrm{TV}(\mathbb P_{\mathrm{sh}},\mathbb P_{\mathrm{ind}})
\le\frac32\zeta M=\delta$.
In particular,
\begin{equation}\label{out:fixed-reference-comparison}
  \mathbb P_{\mathrm{sh}}(E_{\tau,S})
  \le \mathbb P_{\mathrm{ind}}(E_{\tau,S})+\delta.
\end{equation}

In the independent model, we implement the simulation as
a randomized algorithm querying the prefix potential~$V_{i-1}$
formed from~$\Xi_{<i}^{\mathrm{local}}$.
Its enlarged seed
$\hat\omega=(\omega,\Xi_i,\Xi_{<i}^{\mathrm{far}})$
is independent of~$\Xi_{<i}^{\mathrm{local}}$.
The separation checks use only the current and earlier point
coordinates, and every $\mathsf{far}$ reply can be computed
from~$\hat\omega$ and the current point.
Each admitted $\mathsf{local}$ round $k\in S$ requires
one first-order query to the local prefix potential.
The query choices, checks, and output therefore depend only
on~$\hat\omega$ and these oracle replies, without using
the auxiliary degreewise measurements.
This algorithm makes at most $\vert{}S\vert{}<T$
first-order queries and outputs $Z'=x_\tau$ before
the terminal batch is recorded.
Its output passes the gate~$G_{i-1}$ formed from
$\Xi_{<i}^{\mathrm{local}}$ precisely on~$E_{\tau,S}$.
The hypothesis of Lemma~\ref{out:component-charge} gives
$\mathbb P_{\mathrm{ind}}(E_{\tau,S})\le e$.
Combining this with \eqref{out:fixed-reference-comparison}
proves \eqref{out:fixed-reference-bound}.
\end{proof}

\begin{fact}[\bf Covering the bad event]
\label{fact8}
Couple all shared-model simulations to the original run
using the same coefficients~$\Xi$ and seed~$\omega$.
Then
\[
 \mathsf B_i(T)
 \subseteq
 \bigcup_{\tau=1}^q
 \ \bigcup_{\substack{
 S\subseteq\{1,\ldots,\tau-1\}\\ |S|<T}}
 \mathsf E_{\tau,S}.
\]
\end{fact}

\begin{proof}[Proof of Fact~\ref{fact8}]
Fix a realization of~$\mathsf B_i(T)$ and choose an active
component~$\mathcal C$ with fewer than~$T$ queries before
its first active index.
Set $\tau=\tau_{\mathcal C}$ and
$S=\{s\in\mathcal C:s<\tau\}$.
Then $|S|<T$, and every query indexed by~$S$ is inactive,
so its replies from~$V_i$ and~$V_{i-1}$ coincide.

We show inductively that the shared-model simulation for this candidate
$(\tau,S)$ agrees with the original run without failure through
index~$\tau$.
Suppose the two runs have agreed through index $k-1$, where $k\le\tau$,
so that the next proposed point is the original query~$y_k$.
Since $\mathsf{local}$ indices belong to~$\mathcal C$ and $\mathsf{far}$
indices lie outside~$\mathcal C$, no opposite-label pair through
index~$k$ can have positive cutoffs and $d_i$-distance less than~$r$, as
such a pair would form an edge between~$\mathcal C$ and its complement.
Thus the separation check at round~$k$ passes and $x_k=y_k$.
For $k<\tau$, a $\mathsf{local}$ reply agrees with the original reply by
inactivity, and a $\mathsf{far}$ reply agrees by construction,
completing the induction.

In particular, at $k=\tau$, the check passes and $x_\tau=y_\tau$.
Since $\tau=\tau_{\mathcal C}$ is active, we have $A_i(y_\tau)>0$, which
implies $G_{i-1}(x_\tau)=G_{i-1}(y_\tau)>0$.
Thus $E_{\tau,S}$ occurs, proving the inclusion.
\end{proof}

There are at most $\sum_{\tau=1}^q 2^{\tau-1}=2^q-1<2^Q$ candidate
pairs.
Combining Facts~7 and~8 with a union bound yields
$\mathbb P(\mathsf B_i(T))\le 2^Q(e+\delta)$.
\end{proof}

\subsubsection{Proof of Theorem~\ref{hf:output-fails}}
\label{out:recursive-query-bound}
We now combine Lemmas~\ref{out:two-components-lemma}
and~\ref{out:component-charge}.
The thresholds on $n$ in these lemmas and in
Lemma~\ref{out:parameter-bounds} do not depend on $i$, so a single
$n_0$ serves for all levels.
\begin{proof}[Proof of Theorem~\ref{hf:output-fails}]
We show by induction on $i$ that, for all sufficiently large $n$,
every algorithm making fewer than $Q_i$ queries to $V_i$, with a
seed independent of $\Xi_{\le i}$ and an arbitrary output $Z$,
satisfies $\mathbb P\{G_i(Z)>0\}\le e_i$.
By the standing assumptions, we may take $q=Q_i-1$, which is less
than $Q$.

For $i=0$, Lemma~\ref{out:two-components-lemma} shows that,
except on an event of probability at most $\zeta=e_0$, passing
gate $0$ requires two queries in distinct components, hence
$q\ge2$, whereas $q=Q_0-1=1$.
For $i\ge1$, assume the claim at level $i-1$.
By Lemma~\ref{out:two-components-lemma}, except on an event of
probability at most $\zeta$, passing gate $i$ requires two active
queries in distinct components $\mathcal C_1$ and $\mathcal C_2$.
By Lemma~\ref{out:component-charge} with $T=Q_{i-1}$ and
$e=e_{i-1}$, whose hypothesis is the induction hypothesis, except
on an event of probability at most $2^Q(e_{i-1}+\delta)$, every
active component contains at least $Q_{i-1}$ indices preceding
its first active index.
Outside the union of these two events, passing gate $i$ would
give
\[
 q\ge\#\{s\in\mathcal C_1:s<\tau_{\mathcal C_1}\}
 +\#\{s\in\mathcal C_2:s<\tau_{\mathcal C_2}\}
 \ge2Q_{i-1}=Q_i,
\]
contradicting $q=Q_i-1$.
Hence $\mathbb P\{G_i(Z)>0\}\le\zeta+2^Q(e_{i-1}+\delta)=e_i$.

At level $h$, we have $Q_h=Q$, $V_h=V$, and $e_h\le n^{-10}$ by
Lemma~\ref{out:parameter-bounds}(iii).
Since passing the energy test implies $G_h(X)=1$
(Section~\ref{sec:energy-test}), this proves
\eqref{hf:output-test-bound}.
The output was arbitrary and need not have been queried.
\end{proof}

\section{The Target Passes the Energy Test}
\label{sec:target-good-mass}

Throughout this section, expectations over the coefficients are taken
under the original product Gaussian law $\mu_n$, without conditioning
on admissibility. We prove the following estimate for the energy test
$\mathcal T_n$ of Definition~\ref{hf:energy-tests}, where
$\mathcal T_n(\Xi)=\{x:(\Xi,x)\in\mathcal T_n\}$.

\begin{theorem}[\bf The target passes the energy test]
\label{thm:target-good-mass}
For all sufficiently large $n$,
\begin{equation}
  \mathbb E_{\Xi\sim\mu_n}\,
  \pi^\Xi\bigl(\mathcal T_n(\Xi)^c\bigr)
  \le\frac{C}{n},
  \label{eq:target-good-mass-conclusion}
\end{equation}
where $C$ is an absolute constant.
\end{theorem}

The argument proceeds as follows.
In Section~\ref{sec:target-good-event} we
define a good event on which every gate equals one, every energy
truncation is linear, and the energy test passes. On this event the
target weight is the exponential of a Gaussian linear form, so its
expectation over the coefficients can be computed by completing the
square. In Section~\ref{sec:target-first-moments} we show that the
expected weight outside the good event is exponentially small, and in
Section~\ref{sec:target-second-moment} that the weight inside
concentrates around its expectation, by a second-moment computation
that reduces to the overlap of two independent random directions on
the sphere. In Section~\ref{sec:target-proof} we pass from these
estimates on the unnormalized weight to the target probability of the
bad event, which is a ratio of two such weights, by an elementary
inequality.
\subsection{The Good Event}
\label{sec:target-good-event}

Recall that $B(x)=\frac34\|x\|^2$, that $A_i^\Xi=\prod_{j<i}G_j^\Xi$
is the product of the gates below level $i$, and that
$V^\Xi=B+\sum_{i=0}^hA_i^\Xi\,U_i(\langle\Xi_i,\Phi_i\rangle)$. We
abbreviate the fields by
$H_{i,p}^\Xi(x):=\langle\Xi_{i,p},\Phi_{i,p}(x)\rangle$ and
$H_i^\Xi:=\sum_{p\in\mathcal P}H_{i,p}^\Xi$, and suppress $\Xi$ when
convenient. Let $\gamma_n:=\mathcal N(0,\frac23I_{d_n})$ be the target
distribution of $B$. Then the
target of $V^\Xi$ is $\pi^\Xi(dx)=\frac{w(\Xi,x)}{Z(\Xi)}\gamma_n(dx)$
with
\begin{equation}
  w(\Xi,x):=\exp\Bigl[-\sum_{i=0}^hA_i^\Xi(x)\,U_i\bigl(H_i^\Xi(x)\bigr)\Bigr],
  \qquad
  Z(\Xi):=\int w(\Xi,x)\,d\gamma_n(x).
  \label{eq:target-good-mass-weight}
\end{equation}
The bounded truncations and $0\le A_i^\Xi\le1$ give $0<Z(\Xi)<\infty$.

The weight is explicit wherever all gates equal one and all
truncations are inactive. The good event is a set of pairs $(\Xi,x)$
on which this happens.
\begin{definition}[\bf Good event]
\label{def:target-good-event}
Let $\mathcal R$ be the event that
$r_j(x):=\frac{\|x_j\|}{\sqrt n}\in[0.75,0.9]$ for every $0\le j\le h$,
and for each level let
\[
  \mathcal D_i:=\Bigl\{(\Xi,x):
  \max_{p\in\mathcal P}\Bigl|\frac{H_{i,p}^\Xi(x)}{na_{i,p}^2}+\beta_i\Bigr|
  \le\frac{\epsilon_g\beta_i}{4}\Bigr\}.
\]
The good event is
$\mathcal F_n:=\{(\Xi,x):x\in\mathcal R\}\cap\bigcap_{i=0}^h\mathcal D_i$,
with sections $\mathcal F_n(\Xi):=\{x:(\Xi,x)\in\mathcal F_n\}$.
\end{definition}
Set $v_i:=\sum_{p\in\mathcal P}a_{i,p}^2$ and
$t_i:=\frac{n\beta_i}{\lambda}$, so that $U_i(s)=\beta_is$ for
$|s|\le t_i$. By the coefficient normalization,
$v_i\le\frac1{\log8}\sum_{p\ge4}p^{-8}\le\frac1{3\log8}<1$.

\begin{fact}[\bf Properties of the good event]
\label{fact:target-good-event}
On $\mathcal F_n$, every gate equals one, $|H_i|<t_i$ for every $i$,
and the energy test passes. Consequently
\begin{equation}
  \mathcal F_n\subseteq\mathcal T_n,
  \qquad
  w(\Xi,x)=\exp\Bigl[-\sum_{i=0}^h\beta_iH_i^\Xi(x)\Bigr]
  \quad\text{on }\mathcal F_n .
  \label{eq:target-good-mass-linear-on-good}
\end{equation}
\end{fact}

\begin{proof}
On $\mathcal R$ all radial cutoffs $C_i$ and all radial factors $J$
equal one, so the conditions $\mathcal D_i$ are the hypotheses of the
gate activation property, Lemma~\ref{hf:gate-prefix}(2), and
$G_i=1$ for every $i$. Next, on $\mathcal D_i$ each $|H_{i,p}|$ is at
most $na_{i,p}^2\beta_i(1+\frac{\epsilon_g}{4})$, and summing over
$p$,
\[
  |H_i|\le n\beta_iv_i\Bigl(1+\frac{\epsilon_g}{4}\Bigr)
  =t_i\lambda v_i\Bigl(1+\frac{\epsilon_g}{4}\Bigr)<t_i,
\]
because $v_i<1$, $\lambda\le\lambda_*\le\frac12$, and
$\epsilon_g=\frac1{100}$. Finally, $x\in\mathcal R$ gives
$x\in\mathcal A_n$, and since $C_h=1$ on $\mathcal R$ the fields
$H_{h,p}$ are the raw angular fields of the energy test, so
$\mathcal D_h$ is the condition of the test and
$(\Xi,x)\in\mathcal T_n$. The first two assertions give $A_i=1$ and
$U_i(H_i)=\beta_iH_i$ in \eqref{eq:target-good-mass-weight}, which is
the formula for $w$.
\end{proof}

Split the normalizer along the good event,
\[
  Z_{\rm good}(\Xi):=\int w(\Xi,x)\mathbf 1_{\mathcal F_n}(\Xi,x)\,d\gamma_n(x),
  \qquad
  Z_{\rm bad}(\Xi):=\int w(\Xi,x)\mathbf 1_{\mathcal F_n^c}(\Xi,x)\,d\gamma_n(x),
\]
so that $Z=Z_{\rm good}+Z_{\rm bad}$. Since
$\mathcal F_n\subseteq\mathcal T_n$,
\begin{equation}
  \pi^\Xi\bigl(\mathcal T_n(\Xi)^c\bigr)
  \le\pi^\Xi\bigl(\mathcal F_n(\Xi)^c\bigr)
  =\frac{Z_{\rm bad}(\Xi)}{Z(\Xi)},
  \label{eq:target-good-mass-ratio}
\end{equation}
and the theorem follows from estimates on the two masses.

The natural scale for these masses is the expectation of the linear
weight in \eqref{eq:target-good-mass-linear-on-good}. At a fixed $x$,
the fields $H_{i,p}(x)$ are independent centered Gaussians with
variances $nC_i(x)^2a_{i,p}^2$ by Lemma~\ref{hf:covariance}, so
$\sum_i\beta_iH_i(x)$ is a centered Gaussian with variance
$n\sum_i\beta_i^2C_i(x)^2v_i$, and
\begin{equation}
  m(x):=\mathbb E_{\Xi\sim\mu_n}\exp\Bigl[-\sum_{i=0}^h\beta_iH_i^\Xi(x)\Bigr]
  =\exp\Bigl[\frac n2\sum_{i=0}^h\beta_i^2C_i(x)^2v_i\Bigr].
  \label{eq:target-good-mass-scale}
\end{equation}
Put $M:=\int m\,d\gamma_n$ and
$m_*:=\exp[\frac n2\sum_{i=0}^h\beta_i^2v_i]$. Since $0\le C_i\le1$
with $C_i=1$ on $\mathcal R$, we have $1\le m(x)\le m_*$, $M\le m_*$,
and $m(x)=m_*$ on $\mathcal R$. In the following we bound
$\mathbb EZ_{\rm bad}$ and $\mathbb EZ_{\rm good}$ in terms of $M$.

\subsection{First Moments of the Total and Good Masses}
\label{sec:target-first-moments}

We first compare the expected masses with $M$.

\begin{lemma}[\bf First-moment bounds]
\label{lem:target-good-mass-first}
For all sufficiently large $n$,
\begin{equation}
  \mathbb E_{\Xi\sim\mu_n}Z\le M,
  \qquad
  \mathbb E_{\Xi\sim\mu_n}Z_{\rm good}\ge(1-\eta_n)M,
  \qquad
  \mathbb E_{\Xi\sim\mu_n}Z_{\rm bad}\le\eta_nM,
  \label{eq:target-good-mass-first}
\end{equation}
where $0\le\eta_n\le e^{-n^{3/4}}$.
\end{lemma}
\begin{proof}[Proof of Lemma~\ref{lem:target-good-mass-first}]
Recall that the load of level $i$ is $A_i\,U_i(H_i^\Xi)$, where
$U_i(s)=\frac{n\beta_i^2}{\lambda}\varphi\bigl(\frac{\lambda s}{n\beta_i}\bigr)$
is the truncation of the linear load $\beta_is$ and $A_i\in[0,1]$ is
the product of the earlier gates. Since $\varphi$ is odd with
$|\varphi(t)|\le|t|$, the function $U_i$ is odd and
\[
  |U_i(y)|
  =\frac{n\beta_i^2}{\lambda}\Bigl|\varphi\Bigl(\frac{\lambda y}{n\beta_i}\Bigr)\Bigr|
  \le\frac{n\beta_i^2}{\lambda}\cdot\frac{\lambda|y|}{n\beta_i}
  =\beta_i|y|.
\]
The following fact compares the truncated load, with the gate frozen
at a value $a$, to the linear load.

\begin{fact}
\label{fact:target-truncation-comparison}
Let $Y$ be a centered Gaussian variable, possibly degenerate, and let
$a\in[0,1]$. Then
$\mathbb Ee^{-aU_i(Y)}\le\exp[\frac{\beta_i^2}{2}\operatorname{Var}(Y)]$.
\end{fact}

\begin{proof}[Proof of Fact~\ref{fact:target-truncation-comparison}]
Since $Y$ and $-Y$ have the same law and $U_i$ is odd,
$\mathbb Ee^{-aU_i(Y)}=\mathbb E\cosh(aU_i(Y))$. Since $\cosh$ is
increasing in the absolute value of its argument and
$|aU_i(y)|\le\beta_i|y|$, this is at most
$\mathbb E\cosh(\beta_iY)=\exp[\frac{\beta_i^2}{2}\operatorname{Var}(Y)]$,
where the equality is the Gaussian moment generating function.
\end{proof}

\paragraph{Proof of $\mathbb E_{\Xi\sim\mu_n}Z\le M$.}
Fix $x$. For $0\le k\le h$, conditionally on $\Xi_0,\dots,\Xi_{k-1}$
the factors $e^{-A_iU_i(H_i^\Xi)}$ with $i<k$ and the value
$A_k(x)\in[0,1]$ are fixed, while
$H_k^\Xi(x)\sim\mathcal N(0,nC_k(x)^2v_k)$ is independent of them.
Fact~\ref{fact:target-truncation-comparison} with $a=A_k(x)$ and
$Y=H_k^\Xi(x)$ therefore gives
\begin{align*}
  \mathbb E_{\Xi\sim\mu_n}\prod_{i\le k}e^{-A_iU_i(H_i^\Xi)}
  &=\mathbb E_{\Xi\sim\mu_n}\Bigl[\prod_{i<k}e^{-A_iU_i(H_i^\Xi)}\,
    \mathbb E\bigl[e^{-A_kU_k(H_k^\Xi)}\bigm|\Xi_0,\dots,\Xi_{k-1}\bigr]\Bigr]\\
  &\le e^{\frac{\beta_k^2}{2}nC_k(x)^2v_k}\,
    \mathbb E_{\Xi\sim\mu_n}\prod_{i<k}e^{-A_iU_i(H_i^\Xi)} .
\end{align*}
Applying this for $k=h,h-1,\dots,0$ in turn,
\[
  \mathbb E_{\Xi\sim\mu_n}w(\Xi,x)
  =\mathbb E_{\Xi\sim\mu_n}\prod_{i\le h}e^{-A_iU_i(H_i^\Xi)}
  \le\prod_{i=0}^he^{\frac{\beta_i^2}{2}nC_i(x)^2v_i}
  =m(x).
\]
Integration in $x$ and Tonelli's theorem give
$\mathbb E_{\Xi\sim\mu_n}Z\le M$.

\paragraph{Proof of $\mathbb E_{\Xi\sim\mu_n}Z_{\rm good}\ge(1-\eta_n)M$.}
Under $\gamma_n$ the coordinates of $x_j$ are independent
$\mathcal N(0,\frac23)$, so $\frac32\|x_j\|^2\sim\chi_n^2$ and
$\mathbb E_{x\sim\gamma_n}e^{t\cdot\frac32\|x_j\|^2}=(1-2t)^{-\frac n2}$
for $t<\frac12$. Exponential Markov with $t=\frac{a-1}{2a}$ gives
\[
  \gamma_n\Bigl\{\tfrac32\|x_j\|^2\ge an\Bigr\}
  \le\exp\Bigl[-\frac n2(a-1-\log a)\Bigr]\ \ (a>1),
\]
and
\[
  \gamma_n\Bigl\{\tfrac32\|x_j\|^2\le an\Bigr\}
  \le\exp\Bigl[-\frac n2(a-1-\log a)\Bigr]\ \ (0<a<1).
\]
The condition $r_j\in[0.75,0.9]$ reads
$\frac32\|x_j\|^2\in[\frac{27}{32}n,\frac{243}{200}n]$, and
\[
  \frac12\Bigl(\frac{27}{32}-1-\log\frac{27}{32}\Bigr)\ge\frac1{150},
  \qquad
  \frac12\Bigl(\frac{243}{200}-1-\log\frac{243}{200}\Bigr)\ge\frac1{150}.
\]
A union bound over the $h+1$ blocks gives
\begin{equation}
  \gamma_n(\mathcal R^c)\le2(h+1)e^{-n/150}.
  \label{eq:target-good-mass-radii}
\end{equation}

Now fix $x\in\mathcal R$. By the definition of $\mathcal F_n$ we have
$\mathbf 1_{\mathcal F_n}(\Xi,x)=\mathbf 1_{\bigcap_i\mathcal D_i}(\Xi)$,
and $w=e^{-\sum_i\beta_iH_i^\Xi}$ on $\mathcal F_n$ by
\eqref{eq:target-good-mass-linear-on-good}, so
\[
  \mathbb E_{\Xi\sim\mu_n}\bigl[w(\Xi,x)\mathbf 1_{\mathcal F_n}(\Xi,x)\bigr]
  =\mathbb E_{\Xi\sim\mu_n}\Bigl[e^{-\sum_{i,p}\beta_iH_{i,p}^\Xi(x)}
    \mathbf 1_{\bigcap_i\mathcal D_i}\Bigr].
\]
Recall that
$\mathcal D_i=\bigl\{\max_{p\in\mathcal P}\bigl|\frac{H_{i,p}^\Xi(x)}{na_{i,p}^2}+\beta_i\bigr|
\le\frac{\epsilon_g\beta_i}{4}\bigr\}$, and that $C_i(x)=1$ on
$\mathcal R$, so that under $\mu_n$ the $H_{i,p}^\Xi(x)$ are
independent $\mathcal N(0,na_{i,p}^2)$ and their joint density is the
product of the densities of $\mathcal N(0,na_{i,p}^2)$. Completing the
square in each factor and using
$\prod_{i,p}e^{\frac{\beta_i^2na_{i,p}^2}{2}}=m_*$,
\[
  e^{-\sum_{i,p}\beta_iy_{i,p}}
  \prod_{i,p}\frac{e^{-\frac{y_{i,p}^2}{2na_{i,p}^2}}}{\sqrt{2\pi na_{i,p}^2}}
  =m_*\prod_{i,p}
  \frac{e^{-\frac{(y_{i,p}+\beta_ina_{i,p}^2)^2}{2na_{i,p}^2}}}{\sqrt{2\pi na_{i,p}^2}}
  \qquad(y_{i,p}\in\R).
\]
Integrating $\mathbf 1_{\bigcap_i\mathcal D_i}$ against both sides and
substituting $y_{i,p}=z_{i,p}-\beta_ina_{i,p}^2$ on the right, which
turns the condition
$\bigl|\frac{y_{i,p}}{na_{i,p}^2}+\beta_i\bigr|\le\frac{\epsilon_g\beta_i}{4}$
into $|z_{i,p}|\le\frac{\epsilon_g\beta_ina_{i,p}^2}{4}$, we get
\[
  \mathbb E_{\Xi\sim\mu_n}\Bigl[e^{-\sum_{i,p}\beta_iH_{i,p}^\Xi(x)}
    \mathbf 1_{\bigcap_i\mathcal D_i}\Bigr]
  =m_*\Pr_{\Xi\sim\mu_n}\Bigl\{|H_{i,p}^\Xi(x)|\le\frac{\epsilon_g\beta_ina_{i,p}^2}{4}
    \ \text{for all }(i,p)\Bigr\}.
\]
In words, the tilt by $e^{-\sum_i\beta_iH_i^\Xi}$ recentres each
$H_{i,p}^\Xi(x)$ at the centre $-\beta_ina_{i,p}^2$ of the window in
$\mathcal D_i$. Since $H_{i,p}^\Xi(x)\sim\mathcal N(0,na_{i,p}^2)$, the
Gaussian tail bound
$\Pr\{|\mathcal N(0,s^2)|>\lambda\}\le2e^{-\frac{\lambda^2}{2s^2}}$
with $s^2=na_{i,p}^2$ and $\lambda=\frac{\epsilon_g\beta_ina_{i,p}^2}{4}$,
a union bound over $(i,p)$, and $\beta_i\ge\beta_h$ give that for
$x\in\mathcal R$,
\begin{equation}
  \mathbb E_{\Xi\sim\mu_n}\bigl[w(\Xi,x)\mathbf 1_{\mathcal F_n}(\Xi,x)\bigr]
  \ge m_*\Bigl(1-\sum_{i,p}2e^{-\frac{\epsilon_g^2\beta_i^2na_{i,p}^2}{32}}\Bigr)
  \ge m_*\Bigl(1-2(h+1)|\mathcal P|\,
     e^{-\frac{\epsilon_g^2\beta_h^2n\min_{i,p}a_{i,p}^2}{32}}\Bigr).
  \label{eq:target-good-mass-fixed-x}
\end{equation}

Set
\[
  \eta_n:=2(h+1)e^{-n/150}
  +2(h+1)|\mathcal P|\,
   e^{-\frac{\epsilon_g^2\beta_h^2n\min_{i,p}a_{i,p}^2}{32}}.
\]
Integrating \eqref{eq:target-good-mass-fixed-x} over $\mathcal R$,
then using $m_*\ge M$ (as $C_i\le1$) and
\eqref{eq:target-good-mass-radii}, and finally $(1-a)(1-b)\ge1-a-b$,
\begin{align*}
  \mathbb E_{\Xi\sim\mu_n}Z_{\rm good}
  &\ge\int_{\mathcal R}
     \mathbb E_{\Xi\sim\mu_n}\bigl[w(\Xi,x)\mathbf 1_{\mathcal F_n}(\Xi,x)\bigr]\,d\gamma_n(x)
   \ge m_*\Bigl(1-2(h+1)|\mathcal P|\,
     e^{-\frac{\epsilon_g^2\beta_h^2n\min_{i,p}a_{i,p}^2}{32}}\Bigr)
     \gamma_n(\mathcal R)\\
  &\ge M\bigl(1-2(h+1)e^{-n/150}\bigr)
     \Bigl(1-2(h+1)|\mathcal P|\,
     e^{-\frac{\epsilon_g^2\beta_h^2n\min_{i,p}a_{i,p}^2}{32}}\Bigr)
   \ge M(1-\eta_n).
\end{align*}
Finally, Proposition~\ref{hf:parameter-bounds} gives
$\beta_h^2n\ge n^{7/8}$, $\min_{i,p}a_{i,p}^2\ge n^{-1/16}$, and
$(h+1)|\mathcal P|=n^{o(1)}$ for all sufficiently large $n$, so
\[
  \eta_n\le n^{o(1)}e^{-n/150}
  +n^{o(1)}e^{-\frac{\epsilon_g^2}{32}n^{13/16}}
  \le e^{-n^{3/4}}
\]
for all sufficiently large $n$.

\paragraph{Proof of $\mathbb E_{\Xi\sim\mu_n}Z_{\rm bad}\le\eta_nM$.}
Since $Z_{\rm bad}=Z-Z_{\rm good}$, the two bounds just proved give
$\mathbb E_{\Xi\sim\mu_n}Z_{\rm bad}\le M-(1-\eta_n)M=\eta_nM$.
\end{proof}

\subsection{Second Moment of the Good Mass}
\label{sec:target-second-moment}
Recall that $v_i\le\frac1{3\log8}$, and that
$\beta_i\le\beta_0=\eta_*\lambda$ with $\lambda=\frac{\lambda_*}{(h+1)^2}$,
$\eta_*\le1$, and $\lambda_*\le\frac12$. Hence
\begin{equation}
  \alpha_i:=\beta_i^2v_i
  \le\frac{\eta_*^2\lambda_*^2}{(h+1)^4}\cdot\frac1{3\log8}
  \le\frac18,
  \qquad
  \sum_{i=0}^h\alpha_i
  \le\frac{\eta_*^2\lambda_*^2}{(h+1)^3}\cdot\frac1{3\log8}
  \le1.
  \label{eq:target-good-mass-parameters}
\end{equation}
The second moment computation also uses the following spherical
estimate.
\begin{fact}
\label{fact:target-good-mass-overlap}
Let $S$ be the first coordinate of a uniform point on $S^{n-1}$. For
$n\ge6$ and $0\le a\le\frac18$,
\begin{equation}
  \mathbb Ee^{naS^4}\le1+\frac{Ca}{n}.
  \label{eq:target-good-mass-overlap}
\end{equation}
\end{fact}

\begin{proof}[Proof of Fact~\ref{fact:target-good-mass-overlap}]
The density of $S$ is proportional to $(1-s^2)^{\frac{n-3}{2}}$ on
$[-1,1]$. Its normalizing integral over $|s|\le n^{-1/2}$ is at least
$2e^{-1}n^{-1/2}$, by $\log(1-\frac1n)\ge-\frac2n$ on that interval.
Also $\frac{n-3}{2}\ge\frac n4$ and $\log(1-s^2)\le-s^2$, so the
density is at most $C\sqrt n\,e^{-\frac{ns^2}{4}}$. Using
$e^u-1\le ue^u$ for $u\ge0$, and then $as^4\le\frac{s^2}{8}$ for
$|s|\le1$,
\[
  \mathbb E\bigl(e^{naS^4}-1\bigr)
  \le Cna\sqrt n\int_{-1}^1s^4e^{-\frac{ns^2}{4}+nas^4}\,ds
  \le Cna\sqrt n\int_{\mathbb R}s^4e^{-\frac{ns^2}{8}}\,ds
  \le\frac{Ca}{n}.
\]
For the last step, substitute $t=\sqrt n\,s$. The remaining fourth
Gaussian moment is an absolute constant.
\end{proof}

\begin{lemma}[\bf Second moment of the good mass]
\label{lem:target-good-mass-second}
For all sufficiently large $n$,
\begin{equation}
  \mathbb EZ_{\rm good}^2\le\Bigl(1+\frac Cn\Bigr)M^2.
  \label{eq:target-good-mass-second}
\end{equation}
\end{lemma}

\begin{proof}
On the good event the weight is linear in the Gaussian fields by
\eqref{eq:target-good-mass-linear-on-good}. Thus
\begin{equation}
  \mathbb E_{\Xi\sim\mu_n}Z_{\rm good}^2
  \le\iint\mathbb E_{\Xi\sim\mu_n}\,
     e^{-\sum_i\beta_i(H_i^\Xi(x)+H_i^\Xi(y))}
     \,d\gamma_n(x)\,d\gamma_n(y).
  \label{eq:target-good-mass-drop-indicators}
\end{equation}
Since the points with a vanishing block form a $\gamma_n$-null set,
we may take all blocks of $x$ and $y$ nonzero, so that $T_i(x,y)$ is
defined by \eqref{hf:compound-correlation}. Recall from
Lemma~\ref{hf:covariance}(1) that
\[
  \operatorname{Cov}\bigl(H_i^\Xi(x),H_i^\Xi(y)\bigr)
  =nC_i(x)C_i(y)\sum_{p\in\mathcal P}a_{i,p}^2T_i(x,y)^p .
\]
Since different levels are independent, the Gaussian exponential
formula gives
\[
  \mathbb E_{\Xi\sim\mu_n}\,e^{-\sum_i\beta_i(H_i^\Xi(x)+H_i^\Xi(y))}
  =m(x)m(y)e^{k(x,y)},
  \qquad
  k(x,y):=n\sum_{i=0}^h\beta_i^2C_i(x)C_i(y)
  \sum_{p\in\mathcal P}a_{i,p}^2T_i(x,y)^p .
\]
Every degree $p$ is even and at least four, $0\le C_i\le1$, and every
factor in \eqref{hf:compound-correlation} has absolute value at most $1$,
so $|T_i(x,y)|\le\frac{|\langle x_i,y_i\rangle|}{\|x_i\|\|y_i\|}$.
Hence
\begin{equation}
  0\le k(x,y)\le n\sum_{i=0}^h\alpha_i
  \Bigl(\frac{\langle x_i,y_i\rangle}{\|x_i\|\|y_i\|}\Bigr)^4,
  \qquad\alpha_i:=\beta_i^2v_i .
  \label{eq:target-good-mass-kernel}
\end{equation}
Under $\gamma_n\otimes\gamma_n$, condition on all block radii of both
points. The normalized blocks $\frac{x_i}{\|x_i\|}$ and
$\frac{y_i}{\|y_i\|}$ remain independent and uniform on $S^{n-1}$,
so the quotients $\frac{\langle x_i,y_i\rangle}{\|x_i\|\|y_i\|}$,
$0\le i\le h$, are independent and each has the law of $S$ in
Fact~\ref{fact:target-good-mass-overlap}, while $m(x)m(y)$ depends
only on the radii. Therefore the angular integral of $e^{k(x,y)}$ is
at most
\[
  \prod_{i=0}^h\mathbb Ee^{n\alpha_iS^4}
  \le\prod_{i=0}^h\Bigl(1+\frac{C\alpha_i}{n}\Bigr)
  \le\exp\Bigl[\frac Cn\sum_{i=0}^h\alpha_i\Bigr]
  \le1+\frac Cn,
\]
where the first inequality is Fact~\ref{fact:target-good-mass-overlap}
with $a=\alpha_i\le\frac18$ and the last uses
\eqref{eq:target-good-mass-parameters} and $n$ large. The remaining
radial integral of $m(x)m(y)$ is $M^2$. Combining this with
\eqref{eq:target-good-mass-drop-indicators} proves the claim.
Tonelli's theorem applies throughout, and the displayed bounds show
that all integrals are finite.
\end{proof}

\subsection{Proof of Theorem~\ref{thm:target-good-mass}}
\label{sec:target-proof}
The final step uses a deterministic inequality.

\begin{fact}
\label{fact:target-good-mass-normalization}
For $g,b\ge0$ with $g+b>0$,
\begin{equation}
  \frac{b}{g+b}\le b+(1-\sqrt g)^2\le b+(1-g)^2 .
  \label{eq:target-good-mass-normalization}
\end{equation}
\end{fact}

\begin{proof}
Set $t:=g+b>0$. The difference between the middle expression and the
left side is $t+\frac gt-2\sqrt g=\frac{(t-\sqrt g)^2}{t}\ge0$. The
second inequality follows from
$(1-\sqrt g)^2=\frac{(1-g)^2}{(1+\sqrt g)^2}\le(1-g)^2$.
\end{proof}

\begin{proof}[Proof of Theorem~\ref{thm:target-good-mass}]
Lemmas~\ref{lem:target-good-mass-first}
and~\ref{lem:target-good-mass-second} give
\[
  \mathbb E\Bigl(\frac{Z_{\rm good}}{M}-1\Bigr)^2
  =\frac{\mathbb EZ_{\rm good}^2}{M^2}
   -2\,\frac{\mathbb EZ_{\rm good}}{M}+1
  \le\frac Cn+2\eta_n,
  \qquad
  \frac{\mathbb EZ_{\rm bad}}{M}\le\eta_n .
\]
For each realization apply
Fact~\ref{fact:target-good-mass-normalization} with
$g=\frac{Z_{\rm good}}{M}$ and $b=\frac{Z_{\rm bad}}{M}$, whose sum is
$\frac ZM>0$. Since $\mathcal F_n\subseteq\mathcal T_n$,
\[
  \mathbb E\,\pi^\Xi\bigl(\mathcal T_n(\Xi)^c\bigr)
  \le\mathbb E\,\frac{Z_{\rm bad}}{Z}
  \le\frac{\mathbb EZ_{\rm bad}}{M}
   +\mathbb E\Bigl(\frac{Z_{\rm good}}{M}-1\Bigr)^2
  \le\frac Cn+3\eta_n
  \le\frac{C'}{n}.
\]
This proves the theorem under the unconditioned coefficient law
$\mu_n$. Conditioning on admissibility is done in
Section~\ref{sec:parameters-and-lower-bound}.
\end{proof}

\subsection{Proof of Theorem~\ref{thm:main}}
\label{sec:parameters-and-lower-bound}

We combine Theorem~\ref{hf:output-fails} and Theorem~\ref{thm:target-good-mass}. Both are stated under the product law $\mu_n$, so we first pass to the admissible instances by Proposition~\ref{hf:parameter-bounds}(1), which gives the theorem in the dimensions $d_n=(h+1)n$, and then extend it to every sufficiently large dimension by padding with an independent Gaussian factor.

\begin{proof}[Proof of Theorem~\ref{thm:main}]
\smallskip\noindent\emph{The dimensions $d_n$.}
Let $n$ be sufficiently large and put $\mathcal V_{d_n}:=\{V^\Xi:\Xi\in\mathcal E_n\}$. Every $V\in\mathcal V_{d_n}$ is $C^\infty$ and even by Lemma~\ref{hf:smoothness}, and by Proposition~\ref{hf:parameter-bounds}(2) it satisfies $I_{d_n}\preceq\nabla^2V\preceq2I_{d_n}$, has $0$ as its unique minimizer, and has a target with mean zero. This gives (1) and (2).

For (3), fix a randomized algorithm $\mathcal A$ that makes fewer than $Q$ queries on every instance of $\mathcal V_{d_n}$, and let $\mathcal A'$ run $\mathcal A$ but stop and output $0$ if a $Q$-th query is attempted. Then $\mathcal A'$ makes at most $Q-1$ queries on every instance and agrees with $\mathcal A$ on $\mathcal E_n$. Write $\rho_\Xi$ for the output law of $\mathcal A'$ on the instance $\Xi$, including its seed, and put $p_n:=\mu_n(\mathcal E_n)$ and $\mu_n^{\mathcal E}:=\mu_n(\cdot\mid\mathcal E_n)$. For every $\Xi$,
\[
 \operatorname{TV}(\rho_\Xi,\pi^\Xi)\ge1-\pi^\Xi\bigl(\mathcal T_n(\Xi)^c\bigr)-\rho_\Xi\bigl(\mathcal T_n(\Xi)\bigr).
\]
Since $\mathbb E_{\mu_n^{\mathcal E}}f\le\frac1{p_n}\mathbb E_{\mu_n}f$ for $f\ge0$, Theorems~\ref{hf:output-fails} and~\ref{thm:target-good-mass} and Proposition~\ref{hf:parameter-bounds}(1) give
\[
 \mathbb E_{\Xi\sim\mu_n^{\mathcal E}}\operatorname{TV}(\rho_\Xi,\pi^\Xi)
 \ge1-\frac{n^{-10}+\frac Cn}{p_n},\qquad p_n\ge1-2(h+1)Pe^{-n},
\]
and the right-hand side exceeds $\frac1{10}$ for all sufficiently large $n$. Hence $\operatorname{TV}(\rho_\Xi,\pi^\Xi)>\frac1{10}$ for some $\Xi\in\mathcal E_n$, that is, $\mathcal A$ does not sample from $\mathcal V_{d_n}$ with accuracy $\frac1{10}$.

\smallskip\noindent\emph{Arbitrary dimensions.}
Let $d$ be sufficiently large and put
\[
 H_d:=\Bigl\lceil\frac1{32}(\log d)^{\frac13}\Bigr\rceil+1,\qquad n:=\Bigl\lfloor\frac{d}{2H_d}\Bigr\rfloor .
\]
Since $n\le d$, $h(n)+1\le\frac1{32}(\log d)^{\frac13}+1\le H_d$, so $d_n=(h(n)+1)n\le H_d\cdot\frac{d}{2H_d}\le d$. For $\Xi\in\mathcal E_n$ define on $\R^d=\R^{d_n}\times\R^{d-d_n}$
\[
 \widetilde V^\Xi(z,y):=V^\Xi(z)+\frac34\|y\|^2,\qquad\mathcal V_d:=\{\widetilde V^\Xi:\Xi\in\mathcal E_n\}.
\]
Then $\nabla^2\widetilde V^\Xi=\operatorname{diag}\bigl(\nabla^2V^\Xi,\frac32I_{d-d_n}\bigr)$, $\widetilde V^\Xi$ is even with unique minimizer $0$, and $\pi_{\widetilde V^\Xi}=\pi^\Xi\otimes\mathcal N(0,\frac23I_{d-d_n})$ has mean zero, so (1) and (2) hold.

A query to $\widetilde V^\Xi$ at $(z,y)$ is answered by one query to $V^\Xi$ at $z$, since
\[
 \widetilde V^\Xi(z,y)=V^\Xi(z)+\frac34\|y\|^2,\qquad
 \nabla\widetilde V^\Xi(z,y)=\Bigl(\nabla V^\Xi(z),\frac32y\Bigr).
\]
Given an algorithm $\widetilde{\mathcal A}$ for $\mathcal V_d$ with fewer than $Q(n)$ queries, let $\mathcal A$ run $\widetilde{\mathcal A}$ through this simulation and output the $z$-component of its output. The $z$-marginal of $\pi_{\widetilde V^\Xi}$ is $\pi^\Xi$, and total variation does not increase under marginalization, so
\[
 \operatorname{TV}\bigl(\rho^{\mathcal A}_\Xi,\pi^\Xi\bigr)\le\operatorname{TV}\bigl(\rho^{\widetilde{\mathcal A}}_\Xi,\pi_{\widetilde V^\Xi}\bigr),
\]
and the first part gives an instance of $\mathcal V_d$ on which $\widetilde{\mathcal A}$ has error greater than $\frac1{10}$. Finally, $n\ge\frac{d}{2H_d}-1\ge\frac{d}{4H_d}$ and $\log(4H_d)=O(\log\log d)$ give $\log n\ge\frac18\log d$ for all sufficiently large $d$, hence
\[
 \log Q(n)=(h(n)+1)\log2\ge\frac{\log2}{32}(\log n)^{\frac13}\ge\frac{\log2}{64}(\log d)^{\frac13}.
\]
This proves (3) with $c=\frac{\log2}{64}$.
\end{proof}

\begin{remark}[\bf The role of the depth]
The query threshold $Q=2^{h+1}$ is exponential in $h$, while the loss in the last amplitude is $\log\frac1{\beta_h}=O(h^3)$ by Proposition~\ref{hf:parameter-bounds}(3). Taking $h$ of order $(\log n)^{\frac13}$ keeps $\beta_h\ge n^{-\frac1{16}}$, which Theorems~\ref{hf:output-fails} and~\ref{thm:target-good-mass} require, and this is the source of the exponent $\frac13$ in the lower bound.
\end{remark}

\section{Lower Bound for Strongly Log-concave Sampling with Warm Start}
\label{app:slc-warm}

This section extends Theorem~\ref{thm:main} to algorithms initialized at a single draw $X_{\mathrm{init}}\sim\nu$ from an unknown constant-warm starting distribution $\nu$, using only first-order queries to $V$ and independent internal randomness.

We write $\rho^{\mathrm{Alg}}_{V,\nu}$ for the output law and
$\mathbb E_{V,\nu}[N]$ for the expected number of oracle queries.
Both include the randomness of the initial sample and the
algorithm's fresh seed.
Accuracy and query complexity are required uniformly over the
specified family of target--initializer pairs.

\begin{theorem}[\bf Lower bound for strongly log-concave sampling
from a warm start]
\label{thm:slc-warm}
There exist universal constants $c>0$ and $d_0$ such that
for every $d\ge d_0$ there is a family $\mathcal W_d$
of pairs $(V,\nu)$, with $V\in C^\infty(\mathbb R^d)$,
satisfying the following properties:
\begin{enumerate}
    \item
    $I_d\preceq\nabla^2V(x)\preceq2I_d$ for every
    $(V,\nu)\in\mathcal W_d$ and every $x\in\mathbb R^d$;

    \item
    $V(-x)=V(x)$, the unique minimizer of $V$ is $0$,
    and $\pi_V(dx)\propto e^{-V(x)}\,dx$ has mean zero;

    \item
    $\nu$ has a smooth, everywhere positive, even density,
    has mean zero, and satisfies
   $\frac{d\nu}{d\pi_V}\le3$ and $\frac{d\pi_V}{d\nu}\le100$;

    \item
    Sampling from the targets
    $\{\pi_V:(V,\nu)\in\mathcal W_d\}$
    with accuracy $1/10$ in total variation distance,
    uniformly over $(V,\nu)\in\mathcal W_d$,
    requires at least
    $\Omega(\exp(c(\log d)^{1/3}))$
    exact first order queries in worst-case expectation,
    even when one sample $X_{\mathrm{init}}\sim\nu$
    is supplied.
\end{enumerate}
The common minimizer and target mean may be given to the algorithm.
\end{theorem}

\subsection{Construction and Proof of the Warm-Start Lower Bound}
\label{subsec:warm-construction}

\paragraph{The hard family.}
%
For a sufficiently large integer $n$, retain the parameters
\begin{equation}
\begin{aligned}
    h&=\left\lfloor\frac{(\log n)^{1/3}}{32}\right\rfloor,
    &d_n&=(h+1)n,
    &Q&=2^{h+1},\\
    P&=Q^{64},
    &K&=Q^{74},
    &Q_i&=2^{i+1}.
\end{aligned}
\label{eq:warm-parameters}
\end{equation}
Write $x=(x_0,\ldots,x_h)$ with $x_j\in\mathbb R^n$,
and let $\mathcal P=\{4,6,\ldots,P\}$.
Let $\mu_n$ be the original product standard Gaussian law
of the coefficient tensors $\Xi_{i,p}$.
The fields are
\[
    H_{i,p}^\Xi(x)
    =
    \langle\Xi_{i,p},\Phi_{i,p}(x)\rangle,
    \qquad
    H_i^\Xi=\sum_{p\in\mathcal P}H_{i,p}^\Xi.
\]
The prefix potentials are
\[
    V_i^\Xi
    =
    B+\sum_{j=0}^i A_j^\Xi U_j(H_j^\Xi),
    \qquad
    B(x)=\frac34\|x\|^2,
    \qquad
    A_j^\Xi=\prod_{k<j}G_k^\Xi.
\]
All features, amplitudes, cutoffs, gates, and the admissible
event $E_n$ are retained from the original construction.
Proposition~\ref{hf:parameter-bounds} supplies the target regularity on $E_n$.
We write $\pi_\Xi=\pi_{V_h^\Xi}$.

For points with positive level-$i$ cutoffs, define
\begin{equation}
\begin{aligned}
    d_i(x,y)
    &=
    \min\bigl\{
        \|\psi_i(x)-\psi_i(y)\|,
        \|\psi_i(x)+\psi_i(y)\|
    \bigr\},\\
    d_i(x,y)^2
    &=2\bigl(1-|T_i(x,y)|\bigr).
\end{aligned}
\label{eq:warm-distance}
\end{equation}
The tensor structure gives
$|T_{i-1}(x,y)|\le|T_i(x,y)|^K$ for $i\ge1$
whenever the relevant directions are defined.
At level $h$, this is the projective distance between
the normalized last blocks $x_h$ and $y_h$.
Set $\rho_*=P^{-1/16}=Q^{-4}$.
Let
\[
    \mathcal R_n
    =
    \bigl\{
        x:\ 0.75\sqrt n\le\|x_j\|\le0.9\sqrt n
        \text{ for every }0\le j\le h
    \bigr\}.
\]

Let $T_\Xi$ be the energy test of Definition~\ref{hf:energy-tests},
which implies $G_h^\Xi=1$ and
$\|x_h\|\in[0.75,0.9]\sqrt n$.
Theorem~\ref{thm:target-good-mass} and Proposition~\ref{hf:parameter-bounds} give
\begin{equation}
    \mathbb E_{\Xi\sim\mu_n(\cdot\mid E_n)}
    \bigl[\pi_\Xi(T_\Xi^c)\bigr]
    \le\frac{C}{n}
\label{eq:warm-energy-acceptance}
\end{equation}
for a universal constant $C$.

\paragraph{The warm initialization.}
Recall that $\nu$ is $M$-warm with respect to $\pi$
if $\nu\ll\pi$ and $\frac{d\nu}{d\pi}\le M$
almost everywhere.
Fix an even function
$\vartheta\in C^\infty(\mathbb R;[0,1])$
equal to zero for $|t|\le\frac{9}{20}$
and to one for $|t|\ge\frac12$.
Set $f(t)=\frac1{100}+\frac{99}{100}\vartheta(t)$.
For $\Xi\in E_n$ and $v\in S^{n-1}$, define
\begin{equation}
\begin{aligned}
\nu_{\Xi,v}(dx)
    \propto
    {f(v^\top x_h)}
    \pi_\Xi(dx),\quad   \mbox{ with } \quad Z^{\mathrm{init}}_{\Xi,v}
    =
    \int_{\mathbb R^{d_n}}
    f(v^\top x_h)\,\pi_\Xi(dx).
\end{aligned}
\label{eq:warm-initializer}
\end{equation}
Since $\frac1{100}\le f\le1$, we have
$Z^{\mathrm{init}}_{\Xi,v}\in[\frac1{100},1]$,
so $\nu_{\Xi,v}$ is a well-defined probability measure.
The direction $v$ enters only the initialization law,
not the potential or its oracle.
It is not supplied to the algorithm.
For the active dimensions, take
\[
    \mathcal W_{d_n}
    =
    \bigl\{
        (V_h^\Xi,\nu_{\Xi,v})
        :\Xi\in E_n,\ v\in S^{n-1}
    \bigr\}.
\]

Draw $\Xi\sim\mu_n(\cdot\mid E_n)$
and uniform $v\in S^{n-1}$ independently.
Conditional on $(\Xi,v)$, draw
$x^\circ\sim\nu_{\Xi,v}$.
Let $\omega$ be the algorithm's fresh seed,
independent of $(\Xi,v,x^\circ)$.
Denote the resulting joint law by $P_{\mathrm w}$.
The algorithm receives $x^\circ$ and first-order 
oracle access to $V_h^\Xi$.

\paragraph{Warmness and the initialization bias.}
We first verify that the reweighting
in~\eqref{eq:warm-initializer} produces a smooth
constant-warm start.
The same estimate records a uniform deficit of probability
in a central slab, which will be used in the distinguishing test.
Both conclusions hold for every admissible $\Xi$
and every unit direction $v$.

\begin{lemma}[\bf Properties of the warm initialization]
\label{lem:warm-initializer}
For every $\Xi\in E_n$ and $v\in S^{n-1}$,
the law $\nu_{\Xi,v}$ has a smooth, everywhere positive,
even density, a finite first moment, and mean zero.
Its normalizer and density ratios satisfy
\[
    1-\frac1{\sqrt\pi}
    \le Z^{\mathrm{init}}_{\Xi,v}\le1,
    \qquad
    \frac{d\nu_{\Xi,v}}{d\pi_\Xi}\le3,
    \qquad
    \frac{d\pi_\Xi}{d\nu_{\Xi,v}}\le100.
\]
For $b=\frac7{25}$ and
$a_b=\frac{2b e^{-b^2}}{\sqrt{2\pi}}>0.206$,
\[
    \nu_{\Xi,v}\bigl(|v^\top x_h|\le\frac9{20}\bigr)
    \le\frac3{100},
    \qquad
    \pi_\Xi\bigl(|v^\top x_h|\le b\bigr)\ge a_b.
\]
\end{lemma}

\paragraph{Comparison with a Gaussian reference model.}

The supplied sample depends on the instance and therefore
cannot be treated as a random seed independent of
the coefficient tensors.
We introduce a reference model in which conditioning
on the initial sample leaves independent Gaussian
coefficients with an explicit mean shift.
The following lemma compares event probabilities
under this reference model and the actual warm input.

\begin{lemma}[\bf Gaussian comparison]
\label{lem:warm-planting}
There is a probability law $\lambda_n$ on $\mathbb R^{d_n}$,
depending only on the hard family, such that
$\zeta_n:=\lambda_n(\mathcal R_n^c)=o(1)$.
Fix an algorithm whose number of queries is bounded
by a deterministic integer $q$ on every coefficient realization.
Consider the following reference experiment:
\begin{enumerate}
    \item
    Draw $x^\circ\sim\lambda_n$ and uniform
    $v\in S^{n-1}$ independently.

    \item
    Independently of $(x^\circ,v)$, draw independent
    standard Gaussian tensors $\widetilde\Xi_{i,p}$ and set
    \begin{equation}
        \Xi_{i,p}
        =
        \widetilde\Xi_{i,p}
        -\beta_i\Phi_{i,p}(x^\circ),
        \qquad
        0\le i\le h,\quad p\in\mathcal P.
    \label{eq:warm-shift}
    \end{equation}

    \item
    Give the algorithm the initial point $x^\circ$
    and first-order oracle access to $V_h^\Xi$.
    Run it using an independent fresh seed with
    the same law as in the actual warm-input experiment.
    The direction $v$ is not revealed to the algorithm.
\end{enumerate}
Let $\widehat P$ denote the joint law of this experiment,
including the algorithm's transcript and output.
For every measurable event $A$ defined by these variables,
\begin{equation}
    P_{\mathrm w}(A)
    \le
    3\widehat P(A)+Cn^{-1/2},
\label{eq:warm-comparison}
\end{equation}
where $C$ is a universal constant.
\end{lemma}

\paragraph{Gate success far from the initial direction.}
We now control outputs that pass the final gate
while lying far from the initial direction.
In the reference model, conditioning on $x^\circ$
makes the mean shifts deterministic and leaves
the centered coefficient tensors independent standard Gaussians.
The following lemma bounds the probability of producing
such an output with fewer than $Q$ queries,
even when the algorithm may use both $x^\circ$
and the known mean shifts.

\begin{lemma}[\bf Localized query bound at a fixed initial sample]
\label{lem:warm-localized-output}
Fix $x^\circ\in\mathcal R_n$ and use the shifted Gaussian
coefficients in~\eqref{eq:warm-shift},
without conditioning on $E_n$.
Let an algorithm make fewer than $Q$ first-order
queries to $V_h^\Xi$, with a seed independent of
the centered coefficients and possibly containing $x^\circ$.
Its output $Y$, which need not be a queried point, satisfies
\[
    \Pr\bigl\{
        G_h^\Xi(Y)>0,\;
        d_h(Y,x^\circ)\ge\rho_*
    \bigr\}
    \le n^{-10}.
\]
Gate-success events are declared false when their required
directions are undefined.
\end{lemma}

\paragraph{Slab crossing near the initial direction.}
It remains to control outputs that pass the energy test
while staying close to the initial direction.
Such an output can repair the initialization bias only
if it moves from outside the wider slab into the narrower slab.
The following geometric lemma bounds this crossing
probability for a uniform direction $v$.
Under $\widehat P$, the direction $v$ is independent
of the initial sample and the algorithm's run,
so the lemma can be applied after conditioning on them.

\begin{lemma}[\bf A hidden-direction slab bound]
\label{lem:warm-slab-geometry}
Let $x,y\in\mathbb R^n$ satisfy
$\|x\|,\|y\|\in[0.75,0.9]\sqrt n$ and
\[
    \min\left\{
        \left\|
            \frac{x}{\|x\|}-\frac{y}{\|y\|}
        \right\|,
        \left\|
            \frac{x}{\|x\|}+\frac{y}{\|y\|}
        \right\|
    \right\}
    \le\rho,
    \qquad
    0<\rho\le\frac14.
\]
For uniform $v\in S^{n-1}$,
\[
    \Pr_v\left\{
        |v^\top x|>\frac9{20},\;
        |v^\top y|\le\frac7{25}
    \right\}
    \le256\rho^2.
\]
\end{lemma}

The query bound controls successful outputs far from
the initial direction, while the geometric estimate
controls slab crossings near that direction.
We now combine these estimates with the initialization
bias to prove Theorem~\ref{thm:slc-warm}.

\begin{proof}[Proof of Theorem~\ref{thm:slc-warm}]
For $d=d_n$, admissibility and Lemma~\ref{lem:warm-initializer}
ensure that every instance in $\mathcal W_{d_n}$ satisfies
the required target and warm-start conditions.
Consider any algorithm with output $Y$ that makes fewer
than $Q$ queries on $\mathcal W_{d_n}$.
Enforce the same cap on every coefficient realization
by stopping immediately before the $Q$-th query
and returning the origin if necessary.
This does not alter its run on $\mathcal W_{d_n}$.

Define
\begin{equation}
    \mathcal T^{\mathrm{warm}}_{\Xi,v}
    =
    T_\Xi
    \cap
    \bigl\{x:|v^\top x_h|\le b\bigr\},
    \qquad b=\frac7{25}.
\label{eq:warm-final-test}
\end{equation}
Lemma~\ref{lem:warm-initializer}
and~\eqref{eq:warm-energy-acceptance} give
\[
\begin{aligned}
    \mathbb E_{\Xi,v}
    \bigl[\pi_\Xi(\mathcal T^{\mathrm{warm}}_{\Xi,v})\bigr]
    &\ge
    \mathbb E_{\Xi,v}
    \bigl[\pi_\Xi(|v^\top x_h|\le b)\bigr]
    -
    \mathbb E_\Xi[\pi_\Xi(T_\Xi^c)]\ge a_b-\frac{C}{n}.
\end{aligned}
\]

To bound the probability that $Y$ lands in the test set, let
\[
    A=\big\{Y\in\mathcal T^{\mathrm{warm}}_{\Xi,v},\;
    |v^\top x_h^\circ|>\frac{9}{20}\big\},
    \qquad
    \mathcal G=\big\{x^\circ\in\mathcal R_n,\;
    Y\in T_\Xi,\;
    d_h(Y,x^\circ)<\rho_*\big\}.
\]
Under $\widehat P$, the direction $v$ is independent of $\Xi$, $x^\circ$,
and $\omega$, and since the algorithm does not observe $v$, it remains
independent of $Y$ as well.
On $\mathcal G$, the norms $\|x_h^\circ\|$ and $\|Y_h\|$ both lie in
$[0.75,0.9]\sqrt n$ and have projective distance $d_h(Y,x^\circ)<\rho_*$.
Conditioning on $\Xi$, $x^\circ$, $\omega$, and $Y$ and applying
Lemma~\ref{lem:warm-slab-geometry} therefore yields
$\widehat P(A\cap\mathcal G)
\le256\rho_*^2\widehat P(\mathcal G)
\le256\rho_*^2$.

Since $T_\Xi\subseteq\{G_h^\Xi=1\}$,
Lemma~\ref{lem:warm-localized-output},
integrated over $x^\circ\in\mathcal R_n$, yields
\[
\begin{aligned}
    \widehat P(A)
    &\le
    \widehat P(x^\circ\notin\mathcal R_n)
    +
    \widehat P\bigl(
        x^\circ\in\mathcal R_n,\;
        Y\in T_\Xi,\;
        d_h(Y,x^\circ)\ge\rho_*
    \bigr)
    +
    \widehat P(A\cap\mathcal G)\\
    &\le
    \zeta_n+n^{-10}+256\rho_*^2.
\end{aligned}
\]
Using Lemmas~\ref{lem:warm-initializer}
and~\ref{lem:warm-planting}, we obtain
\begin{equation}
\begin{aligned}
    P_{\mathrm w}
    \bigl(Y\in\mathcal T^{\mathrm{warm}}_{\Xi,v}\bigr)
    &\le
    P_{\mathrm w}
    \bigl(|v^\top x_h^\circ|\le\frac9{20}\bigr)
    +P_{\mathrm w}(A)\\
    &\le
    \frac3{100}+3\widehat P(A)+Cn^{-1/2}\\
    &\le
    \frac3{100}
    +3\bigl(\zeta_n+n^{-10}+256\rho_*^2\bigr)
    +Cn^{-1/2}\\
    &=\frac3{100}+o(1).
\end{aligned}
\label{eq:warm-output-acceptance}
\end{equation}

Writing $\rho^{\mathrm{Alg}}_{\Xi,v}$ for the output distribution given $\Xi$ and $v$ averaged over $x^\circ\sim\nu_{\Xi,v}$ and the random seed, we lower bound the total variation distance on the event $\mathcal T^{\mathrm{warm}}_{\Xi,v}$ and average over $\Xi$ and $v$ to obtain
\begin{equation}
\begin{aligned}
    \mathbb E_{\Xi,v}
    \operatorname{TV}
    \bigl(\rho^{\mathrm{Alg}}_{\Xi,v},\pi_\Xi\bigr)
    &\ge
    \mathbb E_{\Xi,v}
    \bigl[
        \pi_\Xi(\mathcal T^{\mathrm{warm}}_{\Xi,v})
        -
        \rho^{\mathrm{Alg}}_{\Xi,v}
        (\mathcal T^{\mathrm{warm}}_{\Xi,v})
    \bigr]\\
    &=
    \mathbb E_{\Xi,v}
    \bigl[\pi_\Xi(\mathcal T^{\mathrm{warm}}_{\Xi,v})\bigr]
    -
    P_{\mathrm w}
    \bigl(Y\in\mathcal T^{\mathrm{warm}}_{\Xi,v}\bigr)\\
    &\ge
    a_b-\frac3{100}-o(1)
    >
    \frac16
\end{aligned}
\label{eq:warm-fixed-budget-gap}
\end{equation}
for all sufficiently large $n$.

Now consider any algorithm with total variation error at most $\frac{1}{10}$
for all $\Xi\in E_n$ and $v\in S^{n-1}$, and assume without loss of generality that
$t=\sup_{\Xi\in E_n,v}\mathbb E_{\Xi,v}[N]<\infty$.
Let $\mathrm{Alg}_{<Q}$ denote the truncated algorithm that stops right before
the $Q$-th query and outputs the origin if it has not yet terminated.
Coupling the two algorithms via the same initial sample and random seed ensures that
their outputs agree whenever $N<Q$, so the coupling, triangle, and Markov inequalities
give, for each $\Xi$ and $v$,
\[
\begin{aligned}
    \operatorname{TV}
    \bigl(\rho^{\mathrm{Alg}_{<Q}}_{\Xi,v},\pi_\Xi\bigr)
    &\le
    \operatorname{TV}
    \bigl(\rho^{\mathrm{Alg}_{<Q}}_{\Xi,v},
          \rho^{\mathrm{Alg}}_{\Xi,v}\bigr)
    +
    \operatorname{TV}
    \bigl(\rho^{\mathrm{Alg}}_{\Xi,v},\pi_\Xi\bigr)\\
    &\le
    \Pr_{\Xi,v}(N\ge Q)+\frac1{10}\\
    &\le
    \frac{\mathbb E_{\Xi,v}[N]}{Q}+\frac1{10}
    \le
    \frac{t}{Q}+\frac1{10}.
\end{aligned}
\]
Applying~\eqref{eq:warm-fixed-budget-gap}
to $\mathrm{Alg}_{<Q}$ therefore forces
$\frac16<\frac1{10}+\frac{t}{Q}$,
and hence $t>\frac{Q}{15}$.

For any sufficiently large $d$, choose $d_n\le d$ with $\log n\ge c\log d$
as in Section~\ref{sec:parameters-and-lower-bound}, and take the product of both the target and initial
distributions with $N\big(0,\frac{2}{3}I_{d-d_n}\big)$.
The common Gaussian factor cancels in the density ratios, preserves smoothness,
evenness, and zero mean, and adds the Hessian block $\frac{3}{2}I_{d-d_n}\in[I,2I]$.
Moreover, an algorithm on $\mathbb R^d$ can be simulated on $\mathbb R^{d_n}$
with the same number of queries by sampling the extra initial coordinates from
its random seed and evaluating the known Gaussian potential directly, while
projecting the output onto $\mathbb R^{d_n}$ can only decrease the total
variation error.
The $d_n$-dimensional lower bound $Q=\exp\big(\Theta(\log^{\frac{1}{3}} n)\big)$
with $\log n\ge c\log d$ therefore proves Theorem~\ref{thm:slc-warm}.
\end{proof}

\subsection{Proof of Lemma~\ref{lem:warm-initializer}}
\label{subsec:warm-initializer-proof}
\begin{proof}[Proof of Lemma~\ref{lem:warm-initializer}]
We first establish one-dimensional marginal bounds for any even potential
$V\in C^2(\mathbb R^d)$ satisfying $I_d\preceq\nabla^2V\preceq2I_d$.
Fix a unit vector $u\in\mathbb R^d$.
We claim that the density $\rho=e^{-\phi}$ of $u^\top X$ under $X\sim\pi_V$
is positive and even, with $\phi\in C^2(\mathbb R)$ and $1\le\phi''\le2$.
When $d=1$, evenness gives $\rho(t)\propto e^{-V(t)}$ and
$\phi''(t)=V''(t)\in[1,2]$.
For $d\ge2$, choose orthonormal coordinates $x=(t,z)$ with $t=u^\top x$
and $z\in\mathbb R^{d-1}$, so that for $Z_V=\int_{\mathbb R^d}e^{-V(x)}\,dx$,
\[
    \rho(t)=\frac{1}{Z_V}
       \int_{\mathbb R^{d-1}}e^{-V(t,z)}\,dz,
\]
and let $\mathbb E_t$ and $\operatorname{Var}_t$ denote expectation
and variance under the conditional distribution of $z$ given $t$.

Since $V$ is even, $\nabla V(0)=0$, and integrating $I_d\preceq\nabla^2V\preceq2I_d$
along the segment from $0$ to $x$ gives $V(x)\ge V(0)+\frac{1}{2}\|x\|^2$ and
$\|\nabla V(x)\|\le2\|x\|$.
The first two $t$-derivatives of $e^{-V(t,z)}$ are $-V_t e^{-V}$ and
$(V_t^2-V_{tt})e^{-V}$, whose absolute values on every bounded interval of $t$
are bounded by a constant times $(1+\|z\|^2)e^{-\frac{\|z\|^2}{2}}$.
Differentiating under the integral sign is therefore valid and gives
$\rho'(t)=-\rho(t)\mathbb E_t[V_t]$ and $\rho''(t)=\rho(t)\mathbb E_t[V_t^2-V_{tt}]$,
showing that $\rho>0$ and $\phi=-\log\rho\in C^2(\mathbb R)$ with
\[
    \phi''(t)
    =-\frac{\rho''(t)}{\rho(t)}
       +\big(\frac{\rho'(t)}{\rho(t)}\big)^2
    =\mathbb E_t[V_{tt}]
       -\mathbb E_t[V_t^2]
       +\big(\mathbb E_t[V_t]\big)^2
    =\mathbb E_t[V_{tt}]-\operatorname{Var}_t(V_t).
\]

For each fixed $t$, the conditional potential $z\mapsto V(t,z)$ is $1$-strongly
convex, so its Poincar\'e inequality gives
$\operatorname{Var}_t(g)\le\mathbb E_t[\|\nabla_zg\|^2]$ for compactly supported
$C^1$ functions $g$.
Let $\chi_R$ be a smooth cutoff equal to $1$ on $\{\|z\|\le R\}$ and $0$ on
$\{\|z\|\ge2R\}$, with $0\le\chi_R\le1$ and $\|\nabla\chi_R\|\le \frac{C}{R}$.
Since $V_t$ grows at most linearly, $V_{zt}$ is bounded, and the conditional
density is bounded by a $t$-dependent constant times $e^{-\frac{\|z\|^2}{2}}$,
as $R\to\infty$ we have $\chi_RV_t\to V_t$ and
$\nabla_z(\chi_RV_t)-V_{zt}=(\chi_R-1)V_{zt}+V_t\nabla\chi_R\to0$ in conditional $L^2$.
Applying the conditional Poincar\'e inequality to $\chi_RV_t$ and letting
$R\to\infty$ thus gives $\operatorname{Var}_t(V_t)\le\mathbb E_t[\|V_{zt}\|^2]$.

Setting $D=\nabla^2V-I_d$, we have $0\preceq D\preceq I_d$ and hence $D^2\preceq D$,
whose first diagonal entry gives $(V_{tt}-1)^2+\|V_{zt}\|^2\le V_{tt}-1$, or
$V_{tt}-\|V_{zt}\|^2\ge1+(V_{tt}-1)^2\ge1$.
Combining this with the variance formula yields
$\phi''(t)\ge\mathbb E_t[V_{tt}-\|V_{zt}\|^2]\ge1$ and
$\phi''(t)\le\mathbb E_t[V_{tt}]\le2$.
Moreover, the change of variables $z\mapsto-z$ with $V(-t,-z)=V(t,z)$ gives
$\rho(-t)=\rho(t)$, so $\phi$ is even with $\phi'(0)=0$.
Integrating $1\le\phi''\le2$ twice from $0$ to $t$ gives
$\frac{t^2}{2}\le\phi(t)-\phi(0)\le t^2$ for all $t\in\mathbb R$, and
integrating the resulting bounds
\[
    \rho(0)e^{-t^2}\le\rho(t)\le\rho(0)e^{-\frac{t^2}{2}}
\]
over $\mathbb R$ yields $\frac{1}{\sqrt{2\pi}}\le\rho(0)\le\frac{1}{\sqrt\pi}$.
Since $\rho(t)\le\rho(0)$, we obtain, for every $b>0$,
\[
\begin{aligned}
    \pi_V\big(|u^\top X|\ge\frac{1}{2}\big)
    &=1-\int_{-\frac{1}{2}}^{\frac{1}{2}}\rho(t)\,dt
      \ge1-\rho(0)
      \ge1-\frac{1}{\sqrt\pi},\\
    \pi_V\big(|u^\top X|\le b\big)
    &=\int_{-b}^b\rho(t)\,dt
      \ge\rho(0)\int_{-b}^b e^{-t^2}\,dt
      \ge2b\,\rho(0)e^{-b^2}
      \ge\frac{2b e^{-b^2}}{\sqrt{2\pi}}.
\end{aligned}
\]

Now fix $\Xi\in E_n$ and $v\in S^{n-1}$.
By admissibility, $V_h^\Xi$ satisfies the above assumptions, so we may apply
these marginal bounds along the unit vector $u$ aligned with $v$ in the
$x_h$-block, for which $u^\top x=v^\top x_h$.
Since $\frac{1}{100}\le f\le1$ and $f(t)=1$ for $|t|\ge\frac{1}{2}$, we have
\[
    1\ge Z^{\mathrm{init}}_{\Xi,v}
    \ge\pi_\Xi\big(|v^\top x_h|\ge\frac{1}{2}\big)
    \ge1-\frac{1}{\sqrt\pi}>\frac{1}{3},
\]
and therefore
\[
\begin{aligned}
    \frac{d\nu_{\Xi,v}}{d\pi_\Xi}(x)
    &=\frac{f(v^\top x_h)}{Z^{\mathrm{init}}_{\Xi,v}}
      \le\frac{1}{1-\frac{1}{\sqrt\pi}}<3,\\
    \frac{d\pi_\Xi}{d\nu_{\Xi,v}}(x)
    &=\frac{Z^{\mathrm{init}}_{\Xi,v}}{f(v^\top x_h)}
      \le100.
\end{aligned}
\]
Because both $\pi_\Xi$ and $x\mapsto f(v^\top x_h)$ are smooth, positive,
even, so is $\nu_{\Xi,v}$, and $\nu_{\Xi,v}\le3\pi_\Xi$ together
with the Gaussian tails of $\pi_\Xi$ implies
$\int\|x\|\,\nu_{\Xi,v}(dx)\le3\int\|x\|\,\pi_\Xi(dx)<\infty$ and
$\int x\,\nu_{\Xi,v}(dx)=0$ by symmetry.

Finally, since $f=\frac{1}{100}$ on $\big[-\frac{9}{20},\frac{9}{20}\big]$
and $Z^{\mathrm{init}}_{\Xi,v}>\frac{1}{3}$, we conclude that for $b=\frac{7}{25}$,
\[
\begin{aligned}
    \nu_{\Xi,v}\big(|v^\top x_h|\le\frac{9}{20}\big)
    &=\frac{\pi_\Xi\big(|v^\top x_h|\le\frac{9}{20}\big)}
            {100Z^{\mathrm{init}}_{\Xi,v}}
      \le\frac{1}{100Z^{\mathrm{init}}_{\Xi,v}}
      <\frac{3}{100},\\
    \pi_\Xi\big(|v^\top x_h|\le b\big)
    &\ge\frac{2b e^{-b^2}}{\sqrt{2\pi}}
      =a_b>0.206.
\end{aligned}
\]
\end{proof}

\subsection{Proof of Lemma~\ref{lem:warm-planting}}
\label{subsec:warm-comparison-proof}
\begin{proof}

\par\medskip
\noindent\textbf{Construction of the reference law.}\quad
Let $\gamma_n=N\big(0,\frac{2}{3}I_{d_n}\big)$ be the Gaussian measure
associated with $B(x)=\frac{3}{4}\|x\|^2$, and define
\begin{equation}
\begin{aligned}
    w_\Xi(x)
    &=e^{-(V_h^\Xi(x)-B(x))},
    &Z_\Xi
    &=\int w_\Xi\,d\gamma_n,\\
    L_\Xi(x)
    &=\exp\big(-\sum_{i=0}^h\beta_iH_i^\Xi(x)\big),
    &\mathfrak M
    &=\mathbb E_{\mu_n}\int L_\Xi\,d\gamma_n.
\end{aligned}
\label{eq:warm-weights}
\end{equation}
The bounded truncations ensure that $0<Z_\Xi<\infty$ and
$\pi_\Xi(dx)=\frac{w_\Xi(x)}{Z_\Xi}\gamma_n(dx)$ for all $\Xi$.
By independence of the Gaussian tensors,
$\|\Phi_{i,p}(x)\|^2=na_{i,p}^2C_i(x)^2$, and $0\le C_i\le1$,
we have
\[
\begin{aligned}
    m(x):=\mathbb E_{\mu_n}L_\Xi(x)
    &=\exp\big(
        \frac{n}{2}\sum_{i=0}^h\sum_{p\in\mathcal P}
        \beta_i^2a_{i,p}^2C_i(x)^2
    \big)\\
    &\le
    \exp\big(
        \frac{n}{2}\sum_{i=0}^h\sum_{p\in\mathcal P}
        \beta_i^2a_{i,p}^2
    \big)=:m_*.
\end{aligned}
\]
Thus $\mathfrak M=\int m\,d\gamma_n\in(0,m_*]$, and
\begin{equation}
    \widehat P^{\Xi,x^\circ}(d\Xi,dx^\circ)
    =
    \frac{L_\Xi(x^\circ)}{\mathfrak M}\,
    \mu_n(d\Xi)\gamma_n(dx^\circ)
\label{eq:warm-planted-law}
\end{equation}
defines a probability measure.

\par\medskip
\noindent\textbf{Conditional Gaussian coefficients.}\quad
Under~\eqref{eq:warm-planted-law}, the conditional density
of $\Xi$ given $x^\circ$ factors over $i$ and $p$ with exponent
\[
\begin{aligned}
    -\frac{1}{2}\|\Xi_{i,p}\|^2
    -\beta_i\langle\Xi_{i,p},\Phi_{i,p}(x^\circ)\rangle
    &=
    -\frac{1}{2}
    \|\Xi_{i,p}+\beta_i\Phi_{i,p}(x^\circ)\|^2\\
    &\quad+
    \frac{\beta_i^2}{2}\|\Phi_{i,p}(x^\circ)\|^2.
\end{aligned}
\]
Consequently, the conditional law of
$\widetilde\Xi_{i,p}
=\Xi_{i,p}+\beta_i\Phi_{i,p}(x^\circ)$
is a product of standard Gaussian laws that does not depend
on $x^\circ$.
These centered tensors are therefore independent of $x^\circ$,
proving~\eqref{eq:warm-shift}.

\par\medskip
\noindent\textbf{The reference initial-point law.}\quad
Integrating~\eqref{eq:warm-planted-law} over $\Xi$ gives
the marginal law
$\lambda_n(dx)=\frac{m(x)}{\mathfrak M}\gamma_n(dx)$.
Since $m=m_*$ on $\mathcal R_n$ and $m\le m_*$ everywhere,
$\mathfrak M\ge m_*\gamma_n(\mathcal R_n)$ and hence
$\lambda_n(\mathcal R_n^c)
\le\frac{\gamma_n(\mathcal R_n^c)}{\gamma_n(\mathcal R_n)}$.
Under $\gamma_n$, each
$S_j=\frac{3}{2}\|x_j\|^2\sim\chi_n^2$, and
$x\in\mathcal R_n$ is equivalent to
$\frac{S_j}{n}\in[a_0,b_0]$ for all $j$,
with $a_0=\frac{27}{32}<1<b_0=\frac{243}{200}$.
Setting $I(s)=\frac{1}{2}(s-1-\log s)$ and
$c=\min\{I(a_0),I(b_0)\}>0$, the $\chi_n^2$
Chernoff bounds and a union bound give
\[
    \gamma_n(\mathcal R_n^c)
    \le(h+1)\big[e^{-nI(a_0)}+e^{-nI(b_0)}\big]
    \le2(h+1)e^{-cn}=o(1),
\]
so $\zeta_n:=\lambda_n(\mathcal R_n^c)=o(1)$.

\par\medskip
\noindent\textbf{Control of the weights and normalizers.}\quad
On the good event $\mathcal F_n$ defined in Definition~\ref{def:target-good-event},
all gates equal $1$ and all truncations are linear,
so $w_\Xi=L_\Xi$.
Set $Z_g=\int w_\Xi\mathbf1_{\mathcal F_n}\,d\gamma_n$
and $Z_b=Z_\Xi-Z_g$.
Under $\Xi\sim\mu_n$, Lemmas~\ref{lem:target-good-mass-first}--\ref{lem:target-good-mass-second} give
$0\le\eta_n\le e^{-n^{\frac{3}{4}}}$ such that
\[
    \mathbb E Z_g\ge(1-\eta_n)\mathfrak M,
    \qquad
    \mathbb E Z_g^2
    \le\big(1+\frac{C}{n}\big)\mathfrak M^2,
    \qquad
    \mathbb E Z_b\le\eta_n\mathfrak M.
\]
Thus
$\mathbb E\big(\frac{Z_g}{\mathfrak M}-1\big)^2
\le\frac{C}{n}+2\eta_n$,
and Cauchy--Schwarz together with
$w_\Xi=L_\Xi$ on $\mathcal F_n$ yields
\[
\begin{aligned}
    \mathbb E\big|\frac{Z_\Xi}{\mathfrak M}-1\big|
    &\le\sqrt{\frac{C}{n}+2\eta_n}+\eta_n,\\
    \mathbb E\int|w_\Xi-L_\Xi|\,d\gamma_n
    &\le\mathbb E Z_b+(\mathfrak M-\mathbb E Z_g)
      \le2\eta_n\mathfrak M.
\end{aligned}
\]

\par\medskip
\noindent\textbf{Comparison with a target sample.}\quad
For
$P_{\mathrm{tar}}(d\Xi,dx^\circ)
=\mu_n(d\Xi)\pi_\Xi(dx^\circ)$,
using $\int w_\Xi\,d\gamma_n=Z_\Xi$
and inserting $w_\Xi/\mathfrak M$ between the two densities gives
\[
\begin{aligned}
    \operatorname{TV}
    (P_{\mathrm{tar}},\widehat P^{\Xi,x^\circ})
    &=
    \frac{1}{2}\mathbb E_{\mu_n}\int
    \big|\frac{w_\Xi}{Z_\Xi}
          -\frac{L_\Xi}{\mathfrak M}\big|\,d\gamma_n\\
    &\le
    \frac{1}{2}\mathbb E_{\mu_n}
    \bigg[
        \big|1-\frac{Z_\Xi}{\mathfrak M}\big|
        +\frac{1}{\mathfrak M}
         \int|w_\Xi-L_\Xi|\,d\gamma_n
    \bigg]\\
    &\le
    \frac12\sqrt{\frac Cn+2\eta_n}
    +\frac32\eta_n
    =O(n^{-\frac12}).
\end{aligned}
\]
For
$P_{\mathrm{tar}}^E(d\Xi,dx^\circ)
=\mu_n(d\Xi\mid E_n)\pi_\Xi(dx^\circ)$,
Proposition~4.2 gives
$\operatorname{TV}(P_{\mathrm{tar}}^E,P_{\mathrm{tar}})
=\mu_n(E_n^c)\le2(h+1)Pe^{-n}=o(n^{-\frac12})$.
The triangle inequality therefore implies
$\operatorname{TV}
(P_{\mathrm{tar}}^E,\widehat P^{\Xi,x^\circ})
=O(n^{-\frac12})$.

\par\medskip
\noindent\textbf{The warm input and the algorithm run.}\quad
Adjoin to both laws a uniform $v\in S^{n-1}$ and a seed
$\omega$ with the same law as the actual algorithm's fresh seed,
with $v$ and $\omega$ mutually independent and independent
of $(\Xi,x^\circ)$.
These common independent factors preserve the total variation
bound between the extended laws $P_{\mathrm{tar}}^E$
and $\widehat P$.
Since
$\frac{dP_{\mathrm w}}{dP_{\mathrm{tar}}^E}
=\frac{f(v^\top x_h^\circ)}{Z^{\mathrm{init}}_{\Xi,v}}\le3$
by Lemma~\ref{lem:warm-initializer},
every measurable event $A$ satisfies
\[
\begin{aligned}
    P_{\mathrm w}(A)
    &\le3P_{\mathrm{tar}}^E(A)\\
    &\le3\widehat P(A)
       +3\operatorname{TV}(P_{\mathrm{tar}}^E,\widehat P)\\
    &\le3\widehat P(A)+Cn^{-\frac12}.
\end{aligned}
\]
Finally, the transcript and output of a globally capped algorithm
are obtained from $(\Xi,v,x^\circ,\omega)$ by the same measurable
run map under $P_{\mathrm w}$ and $\widehat P$.
Applying the preceding inequality to the inverse image
of any event involving the run proves the claimed extension.
\end{proof}

\subsection{Proof of Lemma~\ref{lem:warm-localized-output}}
\label{subsec:warm-localized-proof}
Fix $x^\circ\in\mathcal R_n$ and write
\[
    \Xi_{j,p}
    =
    \widetilde\Xi_{j,p}-\beta_j\Phi_{j,p}(x^\circ),
\]
where the tensors $\widetilde\Xi_{j,p}$ are independent
standard Gaussians.
The point $x^\circ$ and the mean shifts are deterministic,
and no conditioning on $E_n$ is imposed.
All seeds below are independent of the centered coefficients
to which the algorithm has oracle access.
Retain the parameters $\zeta,r,M,\delta,Q_i,e_i$
and the query graph of Section~\ref{out:query-components}.
Their bounds, including $Qr<P^{-1/2}$ and $e_h\le n^{-10}$,
are supplied by Lemma~\ref{out:parameter-bounds}.
Set
\begin{equation}
    \rho_*=P^{-1/16}=Q^{-4},
    \qquad
    p_{\mathrm{step}}=P^{1/4}=Q^{16},
    \qquad
    P_{\mathrm{eff}}=\frac{P}{p_{\mathrm{step}}}=Q^{48}.
\label{eq:warm-subdegrees}
\end{equation}
We first show that high degrees suppress the mean shift
at outputs far from $x^\circ$.

\begin{fact}[\bf The high-degree profile]
\label{fact:warm-reciprocal-profile}
Consider an algorithm making $q<Q$ queries to the value
and gradient of $H_i^\Xi$, with an arbitrary output $Z$.
Except on an event of probability at most $\zeta$,
if $G_i^\Xi(Z)>0$ and $d_i(Z,x^\circ)\ge\rho_*$,
then there is an exponential sum
\[
    g(\ell)=\sum_{j:T_j\ne0}c_je^{-\sigma_j\ell},
    \qquad
    \sigma_j=-p_{\mathrm{step}}\log|T_j|,
\]
satisfying
\begin{equation}
    |\ell g(\ell)-1|<\frac1{10},
    \qquad \ell=4,6,\ldots,P_{\mathrm{eff}}.
\label{eq:warm-rescaled-error}
\end{equation}
Here $T_j=T_i(Z,y_j)$ for positive-cutoff queries,
and $T_j=0$ otherwise.
\end{fact}

\begin{proof}
Each query to the shifted field is simulated by one query to the centered
field minus the known mean function and its gradient.
Since $G_i^\Xi(Z)>0$ and $x^\circ\in\mathcal R_n$ imply $C_i(Z)=C_i(x^\circ)=1$,
Theorem~\ref{out:profile} and the error bound in Proposition~\ref{out:source-separation}
show that outside probability $\zeta$, whenever $G_i^\Xi(Z)>0$,
\begin{equation}
    \frac{H_{i,p}^\Xi(Z)}{na_{i,p}^2}
    =
    p\sum_{j=1}^qB_jT_j^{p-1}+E_p
    -\beta_iT_i(Z,x^\circ)^p,
    \qquad
    \max_p|E_p|\le\frac{\beta_i}{20},
\label{eq:warm-profile}
\end{equation}
with coefficients $B_j$ shared across all degrees.

Moreover, $d_i(Z,x^\circ)\ge\rho_*$ gives $\vert{}T_i(Z,x^\circ)\vert{}\le1-\frac{\rho_*^2}{2}$,
so for $p=p_{\mathrm{step}}\ell$ with $\ell\ge4$,
\begin{equation}
    |T_i(Z,x^\circ)|^p
    \le e^{-\frac{p_{\mathrm{step}}\ell\rho_*^2}{2}}
    \le e^{-2Q^8}.
\label{eq:warm-drift-decay}
\end{equation}
For $T_j\ne0$, set $c_j=-\frac{p_{\mathrm{step}}B_j}{\beta_iT_j}$.
Since $p$ is even,
\[\ell g(\ell) = -\frac{p_{\mathrm{step}}\ell}{\beta_i} \sum_{j:T_j\ne0}\frac{B_j}{T_j}\vert{}T_j\vert{}^p = -\frac{p}{\beta_i}\sum_jB_jT_j^{p-1}.\]
Combining the gate condition of Lemma~\ref{hf:gate-prefix} with
\eqref{eq:warm-profile}--\eqref{eq:warm-drift-decay} yields
\[\vert{}\ell g(\ell)-1\vert{} \le\epsilon_g+\frac{\vert{}E_p\vert{}}{\beta_i} +\vert{}T_i(Z,x^\circ)\vert{}^p \le\frac{1}{100}+\frac{1}{20}+e^{-2Q^8} <\frac{1}{10},\]
so the sum defining $g$ is nonempty.
\end{proof}

We next use the two-window argument from the proof
of Lemma~\ref{out:two-directions}.
The additional conclusion needed here is that both selected
queries are close to the output.

\begin{fact}[\bf Two nearby active queries]
\label{fact:warm-active-queries}
Consider an algorithm making fewer than $Q$ queries
to $V_i^\Xi$, with an arbitrary output $Z$.
Except on an event of probability at most $\zeta$,
if $G_i^\Xi(Z)>0$ and $d_i(Z,x^\circ)\ge\rho_*$,
then two active queries $y,y'$ have positive cutoffs and satisfy
\begin{equation}
    d_i(y,y')>P^{-1/2},
    \qquad
    d_i(y,Z),d_i(y',Z)<(5p_{\mathrm{step}})^{-1/2}.
\label{eq:warm-near-queries}
\end{equation}
\end{fact}

\begin{proof}
Use the active-query simulation from the proof of Lemma~\ref{out:two-components-lemma},
with enlarged seed $(\omega,\widetilde\Xi_{<i})$
and the known mean shifts.
Inactive replies are computed from this seed,
and each active reply requires one query to $H_i^\Xi$.
The seed is independent of $\widetilde\Xi_i$,
so Fact~\ref{fact:warm-reciprocal-profile} applies.

Let $c_0=2^{-20}$ and $C_0=10240$.
Apply Lemma~\ref{out:rate-gap} to $g$ at
$u_1=c_0P_{\mathrm{eff}}/Q^2$ and $u_0=10C_0Q$.
Since $P_{\mathrm{eff}}=Q^{48}$,
both scales are admissible for sufficiently large $n$.
They give two nonzero terms with
\begin{equation}
    \frac1{P_{\mathrm{eff}}}<\sigma_1
    <\frac{C_0Q^3}{c_0P_{\mathrm{eff}}},
    \qquad
    \frac{c_0}{10C_0Q^3}<\sigma_2<\frac1{10}.
\label{eq:warm-rates}
\end{equation}
Thus $\sigma_2>16\sigma_1$ for sufficiently large $n$.
Putting $s_j=\sigma_j/p_{\mathrm{step}}$ gives
$s_1>1/P$, $s_2>16s_1$, and
$s_1,s_2<1/(10p_{\mathrm{step}})$.

For the corresponding queries,
$d_i(y_j,Z)^2=2(1-e^{-s_j})$.
Using $s\le2(1-e^{-s})\le2s$ for $0\le s\le1/10$,
we obtain
\[
\begin{aligned}
    d_i(y_j,Z)^2
    &\le2s_j<\frac1{5p_{\mathrm{step}}},\\
    d_i(y,y')
    &\ge d_i(y',Z)-d_i(y,Z)
      \ge\sqrt{s_2}-\sqrt{2s_1}
      >P^{-1/2}.
\end{aligned}
\]
The selected queries are active because they are queries
of the active-query simulator.
Their cutoffs are positive because their $T_j$ are nonzero.
\end{proof}

The query graph and its diameter bound are unchanged.
The next fact verifies the extra distance condition
needed to apply the localized preceding-level hypothesis.

\begin{fact}[\bf Two preceding-level far components]
\label{fact:warm-two-far-components}
Fix $i\ge1$.
If a distant successful output satisfies the conclusion
of Fact~\ref{fact:warm-active-queries},
the two selected queries belong to distinct active components.
The first active point $x$ of each selected component satisfies
$d_{i-1}(x,x^\circ)>\rho_*$.
\end{fact}

\begin{proof}
By Section~\ref{out:query-components} and Lemma~\ref{out:parameter-bounds}(ii),
each component has diameter less than $Qr<P^{-1/2}$.
The two selected queries therefore belong to distinct components.
Let $y$ be either selected query and $x$ the first active
point of its component.
Then
\[
\begin{aligned}
    d_i(x,x^\circ)
    &\ge d_i(Z,x^\circ)-d_i(Z,y)-d_i(y,x)\\
    &>\rho_*-(5p_{\mathrm{step}})^{-1/2}-Qr
      >\frac{\rho_*}{2}.
\end{aligned}
\]
The last inequality follows from
$(5p_{\mathrm{step}})^{-1/2}=Q^{-8}/\sqrt5$
and $Qr<Q^{-32}$, both of which are $o(\rho_*)$.
Since $x$ is active, it passes $G_{i-1}^\Xi$,
so the preceding-level directions are defined.
Ancestor decay gives
\begin{equation}
    |T_{i-1}(x,x^\circ)|
    \le|T_i(x,x^\circ)|^K
    \le e^{-K\rho_*^2/8}
    =e^{-Q^{66}/8}.
\label{eq:warm-ancestor}
\end{equation}
Hence $d_{i-1}(x,x^\circ)^2
=2(1-|T_{i-1}(x,x^\circ)|)>1>\rho_*^2$
for sufficiently large $n$.
\end{proof}

For $i\ge1$, define the localized bad-component event
\begin{equation}
    B_i^{\mathrm{far}}(T)
    =
    \bigl\{
        \exists\text{ active component }\mathcal C:
        \#\{s\in\mathcal C:s<\tau_{\mathcal C}\}<T,\;
        d_{i-1}(y_{\tau_{\mathcal C}},x^\circ)\ge\rho_*
    \bigr\}.
\label{eq:warm-far-component}
\end{equation}
We adapt the proof of Lemma~\ref{out:component-charge} by adding the terminal
distance condition and retaining the deterministic mean shifts.

\begin{fact}[\bf The localized component charge]
\label{lem:warm-charge}
Suppose every algorithm making fewer than $T$ queries
to $V_{i-1}^\Xi$, with an independent seed
and arbitrary output $Z'$, satisfies
\begin{equation}
    \Pr\bigl\{
        G_{i-1}^\Xi(Z')>0,\;
        d_{i-1}(Z',x^\circ)\ge\rho_*
    \bigr\}
    \le e.
\label{eq:warm-charge-hypothesis}
\end{equation}
Then every level-$i$ algorithm making at most $Q$ queries satisfies
$\Pr(B_i^{\mathrm{far}}(T))\le2^Q(e+\delta)$.
\end{fact}

\begin{proof}
Use the candidate simulations $(\tau,S)$ from the proof
of Lemma~\ref{out:component-charge} , with $|S|<T$.
Local rounds return the shifted $V_{i-1}$ reply,
far rounds return the shifted $V_i$ reply,
and the candidate outputs $x_\tau$
before receiving the terminal reply.
Retain the same separation checks, failure state,
and auxiliary measurement batches.
Replace the original terminal event by
\begin{equation}
    \mathcal E_{\tau,S}^{\mathrm{far}}
    =
    \bigl\{
        G_{i-1}(x_\tau)>0,\;
        d_{i-1}(x_\tau,x^\circ)\ge\rho_*
    \bigr\}.
\label{eq:warm-candidate}
\end{equation}

All underlying measurements act on the centered coefficients
$\widetilde\Xi_{<i}$.
The known mean shifts change only the return maps,
not the measurement vectors or their covariances.
The added distance condition depends only on $x_\tau$
and $x^\circ$, so the original terminal batch still suffices.
Thus the measurement count and separation conditions
verified in the proof of Lemma~\ref{out:component-charge} remain unchanged.
Corollary~\ref{out:ancestor-comparison} and Lemma~\ref{out:parameter-bounds}(i) give
$\operatorname{TV}(P_{\mathrm{sh}},P_{\mathrm{ind}})\le\delta$.

In the independent model, the enlarged seed
$(\omega,\widetilde\Xi_i,\widetilde\Xi_{<i}^{\mathrm{far}})$
is independent of the local centered coefficients.
Every far reply is computable from this seed,
whereas the local rounds use at most $|S|<T$
queries to the shifted local prefix.
The output is chosen before the terminal measurements,
which are not returned to the algorithm.
The terminal gate is evaluated using the local coefficients,
so~\eqref{eq:warm-charge-hypothesis} gives
\[
    P_{\mathrm{sh}}(\mathcal E_{\tau,S}^{\mathrm{far}})
    \le
    P_{\mathrm{ind}}(\mathcal E_{\tau,S}^{\mathrm{far}})
    +\delta
    \le e+\delta.
\]

The covering argument in the proof of Lemma~\ref{out:component-charge} 
also applies to $B_i^{\mathrm{far}}(T)$.
Indeed, for a witnessing component, take
$\tau=\tau_{\mathcal C}$ and
$S=\{s\in\mathcal C:s<\tau\}$.
The earlier local queries are inactive, so their
shifted $V_i$ and $V_{i-1}$ replies coincide.
The component and its complement satisfy the same
separation checks, so the shared candidate agrees
with the original run through $\tau$.
Its terminal point also satisfies the added distance
condition by the definition of $B_i^{\mathrm{far}}(T)$.
There are fewer than $2^Q$ candidates, exactly as in
Lemma~\ref{out:component-charge} , and hence
\[
    \Pr(B_i^{\mathrm{far}}(T))
    \le
    \sum_{\tau,S}
    P_{\mathrm{sh}}(\mathcal E_{\tau,S}^{\mathrm{far}})
    \le2^Q(e+\delta).
\]
\end{proof}

The remaining induction uses the same query thresholds
and error sequence as Section~\ref{out:recursive-query-bound}.

\begin{fact}[\bf The localized prefix induction]
\label{fact:warm-prefix-induction}
Every algorithm making fewer than $Q_i$ queries
to $V_i^\Xi$, with an independent seed
and arbitrary output $Z$, satisfies
\begin{equation}
    \Pr\bigl\{
        G_i^\Xi(Z)>0,\;
        d_i(Z,x^\circ)\ge\rho_*
    \bigr\}
    \le e_i.
\label{prop:warm-localized-prefix}
\end{equation}
\end{fact}

\begin{proof}
For $i=0$, the algorithm makes fewer than two queries.
Fact~\ref{fact:warm-active-queries} requires two distinct
queries for localized success, except on probability
$\zeta=e_0$.

Suppose the claim holds at level $i-1$,
and consider $q<Q_i$ queries to $V_i^\Xi$.
Except on probability $\zeta$,
Facts~\ref{fact:warm-active-queries}
and~\ref{fact:warm-two-far-components}
give two distinct active components whose first active
points are far from $x^\circ$ at level $i-1$.
Outside $B_i^{\mathrm{far}}(Q_{i-1})$,
each contains at least $Q_{i-1}$ queries before
its first active point.
Their index sets are disjoint,
so $q\ge2Q_{i-1}=Q_i$, a contradiction.
Applying Fact~\ref{lem:warm-charge} with
$T=Q_{i-1}$ and $e=e_{i-1}$ therefore gives
\[
\begin{aligned}
    \Pr\bigl\{
        G_i^\Xi(Z)>0,\;
        d_i(Z,x^\circ)\ge\rho_*
    \bigr\}
    &\le\zeta+\Pr(B_i^{\mathrm{far}}(Q_{i-1}))\\
    &\le\zeta+2^Q(e_{i-1}+\delta)
      =e_i.
\end{aligned}
\]
\end{proof}

\begin{proof}[Proof of Lemma~\ref{lem:warm-localized-output}]
Take $i=h$ in Fact~\ref{fact:warm-prefix-induction}.
Since $Q_h=Q$ and $e_h\le n^{-10}$ by Lemma~\ref{out:parameter-bounds}(iii),
the desired bound follows.
All estimates are uniform over the fixed
$x^\circ\in\mathcal R_n$.
\end{proof}

\subsection{Proof of Lemma~\ref{lem:warm-slab-geometry}}
\label{subsec:warm-geometry-proof}
\begin{proof}
Write $u=\frac{x}{\|x\|}$, $w=\frac{y}{\|y\|}$, and $\alpha=\langle u,w\rangle$.
The projective distance bound gives
$\min\{\|u-w\|^2,\|u+w\|^2\}=2(1-|\alpha|)\le\rho^2$,
so $|\alpha|\ge1-\frac{\rho^2}{2}$.
Decompose $y=cx+e$ with $c=\frac{\langle y,x\rangle}{\|x\|^2}$ and $e\perp x$.
The radius bounds give
\[
\begin{aligned}
    |c|
    &=\frac{\|y\|}{\|x\|}|\alpha|
      \ge\frac{5}{6}\big(1-\frac{\rho^2}{2}\big),\\
    \|e\|^2
    &=\|y\|^2(1-\alpha^2)
      \le\|y\|^2
         \big[1-\big(1-\frac{\rho^2}{2}\big)^2\big]
      \le\|y\|^2\rho^2
      \le n\rho^2.
\end{aligned}
\]

When $|v^\top x|>\frac{9}{20}$, $|v^\top y|\le\frac{7}{25}$, and
$\rho\le\frac{1}{4}$, the relation $v^\top e=v^\top y-cv^\top x$ implies
\[
    |v^\top e|
    \ge |c|\,|v^\top x|-|v^\top y|
    >\frac{3}{8}\big(1-\frac{\rho^2}{2}\big)-\frac{7}{25}
    \ge\frac{533}{6400}>\frac{1}{16}.
\]
For a uniform $v\in S^{n-1}$, rotational invariance gives
$\mathbb E_v[vv^\top]=\frac{1}{n}I_n$, so
$\mathbb E_v[(v^\top e)^2]=\frac{\|e\|^2}{n}\le\rho^2$.
Markov's inequality therefore yields
\[
    \Pr_v\big(
        |v^\top x|>\frac{9}{20},\;
        |v^\top y|\le\frac{7}{25}
    \big)
    \le
    \Pr_v\big(|v^\top e|>\frac{1}{16}\big)
    \le256\,\mathbb E_v[(v^\top e)^2]
    \le256\rho^2.
\]
\end{proof}

\section{Lower Bound for LSI Sampling}
\label{sec:lsi-lower-bound}

This section proves a query lower bound of order $\kappa=L/\alpha$
for sampling from $L$-smooth targets satisfying a log-Sobolev
inequality with constant $\alpha$. The bound holds even when the
minimizer is known and a common warm start is available.

\begin{theorem}[\bf Lower bound for smooth LSI sampling]
\label{thm:lsi-sampling-lower-bound}
There are universal constants $c,C,\kappa_0>0$ such that the
following holds. Let $L\ge\alpha>0$, $\kappa:=L/\alpha\ge\kappa_0$,
$d\ge C\log^2(2\kappa)$, and $\varepsilon\in(0,\frac1{10}]$. There
exist even potentials $W_u\in C^\infty(\R^d)$, $u\in S^{d-1}$, with
targets $\mu_u$, such that for every $u$:
\begin{enumerate}
\item $W_u$ is $L$-smooth, $W_u(0)=0$, and $0$ is its unique minimizer.
\item $\mu_u$ satisfies $\mathsf{LSI}(\alpha)$.
\item $\frac{d\nu}{d\mu_u}\le4$ for a fixed centered 
  Gaussian $\nu$ depending only on $(d,L,\alpha)$.
\item Sampling from the class $\{\mu_u:u\in S^{d-1}\}$ with accuracy
  $\varepsilon$ in total variation requires
  $\sup_u\mathbb E[N_u]\ge c\,\kappa$ exact value-and-gradient
  queries.
\end{enumerate}
\end{theorem}
We prove Theorem~\ref{thm:lsi-sampling-lower-bound} using a family
indexed by one hidden direction, defined in
Section~\ref{sec:lsi-family}. As in Sections~\ref{hf:section}--\ref{sec:target-good-mass}, we define a
test, prove that an algorithm with few queries rarely passes it, and
show that a target sample passes it with constant probability
(Section~\ref{sec:lsi-cap-test}). The log-Sobolev constant of the
family is determined in Section~\ref{sec:lsi-constant}, and
Section~\ref{sec:lsi-proof} chooses the parameters and completes the
proof.

Throughout this section, we assume $t\ge4$ and $d\ge Dt^4$,
where $D$ is a sufficiently large universal constant that may
be increased as needed.
We
write $\omega_d$ for the uniform probability measure on $S^{d-1}$.
All constants in this section are independent of $d$, $t$,
and the hidden direction, but may depend on the fixed mollifier
and radial cutoff introduced below.

\subsection{The Hidden-cap Family}
\label{sec:lsi-family}

We perturb a known quadratic potential by a smooth bump supported in a spherical shell. The angular dependence of the perturbation is determined by the latitude relative to a hidden unit vector. We first define the capped hinge that specifies this angular profile.

\begin{definition}[\bf Capped hinge]
\label{def:lsi-hinge}
Fix $\eta:=10^{-3}$ and a nonnegative even function
$\varrho\in C_c^\infty((-1,1))$ with $\int_{\R}\varrho(v)\,dv=1$.
For $t\ge4$, let $w:=\frac{\eta}{t}$ and
$\varrho_w(v):=\frac1w\varrho\bigl(\frac vw\bigr)$, and define
\[
  \widehat h_t(s):=2t\bigl((|s|-t)_+-(|s|-3t)_+\bigr),
  \qquad
  h_t:=\widehat h_t*\varrho_w,
\]
where $a_+:=\max\{a,0\}$.
\end{definition}

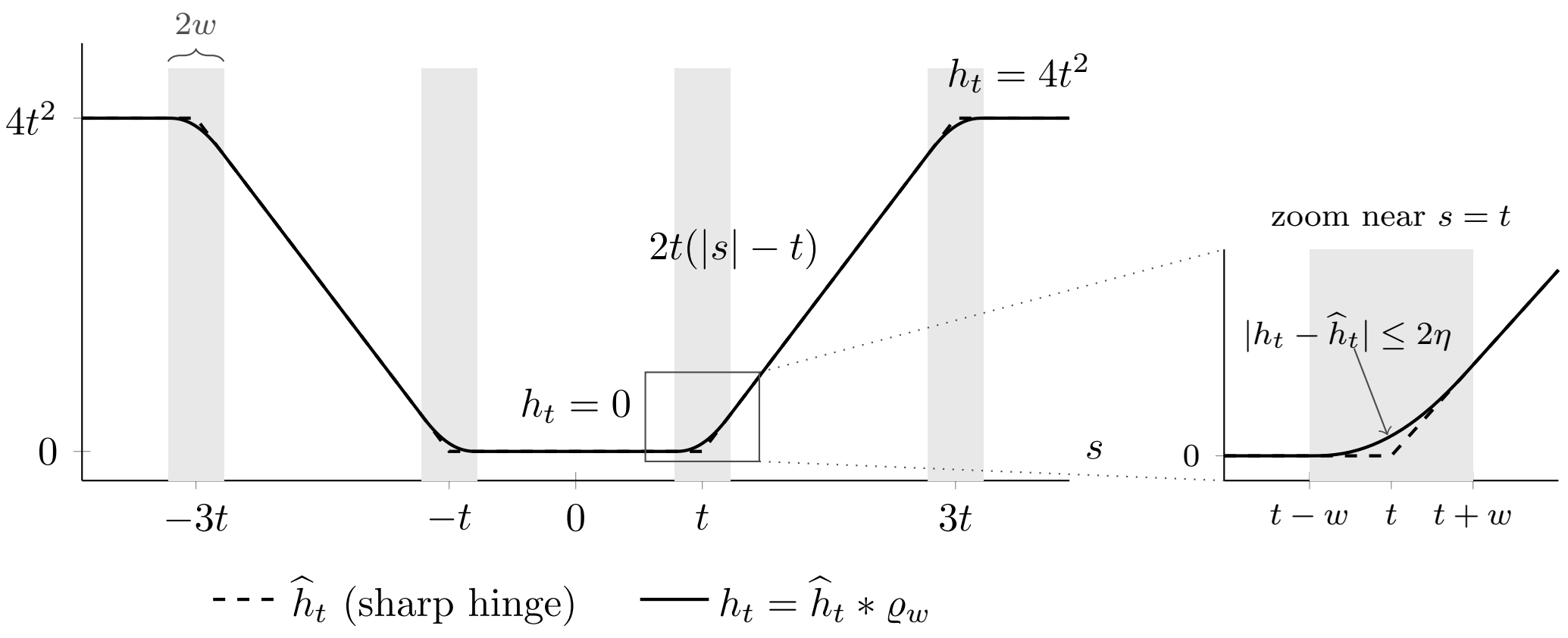
\begin{figure}[t]
\centering
\begin{tikzpicture}
  \begin{axis}[
    name=main,
    width=10.2cm, height=5.4cm,
    xmin=-3.9, xmax=3.9, ymin=-0.35, ymax=4.9,
    axis lines=left, axis line style={-},
    xtick={-3,-1,0,1,3},
    xticklabels={$-3t$,$-t$,$0$,$t$,$3t$},
    ytick={0,4}, yticklabels={$0$,$4t^2$},
    tick label style={font=\small},
    xlabel={$s$},
    xlabel style={at={(axis description cs:1,0.07)},anchor=west,font=\small},
    clip=false,
    legend style={at={(0.5,-0.18)},anchor=north,legend columns=2,font=\small,
      draw=none,/tikz/every even column/.append style={column sep=1.2em}},
    legend cell align=left,
  ]
    \pgfplotsinvokeforeach{-3,-1,1,3}{
      \fill[gray!18] (axis cs:#1-0.22,-0.35) rectangle (axis cs:#1+0.22,4.6);
    }
    \addplot[black, dashed, thick] coordinates
      {(-3.9,4) (-3,4) (-1,0) (1,0) (3,4) (3.9,4)};
    \addlegendentry{$\widehat h_t$ (sharp hinge)}
    \addplot[black, thick] table {\hingeMainData};
    \addlegendentry{$h_t=\widehat h_t*\varrho_w$}
    \node[font=\small] at (axis cs:0,0.55) {$h_t=0$};
    \node[font=\small, anchor=south east] at (axis cs:2.05,2.0) {$2t(|s|-t)$};
    \node[font=\small, anchor=south] at (axis cs:3.5,4.12) {$h_t=4t^2$};
    \draw[gray!60!black, decorate, decoration={brace, amplitude=3pt}]
      (axis cs:-3.22,4.68) -- (axis cs:-2.78,4.68)
      node[midway, above=3pt, font=\scriptsize, gray!60!black] {$2w$};
    \draw[gray!60!black, thin] (axis cs:0.55,-0.12) rectangle (axis cs:1.45,0.95);
    \coordinate (zoomTR) at (axis cs:1.45,0.95);
    \coordinate (zoomBR) at (axis cs:1.45,-0.12);
  \end{axis}
  \begin{axis}[
    name=zoom,
    at={($(main.south east)+(1.35cm,0cm)$)}, anchor=south west,
    width=4.5cm, height=3.6cm,
    xmin=0.55, xmax=1.45, ymin=-0.12, ymax=1.0,
    axis lines=left, axis line style={-},
    xtick={0.78,1,1.22},
    xticklabels={$t-w$,$t$,$t+w$},
    ytick={0}, yticklabels={$0$},
    tick label style={font=\scriptsize},
    title={zoom near $s=t$},
    title style={font=\scriptsize, yshift=-4pt},
    clip=false,
  ]
    \fill[gray!18] (axis cs:0.78,-0.12) rectangle (axis cs:1.22,1.0);
    \addplot[black, dashed, thick] coordinates {(0.55,0) (1,0) (1.45,0.9)};
    \addplot[black, thick] table {\hingeZoomData};
    \node[font=\scriptsize, anchor=west] at (axis cs:0.56,0.6) {$|h_t-\widehat h_t|\le 2\eta$};
    \draw[gray!60!black, thin, ->] (axis cs:0.9,0.52) -- (axis cs:0.99,0.1);
  \end{axis}
  \draw[gray!60!black, thin, dotted] (zoomTR) -- (zoom.north west);
  \draw[gray!60!black, thin, dotted] (zoomBR) -- (zoom.south west);
\end{tikzpicture}
\caption{The capped hinge. The dashed curve shows the sharp hinge $\widehat h_t$ (zero on $[-t,t]$, linear with slope $\pm2t$ for $t < \vert{}s\vert{} < 3t$, and constant at $4t^2$ for $\vert{}s\vert{}\ge3t$), while the solid curve depicts its smoothed counterpart $h_t=\widehat h_t*\varrho_w$. By Lemma~\ref{lem:lsi-hinge}, $h_t$ and $\widehat h_t$ are identical outside the shaded intervals of width $2w$ surrounding the kinks $\pm t$ and $\pm3t$, within which they differ by at most $2\eta$ (with $w=\eta/t$ magnified for visual clarity). Right: zoomed-in view of the smoothing transition at $s=t$.}
\label{fig:lsi-hinge}
\end{figure}

The sharp hinge $\widehat h_t$ vanishes on $[-t,t]$, grows linearly with
$|s|$ between $t$ and $3t$, and is constant beyond $3t$; see
Figure~\ref{fig:lsi-hinge}. Mollification makes it smooth while
preserving these pieces away from their endpoints.

\pgfplotstableread[col sep=space,header=false]{
-3.900 4.0000
-3.250 4.0000
-3.225 4.0000
-3.200 3.9999
-3.175 3.9983
-3.150 3.9941
-3.125 3.9871
-3.100 3.9769
-3.075 3.9635
-3.050 3.9468
-3.025 3.9268
-3.000 3.9035
-2.975 3.8768
-2.950 3.8468
-2.925 3.8135
-2.900 3.7769
-2.875 3.7371
-2.850 3.6941
-2.825 3.6483
-2.800 3.5999
-2.775 3.5500
-2.750 3.5000
-1.250 0.5000
-1.225 0.4500
-1.200 0.4001
-1.175 0.3517
-1.150 0.3059
-1.125 0.2629
-1.100 0.2231
-1.075 0.1865
-1.050 0.1532
-1.025 0.1232
-1.000 0.0965
-0.975 0.0732
-0.950 0.0532
-0.925 0.0365
-0.900 0.0231
-0.875 0.0129
-0.850 0.0059
-0.825 0.0017
-0.800 0.0001
-0.775 0.0000
-0.750 0.0000
0.750 0.0000
0.775 0.0000
0.800 0.0001
0.825 0.0017
0.850 0.0059
0.875 0.0129
0.900 0.0231
0.925 0.0365
0.950 0.0532
0.975 0.0732
1.000 0.0965
1.025 0.1232
1.050 0.1532
1.075 0.1865
1.100 0.2231
1.125 0.2629
1.150 0.3059
1.175 0.3517
1.200 0.4001
1.225 0.4500
1.250 0.5000
2.750 3.5000
2.775 3.5500
2.800 3.5999
2.825 3.6483
2.850 3.6941
2.875 3.7371
2.900 3.7769
2.925 3.8135
2.950 3.8468
2.975 3.8768
3.000 3.9035
3.025 3.9268
3.050 3.9468
3.075 3.9635
3.100 3.9769
3.125 3.9871
3.150 3.9941
3.175 3.9983
3.200 3.9999
3.225 4.0000
3.250 4.0000
3.900 4.0000
}\hingeMainData
\pgfplotstableread[col sep=space,header=false]{
0.550 0.0000
0.750 0.0000
0.770 0.0000
0.790 0.0000
0.810 0.0005
0.830 0.0023
0.850 0.0059
0.870 0.0113
0.890 0.0187
0.910 0.0281
0.930 0.0396
0.950 0.0532
0.970 0.0689
0.990 0.0868
1.010 0.1068
1.030 0.1289
1.050 0.1532
1.070 0.1796
1.090 0.2081
1.110 0.2387
1.130 0.2713
1.150 0.3059
1.170 0.3423
1.190 0.3805
1.210 0.4200
1.230 0.4600
1.250 0.5000
1.450 0.9000
}\hingeZoomData

\begin{lemma}[\bf Properties of the capped hinge]
\label{lem:lsi-hinge}
The function $h_t$ is even and belongs to $C^\infty(\R)$.
It satisfies $0\le h_t\le4t^2$ and
\[
  h_t(s)=
  \begin{cases}
    0, & |s|\le t-w,\\
    2t(|s|-t), & t+w\le |s|\le3t-w,\\
    4t^2, & |s|\ge3t+w.
  \end{cases}
\]
Moreover,
$\|h_t-\widehat h_t\|_\infty\le2\eta$,
$\|h_t'\|_\infty\le2t$, and
$\|h_t''\|_\infty\le Ct^2$,
where $C$ depends only on the fixed mollifier.
\end{lemma}

\begin{proof}
Since $\widehat h_t$ is bounded and $\varrho_w$ is smooth
and compactly supported, their convolution is smooth.
The kernel is nonnegative, even, and has integral one,
so convolution preserves evenness and the range $[0,4t^2]$.
Its support lies in $[-w,w]$, which gives the asserted
constant pieces. Since $\int v\varrho_w(v)\,dv=0$,
convolution also preserves each affine piece wherever
the integration interval remains within that piece.
This proves the formula on $t+w\le|s|\le3t-w$.

The function $\widehat h_t$ is $2t$-Lipschitz.
Consequently,
$|h_t(s)-\widehat h_t(s)|
\le2t\int|v|\varrho_w(v)\,dv
\le2tw=2\eta$.
Convolution preserves the Lipschitz bound, giving
$\|h_t'\|_\infty\le2t$.
Finally, the distributional second derivative is
$\widehat h_t''=
2t(\delta_t+\delta_{-t}-\delta_{3t}-\delta_{-3t})$,
where $\delta_a$ denotes the Dirac mass at $a$.
Convolving with $\varrho_w$ gives
$\|h_t''\|_\infty
\le8t\|\varrho_w\|_\infty
=8t^2\|\varrho\|_\infty/\eta$.
\end{proof}

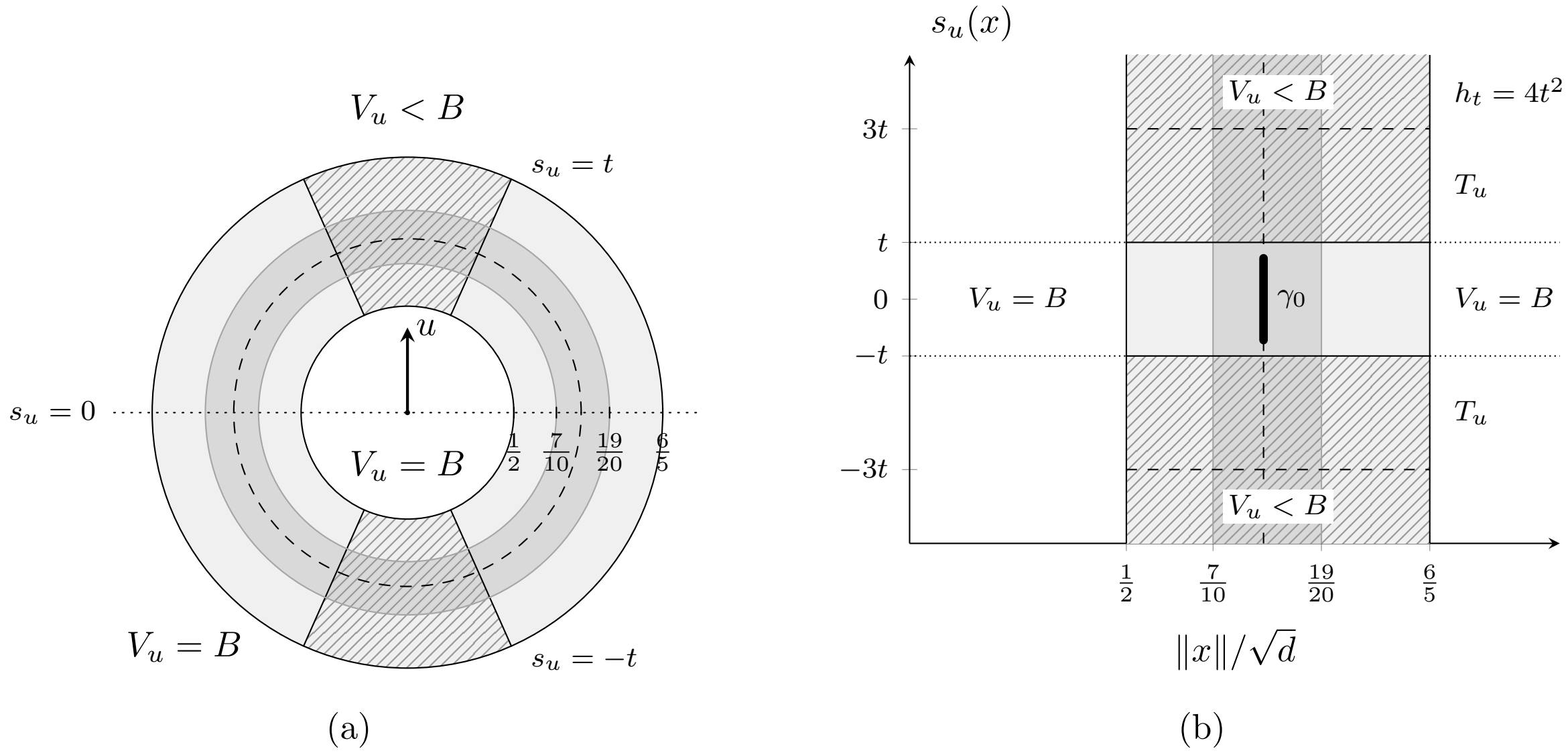
\begin{figure}[t]
\centering
\begin{minipage}[b]{0.45\textwidth}
\centering
\begin{tikzpicture}[scale=2.1,>=stealth,line cap=round,line join=round]
  \def\rin{0.5}\def\rpa{0.7}\def\rpb{0.95}\def\rout{1.2}\def\rg{0.8165}
  \def\capang{24}   
  \fill[gray!12,even odd rule] (0,0) circle[radius=\rin] (0,0) circle[radius=\rout];
  \fill[gray!30,even odd rule] (0,0) circle[radius=\rpa] (0,0) circle[radius=\rpb];
  \foreach \c in {90,270}{
    \fill[pattern=north east lines,pattern color=gray!80]
      (\c-\capang:\rin)
      arc[start angle=\c-\capang,end angle=\c+\capang,radius=\rin]
      -- (\c+\capang:\rout)
      arc[start angle=\c+\capang,end angle=\c-\capang,radius=\rout] -- cycle;
    \draw[thin] (\c-\capang:\rin) -- (\c-\capang:\rout);
    \draw[thin] (\c+\capang:\rin) -- (\c+\capang:\rout);
  }
  \draw[thin,gray!70] (0,0) circle[radius=\rpa];
  \draw[thin,gray!70] (0,0) circle[radius=\rpb];
  \draw[thin] (0,0) circle[radius=\rin];
  \draw[thin] (0,0) circle[radius=\rout];
  \draw[thin,dashed] (0,0) circle[radius=\rg];
  \draw[thin,dotted] (-1.38,0) -- (1.38,0);
  \node[font=\scriptsize,anchor=east] at (-1.4,0) {$s_u=0$};
  \foreach \r/\lab in {\rin/{\tfrac12},\rpa/{\tfrac{7}{10}},\rpb/{\tfrac{19}{20}},\rout/{\tfrac65}}{
    \draw[thin] (\r,0.025) -- (\r,-0.025);
    \node[font=\scriptsize,anchor=north] at (\r,-0.035) {$\lab$};
  }
  \fill (0,0) circle[radius=0.012];
  \draw[->,thick] (0,0) -- (0,0.4) node[font=\small,anchor=west,inner sep=2pt] {$u$};
  \node[font=\small] at (0,-0.25) {$V_u=B$};
  \node[font=\small] at (-1.05,-1.1) {$V_u=B$};
  \node[font=\small,anchor=south] at (0,1.29) {$V_u<B$};
  \node[font=\scriptsize,anchor=west] at (66:1.27) {$s_u=t$};
  \node[font=\scriptsize,anchor=west] at (-66:1.27) {$s_u=-t$};
\end{tikzpicture}\\[2pt]
{\small (a)}
\end{minipage}\hfill
\begin{minipage}[b]{0.53\textwidth}
\centering
\begin{tikzpicture}[>=stealth]
\begin{axis}[
  width=8.0cm, height=6.4cm,
  xmin=0, xmax=1.5, ymin=-4.3, ymax=4.3,
  axis lines=left,
  xtick={0.5,0.7,0.95,1.2},
  xticklabels={$\tfrac12$,$\tfrac{7}{10}$,$\tfrac{19}{20}$,$\tfrac65$},
  ytick={-3,-1,0,1,3},
  yticklabels={$-3t$,$-t$,$0$,$t$,$3t$},
  tick label style={font=\scriptsize},
  xlabel={$\|x\|/\sqrt d$},
  xlabel style={font=\small},
  ylabel={$s_u(x)$},
  ylabel style={font=\small,rotate=-90,at={(axis description cs:0,1)},anchor=south west,xshift=2pt},
  clip=false,
]
  \fill[gray!12] (axis cs:0.5,-4.3) rectangle (axis cs:1.2,4.3);
  \fill[gray!30] (axis cs:0.7,-4.3) rectangle (axis cs:0.95,4.3);
  \fill[pattern=north east lines,pattern color=gray!80]
    (axis cs:0.5,1) rectangle (axis cs:1.2,4.3);
  \fill[pattern=north east lines,pattern color=gray!80]
    (axis cs:0.5,-4.3) rectangle (axis cs:1.2,-1);
  \draw[thin,gray!70] (axis cs:0.7,-4.3) -- (axis cs:0.7,4.3);
  \draw[thin,gray!70] (axis cs:0.95,-4.3) -- (axis cs:0.95,4.3);
  \draw[thin] (axis cs:0.5,-4.3) -- (axis cs:0.5,4.3);
  \draw[thin] (axis cs:1.2,-4.3) -- (axis cs:1.2,4.3);
  \draw[thin] (axis cs:0.5,1) -- (axis cs:1.2,1);
  \draw[thin] (axis cs:0.5,-1) -- (axis cs:1.2,-1);
  \draw[thin,dashed] (axis cs:0.5,3) -- (axis cs:1.2,3);
  \draw[thin,dashed] (axis cs:0.5,-3) -- (axis cs:1.2,-3);
  \draw[thin,dashed] (axis cs:0.8165,-4.3) -- (axis cs:0.8165,4.3);
  \draw[line width=2.4pt,line cap=round] (axis cs:0.8165,-0.72) -- (axis cs:0.8165,0.72);
  \node[font=\scriptsize,anchor=west,inner sep=2pt] at (axis cs:0.83,0) {$\gamma_0$};
    \draw[densely dotted] (axis cs:0,1) -- (axis cs:0.5,1);
  \draw[densely dotted] (axis cs:1.2,1) -- (axis cs:1.5,1);
  \draw[densely dotted] (axis cs:0,-1) -- (axis cs:0.5,-1);
  \draw[densely dotted] (axis cs:1.2,-1) -- (axis cs:1.5,-1);
  \node[font=\scriptsize] at (axis cs:0.25,0) {$V_u=B$};
  \node[font=\scriptsize,anchor=west] at (axis cs:1.225,0) {$V_u=B$};
  \node[font=\scriptsize,fill=white,inner sep=1.5pt] at (axis cs:0.85,3.65) {$V_u<B$};
  \node[font=\scriptsize,fill=white,inner sep=1.5pt] at (axis cs:0.85,-3.65) {$V_u<B$};
  \node[font=\scriptsize,anchor=west] at (axis cs:1.225,3.65) {$h_t=4t^2$};
  \node[font=\scriptsize,anchor=west] at (axis cs:1.225,2.0) {$T_u$};
  \node[font=\scriptsize,anchor=west] at (axis cs:1.225,-2.0) {$T_u$};
\end{axis}
\end{tikzpicture}\\[2pt]
{\small (b)}
\end{minipage}
\caption{The hidden-cap family. (a)~A cross-section through the hidden
direction $u$, with lengths in units of $\sqrt d$ (angles not to
scale). The gray region is the shell $\frac12<\|x\|/\sqrt d<\frac65$, outside
of which the cutoff $\chi(\|x\|/\sqrt d)$ vanishes, darker where it
equals $1$. The dashed circle is the sphere $\|x\|=\sqrt{2d/3}$, around
which $\gamma_0$ concentrates. The hatched region, the intersection of the shell with the caps
$\{|s_u|>t-w\}$, is the only region where $V_u\ne B$. The strip
$t-w<|s_u|\le t$ has width $w=\frac{\eta}{t}$ and is not visible at
this scale. (b)~The same in the coordinates $(\|x\|/\sqrt d,\,s_u(x))$,
which determine $V_u$. The thick segment marks where most of the mass
of $\gamma_0$ lies and the cap test of Definition~\ref{def:lsi-cap-test}
accepts the region $T_u=\{|s_u|\ge t\}$ beyond the dotted lines.}
\label{fig:lsi-family}
\end{figure}

The hinge $h_t$ lowers the quadratic potential $B(x)=\tfrac34\|x\|^2$
inside two antipodal caps centered at $u$ and $-u$, where $u$ is a
hidden direction (Figure~\ref{fig:lsi-family}(a)) and using both caps keeps
the potential even.
A radial
cutoff confines the perturbation to the spherical shell where the
Gaussian target of $B$ concentrates, so the potential stays exactly
quadratic near the origin and outside the shell. The caps carry only a
fraction of order $e^{-t^2/2}/t$ of the Gaussian mass, which is what hides
them from the algorithm.

\begin{definition}[\bf Hidden-cap family]
\label{def:lsi-family}
Fix $\chi\in C_c^\infty(\R)$ with $0\le\chi\le1$,
$\operatorname{supp}\chi\subset\bigl(\frac12,\frac65\bigr)$,
and $\chi=1$ on $\bigl[\frac7{10},\frac{19}{20}\bigr]$.
Let $B(x):=\frac34\|x\|^2$ and $\gamma_0:=\mathcal{N}(0,\frac23 I_d)$.
For $u\in S^{d-1}$, let $  s_u(x):=\sqrt d\,\frac{u\cdot x}{\|x\|}\quad(x\ne0),
$  with  $s_u(0):=0,$
and define the potential
\begin{equation}
  V_u(x):=B(x)-\chi\left(\frac{\|x\|}{\sqrt d}\right)\,h_t(s_u(x)),
  \label{eq:lsi-potential}
\end{equation}
together with
\begin{equation}
  Z_u:=\int_{\R^d}
       e^{\chi\left(\frac{\|x\|}{\sqrt d}\right)\,h_t(s_u(x))}\,\gamma_0(dx),
  \qquad
  \pi_u(dx):=
       \frac{e^{\chi\left(\frac{\|x\|}{\sqrt d}\right)\,h_t(s_u(x))}}{Z_u}\,\gamma_0(dx).
  \label{eq:lsi-target}
\end{equation}
\end{definition}

\paragraph{Basic properties.} Three facts are immediate. First, $V_u\in C^\infty(\R^d)$: the cutoff
$\chi(\|x\|/\sqrt d)$ vanishes for $\|x\|\le\sqrt d/2$, so the
discontinuity of $s_u$ at the origin is immaterial. Second, since
$0\le\chi h_t\le4t^2$, we have $1\le Z_u\le e^{4t^2}$ and since
$e^{-B(x)}\,dx$ is a constant multiple of $\gamma_0(dx)$, the measure
$\pi_u$ is the target distribution of $V_u$. Third, the perturbation
$\chi(\|x\|/\sqrt d)\,h_t(s_u(x))$ vanishes, together with its gradient,
outside the shell $\{\sqrt d/2<\|x\|<6\sqrt d/5\}$ (as
$\operatorname{supp}\chi\subset(1/2,6/5)$) and on the set
$\{|s_u|\le t-w\}$ (as $h_t=0$ there, by Lemma~\ref{lem:lsi-hinge}).
Hence
\begin{equation}
  \bigl(V_u(x),\nabla V_u(x)\bigr)=\bigl(B(x),\nabla B(x)\bigr)
  \qquad\text{whenever }
  \|x\|\notin\bigl(\tfrac{\sqrt d}{2},\tfrac{6\sqrt d}{5}\bigr)
  \ \text{ or }\ |s_u(x)|\le t-w.
  \label{eq:lsi-agree}
\end{equation}
In words, the oracle for $V_u$ returns the same replies as the oracle
for $B$ unless the query lies in the shell and satisfies $|s_u(x)|>t-w$,
the hatched region in Figure~\ref{fig:lsi-family}(b).

The lower bound in Section~\ref{sec:lsi-proof} uses three further
properties of the family, which occupy the rest of this subsection:
(1) $V_u$ is $O(t^2)$-smooth with a known minimizer
(Lemma~\ref{lem:lsi-regularity}); (2) under $\gamma_0$ the latitude $s_u$ is
approximately standard Gaussian, so the caps carry mass of order
$e^{-t^2/2}/t$ (Lemma~\ref{lem:lsi-spherical}); and (3) the perturbation
changes the normalizer by at most a constant factor, so that $\gamma_0$
is a warm start for every $\pi_u$ at once (Lemma~\ref{lem:lsi-normalizer}).

\paragraph{Regularity.}
We first record the smoothness scale of $V_u$ and the location of its minimizer.

\begin{lemma}[\bf Regularity and the known minimizer]
\label{lem:lsi-regularity}
Each $V_u$ is even, with $V_u(0)=0$ and $V_u(x)\ge\|x\|^2/2$ for all
$x\in\R^d$. In particular, the origin is its unique minimizer and
$\pi_u$ has mean zero. Moreover,
$\|\nabla^2V_u\|_{\mathrm{op}}\le C_Lt^2$ on $\R^d$ for a universal
constant $C_L$.
\end{lemma}

\begin{proof}
Away from the origin, every term in \eqref{eq:lsi-potential}
is smooth. Near the origin, the perturbation vanishes
identically, so $V_u=B$ there and $V_u$ is smooth everywhere.
The evenness of $h_t$ and the identity $s_u(-x)=-s_u(x)$
show that $V_u$ is even.

We first bound the Hessian.
For $x\ne0$, write $r=\|x\|$, $\theta=\frac{x}{r}$, and
$b=u\cdot\theta$.
Differentiating $s_u(x)=\sqrt d\,b$ gives
\begin{align*}
  \nabla s_u(x)
    &=\frac{\sqrt d}{r}(u-b\theta),\qquad
  \nabla^2s_u(x)
    =\frac{\sqrt d}{r^2}
      \bigl(-u\theta^\top-\theta u^\top-bI_d
            +3b\theta\theta^\top\bigr).
\end{align*}
Hence $\|\nabla s_u(x)\|\le\frac{\sqrt d}{r}$ and
$\|\nabla^2s_u(x)\|_{\mathrm{op}}\le\frac{6\sqrt d}{r^2}$.
Write $a(x)=\chi\bigl(\frac{r}{\sqrt d}\bigr)$ and
$H(x)=h_t(s_u(x))$.
On the annulus $\frac{\sqrt d}{2}<r<\frac{6\sqrt d}{5}$,
Lemma~\ref{lem:lsi-hinge} and the chain rule give
$\|H\|_\infty\le4t^2$,
$\|\nabla H(x)\|\le\frac{2t\sqrt d}{r}\le4t$, and
\[
  \|\nabla^2H(x)\|_{\mathrm{op}}
  \le
  \|h_t''\|_\infty\frac{d}{r^2}
  +\|h_t'\|_\infty\frac{6\sqrt d}{r^2}
  \le Ct^2.
\]
Radial differentiation gives
$\|\nabla a(x)\|\le\frac{\|\chi'\|_\infty}{\sqrt d}$
and
$\|\nabla^2a(x)\|_{\mathrm{op}}
\le\frac{\|\chi''\|_\infty}{d}
+\frac{\|\chi'\|_\infty}{r\sqrt d}\le\frac{C}{d}$.
Using the product rule and the preceding bounds,
\begin{align*}
  \nabla^2(aH)
  &=a\nabla^2H+H\nabla^2a
    +\nabla a(\nabla H)^\top+\nabla H(\nabla a)^\top,\\
  \|\nabla^2(aH)\|_{\mathrm{op}}
  &\le |a|\|\nabla^2H\|_{\mathrm{op}}
       +|H|\|\nabla^2a\|_{\mathrm{op}}
       +2\|\nabla a\|\|\nabla H\|\\
  &\le Ct^2+\frac{Ct^2}{d}+\frac{Ct}{\sqrt d}
   \le Ct^2.
\end{align*}
Outside the annulus, the perturbation vanishes.
Adding $\nabla^2B=(3/2)I_d$ proves the claimed Hessian bound.
The identity at zero follows from $V_u=B$ near zero.

Finally, wherever the perturbation is nonzero,
$r^2\ge\frac{d}{4}$.
Taking $D$ sufficiently large ensures $d\ge64t^2$, so
\begin{align*}
  4t^2
  &\le\frac{d}{16}\le\frac{r^2}{4},\qquad
  V_u(x)
  \ge\frac{3r^2}{4}-4t^2
   \ge\frac{r^2}{2}.
\end{align*}
At all other points, $V_u(x)=\frac{3r^2}{4}$.
Thus $V_u(x)\ge\frac{\|x\|^2}{2}$ everywhere.
Since $V_u(0)=0$, the origin is the unique minimizer.
The quadratic lower bound implies finite moments of every order,
and evenness then gives $\mathbb E_{\pi_u}X=0$.
\end{proof}

\paragraph{The law of the latitude under $\gamma_0$.}
Under $\gamma_0$, the direction $x/\|x\|$ is uniform on $S^{d-1}$ and
independent of the radius, so the law of the latitude $s_u(x)$ does not
depend on $u$. We write $q_d$ for its density, $p_d(a):=\gamma_0(|s_u|\ge a)$
for its tail ($a\ge0$), and $\varphi(s):=(2\pi)^{-1/2}e^{-s^2/2}$ for the
standard Gaussian density. The latitude is close to standard Gaussian on
the scale of $t$, and its tail is what makes the caps light. The
following estimates are proved in Appendix~\ref{app:lsi-spherical}.

\begin{lemma}[\bf Spherical density and tail comparison]
\label{lem:lsi-spherical}
The latitude $s_u(x)$ has density
\[
  q_d(s)=
  \frac{\Gamma(d/2)}{\sqrt{\pi d}\,\Gamma((d-1)/2)}
  \Bigl(1-\frac{s^2}{d}\Bigr)^{(d-3)/2}
  \qquad(|s|<\sqrt d),
\]
and $q_d(s)=0$ otherwise. Under the standing assumption $d\ge Dt^4$,
\begin{enumerate}
\item[(i)] $q_d(s)\le e^{5/d}\varphi(s)$ for all $s\in\R$;
\item[(ii)] $0.99\,\varphi(s)\le q_d(s)\le1.01\,\varphi(s)$ for $|s|\le3t$;
\item[(iii)] $p_d(t) = \Theta( e^{-t^2/2}/t)$;
\item[(iv)] $p_d(t-w)\le1.002\,p_d(t)$.
\end{enumerate}
\end{lemma}

The quantity $  \mathcal R_t:=\frac{1}{p_d(t)\,t^2} = \Theta(\frac{e^{t^2/2}}{t})$
will serve as the resistance of the caps: it governs both the
log-Sobolev constant of $\pi_u$ (Section~\ref{sec:lsi-constant}) and
the query complexity (Section~\ref{sec:lsi-proof}). Since
$\int s^2q_d(s)\,ds=1$, Chebyshev's inequality gives $p_d(t)t^2\le1$,
that is, $\mathcal R_t\ge1$.
\paragraph{Normalizers.}
The next lemma controls the angular normalizers uniformly
over the strengths introduced by the radial cutoff.

\begin{lemma}[\bf Uniform normalizers and a common warm start]
\label{lem:lsi-normalizer}
For $\beta\in[0,1]$, define
$Z_\beta:=\int_{\R}e^{\beta h_t(s)}q_d(s)\,ds$.
Then $1\le Z_\beta\le3.1$ for every $\beta\in[0,1]$,
and $1\le Z_u\le3.1$ for every $u\in S^{d-1}$.
Furthermore, $d\gamma_0/d\pi_u\le3.1$ everywhere.
\end{lemma}

\begin{proof}
Let $h_{\mathrm{unc}}(s):=2t(|s|-t)_+$.
For $\beta\in[0,1]$, put
$m=2\beta t$ and $\delta=2\beta(1-\beta)t^2\ge0$.
Completing the square gives
\[
  e^{\beta h_{\mathrm{unc}}(s)}\varphi(s)
  =
  \max\bigl\{
    \varphi(s),
    e^{-\delta}\varphi(s-m),
    e^{-\delta}\varphi(s+m)
  \bigr\}.
\]
Indeed, $\beta h_{\mathrm{unc}}(s)$ is the maximum of
$0$, $2\beta t(s-t)$, and $2\beta t(-s-t)$,
and each term gives the corresponding Gaussian density.
Bounding the maximum by the sum shows that its integral
is at most $1+2e^{-\delta}\le3$.
Since $\widehat h_t\le h_{\mathrm{unc}}$ and
$h_t\le\widehat h_t+2\eta$,
Lemma~\ref{lem:lsi-spherical} implies
\[
  Z_\beta
  \le e^{2\eta+C/d}
      \int_{\R}e^{\beta h_{\mathrm{unc}}(s)}\varphi(s)\,ds
  \le3e^{2\eta+C/d}
  \le3.1,
\]
after increasing $D$.
The lower bound follows from $h_t\ge0$.

Under $\gamma_0$, the radius and direction are independent,
and the direction has law $\omega_d$.
Rotational invariance therefore gives
$Z_u=\int Z_{\chi(r/\sqrt d)}\,\nu_0(dr)$,
where $\nu_0$ is the law of the radius under $\gamma_0$.
Thus $1\le Z_u\le3.1$.
Finally, \eqref{eq:lsi-target} gives
$d\gamma_0/d\pi_u
=Z_u e^{-\chi(\|x\|/\sqrt d)h_t(s_u(x))}
\le Z_u\le3.1$.
\end{proof}

\subsection{The Cap Test}
\label{sec:lsi-cap-test}

Under $\pi_u$ the caps carry a constant fraction of the mass, whereas a
point chosen without knowledge of $u$ lands in them with probability at
most $p_d(t)$. The following test turns this discrepancy into a query
lower bound: an algorithm that makes few queries never sees the caps,
so its output is no better than a point chosen without knowledge of
$u$. Throughout, $U$ denotes a uniform random point of $S^{d-1}$ and
$w=\eta/t$ the smoothing width of Lemma~\ref{lem:lsi-hinge}.

\begin{definition}[\bf Cap test]
\label{def:lsi-cap-test}
For $u\in S^{d-1}$, let $T_u:=\{x\in\R^d:|s_u(x)|\ge t\}$. The cap test
associated with $u$ accepts $x$ if $x\in T_u$.
\end{definition}

The test depends on the hidden direction $u$.
It serves only as an analytical tool and is never actually executed by the algorithm.
We use one fact about
it repeatedly. For fixed $y\ne0$, the latitude $s_U(y)$ of $y$ relative
to the random direction $U$ has the same law as the latitude of a point
of $\gamma_0$ relative to a fixed direction, since both are $\sqrt d$
times a coordinate of a uniform point of $S^{d-1}$. Hence
\begin{equation}
  \Pr\bigl(|s_U(y)|\ge a\bigr)=p_d(a)\qquad(y\ne0,\ a\ge0).
  \label{eq:lsi-duality}
\end{equation}

\begin{proposition}[\bf The output fails the cap test]
\label{prop:lsi-output-test}
Let $X_u$ be the output on $V_u$ of a randomized algorithm that makes
at most $q$ queries.
Then
\begin{equation}
  \Pr(X_U\in T_U)
  \le(1.002\,q+1)\,p_d(t).
  \label{eq:lsi-output-test}
\end{equation}
\end{proposition}

\begin{proof}
Consider a \emph{reference run} of the algorithm on the base function $B(x):=\frac{3}{4}\Vert{}x\Vert{}^2$ using the same internal randomness as on $V_U$. 
Let $Y$ denote its output, and let $Y_1,\dots,Y_q$ denote its queries, padded with zeros if it halts early.
The reference run does not involve $U$, and
the internal randomness is independent of $U$, so $Y_1,\dots,Y_q,Y$
are independent of $U$.

Let $H$ be the event that some reference query $Y_i$ lies in the
perturbation region, that is, 
\[
  \frac{\sqrt d}{2}<\|Y_i\|<\frac{6\sqrt d}{5}
  \qquad\text{and}\qquad
  |s_U(Y_i)|>t-w
  \qquad\text{for some } 1\le i\le q.
\]
On $H^c$, the two runs coincide. The first queries match since the internal randomness is shared. Whenever the histories agree so far, the next query equals the corresponding reference query and thus lies outside the perturbation region on $H^c$, ensuring by~\eqref{eq:lsi-agree} that both oracles return the same reply.
By induction, the transcripts, halting times, and outputs coincide. In particular, $X_U=Y$ on $H^c$.

Since each $Y_i$ is independent of $U$ and the origin lies outside the
perturbation region, \eqref{eq:lsi-duality} and a union bound give
$\Pr(H)\le q\,p_d(t-w)$. Likewise, $Y$ is independent of $U$ and
$0\notin T_U$, so $\Pr(Y\in T_U)\le p_d(t)$. Therefore, by
Lemma~\ref{lem:lsi-spherical}(iv),
\[
  \Pr(X_U\in T_U)
  \le\Pr(H)+\Pr(Y\in T_U)
  \le q\,p_d(t-w)+p_d(t)
  \le(1.002\,q+1)\,p_d(t).\qedhere
\]
\end{proof}

\begin{proposition}[\bf The target passes the cap test]
\label{prop:lsi-target-test}
For every $u\in S^{d-1}$, $\pi_u(T_u)>0.55$.
\end{proposition}

\begin{proof}
Let $X\sim\gamma_0$ and $R:=\|X\|$. The coordinates of $X$ are
independent $\mathcal N(0,\frac23)$ variables, so
$\mathbb E R^2=\frac{2d}{3}$ and $\operatorname{Var}(R^2)=\frac{8d}{9}$.
Let $A:=\{\frac{7}{10}\sqrt d\le R\le\frac{19}{20}\sqrt d\}$, the event
on which the radial cutoff equals one. Because $(\frac7{10})^2<\frac12$
and $\frac56<(\frac{19}{20})^2$, the event
$\{|R^2-\frac{2d}{3}|\le\frac d6\}$ is contained in $A$, and
Chebyshev's inequality gives $\Pr(A)\ge1-\frac{32}{d}\ge0.99$, as
$d\ge Dt^4\ge3200$.

For $t+1\le s\le3t-1$ we have $h_t(s)=2t(s-t)$ by
Lemma~\ref{lem:lsi-hinge}, since $w<1$, and completing the square gives
$e^{h_t(s)}\varphi(s)=\varphi(s-2t)$. Under $\gamma_0$ the radius and
the latitude are independent, and $h_t$ and $q_d$ are even. Restricting
the expectation below to $A$ and to $t+1\le|s_u(X)|\le3t-1$, and using
Lemma~\ref{lem:lsi-spherical}(ii), we obtain
\begin{align*}
  Z_u\,\pi_u(T_u)
  &=\mathbb E_{\gamma_0}\Bigl[\mathbf 1_{\{|s_u(X)|\ge t\}}\,
     e^{\chi(R/\sqrt d)\,h_t(s_u(X))}\Bigr]
   \ge2\Pr(A)\int_{t+1}^{3t-1}e^{h_t(s)}q_d(s)\,ds\\
  &\ge2(0.99)^2\int_{t+1}^{3t-1}\varphi(s-2t)\,ds
   =2(0.99)^2\,\Pr\bigl(|G|\le t-1\bigr),
\end{align*}
where $G\sim\mathcal N(0,1)$. By Chebyshev's inequality and $t\ge4$,
$\Pr(|G|\le t-1)\ge1-(t-1)^{-2}\ge\frac89$. With $Z_u\le3.1$ from
Lemma~\ref{lem:lsi-normalizer},
\[
  \pi_u(T_u)\ge\frac{2(0.99)^2}{3.1}\cdot\frac89>0.55.\qedhere
\]
\end{proof}

The two propositions combine through the inequality
$\operatorname{TV}(\rho,\pi)\ge\pi(T)-\rho(T)$, valid for every
measurable set $T$.

\begin{corollary}[\bf Cap separation]
\label{cor:lsi-cap-separation}
Let $\mathcal A$ and $q$ be as in Proposition~\ref{prop:lsi-output-test},
and let $\rho_u$ be the law of its output on $V_u$. Then
\begin{equation}
  \mathbb E_U\operatorname{TV}(\rho_U,\pi_U)
  \ge\mathbb E_U\bigl[\pi_U(T_U)-\rho_U(T_U)\bigr]
  >0.55-(1.002\,q+1)\,p_d(t).
  \label{eq:lsi-cap-separation}
\end{equation}
\end{corollary}

\begin{proof}
The first inequality holds for every value of $U$, and
$\mathbb E_U\rho_U(T_U)=\Pr(X_U\in T_U)$, so
\eqref{eq:lsi-cap-separation} follows from the two propositions.
\end{proof}

The corollary concerns algorithms with a bounded number of queries.
A truncation extends it to the expected number of queries.

\begin{corollary}[\bf Expected number of queries]
\label{cor:lsi-expected-queries}
Let $\mathcal A$ be a randomized algorithm, let $\rho_u$ be the law
of its output on $V_u$, and let $N_u$ be its number of queries on
$V_u$. If $\operatorname{TV}(\rho_u,\pi_u)\le\frac1{10}$ for every
$u$, then $\sup_u\mathbb E[N_u]\ge\frac{1}{50\,p_d(t)}$.
\end{corollary}

\begin{proof}
Let $T:=\mathbb E_U\mathbb E[N_U]\le\sup_u\mathbb E[N_u]$, and
assume $T<\infty$, since otherwise there is nothing to prove. Set
$q:=\max\{1,\lceil10T\rceil\}$ and let $\mathcal A$ halt with output
$0$ as soon as it attempts a $(q+1)$-st query. The truncated
algorithm makes at most $q$ queries, and its output on $V_u$ differs
from that of $\mathcal A$ only on the event $\{N_u>q\}$, so its
output laws $\rho_u'$ satisfy
$\operatorname{TV}(\rho_u',\rho_u)\le\Pr(N_u>q)$ and
\[
  \mathbb E_U\operatorname{TV}(\rho_U',\pi_U)
  \le\mathbb E_U\operatorname{TV}(\rho_U,\pi_U)+\mathbb E_U\Pr(N_U>q)
  \le\frac1{10}+\frac Tq\le\frac15,
\]
where the first inequality is the triangle inequality, the second is
Markov's inequality, and the third uses $q\ge10T$.
Corollary~\ref{cor:lsi-cap-separation} applied to the truncated
algorithm gives $\frac15>0.55-(1.002\,q+1)\,p_d(t)$, that is,
$(1.002\,q+1)\,p_d(t)>0.35$. Since $q\le10T+1$ and
$p_d(t)\le\frac1{t^2}\le\frac1{16}$,
\[
  T>\frac{0.35-2.002\,p_d(t)}{10.02\,p_d(t)}>\frac{1}{50\,p_d(t)}.
  \qedhere
\]
\end{proof}

\subsection{The Log-Sobolev Inequality for the Hidden-cap Family}
\label{sec:lsi-constant}

Recall the standing assumption $t\ge4$, $d\ge Dt^4$, and that
$\mathcal R_t=\frac{1}{p_d(t)t^2}=\Theta(e^{t^2/2}/t)$. By
Lemma~\ref{prop:lsi-bakry-emery}(i), the
Gaussian $\gamma_0$ satisfies $\mathsf{LSI}(\frac32)$, and since
$0\le\chi h_t\le4t^2$, Lemma~\ref{prop:lsi-bounded-perturbation}(i)
gives only $\mathsf{LSI}(\frac32e^{-4t^2})$ for $\pi_u$.
This estimate is far from tight.
We prove the much better rate $\frac{c}{\mathcal R_t}$ in
Theorem~\ref{thm:lsi-family}.


We first prove the global inequality from an $\mathsf{LSI}$ on balls
and a global Poincar\'e inequality
(Section~\ref{sec:lsi-global-proof}). We then establish the
Poincar\'e inequality (Section~\ref{sec:lsi-poincare-proof}), the
local $\mathsf{LSI}$ (Section~\ref{sec:lsi-local-proof}), and the
angular inequality that supplies the resistance scale
(Section~\ref{sec:lsi-angular-proof}). The standard functional
inequalities used throughout are recorded in
Appendix~\ref{app:lsi-functional-tools}.

\subsubsection{The Global Log-Sobolev Inequality}
\label{sec:lsi-global-proof}
The main result of this section is the following.
\begin{theorem}[\bf Log-Sobolev inequality]
\label{thm:lsi-family}
There is a universal constant $c>0$ such that every $\pi_u$
satisfies $\mathsf{LSI}(\frac{c}{\mathcal R_t})$.
\end{theorem}

The proof has two ingredients. The first is a log-Sobolev inequality
for $\pi_u$ conditioned on a ball, whose rate deteriorates with the
radius. Write $\mathbb B_R:=\{x\in\R^d:\|x\|<R\}$.

\begin{restatable}[\bf Local $\mathsf{LSI}$]{lemma}{locallsi}
\label{lem:lsi-local}
For every $R>0$, the conditional law
$\pi_u(\cdot\mid\mathbb B_R)$ satisfies
$\mathsf{LSI}\bigl(\frac{c}{\mathcal R_t+R^2/d}\bigr)$
for the Euclidean gradient on $\mathbb B_R$.
\end{restatable}

The second is a Poincar\'e inequality for $\pi_u$ itself, which
controls the Gaussian tails outside a ball of radius of order
$\sqrt{\mathcal R_td}$.

\begin{restatable}[\bf Global $\mathsf{PI}$]{lemma}{globalpi}
\label{lem:lsi-global-poincare}
Every $\pi_u$ satisfies $\mathsf{PI}(\frac{c}{\mathcal R_t})$.
\end{restatable}

Lemmas~\ref{lem:lsi-local} and~\ref{lem:lsi-global-poincare} are
proved in Sections~\ref{sec:lsi-local-proof}
and~\ref{sec:lsi-poincare-proof}. 
Assuming these two lemmas, we glue
the local inequality to the Gaussian inequality outside the ball via a
partition of unity, which yields a log-Sobolev inequality with a
defect, and then remove the defect using the Poincar\'e inequality.

\begin{proof}[Proof of Theorem~\ref{thm:lsi-family}]
Fix $f\in C_c^\infty(\R^d)$ and set $R:=4\sqrt{\mathcal R_td}$, so
that $R^2/d=16\mathcal R_t$ and $R/2\ge2\sqrt d$. Choose a smooth
radial function $\vartheta$ with $\vartheta=0$ on $\mathbb B_{R/2}$,
$\vartheta=\frac\pi2$ outside $\mathbb B_R$, and
$\|\nabla\vartheta\|\le C/R$, and split
$f^2=f^2\cos^2\vartheta+f^2\sin^2\vartheta$. We bound the entropy of
each piece separately.

The function $f\cos\vartheta$ vanishes outside $\mathbb B_R$. Since
$d\pi_u/d\gamma_0\ge Z_u^{-1}\ge\frac1{3.1}$ by
Lemma~\ref{lem:lsi-normalizer} and $\int\|x\|^2\,d\gamma_0=\frac{2d}3$,
Markov's inequality gives
$\pi_u(\mathbb B_R)\ge\frac1{3.1}\bigl(1-\frac{2d}{3R^2}\bigr)\ge\frac14$,
so Lemma~\ref{lem:lsi-entropy-calculus}(iii) and
Lemma~\ref{lem:lsi-local}, with $\mathcal R_t+R^2/d=17\mathcal R_t$,
give
\begin{align*}
  \operatorname{Ent}_{\pi_u}(f^2\cos^2\vartheta)
  &=\pi_u(\mathbb B_R)
    \operatorname{Ent}_{\pi_u(\cdot\mid\mathbb B_R)}(f^2\cos^2\vartheta)
    +\log\frac1{\pi_u(\mathbb B_R)}\int f^2\cos^2\vartheta\,d\pi_u\\
  &\le C\mathcal R_t\int\|\nabla(f\cos\vartheta)\|^2\,d\pi_u
    +(\log4)\int f^2\cos^2\vartheta\,d\pi_u.
\end{align*}
The function $f\sin\vartheta$ and its gradient vanish on
$\mathbb B_{R/2}$. Outside $\mathbb B_{R/2}\supseteq\mathbb B_{2\sqrt d}$
the perturbation vanishes, so $\pi_u=Z_u^{-1}\gamma_0$ there. By
Lemma~\ref{prop:lsi-bakry-emery}(i) with $U=\frac34\|x\|^2$, the
Gaussian $\gamma_0$ satisfies $\mathsf{LSI}(\frac32)$, so
Lemma~\ref{lem:lsi-entropy-calculus}(iii) with $c=Z_u^{-1}$ and
$Z_u\le3.1$ give
\begin{align*}
  \operatorname{Ent}_{\pi_u}(f^2\sin^2\vartheta)
  &=Z_u^{-1}\operatorname{Ent}_{\gamma_0}(f^2\sin^2\vartheta)
    +(\log Z_u)\int f^2\sin^2\vartheta\,d\pi_u\\
  &\le\frac43\int\|\nabla(f\sin\vartheta)\|^2\,d\pi_u
    +(\log3.1)\int f^2\sin^2\vartheta\,d\pi_u.
\end{align*}

By subadditivity (Lemma~\ref{lem:lsi-entropy-calculus}(ii)), the two
bounds add up to a bound on $\operatorname{Ent}_{\pi_u}(f^2)$.
Moreover,
\[
  \|\nabla(f\cos\vartheta)\|^2+\|\nabla(f\sin\vartheta)\|^2
  =\|\nabla f\|^2+f^2\|\nabla\vartheta\|^2
  \le\|\nabla f\|^2+\frac{C}{R^2}f^2,
\]
and $\mathcal R_t/R^2=\frac1{16d}$. Using $\mathcal R_t\ge1$, we
obtain
\begin{equation}
  \operatorname{Ent}_{\pi_u}(f^2)
  \le C\mathcal R_t\int\|\nabla f\|^2\,d\pi_u+C\int f^2\,d\pi_u.
  \label{eq:lsi-defective}
\end{equation}
Multiplication by smooth spatial cutoffs extends this inequality to
bounded smooth functions of finite energy since boundedness gives
convergence of the entropy terms, and the cutoff contribution to the
energy vanishes.

It remains to remove the last term. The function
$f-\mathbb E_{\pi_u}f$ is bounded and has the same gradient as $f$,
so \eqref{eq:lsi-defective} applies to it, and
Lemma~\ref{lem:lsi-rothaus} together with
Lemma~\ref{lem:lsi-global-poincare} gives
\[
  \operatorname{Ent}_{\pi_u}(f^2)
  \le\operatorname{Ent}_{\pi_u}\bigl((f-\mathbb E_{\pi_u}f)^2\bigr)
    +2\operatorname{Var}_{\pi_u}(f)
  \le C\mathcal R_t\int\|\nabla f\|^2\,d\pi_u
    +C\operatorname{Var}_{\pi_u}(f)
  \le C\mathcal R_t\int\|\nabla f\|^2\,d\pi_u,
\]
which is $\mathsf{LSI}(\frac{c}{\mathcal R_t})$ with $c=2/C$.
\end{proof}
\subsubsection{The Local Log-Sobolev Inequality}
\label{sec:lsi-local-proof}

\locallsi*

The proof of Lemma~\ref{lem:lsi-local} combines a radial curvature
bound with a uniform angular estimate. We state the angular estimate
here and prove it in Section~\ref{sec:lsi-angular-proof}.

Conditionally on $\|x\|=r$, the direction $x/\|x\|$ under $\pi_u$
carries the hinge with strength $\chi\bigl(\frac{r}{\sqrt d}\bigr)$.
For
$\beta\in[0,1]$, define the angular law of strength $\beta$ as the
probability law
$\sigma_\beta(d\theta)\propto e^{\beta h_t(s_u(\theta))}\omega_d(d\theta)$,
where $s_u(\theta)=\sqrt d\,u\cdot\theta$. Since the latitude $s_u(\theta)$ has
density $q_d$ under $\omega_d$, its normalizer is
$\int e^{\beta h_t(s_u(\theta))}\,\omega_d(d\theta)
 =\int e^{\beta h_t(s)}q_d(s)\,ds=Z_\beta$, the quantity of
Lemma~\ref{lem:lsi-normalizer}. 
For $r>0$, let $\mu_{r,\beta}$ be the
law of $r\Theta$ when $\Theta\sim\sigma_\beta$.

\begin{restatable}[\bf Angular LSI]{lemma}{angularlsi}
\label{lem:lsi-angular}
For every $r>0$ and $\beta\in[0,1]$, the law $\mu_{r,\beta}$ satisfies
$\mathsf{LSI}\bigl(\frac{cd}{r^2\mathcal R_t}\bigr)$ for the intrinsic
gradient on the sphere of radius $r$. For $\beta=0$, it satisfies the
sharper $\mathsf{LSI}\bigl(\frac{cd}{r^2}\bigr)$.
\end{restatable}

To combine this estimate with the radial marginal, we use the
following entropy decomposition lemma.

\begin{lemma}[\bf Two-scale LSI]
\label{lem:lsi-two-level}
Let $\mu(dx,dy)=\nu(dx)\mu_x(dy)$, where $\nu$ is a probability
measure on an interval and $\mu_x$ has a positive density $k_x$,
smooth in $x$, with respect to a fixed reference measure. Suppose that
$\nu$ satisfies $\mathsf{LSI}(\alpha_0)$ for an energy
$\int a|g'|^2\,d\nu$ with $a>0$, that each $\mu_x$ satisfies
$\mathsf{LSI}(\alpha_1)$ for a gradient energy
$\int\Gamma_x(f)\,d\mu_x$, and that
$K^2:=\sup_xa(x)\operatorname{osc}(\partial_x\log k_x)^2<\infty$. Then
\begin{equation}
  \operatorname{Ent}_\mu(f^2)
  \le\frac4{\alpha_0}\,\mathbb E_\mu\bigl[a(x)|\partial_xf|^2\bigr]
   +\frac2{\alpha_1}\Bigl(1+\frac{K^2}{2\alpha_0}\Bigr)
    \mathbb E_\mu\Gamma_x(f).
  \label{eq:lsi-two-level}
\end{equation}
If $k_x$ does not depend on $x$, then $K=0$ and $\frac4{\alpha_0}$ can
be replaced by $\frac2{\alpha_0}$.
\end{lemma}

\begin{proof}[Proof of Lemma~\ref{lem:lsi-two-level}]
Set $g(x):=(\mathbb E_{\mu_x}f^2)^{1/2}$ and
$\ell_x:=\partial_x\log k_x$. Differentiating $\int k_x=1$ gives
$\mathbb E_{\mu_x}\ell_x=0$. By Lemma~\ref{lem:lsi-entropy-calculus}(i)
and $\mathsf{LSI}(\alpha_1)$ for each $\mu_x$,
\begin{align*}
  \operatorname{Ent}_\mu(f^2)
  &=\mathbb E_\nu\operatorname{Ent}_{\mu_x}(f^2)
    +\operatorname{Ent}_\nu(g^2)\\
  &\le\frac2{\alpha_1}\,\mathbb E_\mu\Gamma_x(f)
    +\operatorname{Ent}_\nu(g^2),
\end{align*}
so it remains to bound $\operatorname{Ent}_\nu(g^2)$ through the
marginal inequality, which requires a bound on $|g'|$.

Let $m_x$ be the midpoint of the range of $\ell_x$, so that
$|\ell_x-m_x|\le\frac12\operatorname{osc}(\ell_x)$. Differentiating
$g^2=\int f^2k_x$ under the integral gives
\begin{align*}
  2gg'
  &=2\,\mathbb E_{\mu_x}(f\partial_xf)+\mathbb E_{\mu_x}(f^2\ell_x)\\
  &=2\,\mathbb E_{\mu_x}(f\partial_xf)
    +\mathbb E_{\mu_x}\bigl[(f^2-g^2)(\ell_x-m_x)\bigr],
\end{align*}
where the first line uses $\partial_xk_x=\ell_xk_x$ and the second
uses $\mathbb E_{\mu_x}\ell_x=0$ together with
$\mathbb E_{\mu_x}(f^2-g^2)=0$. Moreover,
\begin{align*}
  \mathbb E_{\mu_x}|f^2-g^2|
  &\le\sqrt{\mathbb E_{\mu_x}(f-g)^2}\,\sqrt{\mathbb E_{\mu_x}(f+g)^2}\\
  &=\sqrt{2g^2-2g\,\mathbb E_{\mu_x}f}\,
    \sqrt{2g^2+2g\,\mathbb E_{\mu_x}f}
   =2g\sqrt{\operatorname{Var}_{\mu_x}(f)},
\end{align*}
where the first line is Cauchy--Schwarz and the second uses
$\mathbb E_{\mu_x}f^2=g^2$. Dividing the identity for $2gg'$ by $2g$
and applying Cauchy--Schwarz to its first term, we obtain
\begin{align*}
  |g'|
  &\le\sqrt{\mathbb E_{\mu_x}|\partial_xf|^2}
    +\frac{\operatorname{osc}(\ell_x)}{2}
     \sqrt{\operatorname{Var}_{\mu_x}(f)}\\
  &\le\sqrt{\mathbb E_{\mu_x}|\partial_xf|^2}
    +\frac{\operatorname{osc}(\ell_x)}{2\sqrt{\alpha_1}}
     \sqrt{\mathbb E_{\mu_x}\Gamma_x(f)},
\end{align*}
where the second line is Lemma~\ref{lem:lsi-implies-poincare} for
$\mu_x$. Squaring with $(u+v)^2\le2u^2+2v^2$, multiplying by $a(x)$,
and using the definition of $K$,
\[
  a(x)|g'|^2
  \le2a(x)\,\mathbb E_{\mu_x}|\partial_xf|^2
    +\frac{K^2}{2\alpha_1}\,\mathbb E_{\mu_x}\Gamma_x(f).
\]
Integrating against $\nu$ and using $\mathsf{LSI}(\alpha_0)$ for
$\nu$,
\begin{align*}
  \operatorname{Ent}_\nu(g^2)
  &\le\frac2{\alpha_0}\int a|g'|^2\,d\nu\\
  &\le\frac4{\alpha_0}\,\mathbb E_\mu\bigl[a(x)|\partial_xf|^2\bigr]
    +\frac{K^2}{\alpha_0\alpha_1}\,\mathbb E_\mu\Gamma_x(f).
\end{align*}
Adding the bound from the first display proves
\eqref{eq:lsi-two-level}. If $k_x$ does not depend on $x$, then
$\ell_x=0$, the covariance term is absent, and
$|g'|^2\le\mathbb E_{\mu_x}|\partial_xf|^2$ without the factor $2$,
which gives the coefficient $\frac2{\alpha_0}$. Finally, at zeros of
$g$ the division above is not allowed, so one applies the argument
to $(g^2+\varepsilon)^{1/2}$ and lets $\varepsilon\downarrow0$.
\end{proof}

We can now prove Lemma~\ref{lem:lsi-local}. In polar coordinates,
$\pi_u(\cdot\mid\mathbb B_R)$ disintegrates into a radial marginal
and the angular laws $\sigma_{\beta(r)}$, and the proof consists in
checking the three hypotheses of Lemma~\ref{lem:lsi-two-level} for
this disintegration: a log-Sobolev inequality for the radial
marginal, which comes from the curvature criterion, a uniform
log-Sobolev inequality for the angular laws, which is
Lemma~\ref{lem:lsi-angular}, and a bound on the coupling constant $K$,
which is where the standing assumption $d\ge Dt^4$ enters.

\begin{proof}[Proof of Lemma~\ref{lem:lsi-local}]
Write $x=r\theta$ and $\beta(r):=\chi\left(\frac{r}{\sqrt{d}}\right)$. By
\eqref{eq:lsi-target}, the conditional law $\pi_u(\cdot\mid\mathbb B_R)$
decomposes as $\nu_R(dr)\,\sigma_{\beta(r)}(d\theta)$ with radial
marginal
\[
  \nu_R(dr)\propto
  \mathbf 1_{(0,R)}(r)\,r^{d-1}e^{-3r^2/4}Z_{\beta(r)}\,dr.
\] We
apply Lemma~\ref{lem:lsi-two-level} with marginal variable $r$ and
weight $a\equiv1$.

For the radial marginal, consider the baseline law on $(0,R)$ with
density proportional to $r^{d-1}e^{-3r^2/4}$. Its potential
$U_0(r):=\frac34r^2-(d-1)\log r$ satisfies
$U_0''(r)=\frac32+\frac{d-1}{r^2}\ge\frac32$, so
Lemma~\ref{prop:lsi-bakry-emery} (applied on $[\varepsilon,R]$
as $\varepsilon\downarrow0$) yields $\mathsf{LSI}(\frac32)$. Since the
density of $\nu_R$ differs from this baseline by the factor
$Z_{\beta(r)}\in[1,3.1]$ from Lemma~\ref{lem:lsi-normalizer},
Proposition~\ref{prop:lsi-bounded-perturbation} gives
$\mathsf{LSI}(\alpha_0)$ for $\nu_R$ with
$\alpha_0=\frac{3}{2\cdot3.1}=\frac{15}{31}$.

For the angular laws $\sigma_{\beta(r)}$, we equip $S^{d-1}$ with the
rescaled energy $r^{-2}\|\nabla_{S^{d-1}}f\|^2$, which is the angular
part of the Euclidean gradient energy at radius~$r$. When
$\beta(r)\ne0$, we have $r\le\frac65\sqrt d$, so
Lemma~\ref{lem:lsi-angular} gives an $\mathsf{LSI}$ rate of at least
$\frac{cd}{r^2\mathcal R_t}\ge\frac{c'}{\mathcal R_t}$. When
$\beta(r)=0$, $\sigma_0=\omega_d$ is the uniform measure on $S^{d-1}$,
and under the rescaled energy Proposition~\ref{prop:lsi-bakry-emery}
gives rate $\frac{d-2}{r^2}\ge\frac{c'd}{R^2}$. Combining the two
cases, every $\sigma_{\beta(r)}$ with $r<R$ satisfies
$\mathsf{LSI}(\alpha_1)$ with $\alpha_1=\frac{c''}{\mathcal R_t+R^2/d}$.

For the coupling, the density of $\sigma_{\beta(r)}$ with respect to
$\omega_d$ is $Z_{\beta(r)}^{-1}e^{\beta(r)h_t(s_u(\theta))}$, and
$\partial_\beta\log Z_\beta=\mathbb E_{\sigma_\beta}h_t(s_u(\Theta))$,
so its logarithmic derivative in $r$ is
\[
  \ell_r(\theta)
  =\frac{\chi'(r/\sqrt d)}{\sqrt d}
   \Bigl[h_t(s_u(\theta))-\mathbb E_{\sigma_{\beta(r)}}h_t(s_u(\Theta))\Bigr].
\]
Since $0\le h_t\le4t^2$, the bracket has oscillation at most $4t^2$,
and therefore
\[
  K^2=\sup_r\operatorname{osc}(\ell_r)^2
  \le\frac{16\|\chi'\|_\infty^2t^4}{d}\le C,
\]
where the last step uses $d\ge Dt^4$.

Lemma~\ref{lem:lsi-two-level} now gives, for $f$ smooth on
$\mathbb B_R$,
\begin{align*}
  \operatorname{Ent}_{\pi_u(\cdot\mid\mathbb B_R)}(f^2)
  &\le\frac4{\alpha_0}\,\mathbb E\bigl[|\partial_rf|^2\bigr]
    +\frac2{\alpha_1}\Bigl(1+\frac{K^2}{2\alpha_0}\Bigr)
     \mathbb E\bigl[r^{-2}\|\nabla_{S^{d-1}}f\|^2\bigr]\\
  &\le C\Bigl(\mathcal R_t+\frac{R^2}{d}\Bigr)
     \mathbb E\bigl[|\partial_rf|^2+r^{-2}\|\nabla_{S^{d-1}}f\|^2\bigr]
   =C\Bigl(\mathcal R_t+\frac{R^2}{d}\Bigr)
     \mathbb E\|\nabla f\|^2,
\end{align*}
where the expectations are under $\pi_u(\cdot\mid\mathbb B_R)$, the
second line uses that $\alpha_0$ and $K$ are universal constants and
$\mathcal R_t\ge1$, and the last equality is the polar decomposition
of the Euclidean gradient. This is the stated local $\mathsf{LSI}$.
\end{proof}

\subsubsection{The Angular Log-Sobolev Inequality}
\label{sec:lsi-angular-proof}

Recall from Section~\ref{sec:lsi-local-proof} that
$\sigma_\beta(d\theta)\propto e^{\beta h_t(s_u(\theta))}\,\omega_d(d\theta)$
is the angular law of strength $\beta\in[0,1]$ on the unit sphere,
with $s_u(\theta)=\sqrt d\,u\cdot\theta$ and normalizer $Z_\beta$, and
that $\mu_{r,\beta}$ is the law of $r\Theta$ for $\Theta\sim\sigma_\beta$.
In particular, $\mu_{r,0}$ is the uniform law on the sphere of radius
$r$.

\angularlsi*

The density of $\mu_{r,\beta}$ depends on $\theta$ only through the
latitude $s_u(\theta)$, so the problem is essentially one-dimensional.
We prove a log-Sobolev inequality for the law of the latitude with the
weighted energy inherited from the sphere, by a Hardy-type criterion,
and then recover the full sphere by Lemma~\ref{lem:lsi-two-level}.
The estimate must hold uniformly in $\beta\in[0,1]$, since the radial
cutoff takes every value between zero and one.

The criterion is the following form of the Bobkov--G\"otze
characterization of one-dimensional log-Sobolev inequalities.
%

\begin{proposition}[\textbf{Weighted Hardy criterion~\cite[Theorem~1]{BartheRoberto2003}}]
\label{prop:lsi-hardy}
Let $\mu(ds)=\rho(s)\,ds$ be a probability measure on
$I=(\ell_-,\ell_+)$ with positive density, let $m$ be a median, and
let $a:I\to(0,\infty)$ be continuous. Define
\begin{align*}
  B_+&:=\sup_{m<x<\ell_+}
        \mu([x,\ell_+))\log\frac1{\mu([x,\ell_+))}
        \int_m^x\frac{ds}{a(s)\rho(s)},\\
  B_-&:=\sup_{\ell_-<x<m}
        \mu((\ell_-,x])\log\frac1{\mu((\ell_-,x])}
        \int_x^m\frac{ds}{a(s)\rho(s)}.
\end{align*}
Then $\operatorname{Ent}_\mu(f^2)\le468(B_++B_-)\int a|f'|^2\,d\mu$
for every locally absolutely continuous $f\in L^2(\mu)$ of finite
energy. Conversely, if
$\operatorname{Ent}_\mu(f^2)\le\frac2\alpha\int a|f'|^2\,d\mu$ for
all such $f$, then $\max(B_+,B_-)\le\frac{300}{\alpha}$.
\end{proposition}

For the latitude law, the hinge tilts a Gaussian, so the Hardy
quantities involve tails of shifted Gaussians and integrals of
$e^{y^2/2}$ over shifted intervals.
The following Gaussian estimates
are proved in Appendix~\ref{app:lsi-shifted-gaussian}.

\begin{restatable}[\bf Gaussian estimates]{lemma}{gaussianest}
\label{lem:lsi-gaussian-estimates}Write
$\overline\Phi(y):=\int_y^\infty\varphi(s)\,ds$ for the standard
Gaussian tail.
\begin{enumerate}
\item[(i)] For $M\ge1$,
  $\int_0^Me^{\frac{y^2}{2}}\,dy\le\frac{2e^{\frac{M^2}{2}}}{M}$.
\item[(ii)] Let $t\ge4$, $y_-\in[-t,t]$, $y_+\ge y_-$, and
  $\delta:=\frac{t^2-y_-^2}{2}$. Then
  \[
    \overline\Phi(y_+)
    \Bigl(1+\delta+\log\frac1{\overline\Phi(y_+)}\Bigr)
    \int_{y_-}^{y_+}e^{\frac{y^2}{2}}\,dy
    \le C\frac{e^{\frac{t^2}{2}}}{t}.
  \]
\end{enumerate}
\end{restatable}

Now we are ready to prove Lemma~\ref{lem:lsi-angular}.
The proof verifies the Hardy condition for the sharp hinge $\widehat h_t$, which
vanishes on $[0,t]$, is linear on $[t,3t]$, and remains constant at
$4t^2$ beyond $3t$. The analysis splits into these three regimes. On
$[0,t]$, the density is unperturbed, and the Hardy quantity at $x=t$
is already of order $\frac{r^2}{d}\mathcal R_t$, which dictates the
overall rate. On $[t,3t]$, the linear hinge tilts the density into a
shifted Gaussian, which we bound using
Lemma~\ref{lem:lsi-gaussian-estimates}. Beyond $3t$, the hinge merely
rescales the density by a constant factor, reducing the Hardy quantity
to the unperturbed case $\beta=0$. Finally, this one-dimensional
latitude inequality lifts to the full sphere because the conditional
measures on the latitude fibers are uniform on spheres.

\begin{proof}[Proof of Lemma~\ref{lem:lsi-angular}]
A function on the sphere of radius $r$ that depends only on the
latitude has the form $f(r\theta)=F(s_u(\theta))$ with $F$ a function
on $(-\sqrt d,\sqrt d)$, and its intrinsic energy is
$a_r(s)|F'(s)|^2$ with $a_r(s):=\frac{d-s^2}{r^2}$, because
$\|\nabla_{S^{d-1}}s_u\|^2=d-s_u^2$ on the unit sphere.

For $\beta=0$, the law $\mu_{r,0}$ is uniform on the sphere of radius
$r$, whose Ricci curvature is $\frac{d-2}{r^2}$ times the metric, so
Lemma~\ref{prop:lsi-bakry-emery}(i) gives
$\mathsf{LSI}(\frac{d-2}{r^2})$. 
 This is the second claim. 

Now let $F$ be a function on
$(-\sqrt d,\sqrt d)$ and $f=F(s_u)$. Under $\mu_{r,0}$ the latitude
has density $q_d$, so $\operatorname{Ent}_{\mu_{r,0}}(f^2)$ is the
entropy of $F^2$ under the law $q_d(s)\,ds$, and the energy of $f$ is
$\int a_r|F'|^2q_d\,ds$. Writing $q_d$ also for this law,
\[
  \operatorname{Ent}_{q_d}(F^2)\le\frac{2r^2}{d-2}\int a_r|F'|^2\,q_d\,ds,
\]
that is, the latitude law with $\beta=0$ satisfies
$\mathsf{LSI}(\frac{d-2}{r^2})$ for the energy $\int a_r|F'|^2$.

We first work with the sharp hinge $\widehat h_t$, for which the
Hardy quantities are explicit, and return to $h_t$ at the end. For
$\beta\in[0,1]$, write
$\widehat Z_\beta:=\int e^{\beta\widehat h_t(s)}q_d(s)\,ds$. By
Lemmas~\ref{lem:lsi-hinge} and~\ref{lem:lsi-normalizer},
\[
  1\le\widehat Z_\beta\le e^{2\eta}Z_\beta\le3.1e^{2\eta}\le4.
\]
Here the lower bound uses $\widehat h_t\ge0$, and the second
inequality uses $\widehat h_t\le h_t+2\eta$.

Under the sharp-hinge law, the latitude has density
$\frac{e^{\beta\widehat h_t(s)}q_d(s)}{\widehat Z_\beta}$ on
$(-\sqrt d,\sqrt d)$. This density is even, so zero is a median. For
$\beta\in[0,1]$ and $0<x<\sqrt d$, define
\[
  T_\beta(x):=\int_x^{\sqrt d}e^{\beta\widehat h_t(s)}q_d(s)\,ds,
  \qquad
  J_\beta(x):=\int_0^x\frac{e^{-\beta\widehat h_t(s)}}{a_r(s)q_d(s)}\,ds,
  \qquad
  \mathcal B_\beta(x):=T_\beta(x)\log\frac{e\widehat Z_\beta}{T_\beta(x)}\,J_\beta(x).
\]
In Proposition~\ref{prop:lsi-hardy}, the tail $\mu([x,\sqrt d))$ and
the integral $\int_0^x\frac{ds}{a\rho}$ are
$\frac{T_\beta(x)}{\widehat Z_\beta}$ and $\widehat Z_\beta J_\beta(x)$,
so the normalizing factors cancel outside the logarithm, and
$B_+=B_-$ by symmetry. The factor $e$ in the logarithm of
$\mathcal B_\beta$ only increases the Hardy quantity, so
Proposition~\ref{prop:lsi-hardy} shows that the latitude law
satisfies $\mathsf{LSI}(\alpha)$ for the energy $\int a_r|F'|^2$ with
$\frac2\alpha\le 936\sup_x\mathcal B_\beta(x)$. Fix $\beta\in[0,1]$.
It therefore suffices to prove
\begin{equation}
  \mathcal B_\beta(x)\le C\frac{r^2}{d}\mathcal R_t
  \qquad(0<x<\sqrt d).
  \label{eq:lsi-angular-hardy}
\end{equation}
We will repeatedly use that $v\mapsto v\log\frac{eZ}{v}$ is
increasing on $(0,Z]$ and bounded above by $Z$ there.

We establish \eqref{eq:lsi-angular-hardy} through three facts.
 The
first records what the case $\beta=0$ provides: a bound on
$\mathcal B_0(x)$ for every $x$, and a bound on $J_0(t)$, the
integral up to the kink of the hinge, which is where $\mathcal R_t$
appears.

\begin{fact}[\bf Baseline estimates]
\label{fact:lsi-angular-baseline}
For every $0<x<\sqrt d$,
\begin{equation}
  \mathcal B_0(x)=T_0(x)\log\frac e{T_0(x)}\,J_0(x)\le C\frac{r^2}{d}.
  \label{eq:lsi-baseline-hardy}
\end{equation}
Moreover,
\begin{equation}
  J_0(t)\le C\frac{r^2}{d}\frac{e^{\frac{t^2}{2}}}{t}
  \le C\frac{r^2}{d}\mathcal R_t.
  \label{eq:lsi-boundary-resistance}
\end{equation}
\end{fact}
\begin{proof}
The latitude law with $\beta=0$ satisfies $\mathsf{LSI}(\frac{d-2}{r^2})$
for the energy $\int a_r|F'|^2$, as shown at the beginning of the
proof. By the extension to $H^1$ recalled in
Appendix~\ref{app:lsi-functional-tools}, the hypothesis of the
converse in Proposition~\ref{prop:lsi-hardy} is met, and since
$\widehat Z_0=1$ that converse gives
$T_0(x)\log\frac1{T_0(x)}J_0(x)\le\frac{300r^2}{d-2}$ for every $x$.
This is \eqref{eq:lsi-baseline-hardy}, because $T_0(x)\le\frac12$
gives $\log\frac e{T_0(x)}\le\bigl(1+\frac1{\log2}\bigr)\log\frac1{T_0(x)}$.

On $[0,t]$, $a_r(s)=\Theta(\frac d{r^2})$ because $t^2\le\frac dD$,
and $q_d(s)=\Theta(\varphi(s))$ by Lemma~\ref{lem:lsi-spherical}(ii).
Consequently,
\[
  J_0(t)\le C\frac{r^2}{d}\int_0^te^{\frac{s^2}{2}}\,ds
  \le C\frac{r^2}{d}\frac{e^{\frac{t^2}{2}}}{t},
\]
where the last step is Lemma~\ref{lem:lsi-gaussian-estimates}(i).
Lemma~\ref{lem:lsi-spherical}(iii) then gives
\eqref{eq:lsi-boundary-resistance}.
\end{proof}

The second fact bounds $\mathcal B_\beta(x)$ on the range where the
hinge is zero or linear.
\begin{fact}[\bf Hardy estimate up to the plateau]
\label{fact:lsi-angular-local-hardy}
For every $0<x\le3t$,
$\mathcal B_\beta(x)\le C\frac{r^2}{d}\mathcal R_t$.
\end{fact}
\begin{proof}
If $0<x\le t$, then $\widehat h_t=0$ on $[0,x]$, so
$J_\beta(x)=J_0(x)\le J_0(t)$. Since
$T_\beta(x)\log\frac{e\widehat Z_\beta}{T_\beta(x)}\le\widehat Z_\beta\le4$,
the claim follows from Fact~\ref{fact:lsi-angular-baseline}.

Now suppose $t\le x\le3t$. Put $y_-:=(1-2\beta)t$,
$y_+:=x-2\beta t$, and
$\delta:=2\beta(1-\beta)t^2=\frac{t^2-y_-^2}{2}$. Completing the
square gives
\[
\begin{split}
  e^{2\beta t(s-t)}\varphi(s)&=e^{-\delta}\varphi(s-2\beta t),\\
  e^{-2\beta t(s-t)}e^{\frac{s^2}{2}}&=e^{\delta}e^{\frac{(s-2\beta t)^2}{2}}.
\end{split}
\]
For $s\ge t$, we have $\widehat h_t(s)\le2t(s-t)$, with equality on
$[t,3t]$. Hence
\begin{align*}
  T_\beta(x)
  &\le e^{\frac5d}\int_x^\infty e^{2\beta t(s-t)}\varphi(s)\,ds
   \le Ce^{-\delta}\overline\Phi(y_+),\\
  J_\beta(x)
  &=J_0(t)+\int_t^x\frac{e^{-2\beta t(s-t)}}{a_r(s)q_d(s)}\,ds
   \le J_0(t)+C\frac{r^2}{d}e^{\delta}\int_{y_-}^{y_+}e^{\frac{y^2}{2}}\,dy.
\end{align*}
The first line uses $q_d\le e^{\frac5d}\varphi$ from
Lemma~\ref{lem:lsi-spherical}(i) and the first identity above. The
second line uses $a_r(s)q_d(s)\ge c\frac d{r^2}\varphi(s)$ on $[t,3t]$
from Lemma~\ref{lem:lsi-spherical}(ii), the second identity above,
and the substitution $y=s-2\beta t$.

To control the entropy factor, set
$w:=e^{-\delta}\overline\Phi(y_+)\in(0,1]$ and
$v:=\frac{T_\beta(x)}{w}$, so that $v\le C$ by the bound on
$T_\beta(x)$. Since $v\log\frac1v\le\frac1e$ for all $v>0$ and
$1\le\widehat Z_\beta\le4$,
\begin{align*}
  T_\beta(x)\log\frac{e\widehat Z_\beta}{T_\beta(x)}
  &=wv\log\frac{e\widehat Z_\beta}{w}+wv\log\frac1v\\
  &\le Cw\Bigl(1+\log4+\log\frac1w\Bigr)+\frac we
   \le Cw\Bigl(1+\log\frac1w\Bigr).
\end{align*}
That is,
\begin{equation}
  T_\beta(x)\log\frac{e\widehat Z_\beta}{T_\beta(x)}
  \le Ce^{-\delta}\overline\Phi(y_+)
      \Bigl(1+\delta+\log\frac1{\overline\Phi(y_+)}\Bigr).
  \label{eq:lsi-tail-entropy-comparison}
\end{equation}
Multiplying by the bound on $J_\beta(x)$, and using the bound $4$ for
the entropy factor in front of $J_0(t)$, gives
\begin{align*}
  \mathcal B_\beta(x)
  &\le4J_0(t)
    +C\frac{r^2}{d}\,\overline\Phi(y_+)
     \Bigl(1+\delta+\log\frac1{\overline\Phi(y_+)}\Bigr)
     \int_{y_-}^{y_+}e^{\frac{y^2}{2}}\,dy\\
  &\le4J_0(t)+C\frac{r^2}{d}\frac{e^{\frac{t^2}{2}}}{t}
   \le C\frac{r^2}{d}\mathcal R_t.
\end{align*}
In the first line the factors $e^{\delta}$ and $e^{-\delta}$ cancel.
The second line is Lemma~\ref{lem:lsi-gaussian-estimates}(ii), whose
hypotheses hold because $y_-\in[-t,t]$, $y_+\ge y_-$, and
$\delta=\frac{t^2-y_-^2}{2}$. The last inequality follows from
Fact~\ref{fact:lsi-angular-baseline} and
Lemma~\ref{lem:lsi-spherical}(iii).
\end{proof}

Beyond $3t$ the hinge is constant, so the tilt only rescales the
density there, and the third fact reduces this range to what is
already known.
\begin{fact}[\bf Extension across the plateau]
\label{fact:lsi-angular-tail-extension}
For every $3t\le x<\sqrt d$,
$\mathcal B_\beta(x)\le\mathcal B_\beta(3t)+\mathcal B_0(x)$.
\end{fact}
\begin{proof}
Since $\widehat h_t=4t^2$ on $[3t,\sqrt d)$,
\[
  T_\beta(x)=e^{4\beta t^2}T_0(x),
  \qquad
  J_\beta(x)=J_\beta(3t)+e^{-4\beta t^2}\bigl(J_0(x)-J_0(3t)\bigr).
\]
The function $v\mapsto v\log\frac{e\widehat Z_\beta}{v}$ is
increasing on $(0,\widehat Z_\beta]$, while $T_\beta$ is decreasing,
so the contribution of $J_\beta(3t)$ to $\mathcal B_\beta(x)$ is at
most $\mathcal B_\beta(3t)$. Moreover, $\widehat h_t\le4t^2$
everywhere gives $\widehat Z_\beta\le e^{4\beta t^2}$. After the
exponential factors cancel, the remaining contribution is
\[
  T_0(x)\log\frac{e\widehat Z_\beta e^{-4\beta t^2}}{T_0(x)}
  \bigl(J_0(x)-J_0(3t)\bigr)
  \le T_0(x)\log\frac e{T_0(x)}\,J_0(x)
  =\mathcal B_0(x).
\]
This proves the claim.
\end{proof}

Fact~\ref{fact:lsi-angular-local-hardy} proves
\eqref{eq:lsi-angular-hardy} for $x\le3t$. For $3t\le x<\sqrt d$,
Fact~\ref{fact:lsi-angular-tail-extension},
Fact~\ref{fact:lsi-angular-local-hardy} at $x=3t$, and
Fact~\ref{fact:lsi-angular-baseline} give
$\mathcal B_\beta(x)\le C\frac{r^2}{d}\mathcal R_t+C\frac{r^2}{d}
\le C\frac{r^2}{d}\mathcal R_t$, using $\mathcal R_t\ge1$. Hence
the latitude law satisfies $\mathsf{LSI}(\frac{cd}{r^2\mathcal R_t})$
for the energy $\int a_r|F'|^2$.

We now lift this inequality to the sphere. Write
$\theta=\frac{s}{\sqrt d}u+\sqrt{1-\frac{s^2}{d}}\,\omega$, where
$\omega$ lies on the unit sphere of $u^\perp$. Under the sharp-hinge
law, $\omega$ is uniform on that sphere and independent of $s$, and
the intrinsic energy decomposes as
\[
  \|\nabla_{S_r}f\|^2
  =a_r(s)|\partial_sf|^2
   +\frac{\|\nabla_\omega f\|^2}{r^2(1-\frac{s^2}{d})}.
\]
The second term is the intrinsic energy of the fiber at latitude $s$,
a sphere of dimension $d-2$ and radius $r\sqrt{1-\frac{s^2}{d}}$,
whose Ricci curvature is $\frac{d-3}{r^2(1-s^2/d)}$ times the metric.
Lemma~\ref{prop:lsi-bakry-emery}(i) therefore gives rate at least
$\frac{d-3}{r^2}$ for the fiber energy. Lemma~\ref{lem:lsi-two-level}
applies with marginal variable $s$, weight $a_r$, and $K=0$, because
the conditional law of $\omega$ does not depend on $s$. With all
expectations under the sharp-hinge law on the sphere of radius $r$,
it gives
\[
  \operatorname{Ent}(f^2)
  \le\frac{2r^2\mathcal R_t}{cd}\,\mathbb E\bigl[a_r(s)|\partial_sf|^2\bigr]
   +\frac{2r^2}{d-3}\,
    \mathbb E\biggl[\frac{\|\nabla_\omega f\|^2}{r^2(1-\frac{s^2}{d})}\biggr]
  \le\frac{2r^2\mathcal R_t}{cd}\,\mathbb E\|\nabla_{S_r}f\|^2,
\]
where the last step uses $\mathcal R_t\ge1$ and $d-3\ge cd$. This is
$\mathsf{LSI}(\frac{cd}{r^2\mathcal R_t})$ for the sharp-hinge law.

Finally, the angular laws defined by $h_t$ and $\widehat h_t$ have
densities proportional to $e^{\beta h_t(s_u(\theta))}$ and
$e^{\beta\widehat h_t(s_u(\theta))}$, and
$\operatorname{osc}[\beta(h_t-\widehat h_t)]\le4\eta$ by
Lemma~\ref{lem:lsi-hinge}. Lemma~\ref{prop:lsi-bounded-perturbation}(i)
therefore transfers the bound to $h_t$ at the cost of the factor
$e^{4\eta}$.
\end{proof}

\subsubsection{The Global Poincar\'e Inequality}
\label{sec:lsi-poincare-proof}

\globalpi*

We prove Lemma~\ref{lem:lsi-global-poincare} by combining the local
estimate at radius $2\sqrt d$ with a Lyapunov estimate for the
quadratic tails.

\begin{proof}[Proof of Lemma~\ref{lem:lsi-global-poincare}]
Write $\mathbb B:=\mathbb B_{2\sqrt d}$. Let $f\in C_c^\infty(\R^d)$
and $g:=f-\mathbb E_{\pi_u(\cdot\mid\mathbb B)}f$. Since the variance
is the smallest mean square deviation from a constant,
\[
  \operatorname{Var}_{\pi_u}(f)
  \le\int g^2\,d\pi_u
  =\int_{\mathbb B}g^2\,d\pi_u+\int_{\mathbb B^c}g^2\,d\pi_u .
\]
By Lemma~\ref{lem:lsi-local} with $R=2\sqrt d$ and
Lemma~\ref{lem:lsi-implies-poincare}, the law
$\pi_u(\cdot\mid\mathbb B)$ satisfies
$\mathsf{PI}(\frac{c}{\mathcal R_t+4})$, and
$\mathcal R_t+4\le5\mathcal R_t$ because $\mathcal R_t\ge1$. Since
$g$ is centered under this law, multiplying its Poincar\'e
inequality by $\pi_u(\mathbb B)$ gives
\begin{equation}
  \int_{\mathbb B}g^2\,d\pi_u
  \le C\mathcal R_t\int_{\mathbb B}\|\nabla f\|^2\,d\pi_u .
  \label{eq:lsi-pi-inside}
\end{equation}
It therefore suffices to control the mass of $g^2$ outside
$\mathbb B$ by its mass inside and the energy. We prove that every
bounded smooth $g$ whose gradient has compact support satisfies
\begin{equation}
  \int_{\mathbb B^c}g^2\,d\pi_u
  \le\frac1d\int\|\nabla g\|^2\,d\pi_u
   +C\int_{\mathbb B}g^2\,d\pi_u .
  \label{eq:lsi-lyapunov}
\end{equation}
Granting \eqref{eq:lsi-lyapunov}, and using $\nabla g=\nabla f$,
\eqref{eq:lsi-pi-inside}, and $\frac1d\le1\le\mathcal R_t$,
\[
  \operatorname{Var}_{\pi_u}(f)
  \le\frac1d\int\|\nabla f\|^2\,d\pi_u
   +(C+1)\int_{\mathbb B}g^2\,d\pi_u
  \le C\mathcal R_t\int\|\nabla f\|^2\,d\pi_u,
\]
which is $\mathsf{PI}(\frac{c}{\mathcal R_t})$ with $c=\frac1C$.

To prove \eqref{eq:lsi-lyapunov}, let
$\mathcal L:=\Delta-\nabla V_u\cdot\nabla$ be the generator of
$\pi_u$ and $W(x):=e^{\frac{\|x\|^2}{8}}$. We bound
$\int-\frac{\mathcal LW}{W}g^2\,d\pi_u$ from above by the energy and
from below by the two sides of \eqref{eq:lsi-lyapunov}.

For the upper bound, let $g\in C_c^\infty(\R^d)$. Integration by
parts and the product rule give
\[
  \int-\frac{\mathcal LW}{W}\,g^2\,d\pi_u
  =\int\nabla W\cdot\nabla\Bigl(\frac{g^2}{W}\Bigr)d\pi_u
  =\int\Bigl[\frac{2g}{W}\nabla W\cdot\nabla g
     -\frac{g^2}{W^2}\|\nabla W\|^2\Bigr]d\pi_u
  \le\int\|\nabla g\|^2\,d\pi_u,
\]
where the inequality is $2a\cdot b-\|b\|^2\le\|a\|^2$ with
$a=\nabla g$ and $b=\frac gW\nabla W$.

For the lower bound, write $r=\|x\|$, so that $\nabla W=\frac x4W$
and $\Delta W=(\frac d4+\frac{r^2}{16})W$. Since $s_u$ is homogeneous
of degree zero, $x\cdot\nabla s_u=0$ away from the origin, and the
perturbation in \eqref{eq:lsi-potential} vanishes near the origin,
so
$x\cdot\nabla V_u=\frac32r^2-\frac{r}{\sqrt d}\chi'\bigl(\frac{r}{\sqrt d}\bigr)h_t(s_u(x))$.
Hence
\[
  -\frac{\mathcal LW}{W}
  =-\frac d4-\frac{r^2}{16}+\frac{x\cdot\nabla V_u}{4}
  =\frac{5r^2}{16}-\frac d4
   -\frac{r}{4\sqrt d}\chi'\Bigl(\frac{r}{\sqrt d}\Bigr)h_t(s_u(x)).
\]
Outside $\mathbb B$ the cutoff derivative vanishes and
$\frac{5r^2}{16}-\frac d4\ge d$. Inside,
$\frac{5r^2}{16}-\frac d4\ge-\frac d4$, and the last term has
absolute value at most $2\|\chi'\|_\infty t^2\le Cd$ because
$r\le2\sqrt d$, $h_t\le4t^2$, and $t^2\le d$. Therefore
\[
  -\frac{\mathcal LW}{W}\ge d\,\mathbf 1_{\mathbb B^c}-Cd\,\mathbf 1_{\mathbb B},
  \qquad\text{so}\qquad
  \int-\frac{\mathcal LW}{W}\,g^2\,d\pi_u
  \ge d\int_{\mathbb B^c}g^2\,d\pi_u-Cd\int_{\mathbb B}g^2\,d\pi_u .
\]
Combining the two bounds and dividing by $d$ proves
\eqref{eq:lsi-lyapunov} for $g\in C_c^\infty(\R^d)$.

Finally, let $g$ be bounded and smooth with compactly supported
gradient. Fix $\zeta\in C_c^\infty(\R^d)$ with $\zeta=1$ on
$\mathbb B_1$ and apply \eqref{eq:lsi-lyapunov} to $\zeta(\frac xn)g$.
For large $n$, $\nabla\zeta(\frac xn)$ vanishes on the support of
$\nabla g$, so the energy of $\zeta(\frac xn)g$ exceeds that of $g$
by at most $\frac{\|g\|_\infty^2\|\nabla\zeta\|_\infty^2}{n^2}$,
while the two integrals of $\zeta(\frac xn)^2g^2$ converge to those
of $g^2$ by dominated convergence. Letting $n\to\infty$ gives
\eqref{eq:lsi-lyapunov} for $g$.
\end{proof}
\subsection{Proof of Theorem~\ref{thm:lsi-sampling-lower-bound}}
\label{sec:lsi-proof}

The hidden-cap family with parameter $t$ enters the proof through
three properties. Let $C_L$ be the constant of
Lemma~\ref{lem:lsi-regularity}, $c_{\mathrm{LS}}$ the constant $c$ of
Theorem~\ref{thm:lsi-family}, and $c_-,c_+$ the constants implicit in
Lemma~\ref{lem:lsi-spherical}(iii), so that whenever $t\ge4$ and
$d\ge Dt^4$,
\begin{equation}
  \sup_x\|\nabla^2V_u(x)\|_{\mathrm{op}}\le C_Lt^2,
  \qquad
  \pi_u\text{ satisfies }\mathsf{LSI}\bigl(c_{\mathrm{LS}}\,p_d(t)t^2\bigr),
  \qquad
  c_-\frac{e^{-\frac{t^2}{2}}}{t}\le p_d(t)\le c_+\frac{e^{-\frac{t^2}{2}}}{t}.
  \label{eq:lsi-inputs}
\end{equation}
The ratio of the smoothness to the log-Sobolev rate is of order
$\frac1{p_d(t)}$, which is also the lower bound of
Corollary~\ref{cor:lsi-expected-queries}. We choose $t$ so that
$\frac1{p_d(t)}$ is of order $\kappa$ and rescale the family so that the potentials are $L$-smooth.

\begin{proof}[Proof of Theorem~\ref{thm:lsi-sampling-lower-bound}]
Set $a_0:=\max\{1,\frac{C_L}{c_{\mathrm{LS}}c_-}\}$,
$\kappa_0:=4e^8a_0$, $C:=4D$, and $c:=\frac{1}{50c_+a_0}$. Let
$L\ge\alpha>0$ with $\kappa=L/\alpha\ge\kappa_0$ and
$d\ge C\log^2(2\kappa)$.

Since $t\mapsto\frac{e^{-t^2/2}}{t}$ decreases continuously from
$\infty$ to $0$ on $(0,\infty)$, there is a unique $t>0$ with
\begin{equation}
  \frac{e^{-\frac{t^2}{2}}}{t}=\frac{a_0}{\kappa}.
  \label{eq:lsi-parameter-choice}
\end{equation}
Then $t\ge4$, because the left side is decreasing and at $t=4$ equals
$\frac{e^{-8}}{4}=\frac{a_0}{\kappa_0}\ge\frac{a_0}{\kappa}$.
Moreover $e^{\frac{t^2}{2}}=\frac{\kappa}{a_0t}\le\frac{\kappa}{a_0}$,
so $t^2\le2\log\frac{\kappa}{a_0}\le2\log\kappa$ and
$Dt^4\le4D\log^2\kappa\le C\log^2(2\kappa)\le d$. Thus
\eqref{eq:lsi-inputs} applies, and its last part becomes
\begin{equation}
  \frac{c_-a_0}{\kappa}\le p_d(t)\le\frac{c_+a_0}{\kappa}.
  \label{eq:lsi-pd-kappa}
\end{equation}

Set $b:=\sqrt{L/(C_Lt^2)}$, $W_u(z):=V_u(bz)$, and let $\mu_u$ be the
target of $W_u$, which is the law of $X/b$ for $X\sim\pi_u$. Evenness,
smoothness, $W_u(0)=0$, and the minimizer are inherited from $V_u$ by
Lemma~\ref{lem:lsi-regularity}, and
$\nabla^2W_u(z)=b^2\nabla^2V_u(bz)$ gives
$\|\nabla^2W_u\|_{\mathrm{op}}\le b^2C_Lt^2=L$. This is property 1.
For a test function $g$, put $f(x):=g(x/b)$, so that
$\operatorname{Ent}_{\mu_u}(g^2)=\operatorname{Ent}_{\pi_u}(f^2)$ and
$\int\|\nabla f\|^2\,d\pi_u=\frac1{b^2}\int\|\nabla g\|^2\,d\mu_u$.
Hence $\mu_u$ satisfies $\mathsf{LSI}$ with rate
\[
  b^2c_{\mathrm{LS}}\,p_d(t)t^2
  =\frac{c_{\mathrm{LS}}L}{C_L}\,p_d(t)
  \ge\frac{c_{\mathrm{LS}}c_-a_0}{C_L}\cdot\frac{L}{\kappa}
  \ge\alpha,
\]
where the first inequality is \eqref{eq:lsi-pd-kappa} and the second
is the choice of $a_0$ together with $\frac L\kappa=\alpha$. This is
property 2. The image of $\gamma_0$ under $x\mapsto x/b$ is
$\nu:=\mathcal N(0,\frac{2}{3b^2}I_d)$, which depends only on
$(d,L,\alpha)$ because $b$ is determined by $L$ and $t$, and $t$ by
$\kappa$. By Lemma~\ref{lem:lsi-normalizer},
$\frac{d\nu}{d\mu_u}(z)=\frac{d\gamma_0}{d\pi_u}(bz)\le3.1<4$, which
is property 3.

For property 4, let $\mathcal A$ be an algorithm whose total
variation error on every $\mu_u$ is at most $\varepsilon\le\frac1{10}$.
Answering each of its queries $z$ by the single query $bz$ to the
oracle of $V_u$, which returns $W_u(z)=V_u(bz)$ and
$\nabla W_u(z)=b\,\nabla V_u(bz)$, and multiplying its output by $b$
gives an algorithm for $\{\pi_u\}$ with the same numbers of queries
$N_u$. Its output law on $V_u$ is the image under $z\mapsto bz$ of
the output law of $\mathcal A$ on $W_u$, and this map carries $\mu_u$
to $\pi_u$ and preserves total variation, so the errors are the same.
Corollary~\ref{cor:lsi-expected-queries} and \eqref{eq:lsi-pd-kappa}
give
\[
  \sup_u\mathbb E[N_u]\ge\frac{1}{50\,p_d(t)}
  \ge\frac{\kappa}{50\,c_+a_0}=c\kappa.
  \qedhere
\]
\end{proof}

\section{Conclusion}
\label{sec:con}
We established first-order query lower bounds for strongly
log-concave sampling and sampling under a log-Sobolev inequality.
At constant condition number and accuracy, strongly log-concave
sampling requires superpolylogarithmically many queries in the
dimension, separating it from Gaussian sampling.
For log-Sobolev targets, we proved an $\Omega(\kappa)$ lower bound
when $d\gtrsim\log^2\kappa$, even from a constant warm start.
Closing the gap between the dimension-dependent upper and lower
bounds, and determining the joint dependence on dimension,
condition number, and accuracy, remain open directions.

\section*{AI Disclosure}
AI tools were used to assist with literature searches, the development and verification of proof
arguments, the presentation of proofs, the polishing of the exposition, and typesetting. The authors
independently verified all mathematical arguments and take full responsibility for the content of the
paper.

\section*{Acknowledgment}
We thank Sinho Chewi and Anming Gu for helpful discussions. TS was partially supported by JST
CREST (JPMJCR2115) and JSPS KAKENHI (24K02905). This work was partially supported by
JST ERATO (JPMJER2601). This research is supported by the National Research Foundation,
Singapore and the Ministry of Digital Development and Information under the AI Visiting Professorship Programme (award number AIVP-2024-004). Any opinions, findings and conclusions or
recommendations expressed in this material are those of the author(s) and do not reflect the views of
National Research Foundation, Singapore and the Ministry of Digital Development and Information.

\newpage
\bibliographystyle{alpha}
\bibliography{references}
\newpage
\appendix
\section{Tensor Estimations}
\label{out:tensor-spaces}
We establish projection and angle bounds for subspaces
spanned by tensor powers and their first derivatives.
Let $E$ be a finite-dimensional Euclidean space and $N\ge1$
an integer. Equip
$E^{\otimes N}=\underbrace{E\otimes\cdots\otimes E}_{N\text{ factors}}$
with its Euclidean tensor inner product.
Let $\mathfrak S_N$ be the set of permutations of
$\{1,\ldots,N\}$.
Define the linear map
$\operatorname{Sym}:E^{\otimes N}\to E^{\otimes N}$ by
\[
 \operatorname{Sym}(z_1\otimes\cdots\otimes z_N)
 =\frac1{N!}\sum_{\sigma\in\mathfrak S_N}
 z_{\sigma(1)}\otimes\cdots\otimes z_{\sigma(N)},
 \qquad z_1,\ldots,z_N\in E.
\]
This is the orthogonal projection onto the subspace of
tensors invariant under every permutation of the tensor factors.
For a nonempty finite set $A$ of unit vectors in $E$, define
\[
 W_A^N
 =\operatorname{span}\bigl\{
 v^{\otimes N},
 \operatorname{Sym}(v^{\otimes(N-1)}\otimes e):
 v\in A,\ e\in E
 \bigr\}.
\]
Since
$D_e(v^{\otimes N})
=N\operatorname{Sym}(v^{\otimes(N-1)}\otimes e)$,
this is the space spanned by the values and first derivatives
of $v\mapsto v^{\otimes N}$ at the points of $A$.
We use the convention $v^{\otimes0}=1$.

The following lemma bounds the projection of a tensor power
onto $W_A^N$, without any separation assumption on the
points within $A$.

\begin{lemma}[\bf Projection of a tensor power]
\label{out:one-jet}
Let $A=\{v_1,\ldots,v_s\}$ be a nonempty set of unit vectors
in $E$, and let $u\in E$ be a unit vector.
Suppose that $N\ge2s$ is an integer and
$\alpha:=\max_{1\le j\le s}|\langle u,v_j\rangle|<1$.
Then
\[
 \|P_{W_A^N}u^{\otimes N}\|^2
 \le2s\alpha^{2\lfloor N/(2s)\rfloor}.
\]
\end{lemma}

\begin{proof}
Put $k=\lfloor N/(2s)\rfloor$ and
$t_j=\langle u,v_j\rangle$.
We construct a tensor orthogonal to $W_A^N$ whose inner
product with $u^{\otimes N}$ is one.

For each $j$, define
\[
 A_j
 =\frac{u^{\otimes k}-t_j^kv_j^{\otimes k}}
        {1-t_j^{2k}}.
\]
The denominator is positive because $|t_j|<1$.
Using
$\langle u^{\otimes k},v_j^{\otimes k}\rangle=t_j^k$,
we obtain
\[
 \langle A_j,u^{\otimes k}\rangle=1,
 \qquad
 \langle A_j,v_j^{\otimes k}\rangle=0,
 \qquad
 \|A_j\|^2=\frac1{1-t_j^{2k}}.
\]
Since $2ks\le N$, the tensor
\[
 T=\operatorname{Sym}\!\left(
 u^{\otimes(N-2ks)}
 \otimes\bigotimes_{j=1}^s(A_j\otimes A_j)
 \right)
\]
is well defined. Its associated polynomial is
\[
 \langle T,z^{\otimes N}\rangle
 =\langle u,z\rangle^{N-2ks}
   \prod_{j=1}^s\langle A_j,z^{\otimes k}\rangle^2.
\]
At $z=v_j$, the $j$th squared factor and its first
derivatives vanish. Thus this polynomial and all its
first derivatives vanish at every point of $A$.
By the definition of $W_A^N$, this means $T\perp W_A^N$.

The identities for $A_j$ give
$\langle T,u^{\otimes N}\rangle=1$.
Moreover, orthogonal symmetrization does not increase
the norm, so
$\|T\|^2\le\prod_{j=1}^s(1-t_j^{2k})^{-2}$.
Consequently, Cauchy--Schwarz gives
\[
 \begin{aligned}
 1
 &=\left|
   \left\langle T,
   P_{(W_A^N)^\perp}u^{\otimes N}\right\rangle
   \right|^2\\
 &\le\|T\|^2
   \left(1-\|P_{W_A^N}u^{\otimes N}\|^2\right).
 \end{aligned}
\]
Rearranging and applying the norm bound yields
\[
 \begin{aligned}
 \|P_{W_A^N}u^{\otimes N}\|^2
 &\le1-\frac1{\|T\|^2}\\
 &\le1-\prod_{j=1}^s(1-t_j^{2k})^2\\
 &\le2\sum_{j=1}^s t_j^{2k}
 \le2s\alpha^{2k}.
 \end{aligned}
\]
The third inequality follows from
$1-\prod_j b_j\le\sum_j(1-b_j)$ for $b_j\in[0,1]$,
applied with $b_j=(1-t_j^{2k})^2$, and
$1-(1-x)^2\le2x$ for $x\in[0,1]$.
This proves the claim.
\end{proof}

We now use Lemma~\ref{out:one-jet} to compare the spaces
associated with two groups of directions.

\begin{lemma}[\bf Angle between two tensor subspaces;
Lemma~\ref{out:two-jet} restated]
\label{app:two-jet}
Let $A,B$ be nonempty sets of $s,t$ unit vectors in $E$,
respectively, and let $r>0$.
Suppose that
$\|a-b\|\ge r$ and $\|a+b\|\ge r$
for every $a\in A$ and $b\in B$.
Then, for every integer $N\ge8st$,
\[
 \|P_{W_A^N}P_{W_B^N}\|_{\mathrm{op}}
 \le2\sqrt{st}\exp\!\left(-\frac{Nr^2}{16st}\right).
\]
\end{lemma}

\begin{proof}
Put
$\alpha=\max_{a\in A,b\in B}|\langle a,b\rangle|$.
The separation assumptions and the unit norms give
$\alpha\le1-r^2/2<1$.
Set $k=\lfloor N/(2t)\rfloor$, so that
$2tk\le N$ and $k\ge2s$.

For each $b\in B$, assign two sets of $k$ tensor factors,
with all $2t$ sets pairwise disjoint.
This is possible because $2tk\le N$.
For $\nu\in\{1,2\}$, let $R_{b,\nu}$ be the orthogonal
projection onto $\operatorname{span}\{b^{\otimes k}\}$
in the $\nu$th set assigned to $b$, tensored with the
identity on the remaining factors.
These operators act on the full space $E^{\otimes N}$.
Their assigned factors are disjoint, so they commute.
Consequently,
\[
 \Pi=\prod_{b\in B}\prod_{\nu=1}^2(I-R_{b,\nu})
\]
is an orthogonal projection.

We first show that $\Pi W_B^N=\{0\}$.
For each $b\in B$, both assigned sets in
$b^{\otimes N}$ contain only copies of $b$, so
$(I-R_{b,1})b^{\otimes N}=0$.
Each summand of $D_e(b^{\otimes N})$ replaces just one
tensor factor $b$ by $e$.
At least one of the two sets assigned to $b$ therefore
still contains only copies of $b$.
The corresponding factor $I-R_{b,\nu}$ annihilates
that summand.
Since all the projections commute, $\Pi$ annihilates
both $b^{\otimes N}$ and $D_e(b^{\otimes N})$.
These vectors span $W_B^N$, proving the assertion.

We next bound $R_{b,\nu}$ on $W_A^N$.
Fix one assigned set and permute its $k$ factors to
the first $k$ positions.
For each $a\in A$, we have
$a^{\otimes N}=a^{\otimes k}\otimes a^{\otimes(N-k)}$.
The product rule also gives
\[
 \begin{aligned}
 D_e(a^{\otimes N})
 &=
 D_e(a^{\otimes k})\otimes a^{\otimes(N-k)}
 +a^{\otimes k}\otimes D_e(a^{\otimes(N-k)}).
 \end{aligned}
\]
In each term, the first tensor factor belongs to $W_A^k$.
Thus, under this grouping,
$W_A^N\subseteq W_A^k\otimes E^{\otimes(N-k)}$.

Take a unit vector $Z\in W_A^N$.
Choose an orthonormal basis $(f_\ell)_\ell$ of
$E^{\otimes(N-k)}$ and write
$Z=\sum_\ell z_\ell\otimes f_\ell$, where
$z_\ell\in W_A^k$ and $\sum_\ell\|z_\ell\|^2=1$.
Then
\[
 \begin{aligned}
 \|R_{b,\nu}Z\|^2
 &=\sum_\ell
   |\langle b^{\otimes k},z_\ell\rangle|^2\\
 &\le
 \|P_{W_A^k}b^{\otimes k}\|^2
 \sum_\ell\|z_\ell\|^2\\
 &\le2s\alpha^{2\lfloor k/(2s)\rfloor}.
 \end{aligned}
\]
The first inequality is Cauchy--Schwarz, using
$z_\ell\in W_A^k$.
The second follows from Lemma~\ref{out:one-jet},
applied with $u=b$ and tensor degree $k$;
its hypotheses hold because $k\ge2s$ and
$\max_{a\in A}|\langle a,b\rangle|\le\alpha<1$.

Since $\Pi W_B^N=\{0\}$, the range of $\Pi$ is
orthogonal to $W_B^N$.
Moreover, the commuting projections satisfy
\[
 P_{W_B^N}\preceq I-\Pi
 \preceq\sum_{b\in B}\sum_{\nu=1}^2R_{b,\nu}.
\]
For the second inequality, simultaneously diagonalize
the commuting projections and use
$1-\prod_j(1-\lambda_j)\le\sum_j\lambda_j$
for $\lambda_j\in\{0,1\}$.
Therefore, for every unit $Z\in W_A^N$,
\[
 \begin{aligned}
 \|P_{W_B^N}Z\|^2
 &\le
 \sum_{b\in B}\sum_{\nu=1}^2\|R_{b,\nu}Z\|^2\\
 &\le4st\alpha^{2\lfloor k/(2s)\rfloor}.
 \end{aligned}
\]
Taking the supremum over such $Z$, taking square roots,
and using equality of the operator norms of an operator
and its adjoint gives
\[
 \|P_{W_A^N}P_{W_B^N}\|_{\mathrm{op}}
 \le2\sqrt{st}\,
 \alpha^{\lfloor N/(4st)\rfloor}.
\]
Here we used
$\lfloor k/(2s)\rfloor
=\lfloor\lfloor N/(2t)\rfloor/(2s)\rfloor
=\lfloor N/(4st)\rfloor$.

Finally,
$\alpha\le1-r^2/2\le e^{-r^2/2}$, and
$N\ge8st$ implies
$\lfloor N/(4st)\rfloor\ge N/(8st)$.
Consequently,
\[
 \begin{aligned}
 \|P_{W_A^N}P_{W_B^N}\|_{\mathrm{op}}
 &\le2\sqrt{st}\,
 \exp\!\left(
 -\frac{r^2}{2}\left\lfloor\frac{N}{4st}\right\rfloor
 \right)\\
 &\le2\sqrt{st}\exp\!\left(-\frac{Nr^2}{16st}\right),
 \end{aligned}
\]
as required.
\end{proof}

\section{Proof of Lemma~\ref{out:rate-gap}}
\label{app:rate-window}

This appendix proves Lemma~\ref{out:rate-gap}, which
Section~\ref{out:approximation} uses to turn the degree profile of
Theorem~\ref{out:profile} into two separated query directions.
The proof follows the outline given there.
Two auxiliary estimates are established first
(Section~\ref{app:rate-window-aux}): a discrete statement that the
fast part of a bounded power-sum sequence decays geometrically after
$Q$ steps, and a Markov-type derivative bound for exponential sums
with slow rates.
Both are stated for arbitrary real coefficients and allow repeated
rates, and neither involves the size of the coefficients; this is
what the application to Theorem~\ref{out:profile} requires, since
the coefficients there may have arbitrary signs and magnitudes.
The main argument (Section~\ref{app:rate-window-main}) assumes that
no rate lies in the window, samples $f$ on an arithmetic progression
of even integers chosen so that the first estimate separates the
slow and fast rates, shows that beyond $p_0$ the slow part alone
satisfies the relative tolerance, and derives two incompatible
bounds on the derivative of the slow part, one from the second
estimate and one from the decay of $1/p$.



\begin{lemma}[\bf Rate window; the case $Q=q$ of Lemma~\ref{out:rate-gap} restated]
\label{app:rate-window-lemma}
There exist absolute constants $c,C>0$ such that the following holds.
Let $P\ge8$ be even, and let
$f(p)=\sum_{j=1}^{q}c_je^{-s_jp}$, where $q\ge1$ is an integer,
$c_j\in\mathbb R$, and $s_j\ge0$.
Suppose that $|pf(p)-1|\le1/10$ for $p=4,6,\ldots,P$.
Then, for every $p_0\in\mathbb R$ satisfying
\begin{equation}\label{app:p0-range}
 Cq\le p_0\le\frac{cP}{q^2},
\end{equation}
there exists $j\in\{1,\ldots,q\}$ with $c_j\ne0$ such that
\[
 \frac{c}{q^2p_0}<s_j<\frac{Cq}{p_0}.
\]
\end{lemma}

\subsection{Estimates for the Fast and Slow Parts}
\label{app:rate-window-aux}

We first establish two estimates, one for the fast part
(Lemma~\ref{app:discrete-separation}) and one for the slow part
(Lemma~\ref{app:slow-derivative}).
Both estimates are uniform in the coefficients $c_j$ and allow repeated
rates $s_j$.
For an interval $I$, write $\|g\|_I=\sup_{x\in I}|g(x)|$.

The first estimate concerns a power sum whose nodes split into a
fast group in $(0,1/4]$ and a slow group in $[3/4,1]$.
It shows that the fast group decays geometrically once the first
$2Q-1$ values of the whole sum are bounded.


\begin{lemma}[\bf Decay of the fast part from bounded samples]
\label{app:discrete-separation}
Let $Q\ge1$ be an integer and $B\ge0$.
Let $d_1,\ldots,d_Q\in\mathbb R$ and
$x_1,\ldots,x_Q\in(0,1/4]\cup[3/4,1]$,
with repetitions allowed.
For integers $m\ge0$, define
\[
 F_m=\sum_{j=1}^{Q}d_jx_j^m,
 \qquad
 H_m=\sum_{x_j\le1/4}d_jx_j^m.
\]
If $|F_\ell|\le B$ for $0\le\ell<Q$, then
\begin{equation}\label{app:discrete-tail}
 |H_m|\le B24^Q3^{-m}
 \qquad(m\ge Q).
\end{equation}
\end{lemma}

\begin{proof}
Fix an integer $m\ge Q$, and list the distinct nodes among
$x_1,\ldots,x_Q$ as $y_1<\cdots<y_r$, where $r\le Q$.
Assign the interpolation data
\[
 v_i=
 \begin{cases}
 y_i^m,&y_i\le1/4,\\
 0,&y_i\ge3/4.
 \end{cases}
\]
Set $D_{i,0}=v_i$ and define the divided differences recursively by
\[
 D_{i,k}
 =\frac{D_{i+1,k-1}-D_{i,k-1}}{y_{i+k}-y_i}
 \qquad(1\le k<r,\ 1\le i\le r-k).
\]

\begin{fact}\label{app:divided-difference-bound}
For every $0\le k<r$ and $1\le i\le r-k$,
\[
 |D_{i,k}|\le12^k3^{-m}.
\]
\end{fact}

\begin{proof}[Proof of Fact~\ref{app:divided-difference-bound}]
We argue by induction on $k$.
For $k=0$, the bound follows from
$|v_i|\le4^{-m}\le3^{-m}$.
Suppose $k\ge1$ and the bound holds at order $k-1$.

If all nodes $y_i,\ldots,y_{i+k}$ lie in $(0,1/4]$,
the divided-difference mean value theorem gives
\[
 |D_{i,k}|
 \le\binom mk4^{k-m}
 \le12^k3^{-m}.
\]
The second inequality follows from
$\binom mk3^{-k}\le(1+1/3)^m$, by the binomial theorem.

If all these nodes lie in $[3/4,1]$, their interpolation data
are zero, so $D_{i,k}=0$.
Otherwise, $y_{i+k}-y_i\ge1/2$, and the recursive formula
and induction hypothesis give
\[
 \begin{aligned}
 |D_{i,k}|
 &\le2\bigl(|D_{i+1,k-1}|+|D_{i,k-1}|\bigr)\\
 &\le4\cdot12^{k-1}3^{-m}
 \le12^k3^{-m}.
 \end{aligned}
\]
This completes the induction.
\end{proof}

Let $P_m$ be the polynomial interpolating the data.
Its Newton and monomial forms are
\[
 P_m(z)
 =\sum_{k=0}^{r-1}D_{1,k}\prod_{j=1}^{k}(z-y_j)
 =\sum_{\ell=0}^{r-1}b_\ell z^\ell.
\]
Since $0<y_j\le1$, the sum of the absolute coefficients of
$\prod_{j=1}^{k}(z-y_j)$ is at most
$\prod_{j=1}^{k}(1+y_j)\le2^k$.
Fact~\ref{app:divided-difference-bound} therefore implies
\[
 \begin{aligned}
 \sum_{\ell=0}^{r-1}|b_\ell|
 &\le\sum_{k=0}^{r-1}|D_{1,k}|\,2^k\le3^{-m}\sum_{k=0}^{r-1}24^k=3^{-m}\frac{24^r-1}{23}
 \le24^Q3^{-m},
 \end{aligned}
\]
where the last inequality uses $r\le Q$.

By construction, $P_m(x_j)=x_j^m$ on the fast nodes and
$P_m(x_j)=0$ on the slow nodes. Thus
\[
 \begin{aligned}
 H_m
 &=\sum_{j=1}^{Q}d_jP_m(x_j)=\sum_{\ell=0}^{r-1}b_\ell
   \sum_{j=1}^{Q}d_jx_j^\ell=\sum_{\ell=0}^{r-1}b_\ell F_\ell.
 \end{aligned}
\]
Since $r\le Q$, every sample on the right satisfies
$|F_\ell|\le B$ by assumption. Consequently,
\[
 |H_m|
 \le\sum_{\ell=0}^{r-1}|b_\ell|\,|F_\ell|
 \le B\sum_{\ell=0}^{r-1}|b_\ell|
 \le B24^Q3^{-m},
\]
as required.
\end{proof}

The second estimate concerns an exponential sum whose rates all lie
in a short interval $[0,a]$.
It is a Markov-type inequality stating that on an interval of length $D$, the
derivative of such a sum is at most $(4e^{aD}Q^2/D+a)$ times its
supremum, so a sum with slow rates cannot vary quickly without being
large somewhere on the interval.

\begin{lemma}[\bf Derivative bound for slow rates]
\label{app:slow-derivative}
Let $Q\ge1$ be an integer and $a\ge0$.
Let
$E(x)=\sum_{j=1}^{r}d_je^{-s_jx}$, where $1\le r\le Q$,
$d_j\in\mathbb R$, and $0\le s_j\le a$.
On every interval $I=[u,u+D]$ of length $D>0$,
\begin{equation}\label{app:slow-Markov}
 \|E'\|_I
 \le\left(\frac{4e^{aD}Q^2}{D}+a\right)\|E\|_I.
\end{equation}
The same conclusion holds for the empty sum.
\end{lemma}

\begin{proof}
We use the endpoint inequality of Borwein and
Erd\'elyi~\cite[Theorem~1]{BorweinErdelyi1995Upper}:
if $g(t)=b_0+\sum_{j=1}^{r}b_je^{\lambda_jt}$ has real
coefficients and nonnegative exponents, then
\begin{equation}\label{app:BE-endpoint}
 |g'(\alpha)|
 \le\frac{2r^2}{\beta-\alpha}\|g\|_{[\alpha,\beta]}
 \qquad \mbox{for }\alpha<\beta.
\end{equation}
Repeated exponents and zero coefficients are allowed by combining
or deleting terms.

For $x\in[u,u+D/2]$, we apply \eqref{app:BE-endpoint} to
$g_x(t)=e^{a(t-x)}E(t)$ on $[x,u+D]$.
Its exponents $a-s_j$ are nonnegative, and
$\|g_x\|_{[x,u+D]}\le e^{aD}\|E\|_I$.
The interval has length at least $D/2$, while
$g_x'(x)=E'(x)+aE(x)$.
Therefore,
\[
 |E'(x)+aE(x)|
 \le\frac{4e^{aD}Q^2}{D}\|E\|_I.
\]
The triangle inequality gives
$|E'(x)|\le(4e^{aD}Q^2/D+a)\|E\|_I$.

For $x\in[u+D/2,u+D]$, we apply \eqref{app:BE-endpoint}
to $h_x(t)=E(x-t)$ on $[0,x-u]$.
Its exponents are $s_j\ge0$, its norm is at most $\|E\|_I$,
and the interval has length at least $D/2$.
Since $h_x'(0)=-E'(x)$, we obtain
$|E'(x)|\le4Q^2\|E\|_I/D$.
Combining the two halves proves \eqref{app:slow-Markov}.
\end{proof}

\subsection{Proof of the Rate Window Lemma}
\label{app:rate-window-main}

\begin{proof}[Proof of Lemma~\ref{app:rate-window-lemma}]
Set 
\begin{equation}\label{app:rate-window-constants}
 c=2^{-20},
 \qquad
 C=10240.
\end{equation}
Fix $p_0\in\mathbb R$ satisfying Eq.~\eqref{app:p0-range}, that is,
$Cq\le p_0\le cP/q^2$, and set
$a=c/(q^2p_0)$ and $b=Cq/p_0$.
Then
\begin{equation}\label{app:standing}
 p_0\ge10240,\qquad
 b\le1,\qquad
 \frac ab=\frac{c}{Cq^3}<1,\qquad
 \frac1a=\frac{q^2p_0}{c}\le P.
\end{equation}

Suppose, for a contradiction, that no term with a nonzero
coefficient has its rate in $(a,b)$.
Set $s_j=0$ whenever $c_j=0$, which does not change $f$.
Every rate then belongs to $[0,a]\cup[b,\infty)$, so
\[
 f=L+H,\qquad
 L(p)=\sum_{s_j\le a}c_je^{-s_jp},
 \qquad
 H(p)=\sum_{s_j\ge b}c_je^{-s_jp}.
\]

We consider an even grid starting near $p_0/2$:
\[
 p_{\mathrm b}=2\left\lceil\frac{p_0}{4}\right\rceil,
 \qquad
 \Delta=2\left\lceil\frac{\log4}{2b}\right\rceil,
 \qquad
 p_m=p_{\mathrm b}+m\Delta\quad(m\ge0).
\]
Our first fact bounds the fast part using the initial samples
of the whole sum.

\begin{fact}[\bf A bound for the sampled fast part]
\label{app:sampled-fast-bound}
For every integer $m\ge q$,
\begin{equation}\label{app:fast-decay}
 |H(p_m)|\le\frac{2.2}{p_0}\,24^q3^{-m}.
\end{equation}
\end{fact}

\begin{proof}[Proof of Fact~\ref{app:sampled-fast-bound}]
The definitions of $p_{\mathrm b}$ and $\Delta$ give
\begin{align}
 \frac{p_0}{2}
 &\le p_{\mathrm b}<\frac{p_0}{2}+2
 \le\frac{3p_0}{5},
 \label{app:base-point}\\
 \log4
 &\le b\Delta<\log4+2b<4,
 \notag\\
 \Delta
 &<\frac4b=\frac{4p_0}{Cq}
 \le\frac{p_0}{100q}.
 \label{app:Delta}
\end{align}
Here the first line uses $p_0\ge20$, the second uses $b\le1$,
and the third uses $C\ge400$.

The first $q$ sampling points are even integers at least four,
and satisfy
$p_\ell\le p_{q-1}\le3p_0/5+p_0/100
=61p_0/100<p_0\le P$.
The approximation assumption and $p_\ell\ge p_0/2$ therefore imply
\begin{equation}\label{app:initial-samples}
 |f(p_\ell)|\le\frac{1.1}{p_\ell}
 \le\frac{2.2}{p_0}
 \qquad(0\le\ell<q).
\end{equation}

Set $x_j=e^{-s_j\Delta}$ and
$d_j=c_je^{-s_jp_{\mathrm b}}$, so that
$f(p_m)=\sum_{j=1}^{q}d_jx_j^m$.
For $s_j\ge b$, we have
$x_j\le e^{-b\Delta}\le1/4$.
For $s_j\le a$, Eq.~\eqref{app:Delta} gives
$s_j\Delta\le ap_0/(100q)=c/(100q^3)<\log(4/3)$,
and hence $x_j>3/4$.
Thus all nodes belong to $(0,1/4]\cup[3/4,1]$,
and the fast part of this power sum is $H(p_m)$.

By Eq.~\eqref{app:initial-samples},
Lemma~\ref{app:discrete-separation} applies with
$Q=q$ and $B=2.2/p_0$. It yields
\[
 |H(p_m)|\le\frac{2.2}{p_0}24^q3^{-m}
 \qquad(m\ge q),
\]
as required.
\end{proof} 
Let $p_*=\min\{p_m:m\ge0,\ p_m\ge p_0\}$.
The next fact transfers the approximation from $f$ to $L$.

\begin{fact}[\bf Approximation by the slow part]
\label{app:slow-approximation}
We have $p_0\le p_*\le1.01p_0<P$, and
\begin{equation}\label{app:grid-estimate}
 \frac{0.89}{p_m}\le L(p_m)\le\frac{1.11}{p_m}
 \qquad(p_m\in[p_*,P]).
\end{equation}
\end{fact}

\begin{proof}[Proof of Fact~\ref{app:slow-approximation}]
Write
\[
 m_0=\left\lceil\frac{p_0-p_{\mathrm b}}{\Delta}\right\rceil,
 \qquad p_*=p_{m_0}.
\]
By Eqs.~\eqref{app:base-point} and \eqref{app:Delta},
\[
 \begin{aligned}
 m_0
 &\ge\frac{p_0-p_{\mathrm b}}{\Delta}\ge\frac{2p_0/5}{p_0/(100q)}
 =40q
 \ge8q,\\
 p_0
 &\le p_*<p_0+\Delta\le1.01p_0.
 \end{aligned}
\]
Also,
$p_*\le1.01p_0\le1.01cP/q^2<P$,
using the upper bound on $p_0$ and $c=2^{-20}$.

The sequence $p_m3^{-m}$ is decreasing, because
\[
 \begin{aligned}
 \frac{p_{m+1}3^{-(m+1)}}{p_m3^{-m}}
 &=\frac13\left(1+\frac{\Delta}{p_m}\right)\\
 &\le\frac13\left(1+\frac1{50q}\right)
 <1.
 \end{aligned}
\]
The inequality uses $p_m\ge p_0/2$ and
$\Delta\le p_0/(100q)$.
Thus Fact~\ref{app:sampled-fast-bound} gives, for $m\ge m_0$,
\[
 \begin{aligned}
 p_m|H(p_m)|
 &\le\frac{2.2}{p_0}24^q p_m3^{-m}\\
 &\le\frac{2.2p_*}{p_0}24^q3^{-m_0}\\
 &\le2.2\cdot1.01
       \left(\frac{24}{3^8}\right)^q\\
 &<\frac52\frac{24}{3^8}
 =\frac{60}{6561}
 <\frac1{100}.
 \end{aligned}
\]
The third line uses $p_*\le1.01p_0$ and $m_0\ge8q$;
the fourth uses $q\ge1$ and $24/3^8<1$.
Therefore
\[
 |H(p_m)|\le\frac1{100p_m}
 \qquad(m\ge m_0).
\]

For every sampling point $p_m\in[p_*,P]$, we now have
\[
 \begin{aligned}
 \left|L(p_m)-\frac1{p_m}\right|
 &\le
 \left|f(p_m)-\frac1{p_m}\right|+|H(p_m)|\\
 &\le\frac1{10p_m}+\frac1{100p_m}\\
 &=\frac{0.11}{p_m}.
 \end{aligned}
\]
This is equivalent to Eq.~\eqref{app:grid-estimate}.
\end{proof}

To control the derivative of $L$, choose the long interval
\[
 D=\Delta\left\lfloor\frac1{2a\Delta}\right\rfloor,
 \qquad I=[p_*,p_*+D].
\]

\begin{fact}[\bf A derivative bound for the slow part]
\label{app:slow-derivative-bound}
The interval $I$ is contained in $[p_*,P]$, and
\begin{equation}\label{app:L-prime-upper}
 \|L'\|_I<\frac1{4p_0^2}.
\end{equation}
\end{fact}

\begin{proof}[Proof of Fact~\ref{app:slow-derivative-bound}]
By Eq.~\eqref{app:Delta},
$a\Delta\le c/(100q^3)<1/4$.
The inequalities $u-1\le\lfloor u\rfloor\le u$ therefore give
\begin{equation}\label{app:long-interval}
 \frac1{4a}
 \le\frac1{2a}-\Delta
 \le D\le\frac1{2a},
 \qquad aD\le\frac12.
\end{equation}
Moreover,
\[
 \begin{aligned}
 p_*+D
 &\le1.01p_0+\frac1{2a}\\
 &=\frac1a\left(\frac{1.01c}{q^2}+\frac12\right)\\
 &<\frac1a\le P.
 \end{aligned}
\]
The equality uses $ap_0=c/q^2$;
the strict inequality uses $q\ge1$ and $c=2^{-20}$.
Hence $I\subset[p_*,P]$.
Both endpoints of $I$ are sampling points because $D$ is a
multiple of $\Delta$.

By Lemma~\ref{app:slow-derivative},
\[
 \begin{aligned}
 \|L'\|_I
 &\le
 \left(\frac{4e^{aD}q^2}{D}+a\right)\|L\|_I\\
 &\le
 \left(16e^{1/2}aq^2+a\right)\|L\|_I\\
 &=\left(16e^{1/2}+\frac1{q^2}\right)
       \frac{c}{p_0}\|L\|_I\\
 &\le\frac{40c}{p_0}\|L\|_I.
 \end{aligned}
\]
The second line uses Eq.~\eqref{app:long-interval},
the third uses $aq^2=c/p_0$, and
the last uses $q\ge1$ and $e^{1/2}<2$.

For each $x\in I$, choose a sampling point $p_m\in I$
with $|x-p_m|\le\Delta/2$.
Such a point exists because the endpoints of $I$ belong to the
sampling grid.
The mean value theorem and Fact~\ref{app:slow-approximation} give
\[
 |L(x)|
 \le |L(p_m)|+\frac{\Delta}{2}\|L'\|_I
 \le\frac{1.11}{p_0}+\frac{\Delta}{2}\|L'\|_I.
\]
Taking the supremum and applying the derivative estimate,
\[
 \begin{aligned}
 \|L\|_I
 &\le\frac{1.11}{p_0}+\frac{\Delta}{2}\|L'\|_I\\
 &\le\frac{1.11}{p_0}
       +\frac{20c\Delta}{p_0}\|L\|_I\\
 &\le\frac{1.11}{p_0}+\frac c5\|L\|_I.
 \end{aligned}
\]
The last inequality uses $\Delta\le p_0/(100q)\le p_0/100$.
Since $c\le1$, rearranging yields
\[
 \|L\|_I
 \le\frac{1.11}{p_0(1-c/5)}
 \le\frac{1.11}{0.8p_0}
 <\frac2{p_0}.
\]
Substituting this bound into the derivative estimate gives
\[
 \|L'\|_I
 \le\frac{80c}{p_0^2}
 <\frac1{4p_0^2},
\]
where the last inequality uses $c=2^{-20}<1/320$.
\end{proof}

We finish by showing that the approximation in
Fact~\ref{app:slow-approximation} forces a larger derivative.
Let $p_{**}=\min\{p_m:m\ge0,\ p_m\ge2p_*\}$.
Then
\[
 \begin{aligned}
 2p_*&\le p_{**}<2p_*+\Delta,\\
 p_{**}-p_*
 &<p_*+\Delta\\
 &\le1.02p_0\\
 &<\frac{p_0}{4c}
 \le\frac1{4a}
 \le D.
 \end{aligned}
\]
Here we used $p_*\le1.01p_0$, $\Delta\le0.01p_0$,
$c=2^{-20}$, $1/a=q^2p_0/c$, and
Eq.~\eqref{app:long-interval}.
Thus $p_{**}\in I$.

By Fact~\ref{app:slow-approximation},
\[
 \begin{aligned}
 L(p_*)-L(p_{**})
 &\ge\frac{0.89}{p_*}-\frac{1.11}{p_{**}}\\
 &\ge\frac{0.89}{p_*}-\frac{1.11}{2p_*}\\
 &=\frac{0.335}{p_*},
 \end{aligned}
\]
where the second inequality uses $p_{**}\ge2p_*$.
The mean value theorem therefore implies
\[
 \begin{aligned}
 \|L'\|_I
 &\ge\frac{L(p_*)-L(p_{**})}{p_{**}-p_*}\\
 &\ge\frac{0.335}{p_*(p_{**}-p_*)}\\
 &\ge\frac{0.335}{1.01\cdot1.02\,p_0^2}\\
 &>\frac1{4p_0^2}.
 \end{aligned}
\]
The third line uses $p_*\le1.01p_0$ and
$p_{**}-p_*\le1.02p_0$.
This contradicts Fact~\ref{app:slow-derivative-bound}.

Hence some term with a nonzero coefficient has its rate in
$(a,b)$, which is the asserted interval.
\end{proof}

\section{Missing Proofs in Section~\ref{sec:lsi-lower-bound}}
\subsection{Spherical density and tail estimates}
\label{app:lsi-spherical}

\begin{proof}[Proof of Lemma~\ref{lem:lsi-spherical}]
Throughout, $t\ge4$ and $d\ge Dt^4$.
For this lemma, it suffices to take $D\ge10^4$.
In part~(iv), the smoothing shift is $\frac{10^{-3}}{t}$.

\paragraph{The density and its normalizing constant.}
Let $U$ be uniform on $S^{d-1}$.
The latitude under $\gamma_0$ has the same law as $\sqrt d\,U_1$.
In the parametrization
$U=(v,\sqrt{1-v^2}\,\omega)$, with $-1<v<1$ and
$\omega\in S^{d-2}$, the surface element is proportional to
$(1-v^2)^{(d-3)/2}\,dv$ times the surface element on $S^{d-2}$ since 
the latitude direction contributes $(1-v^2)^{-1/2}$,
and the remaining $d-2$ directions contribute
$(1-v^2)^{(d-2)/2}$.
The beta integral gives
\[
  \int_{-1}^1(1-v^2)^{(d-3)/2}\,dv
  =\frac{\sqrt\pi\,\Gamma((d-1)/2)}{\Gamma(d/2)}.
\]
Changing variables from $v$ to $s=\sqrt d\,v$ therefore yields
the stated density, with normalizing factor
$c_d:=\frac{\Gamma(d/2)}
{\sqrt{\pi d}\,\Gamma((d-1)/2)}$.

We first obtain explicit bounds on $A_d:=\sqrt{2\pi}\,c_d$.
For $z>0$, Cauchy--Schwarz applied to the gamma integrals,
together with $\Gamma(z+1)=z\Gamma(z)$, gives
\begin{align*}
  \Gamma\bigl(z+\tfrac12\bigr)^2
  &\le \Gamma(z)\Gamma(z+1)
   =z\Gamma(z)^2,\\
  \Gamma(z+1)^2
  &\le \Gamma\bigl(z+\tfrac12\bigr)
        \Gamma\bigl(z+\tfrac32\bigr)
   =\bigl(z+\tfrac12\bigr)
        \Gamma\bigl(z+\tfrac12\bigr)^2.
\end{align*}
Thus
$\frac{z}{\sqrt{z+1/2}}
\le\frac{\Gamma(z+1/2)}{\Gamma(z)}
\le\sqrt z$.
Taking $z=\frac{d-1}{2}$ gives
\begin{equation}
  1-\frac1d\le A_d\le\sqrt{1-\frac1d}\le1,
  \qquad
  -\frac2d\le\log A_d\le0.
  \label{eq:lsi-spherical-normalizer}
\end{equation}
For the logarithmic lower bound, we used
$\log(1-x)\ge-\frac{x}{1-x}$ with $x=\frac1d$
and $d\ge2$.

\paragraph{The global density bound.}
For $|s|<\sqrt d$, put $v=\frac{s^2}{d}$ and define
$F_d(v):=\frac{d-3}{2}\log(1-v)+\frac d2v$.
Then
\[
  \log\frac{q_d(s)}{\varphi(s)}
  =\log A_d+F_d(v),
  \qquad
  F_d'(v)=\frac{3-dv}{2(1-v)}.
\]
Since $d>3$, the maximum of $F_d$ on $[0,1)$
is attained at $v=\frac3d$.
Using $\log(1-v)\le-v$ and
\eqref{eq:lsi-spherical-normalizer}, we obtain
\[
  \log\frac{q_d(s)}{\varphi(s)}
  \le F_d\left(\frac3d\right)
  =\frac{d-3}{2}\log\left(1-\frac3d\right)+\frac32
  \le\frac{9}{2d}\le\frac5d.
\]
For $|s|\ge\sqrt d$, the claimed bound follows from $q_d(s)=0$.
This proves~(i).

\paragraph{The local density comparison.}
For $0\le v\le\frac12$,
\[
  0\le-\log(1-v)-v
  =\int_0^v\frac{y}{1-y}\,dy
  \le\frac{v^2}{2(1-v)}
  \le v^2.
\]
Writing
$F_d(v)=\frac32v+\frac{d-3}{2}(\log(1-v)+v)$
therefore gives
\[
  -\frac{s^4}{2d}\le F_d(v)\le\frac{3s^2}{2d},
  \qquad
  \left|\log\frac{q_d(s)}{\varphi(s)}\right|
  \le\frac{2+\frac32s^2+\frac12s^4}{d}.
\]
If $|s|\le3t$, then
$v\le\frac{9t^2}{d}\le\frac{9}{16D}\le\frac12$,
and $t\ge4$ implies
\[
  \left|\log\frac{q_d(s)}{\varphi(s)}\right|
  \le
  \frac{2+\frac{27}{2}t^2+\frac{81}{2}t^4}{d}
  \le\frac{42t^4}{d}
  \le\frac{42}{D}.
\]
For $D\ge10^4$, we have
$e^{-42/D}>0.99$ and $e^{42/D}<1.01$.
Exponentiating proves~(ii).

\paragraph{The tail estimate.}
For $0<a<\sqrt d$, define
$v_d(s):=(1-\frac{s^2}{d})^{(d-3)/2}$,
$b_d(s):=(1-\frac{s^2}{d})^{(d-1)/2}$,
and $I_d(a):=\int_a^{\sqrt d}v_d(s)\,ds$.
Since $b_d'(s)=-\frac{d-1}{d}s\,v_d(s)$
and $b_d(\sqrt d)=0$, integration by parts gives
\[
  I_d(a)
  =\frac{d\,b_d(a)}{(d-1)a}
   -\frac{d}{d-1}
       \int_a^{\sqrt d}\frac{b_d(s)}{s^2}\,ds.
\]
The subtracted term is nonnegative.
Moreover, $0\le b_d(s)\le v_d(s)$ and $s\ge a$, so
\[
  0\le
  \frac{d}{d-1}\int_a^{\sqrt d}\frac{b_d(s)}{s^2}\,ds
  \le\frac{d}{(d-1)a^2}\,I_d(a).
\]
Consequently,
\[
  \frac{d\,b_d(a)}{(d-1)a}
  \left(1+\frac{d}{(d-1)a^2}\right)^{-1}
  \le I_d(a)
  \le\frac{d\,b_d(a)}{(d-1)a}.
\]
Using
$b_d(a)=(1-\frac{a^2}{d})v_d(a)$,
$p_d(a)=2c_dI_d(a)$, and $q_d(a)=c_dv_d(a)$,
we obtain the two-sided bound
\begin{equation}
  \frac{d-a^2}{(d-1)a+\frac da}
  \le \frac{p_d(a)}{2q_d(a)}
  \le \frac{d-a^2}{(d-1)a}.
  \label{eq:lsi-spherical-mills}
\end{equation}

At $a=t$, the assumptions $t\ge4$ and $d\ge t^4$ imply
$d-t^2\ge\frac{15d}{16}$,
$(d-1)t+\frac dt\le\frac{17dt}{16}$,
and $d-t^2\le d-1$.
Thus
\[
  \frac{15}{17t}
  \le\frac{p_d(t)}{2q_d(t)}
  \le\frac1t.
\]
By~(ii), $0.99\varphi(t)\le q_d(t)\le1.01\varphi(t)$.
It follows that
\[
  \frac{2(0.99)\cdot15}{17\sqrt{2\pi}}\,
  \frac{e^{-t^2/2}}{t}
  \le p_d(t)
  \le
  \frac{2(1.01)}{\sqrt{2\pi}}\,
  \frac{e^{-t^2/2}}{t},
\]
which proves~(iii).

\paragraph{Stability under the smoothing shift.}
For $0<a<\sqrt d$, differentiating the tail gives
$p_d'(a)=-2q_d(a)$.
The lower bound in \eqref{eq:lsi-spherical-mills}
therefore implies
\[
  -\frac{d}{da}\log p_d(a)
  =\frac{2q_d(a)}{p_d(a)}
  \le\frac{(d-1)a+\frac da}{d-a^2}.
\]
For $t-\frac{10^{-3}}{t}\le a\le t$, we have
\begin{align*}
  \frac{(d-1)a+\frac da}{d-a^2}
  &\le
  \frac{t+\frac{1}{t-\frac{10^{-3}}{t}}}{1-\frac{t^2}{d}}\\
  &=
  t\,\frac{1+\frac{1}{t^2-10^{-3}}}{1-\frac{t^2}{d}}\\
  &\le\frac{256}{225}\,t
   <\frac65\,t.
\end{align*}
Here we used $t^2-10^{-3}\ge15$ and
$\frac{t^2}{d}\le\frac1{16}$.
Integrating over this interval yields
\[
  \log\frac{p_d(t-\frac{10^{-3}}{t})}{p_d(t)}
  =
  \int_{t-\frac{10^{-3}}{t}}^t
       \frac{2q_d(a)}{p_d(a)}\,da
  \le\frac65\,t\cdot\frac{10^{-3}}{t}
  =1.2\times10^{-3}.
\]
Since $e^{1.2\times10^{-3}}<1.002$, this proves~(iv).

All dimension requirements used above follow from
$t\ge4$, $D\ge10^4$, and $d\ge Dt^4$.
\end{proof}

\subsection{Proof of Lemma~\ref{lem:lsi-gaussian-estimates}}
\label{app:lsi-shifted-gaussian}

\gaussianest*

\begin{proof}[Proof of Lemma~\ref{lem:lsi-gaussian-estimates}]For (i), $M^2-y^2=(M-y)(M+y)\ge(M-y)M$ for $0\le y\le M$, so
\[
  \int_0^Me^{\frac{y^2}{2}}\,dy
  \le e^{\frac{M^2}{2}}\int_0^Me^{-\frac{(M-y)M}{2}}\,dy
  \le\frac{2e^{\frac{M^2}{2}}}{M}.
\]

For (ii), we first extend (i) to all $M\ge0$ in the form
$\int_0^Me^{\frac{y^2}{2}}\,dy\le\frac{4e^{\frac{M^2}{2}}}{1+M}$. For
$M\ge1$ this is (i) with $1+M\le2M$, and for $M\le1$ the integral is
at most $e^{\frac12}\le\frac4{1+M}$. Next, we use the standard tail
bounds $\frac{\varphi(y)}{2y}\le\overline\Phi(y)\le\frac{\varphi(y)}{y}$
for $y\ge1$ in the form
\[
  \overline\Phi(y)\,e^{\frac{y^2}{2}}\le\frac{C}{1+y},
  \qquad
  \log\frac1{\overline\Phi(y)}\le y^2+2
  \qquad(y\ge0).
\]
For $0\le y\le1$ these follow from $\overline\Phi(y)\le1$ and
$\overline\Phi(y)\ge\overline\Phi(1)\ge e^{-2}$. For $y\ge1$ they
follow from the tail bounds, since
$\varphi(y)e^{\frac{y^2}{2}}=\frac1{\sqrt{2\pi}}$ and
$\log\frac{2y}{\varphi(y)}=\frac{y^2}{2}+\log(2\sqrt{2\pi}\,y)\le y^2+2$.
For all $y$, $\overline\Phi(y)\le1$ and
$\overline\Phi(y)\log\frac1{\overline\Phi(y)}\le\frac1e$. By the
definition of $\delta$, $0\le\delta\le\frac{t^2}{2}$ and
$e^{\frac{y_-^2}{2}}=e^{\frac{t^2}{2}}e^{-\delta}$. Finally
$t^3\le e^{\frac{t^2}{2}}$ for $t\ge4$, so $Ct^2\le C\frac{e^{t^2/2}}{t}$.

Write $E:=1+\delta+\log\frac1{\overline\Phi(y_+)}$. The integrand is
positive and even, so
\[
  \int_{y_-}^{y_+}e^{\frac{y^2}{2}}\,dy
  \le\int_0^{|y_-|}e^{\frac{y^2}{2}}\,dy+\int_0^{y_+}e^{\frac{y^2}{2}}\,dy,
\]
where the first integral is omitted if $y_-\ge0$ and the second if
$y_+\le0$. We bound the two contributions separately.

For the second, let $y_+>0$. By the extended form of (i) and the two
bounds for $y\ge0$,
\[
  \overline\Phi(y_+)E\int_0^{y_+}e^{\frac{y^2}{2}}\,dy
  \le\frac{C}{(1+y_+)^2}\,(y_+^2+3+\delta)
  \le C(1+\delta)\le Ct^2 .
\]

For the first, let $y_-<0$. Since
$\overline\Phi(y_+)E\le1+\delta+\frac1e$, the extended form of (i)
gives
\[
  \overline\Phi(y_+)E\int_0^{|y_-|}e^{\frac{y^2}{2}}\,dy
  \le\frac{4e^{\frac{t^2}{2}}}{1+|y_-|}\,
    e^{-\delta}\Bigl(1+\delta+\frac1e\Bigr).
\]
The function $\delta\mapsto e^{-\delta}(1+\delta+\frac1e)$ is
decreasing, with value $1+\frac1e\le2$ at $\delta=0$. If
$\delta\le\frac{t^2}{4}$, then $|y_-|=\sqrt{t^2-2\delta}\ge\frac{t}{\sqrt2}$
and the right side is at most $\frac{8\sqrt2\,e^{t^2/2}}{t}$. If
$\delta\ge\frac{t^2}{4}$, then
$e^{-\delta}(1+\delta+\frac1e)\le e^{-\frac{t^2}{4}}(2+\frac{t^2}{4})\le\frac Ct$,
because $te^{-\frac{t^2}{4}}(2+\frac{t^2}{4})$ is bounded on
$[4,\infty)$, and the right side is at most $C\frac{e^{t^2/2}}{t}$
again. Adding the two contributions proves (ii).
\end{proof}

\subsection{Functional Inequalities used in Section~\ref{sec:lsi-constant}}
\label{app:lsi-functional-tools}

Recall that we say a probability measure $\mu$ on $\R^d$ satisfies
$\mathsf{LSI}(\alpha)$, if
$\operatorname{Ent}_\mu(f^2)\le\frac2\alpha\int\|\nabla f\|^2\,d\mu$
for all $f\in C_c^\infty(\R^d)$ and  $\mu$ satisfies the
$\mathsf{PI}(\alpha)$ if $  \operatorname{Var}_\mu(f)\le\frac1\alpha\int\|\nabla f\|^2\,d\mu$.
The same definition applies on a closed ball, a sphere, or an interval,
using the corresponding gradient and taking $f$ smooth up to the
boundary without boundary conditions. By standard approximation, both
$\mathsf{LSI}(\alpha)$ and $\mathsf{PI}(\alpha)$ extend to the
Sobolev space $H^1(\mu)$ of locally Lipschitz $f\in L^2(\mu)$ with
$\int\|\nabla f\|^2\,d\mu<\infty$. We use this extension freely.
For a bounded function $F$, we write
$\operatorname{osc}F:=\sup F-\inf F$.

We begin with three elementary facts about entropy.  Recall  that for a probability measure $\mu$ and a
nonnegative function $g$ with $\int g\,d\mu<\infty$, 
\[
  \operatorname{Ent}_\mu(g)
  :=\int g\log g\,d\mu-\Bigl(\int g\,d\mu\Bigr)\log\Bigl(\int g\,d\mu\Bigr),
  \qquad 0\log0:=0.
\]
\begin{lemma}[\bf Entropy identities and subadditivity]
\label{lem:lsi-entropy-calculus}
The following statements hold for bounded nonnegative
functions, with $0\log0:=0$.
They also hold whenever all displayed integrals are finite.

\begin{enumerate}
\item[(i)] \textbf{(Chain rule.)}
If $\mu(dx,dy)=\nu(dx)\mu_x(dy)$ and
$G(x):=\int g(x,y)\,\mu_x(dy)$, then
\[
  \operatorname{Ent}_\mu(g)
  =
  \int\operatorname{Ent}_{\mu_x}(g(x,\cdot))\,\nu(dx)
  +\operatorname{Ent}_\nu(G).
\]

\item[(ii)]\textbf{ (Homogeneity and subadditivity.)}
For $\lambda\ge0$,
$\operatorname{Ent}_\mu(\lambda g)
=\lambda\operatorname{Ent}_\mu(g)$.
Moreover,
$\operatorname{Ent}_\mu(g+h)
\le\operatorname{Ent}_\mu(g)+\operatorname{Ent}_\mu(h)$.
Consequently, entropy is convex on nonnegative functions.

\item[(iii)]\textbf{(Change of measure on a set.)}
Let $\nu$ be another probability measure and let $A$ be
measurable. Suppose that $\mu|_A=c\nu|_A$ for some $c>0$.
If $g$ vanishes outside $A$, then
\[
  \operatorname{Ent}_\mu(g)
  =
  c\operatorname{Ent}_\nu(g)
  -(\log c)\int g\,d\mu.
\]
In particular, if $p:=\mu(D)>0$ and $g$ vanishes outside $D$,
then
\[
  \operatorname{Ent}_\mu(g)
  =
  p\operatorname{Ent}_{\mu(\cdot\mid D)}(g)
  +\log\frac1p\int g\,d\mu.
\]
\end{enumerate}
\end{lemma}

\begin{proof}
For~(i), put $m:=\int g\,d\mu=\int G\,d\nu$.
Expanding the conditional entropy gives
\[
  \int\operatorname{Ent}_{\mu_x}(g(x,\cdot))\,\nu(dx)
  =
  \int g\log g\,d\mu-\int G\log G\,d\nu.
\]
Adding
$\operatorname{Ent}_\nu(G)=\int G\log G\,d\nu-m\log m$
proves the identity.

For~(ii), homogeneity follows by expanding the definition;
the terms $\lambda\log\lambda\int g\,d\mu$ cancel.
For subadditivity, write $a:=\int g\,d\mu$ and
$b:=\int h\,d\mu$.
If $a=0$ or $b=0$, the claim is immediate.
Otherwise, put $\theta:=\frac{a}{a+b}$ and
$\Phi(z):=z\log z$.
The identity
$\frac{g+h}{a+b}
=\theta\frac ga+(1-\theta)\frac hb$
and convexity of $\Phi$ give
\begin{align*}
  \operatorname{Ent}_\mu(g+h)
  &=(a+b)\int\Phi\left(\frac{g+h}{a+b}\right)d\mu\\
  &\le
  a\int\Phi\left(\frac ga\right)d\mu
  +b\int\Phi\left(\frac hb\right)d\mu\\
  &=\operatorname{Ent}_\mu(g)+\operatorname{Ent}_\mu(h).
\end{align*}
Applying homogeneity and subadditivity to
$\lambda g+(1-\lambda)h$, for $0\le\lambda\le1$,
also proves convexity.

For~(iii), the support condition gives
$\int g\,d\mu=c\int g\,d\nu$ and
$\int g\log g\,d\mu=c\int g\log g\,d\nu$.
If $m:=\int g\,d\nu>0$, then
\begin{align*}
  \operatorname{Ent}_\mu(g)
  &=c\int g\log g\,d\nu-cm\log(cm)\\
  &=c\operatorname{Ent}_\nu(g)-cm\log c.
\end{align*}
If $m=0$, both sides vanish.
The conditional-measure identity follows by taking
$\nu=\mu(\cdot\mid D)$, $A=D$, and $c=p$.
\end{proof}

Next we recall the stability of log-Sobolev and Poincar\'e
inequalities under bounded perturbations of the measure. 
\begin{lemma}[\bf Holley--Stroock perturbation lemma~\cite{HolleyStroock1987,MenzSchlichting2014}]
\label{prop:lsi-bounded-perturbation}
Let $F$ be a bounded measurable function and let
$\nu(dx)\propto e^{F(x)}\mu(dx)$ be a probability measure.
\begin{enumerate}
\item[(i)] If $\mu$ satisfies $\mathsf{LSI}(\alpha)$, then $\nu$
  satisfies $\mathsf{LSI}(\alpha e^{-\operatorname{osc}F})$, that is,
  \[
    \operatorname{Ent}_\nu(f^2)
    \le\frac{2e^{\operatorname{osc}F}}{\alpha}
       \int\|\nabla f\|^2\,d\nu .
  \]
\item[(ii)] If $\mu$ satisfies $\mathsf{PI}(\alpha)$, then $\nu$
  satisfies $\mathsf{PI}(\alpha e^{-\operatorname{osc}F})$, that is,
  \[
    \operatorname{Var}_\nu(f)
    \le\frac{e^{\operatorname{osc}F}}{\alpha}
       \int\|\nabla f\|^2\,d\nu .
  \]
\end{enumerate}
\end{lemma}

The Bakry--\'Emery curvature criterion gives a log-Sobolev inequality
for a measure with density proportional to $e^{-U}$ whenever the
Hessian of $U$, together with the Ricci curvature of the underlying
space, is bounded below by a positive constant. We state it on a
manifold without boundary and on a compact interval, where the
relevant version is the one with reflecting boundary, in which no
condition is imposed on the test functions.

\begin{lemma}[\bf Bakry--\'Emery curvature criterion~\cite{BakryEmery1985,KolesnikovMilman2016}]
\label{prop:lsi-bakry-emery}
Let $\mu(dx)\propto e^{-U(x)}\,\mathrm{vol}(dx)$ be a probability
measure, where $\mathrm{vol}$ is the Riemannian volume of the
underlying space and $U$ is a $C^2$ potential, and let $\alpha>0$.
\begin{enumerate}
\item[(i)] Let the underlying space be a connected complete
  Riemannian manifold $(M,g_M)$ without boundary. If
  \[
    \operatorname{Ric}+\nabla^2U\succeq\alpha\,g_M
    \qquad\text{on }M,
  \]
  then $\mu$ satisfies $\mathsf{LSI}(\alpha)$ for the Riemannian
  gradient.
\item[(ii)] Let the underlying space be a compact interval
  $[a,b]\subset\R$. If
  \[
    U''\ge\alpha\qquad\text{on }[a,b],
  \]
  then $\mu$ satisfies $\mathsf{LSI}(\alpha)$ for all $f$ smooth on
  $[a,b]$, with no condition at the endpoints.
\end{enumerate}
\end{lemma}



Linearizing the log-Sobolev inequality around the constant function
$1$ gives the Poincar\'e inequality with the same rate.
\begin{lemma}[\bf $\mathsf{LSI}$ implies $\mathsf{PI}$]
\label{lem:lsi-implies-poincare}
If $\mu$ satisfies $\mathsf{LSI}(\alpha)$, then $\mu$ satisfies
$\mathsf{PI}(\alpha)$, that is,
\[
  \operatorname{Var}_\mu(f)\le\frac1\alpha\int\|\nabla f\|^2\,d\mu
  \qquad\text{for all }f\in C_c^\infty(\R^d).
\]
\end{lemma}

\begin{proof}
For bounded smooth $f$ of finite energy, apply LSI to
$1+\varepsilon f$.
Taylor expansion at $\varepsilon=0$ gives
$\operatorname{Ent}_\mu((1+\varepsilon f)^2)
=2\varepsilon^2\operatorname{Var}_\mu(f)
+o(\varepsilon^2)$,
whereas the gradient energy is
$\varepsilon^2\int\|\nabla f\|^2\,d\mu$.
Dividing the inequality by $2\varepsilon^2$ and letting
$\varepsilon\to0$ proves the claim.
Truncation and approximation extend it to the Sobolev domain.
\end{proof}

Finally, Rothaus' lemma compares the entropy of $f^2$ with that of a
shifted copy of $f$. Combined with a Poincar\'e inequality, it turns
a defective log-Sobolev inequality, which carries an extra term
proportional to $\int f^2\,d\mu$, into a true one.

\begin{lemma}[\bf Rothaus' lemma~\cite{BartheRoberto2003}]
\label{lem:lsi-rothaus}
Let $\mu$ be a probability measure and let $f\in L^2(\mu)$.
For every constant $a\in\R$,
\[
  \operatorname{Ent}_\mu(f^2)
  \le
  \operatorname{Ent}_\mu((f-a)^2)
  +2\int(f-a)^2\,d\mu.
\]
In particular, with $a=\int f\,d\mu$,
$\operatorname{Ent}_\mu(f^2)
\le\operatorname{Ent}_\mu((f-\mathbb E_\mu f)^2)
+2\operatorname{Var}_\mu(f)$.
\end{lemma}

\end{document}